# Untersuchungen zur Implementierung von Bildverarbeitungsalgorithmen mittels pulsgekoppelter neuronaler Netze

Christian Georg Mayr

Dissertation Thesis

# Abstract

This thesis deals with the study of image processing algorithms which can be implemented by pulse-coupled neural nets. The inspiration for this choice is taken from biological image processing, which achieves with little computational effort in highly parallel processes image analysis tasks such as object recognition, image segmentation, velocity and distance estimation, etc.

Conventional, serially implemented algorithms either cannot realize those tasks at all or will expend significantly more effort. Because the first stages of the visual system comprise a sensor interface, they are comparatively accessible with respect to defining their transfer or processing function. Some of those processing functions or principles are to be used in hardware implementations, with the focus on duplicating especially the highly parallel processing.

This work is structured as follows: As introduction, the development and aims of technical and biological image processing are compared, including a short summary of the first stages of mammalian visual information processing. Following that, the current state-of-the-art concerning biological and information theoretical research and modelling of these stages is given, which shall serve as a theoretical basis for the subsequent chapters and summary of their information processing functions. The following chapters deal with adapting single aspects of biological image processing for technical implementations. The first few chapters are primarily concerned with technical realization and applicability, in part causing the biological processing principles to be heavily modified. In contrast, especially the work discussed in the last chapter aims at pure research, i.e. biological information processing principles are to be transferred to hard- and software faithfully to develop a better understanding of the processing steps carried out in the respective biological neural nets.

# References: Main Author

# References: Coauthor

TECHNISCHE UNIVERSITÄT DRESDEN

# Untersuchungen zur Implementierung von Bildverarbeitungs-algorithmen mittels pulsgekoppelter neuronaler Netze

**Christian Georg Mayr**

geb. am 20.04.1977

von der Fakultät Elektrotechnik und Informationstechnik
der Technischen Universität Dresden

zur Erlangung des akademischen Grades eines

**Doktoringenieurs**

(Dr.-Ing.)

genehmigte Dissertation

# Kurzfassung

Die vorliegende Arbeit befasst sich mit der Studie von Bildverarbeitungsalgorithmen, die mittels pulsgekoppelter neuronaler Netze realisiert werden. Inspiration ist hierbei die biologische Bildverarbeitung, die mit wenig Rechenaufwand hochparallel Bildanalysen wie Objekterkennung, Bildsegmentierung, Geschwindigkeits- und Entfernungsabschätzungen, etc. realisiert, zu denen herkömmliche, seriell arbeitende Algorithmen entweder nicht in der Lage sind oder unverhältnismäßig mehr Aufwand betreiben. Aufgrund der hohen Zugänglichkeit als Sensorschnittstelle sind die ersten Stufen der biologischen Bildverarbeitung hinsichtlich ihrer Übertragungsfunktionen bzw. Verarbeitungsprinzipien (relativ) gut erforscht. Manche dieser Verarbeitungsprinzipien sollen in technische Implementierungen übertragen werden, um insbesondere die hohe Parallelität zu duplizieren.
Die Arbeit gliedert sich wie folgt: Als erstes werden die Entwicklung und Zielsetzung der technischen und biologischen Bildverarbeitung gegenübergestellt, wobei insbesondere über die ersten Stufen der visuellen Informationsverarbeitung bei Säugetieren ein kurzer Überblick gegeben wird. Im weiteren wird der aktuelle Stand der biologischen und informationstheoretischen Forschung und Modellierung bzgl. dieser Stufen wiedergegeben, was als theoretische Grundlage für die folgenden Kapitel und Zusammenfassung der in ihnen stattfindenden Informationsverarbeitung dienen soll. Diese Kapitel befassen sich dann mit der Adaption einzelner Aspekte der biologischen Bildverarbeitung für technische Implementierungen. In den ersten Kapiteln steht die technische Realisierung und Anwendbarkeit im Vordergrund, wobei hier diese Verarbeitungsprinzipien zum Teil stark angepasst werden. Hingegen liegt v.a. im letzten Kapitel der Fokus auf der Forschung, das heißt, biologische Prinzipien werden möglichst unverfälscht in Hard- und Software übertragen, um ein besseres Verständnis für die in den entsprechenden biologischen neuronalen Netzen stattfindenden Verarbeitungsschritte zu erlangen.

# Summary

This thesis deals with the study of image processing algorithms which can be implemented by pulse-coupled neural nets. The inspiration for this choice is taken from biological image processing, which achieves with little computational effort in highly parallel processes image analysis tasks such as object recognition, image segmentation, velocity and distance estimation, etc. Conventional, serially implemented algorithms either cannot realize those tasks at all or will expend significantly more effort. Because the first stages of the visual system comprise a sensor interface, they are comparatively accessible with respect to defining their transfer or processing function. Some of those processing functions or principles are to be used in hardware implementations, with the focus on duplicating especially the highly parallel processing.
This work is structured as follows: As introduction, the development and aims of technical and biological image processing are compared, including a short summary of the first stages of mammalian visual information processing. Following that, the current state-of-the-art concerning biological and information theoretical research and modelling of these stages is given, which shall serve as a theoretical basis for the subsequent chapters and summary of their information processing functions. The following chapters deal with adapting single aspects of biological image processing for technical implementations. The first few chapters are primarily concerned with technical realization and applicability, in part causing the biological processing principles to be heavily modified. In contrast, especially the work discussed in the last chapter aims at pure research, i.e. biological information processing principles are to be transferred to hard- and software faithfully to develop a better understanding of the processing steps carried out in the respective biological neural nets.

# Vorwort

Fachlicher Dank geht an alle Mitarbeiter des Stiftungslehrstuhl hochparallele VLSI-Systeme und Neuromikroelektronik, die durch eine angenehme Arbeitsatmosphäre und intellektuellen Diskurs zum Gelingen dieser Arbeit beigetragen haben. Im Besonderen sei hier die Zusammenarbeit mit Herrn Dipl. Ing. H. Eisenreich und Dr. Ing. S. Henker an den in Kapitel III und V vorgestellten Arbeiten erwähnt, sowie die Diskussionen mit Herrn Dipl.-Ing. J. Partzsch zu verschiedensten Aspekten neuronaler Signalverarbeitung. Herrn Prof. Schüffny möchte ich für die fachliche und persönliche Unterstützung während der Durchführung meiner Promotion danken. Für Assistenz im organisatorischen Bereich (Dienstreisen, Verwaltungsvorgänge, Materialbeschaffung, etc.) zeichnete Herr Dipl. Ing. J. Dohndorf verantwortlich.
Meine Arbeiten wurden finanziell unterstützt durch Infineon München, Division Corporate Research (System Technology), die Deutschen Forschungsgemeinschaft (DFG) im Rahmen des Sonderforschungsbereiches SFB 358 „Automatisierter Systementwurf“, und die Europäische Union, 6th Framework Programme: Information Society Technologies, FP6-IST, Projekt FACETS.
Privat danken möchte ich meiner Familie und meinem Freundeskreis, die mich auf meinem bisherigen wissenschaftlichen und persönlichen Weg unterstützt haben, und ohne die diese Arbeit vermutlich nicht zustande gekommen wäre.

*“A scientist builds in order to learn; an engineer learns in order to build.”*

*Fred Brooks*

# Inhaltsverzeichnis

# Verwendete Abkürzungen

| | |
|---|---|
| AD | Analog-Digital (Konverter) |
| AER | Address-Event-Representation (Methode, in pulsenden neuronalen Netzen Pulsereignisse nur mit Identifikations- und Zeitmarke zu übertragen) |
| AHDL | analog hardware description language |
| ANC | Analog Network Chip (Teil der FACETS Hardware, enthält eigentliche neuronale Funktionalität, d.h. Neuronen und Synapsen) |
| AP | Aktionspotential |
| AS | Auswahlschalter |
| ASIC | Application specific integrated circuit |
| BCM | Bienenstock-Cooper-Monroe (Ratenbasierte, biologisch motivierte LTP/LTD Lernregel) |
| BSA | Bens Spiker Algorithm |
| CB | Crossbar |
| CMOS | Complementary Metal Oxide Semiconductor |
| CNN | cellular neural network |
| CV | Coefficient of Variation (Maß für die Variabilität des ISI in einem Spike Train) |
| DA | Digital-Analog (Konverter) |
| DGL | Differentialgleichung |
| DNC | Digital Network Chip (Teil der FACETS Hardware, zuständig für digitales, paketbasiertes Routing oberhalb des WSS) |
| DoG | Difference-of-Gaussian |
| DSM | Delta-Sigma-Modulator |
| Dxx | Deliverable (Nummer xx), an EU zu liefernder Zwischenstand im Rahmen von FACETS, z.B. D21 |
| EPSC | exzitatorischer postsynaptischer Strom |
| EPSP | “ “ Potential |
| FACETS | Fast Analog Computing of Emergent Transient States |
| FF | Flipflop (in verschiedenen Ausprägungen, etwa D-FF, RS-FF, etc.) |
| FIFO | First In, First Out (serielle Speicherorganisation) |
| Flops | floating point operations per second |
| FPGA | Field programmable Gate Array |
| FTD | Fast Time constant Dynamics |
| GA | Genetic Algorithm |
| GABA | Gamma-Aminobuttersäure (Neurotransmitter) |
| HH | Hodgkin-Huxley (membrane model) |
| HICANN | High Input Count Analog Neural Network (ANC-Prototyp) |
| IAF | Integrate and Fire (Neuron) |
| IC | Integrated Circuit |
| In-vivo | neurobiologische Messungen am lebenden Organismus, z.B. über Elektroden im V1, bei gleichzeitiger Stimulierung der Retina über dem Tier vorgeführte Bilder. |
| In-vitro | Messungen in einer Neuronenkultur (Petrischale), zum Einen in kompletten Ausschnitten/Scheiben, bei denen die originale Verbindungsstruktur erhalten wird, oder als auspräparierte einzelne Neuronen, die über biologische oder emulierte Synapsen neu verbunden werden. |
| I/O | Input-Output |
| IPSP | inhibitorisches postsynaptisches Potential |
| ISI | interspike interval (Zeitabstand zwischen Pulsen in einem spike train) |
| JTAG | Joint Test Action Group (Akronym für den IEEE 1149.1 Standard für IC-Testschnittstellen) |
| LGN | Lateral Geniculate Nucleus |
| LIAF | Leaky Integrate and Fire (Neuron) |
| LOC | Local Orientation Coding |
| LoG | Laplacian of Gaussian |
| LSB | least significant bit |
| LSME | Least Mean Squared Error |
| LTD | Long Term Depression |
| LTP | Long Term Potentation |
| LUT | Look-up-Table |
| LVDS | Low Voltage Differential Signalling |
| MCM | Multi-compartment-model |
| MCO | Multiobjective combinatorial optimization |
| MPW | Multi-Project-Wafer |
| MSB | most significant bit |
| Mxx | Milestone (Nummer xx), an EU zu liefernder Meilenstein im Rahmen von FACETS, z.B. D21 |
| NCSIM | Cadence environment mixed signal simulator |
| NP | Nicht-deterministische Polynomzeit |
| NPU | Neural Processing Unit |
| NTF | Noise Transfer Function |
| OSR | Oversampling Ratio |
| PCA | Principal Component Analysis |
| PCNN | pulse coupled neural network |
| PLL | Phase-locked-loop |
| PLOC | Pulsed Local Orientation Coding |
| PSTH | Peri-Stimulus Time Histogram |
| pt | points (Bildpunkte/Pixel, z.B. bei Ausdehnung einer Filtermaske) |
| PTP | posttetanic potentiation |
| PWM | Pulsweitenmodulation |
| QIAF | Quadratic Integrate and Fire (Neuron) |
| RF | Rezeptives Feld |
| RNN | reduced nearest neighbor |
| ROC | Rank Order Coding |
| ROI | Region of Interest |
| RS | Regular Spiking (Neuron) |
| SFA | spike frequency adaptation |
| SNR | Signal to Noise Ratio |
| STDP | Spike Timing Dependent Plasticity |
| STF | signal transfer function |
| STP | Short Term Potentation |
| SUSAN | Smallest Univalue Segment Assimilating Nucleus |
| VHDL | Very High Speed Integrated Circuit Hardware Description Language |
| VLSI | Very Large Scale Integration |
| WP | Workpackage |
| WTA | Winner-take-all |
| XOR | Exclusive Or |

# Verzeichnis der Formelzeichen

| | |
|---|---|
| $A$ | Umrechnungsfaktor von postsynaptischer Ausschüttungsmenge auf im Mittel dadurch hervorgerufenen Strom |
| $A(m+i,n+i)$ | Einzelantwort der Korrelationsnachbearbeitung des PLOC-Operators |
| $a$ | Akkumulatorstand im IAF-Neuronenmodell, auch Maß für den (Frequenz-) Durchlassbereich eines DoG-Filters |
| $a,b$ | Feedforward bzw. Feedback Koeffizienten im DSM |
| $\alpha,\beta$ | Übergangsraten im HH-Modell, z.B. $\alpha_m$ oder $\beta_m$ |
| $B$ | Menge aller biologischen Elemente beim Mapping von neuronalen Benchmarks auf die FACETS-Hardware, auch $B_p$ als erfolgreich abgebildete Teilmenge aus $B$ |
| $b(m,n)$ | Pixelgrauwert an den Koordinaten (m,n), auch b(x,y) |
| $b'(m,n)$ | Antwort des LOC-Operators an den Koordinaten (m,n), auch Gesamtantwort des PLOC-Operators |
| $b'_k(m,n)$ | Antwort des PLOC-Operators für ein einzelnes Merkmal k an den Koordinaten (m,n) |
| $C_{mem}$ | Membrankapazität (z.T. flächenbezogen) |
| $C_{xx}$ | Korrelation zwischen Neuronen, z.B. $C_{13}$ |
| $C$ | Skalierungskoeffizient, z.B. bei LOC-Operator oder bei Sigmoid-Übertragungsfunktion. |
| $c_X$ | Nebenbedingung beim Mapping, etwa $c_B$ als gute Abbildung von Benchmark-Parametern oder $c_H$ als effiziente Auslastung der FACETS-Hardware durch eine Benchmark. |
| $\delta$ | Diracimpuls |
| $d$ | Exzentrizität der Gabormaske, auch Axondurchmesser |
| $d(x,y)$ | Maskenkoeffizienten eines DoG-Filters |
| $D(x,y)$ | Antwort eines Bildes auf Faltung mit einem DoG-Filters |
| $\Delta$ | Sample-Wegstrecke beim Stochastic Universal Sampling |
| $\Delta_{OS}$, $\Delta_{US}$ | Delta-Pulsanzahlen, obere und untere Schranke bei Ratencode |
| $\Delta t, T_{ISI}$ | Zeit zwischen Pulsen |
| $e_x, i_x$ | Exzitatorische und inhibitorische Eingänge im Dendritenmodell |
| $E(...)$ | Erwartungswert, allgemein |
| $EPSC_n$ | mittlerer EPSC, der beim n-ten Puls in der quantalen synaptischen Kurzzeit-adaption auftritt |
| $\varepsilon_{m,n}(i,j)$ | Antwort des LOC-Operators |
| $f(...)$ | Wahrscheinlichkeitsdichtefunktion |
| $f_m$ | Modulationsfrequenz einer Pulsrate |
| G | Zeitinvarianter Leitwert, z.B. $\hat{G}_{Na}$ Spitzenleitwert des Natriumkanals im HH-Modell |
| $g_{XX}(t)$ | zeitveränderlicher Leitwert von Ionenkanälen im HH-Modell, z.B. $g_{Na}(t)$ |
| $G(m,n)$ | Antwort eines Bildes auf Faltung mit einem Gaußfilter im modifizierten LOC |
| $g(x,y)$ | komplexe Maskenkoeffizienten des biologisch motivierten Gaborfilters, auch z.B. $g_{real}(x,y)$ als Realteil dieser Antwort |
| $gauss(x,y)$ | Maskenkoeffizienten einer Gaußschen Glättung |
| $H$ | Shannon-Entropie |
| $I$ | Strom, z.B. $I_{PS}$ postsynaptischer Strom im HH-Modell, $I_{photo}$ Photostrom einer Pixelzelle |
| $I_{AB}$ | Ähnlichkeitsmaß für zwei DoG-Charakteristiken A und B der Retina |
| $j$ | imaginäre Einheit, auch $(i,j)$ als Relativkoordinatenpaar beim LOC-Operator |
| $i,j,k$ | Laufvariablen in Summenformeln |
| $\kappa$ | Rechenaufwand (Anzahl Operationen) von Bildfaltungen, z.B. $\kappa_{DoG}$ |
| $k$ | Boltzmannkonstante, spatiale Ausdehnung einer Gabormaske, Mengenindex bei PLOC-Merkmalen |
| $k(i,j)$ | Orientierungskoeffizient des LOC-Operators |
| $k_1,k_2$ | Anteil verworfener Pulse im rauschbehafteten ROC-Modell |
| $l_B$ | Verlust an nicht abbildbaren Elementen aus $B$ |
| $\lambda$ | Feuerrate von Neuronen (einzeln oder Population) |
| $h$ | Inaktivierungspartikel im Natriumkanal des HH-Modells |
| $H$ | Menge aller Hardware-Elemente beim Mapping von neuronalen Benchmarks, auch $H_p$ als erfolgreich abgebildete Teilmenge aus $H$ |
| $m$ | Aktivierungspartikel im Natriumkanal des HH-Modells, Auch Mapping-Abbildung von biologischen Benchmarks auf die FACETS-Hardware |
| $n$ | Generelle Zählvariable; für Synapsen: Anzahl der Neurotransmitter-Ausschüttungsstellen, im HH-Modell: Aktivierungspartikel im Kaliumkanal |
| $\eta$ | Plastizitätsskalierung (Lernrate) in verschiedenen Modellen synaptischer Plastizität. |
| $(m,n),(x,y)$ | Bildkoordinaten (Pixel), z.B. beim LOC-Operator oder der Retinamodellierung |
| $N_I,N_C$ | Ionenkonzentration im Intrazellulärraum $N_I$ und im Cytoplasma $N_C$ |
| $N_+,N_-$ | Anzahl der positiven/negativen Eingangszugriffe einer Mikroschaltung beim pulsbasierten Gaborfilter |
| $N$ | Generell als Zählvariable, z.B. $N_G$ als Anzahl an Gabormasken in einer Bildfilterung |
| $\Phi$ | Bewertungsfunktion für die Gewichtsänderung in der BCM-Regel |
| $p$ | Wahrscheinlichkeit für eine postsynaptische Neurotransmitterausschüttung bei einer einzelnen Synapse für ein eingehendes postsynaptisches Aktionspotential |
| $p(...),P(...)$ | Allgemein Wahrscheinlichkeit von (...) |
| $p_+,p_-$ | Wahrscheinlichkeiten für eines positiven/negativen Eingangszugriff einer Mikroschaltung beim pulsbasierten Gaborfilter |
| $p_{sum}$ | Summe über entweder $p_+$ oder $p_-$ |
| $\Phi_{1,2}$ | Taktsignal |
| $q$ | Im Synapsenmodell Menge der postsynaptisch ausgeschütteten Neurotransmitter, auch Ionenladung im HH-Modell (Berechnung der Nernstspannung) |
| $R$ | synaptische Übertragungseffizienz, auch Menge der reellen Zahlen |
| $R_+,R_-$ | positive bzw. negative Maskenantwort des pulsbasierten Gaborfilters |
| $R_i$ | spezifischer elektrischer Widerstand im Membranmodell |
| $R_n$ | (iterative) freie synaptische Ausschüttungsmenge |
| $R(t)$ | zeitkontinuierliche Näherungsformel für $R_n$ |
| $R_k(\lambda)$ | Konvergenzwert von $R_n$ für eine feste Pulsrate λ |
| $r(t)$ | Zeitfunktion einer Folge von Aktionspotentialen, in Form von Dirac-Impulsen oder als $r_X(t)$ in abgewandelter, durch $X$ bezeichneter Form, etwa Gaußfunktionen oder Aktionspotentiale. |
| $R(\omega)$ | Fouriertransformation von $r(t)$ |
| $S(z)$ | äquivalentes Quantisierungsrauschen im Frequenzbereich beim DSM |
| $\theta_M$ | Schwellwert bei synaptischen Adaptionsregeln |
| $\theta_S$ | Signifikanz-Schwellwert der Merkmale beim PLOC-Operator |
| $\theta_{korr}$ | Korrelations-Schwellwert der Merkmale beim PLOC-Operator |
| $\sigma$ | Standardabweichung, z.B. im Gaussfilter, in Dichtefunktionen, etc. |
| $T$ | Zeitraum, z.B. $T_{abs}$ oder $T_{rel}$ als Kennzeichen einer Refraktärzeit, $T_{puls}$ als mittlere Dauer eines Aktionspotentials, auch Temperatur bei Nernstspannung |
| $\tau$ | Zeitkonstante, z.B. $\tau_{reset}$ Resetzeit, $\tau_{int}$ Integrationszeit, $\tau_{facil}$ und $\tau_{rec}$ Zeitkonstanten der quantalen Kurzzeitadaption, $\tau_{konv,u}$ und $\tau_{konv,R}$ Ersatzzeit- |
| | konstanten für zeitkontinuierliche *u(t)* und *R(t)* und Annäherung auf eine feste Rate $\lambda$, $\tau_{u,\lambda 1}$ und $\tau_{R,\lambda 1}$ wie vorhergehend, jedoch mit Bezeichnung der Rate. |
| *t(m,n)* | ortsabhängiger(Signifikanz-) Schwellwert im Grundmodell LOC |
| $\mu$ | Erwartungswert für die Anzahl an Pulsen in einem Intervall |
| $u_n$ | (momentane, iterative) verwendete synaptische Ausschüttungsmenge |
| *u(t)* | zeitkontinuierliche Näherungsformel für $u_n$ |
| $u_{Mem}(x,t)$ | Membranspannung im HH-Modell nach Ort und Zeit |
| $u_k(\lambda)$ | Konvergenzwert von $u_n$ für eine feste Pulsrate λ |
| $u_{x,\lambda 2}$ | Startwerte für *u(t)* zu Beginn des Taktes der Ratenmodulation, der auf die Rate $\lambda_2$ hinführt |
| *U* | verwendete synaptische Ausschüttungsmenge, auch Ausdehnung einer diskretisierten Faltungsmaske in x-Richtung |
| $U_x$ | Spannung, z.B. $U_{mem}$ Membranspannung Neuron, $U_{Na}$ Nernstspannung Natriumkanal, $U_K$ Nernstspannung Kaliumkanal, $U_L$ Nernstspannung übrige passive Ionenkanäle im HH-Modell |
| *U(z)* | komplexe frequenzabhängige Signalspannung im zeitgetakteten System |
| $\overline{UR}_{xx}$ | mittlere synaptische Ausschüttungsmenge |
| *V* | Ausdehnung einer diskretisierten Faltungsmaske in y-Richtung |
| *V(...)* | Varianz |
| *v* | Fortpflanzungsgeschwindigkeit eines APs im HH-Modell |
| $W_{xx}$ | synaptisches Gewicht, z.B. $W_{41}$ Gewicht, das Neuron 1 mit Neuron 4 verbindet, evtl. auch mit Typbezeichnung, |
| $\omega$ | Kreisfrequenz (z.B. bei Fouriertransformation) |
| $\omega_0$ | spatiale Grundfrequenz des Gaborfilters |
| $\gamma$ | Gewichtsabklingkonstante in synaptischen Lernregeln |
| $\chi$ | Indikatorfunktion |
| *X* | Bildausdehnung in x-Richtung |
| $X_1$, *etc.* | Neuronenausgang |
| $(x_d, y_d)$ | diskrete Bildkoordinaten beim pulsbasierten Gaborfilter |
| *Y* | Bildausdehnung in y-Richtung |
| *Y(z)* | komplexes frequenzabhängiges getaktetes Ausgangssignal des DSM |
| *z* | komplexe diskrete Frequenz (z.B. Abtastfrequenz) in getakteten Systemen |

# I Einleitung

Eine Arbeit zum Thema „Untersuchungen zur Implementierung von Bildverarbeitungs-Algorithmen mittels pulsgekoppelter neuronaler Netze“ befasst sich naturgemäß mit einem breiten und nicht klar definierten Wissenschaftsgebiet. Zielsetzung ist im weitesten Sinne, wie in der Einleitung erwähnt, Vorgänge in der biologischen Bildverarbeitung (Abschnitt I.2) zu emulieren und sich damit Verarbeitungsprinzipien nutzbar zu machen, die sich deutlich von konventioneller, algorithmisch orientierter Bildverarbeitung (Abschnitt I.1) unterscheiden. Als Grundlage für nachfolgende Kapitel wird in den auf I.1 folgenden Abschnitten eine kurze Übersicht der visuellen Verarbeitung im Säugetier (Abschnitt I.3), der neuronalen Bausteine und Baugruppen (Abschnitt II.1), sowie informationstheoretischer Aspekte (Abschnitt II.2) gegeben. Als Abschluss der grundlegenden Betrachtungen gibt Sektion II.4 einen Überblick über die vom Einsatz dieser neuronalen Prinzipien erhofften Vorteile. In diesem Abschnitt und in den im Hauptteil dokumentierten Forschungsarbeiten wird dabei das behandelte Themengebiet wiederholt in Richtung allgemeiner neuronaler Informationsverarbeitung erweitert. Dies wird dadurch motiviert, dass die Strukturen anderer informationsverarbeitender Systeme im Kortex starke Parallelen zur Verarbeitung visueller Information aufweisen [Shepherd04] und sich manche Aspekte der Verarbeitungsfunktion besser an nicht optisch basierter Information aufzeigen lassen.
Das Spektrum der im anschließenden Hauptteil (Kapitel III bis V) dokumentierten Arbeiten deckt einen weiten Bereich an Aspekten der neuronalen Bildverarbeitung ab. Die in Kapitel III geschilderten Arbeiten setzen eine starke Betonung auf Vernetzungsaspekte und die dadurch möglichen Verarbeitungsfunktionen. In Kapitel IV wird anhand einzelner Teilbereiche neuronaler Verarbeitung deren technische Anwendung thematisiert, wobei hier auch dokumentiert wird, inwieweit diese neuronale Funktionalität angepasst werden muss, um in der technischen Anwendung einsetzbar zu sei. Das abschließende Kapitel V dokumentiert Arbeiten im Rahmen des EU Projekts Fast Analog Computing of Emergent Transient States (FACETS), das zum Ziel hat, die Verarbeitungsfunktionen großräumiger neuronaler Areale (z.B. V1) nachzubilden. In gewissem Sinne stellt dies eine Synthese der vorhergehenden Kapitel zu Vernetzung und Einzelaspekten neuronaler Verarbeitung dar, da komplexe Netztopologien in Verbindung mit möglichst realistischem Gesamtverhalten der einzelnen Bausteine realisiert werden sollen. Die Hauptgesichtspunkte in diesem Kapitel sind die oben angeführte detailgetreue Nachbildung biologischer Systeme und ihre technische Implementierung. Kapitel VI gibt eine Zusammenfassung der hier dokumentierten Arbeiten und ihre Einordnung in das vom Dissertationsthema abgesteckte Wissenschaftsgebiet.

## I.1 Technische Bildverarbeitung

Das Sprichwort ‚ein Bild sagt mehr als tausend Worte“, enthält die Erkenntnis, dass Informationsverarbeitung und, -darstellung, die Menschen zugänglich sein soll hauptsächlich optisch orientiert ist. Besonders in der Wissenschaft wurde bereits zur Zeit der ersten neuzeitlichen Entdeckungsreisen von Kolumbus, Magellan, etc., stets hochqualifizierte Zeichner und Illustratoren mit auf Reisen geschickt, um sowohl für die Wissenschaft als auch für interessierte Laien die neuen naturalistischen Erkenntnisse visuell begreifbar zu machen. Abgesehen vom Wiedergeben dieser Exotik, wurden Illustrationen vermehrt auch in der heimischen Wissenschaft eingesetzt, um Versuchsaufbauten und –ergebnisse zu dokumentieren [Lardner52]. Mit dem Aufkommen der Photographie eröffneten sich für den wissenschaftlichen Einsatz von bildverarbeitenden Systemen viele neue Anwendungsfelder, weg von der reinen Dokumentation zum Gebrauch als Analyseinstrument, mit dem z.B. erstmals sehr schnelle Vorgänge erforscht werden konnten [Worthington00]. Ebenso eignete sich die Photographie, durch lange Verschlusszeiten langsame oder wenig Licht liefernde Vorgänge zu analysieren [Wood08]. In diesem Zusammenhang wird absichtlich bereits der Begriff der Bildverarbeitung im Gegensatz zur Bildaufnahme verwandt, da zum Einen Versuchsaufbauten bereits so gewählt wurden, dass nur bestimmte Aspekte der optischen Information verwertet wurden. Zum Anderen erschloss sich die Versuchserkenntnis erst aus dem Studium der entstandenen Photographien, also bei ihrer Auswertung oder (neuzeitlich) Verarbeitung:

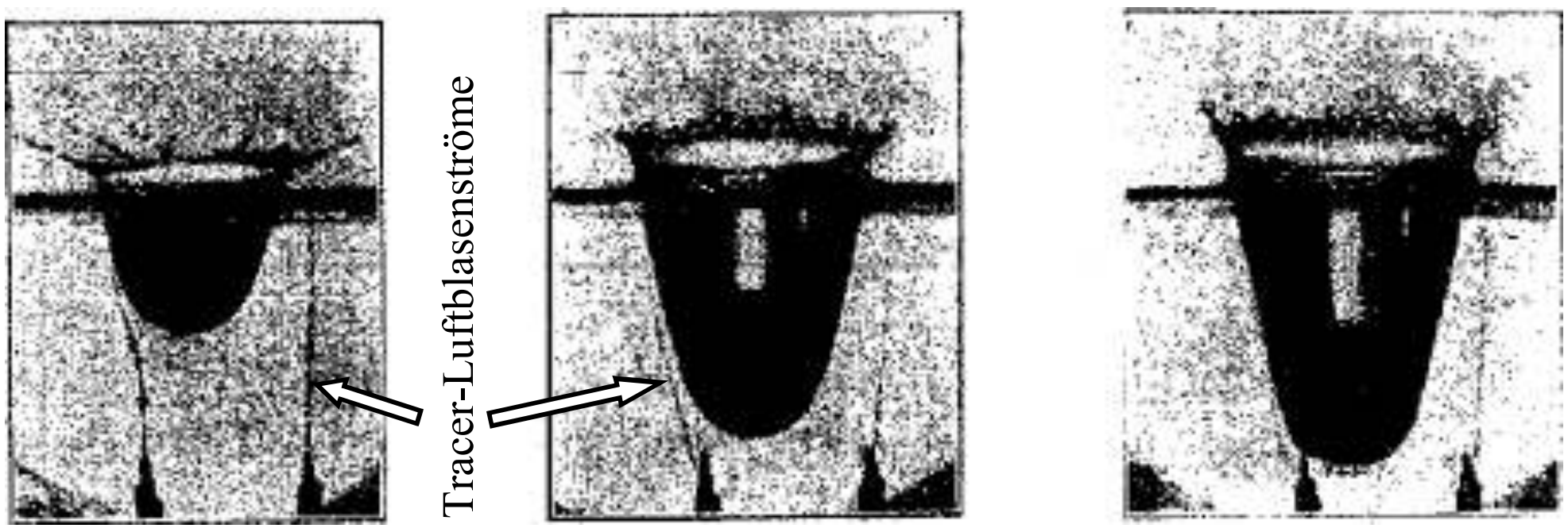

**Abbildung I.1.: Auswertung von Wiederauffüllvorgängen in Wassertropfen über die Vermessung von Luftblasenströmen in photographischen Aufnahmen der Versuchsreihe [Worthington00]**

Die Detailtreue von optischen Aufnahmen konnte u.a. in der Kartographie verwendet werden, um bisher unerreichte Akkuratheit in der Vermessung der Erde [Thomas20] oder anderer Himmelskörper zu liefern [Hale20]. Der Photographie wurden ständig neue Betätigungsfelder erschlossen, sei es in der Rüstungsforschung [Smith25] oder bei der Erforschung kleinster Materiestrukturen [Buerger39].
Die Auswertung der bis jetzt angesprochenen Anwendungen erfolgte jedoch immer manuell, d.h. Datenextraktion, Bildselektion, Bildanalyse wurden wieder von Menschen vollzogen, die sich einzeln mit dem Bildmaterial befassen mussten. Mit dem Fortschritt in der Informationstheorie wurden zunehmend automatisierte Verfahren zur Bildanalyse eingesetzt, die aber mit den heute eingesetzten digitalen Methoden abgesehen von der Theorie wenig gemeinsam hatten, z.B. realisiert man Kontrastvergrößerung oder Kantenfilterung mit einer optischen Bank [Oneill56]. Diese Technologie wurde bis zu relativ komplexen Bildanalysen wie z.B. Buchstabenerkennung weiterentwickelt [Lugt64]. Zunehmender Fortschritt und Verfügbarkeit der digitalen Rechentechnik lieferte hier neue Impulse, bei der zum Einen Computer zum Design einer optischen Bank herangezogen wurden, aber auch zum Auswerten der mit ihr erhaltenen Ergebnisse [Huang67]. Gleichzeitig machten die Fortschritte in der Halbleitertechnik u.a. Bildsensoren möglich, mit denen die gesamte Kette der Bildinformationsverarbeitung elektronisch realisiert werden konnte, wenn auch mit geringerer Auflösung als die der optischen Bank [Alt62]. Getragen von Trends zu militärischer automatisierter Fernerkundung, gehörte die digitale, elektronische Bildaufnahme und –verarbeitung ein Jahrzehnt später zu den Standardmethoden in einem breiten Spektrum wissenschaftlicher Applikationen [Nagy72]. Algorithmen und Hardware wurden permanent weiterentwickelt, um mehr und mehr Information automatisiert aus dem Bildmaterial extrahieren zu können, bis hin zu Robotersteuerungen, die sich in zweifacher Hinsicht ‚ein Bild ihrer Umgebung machen' [Marek02]. Einen Überblick moderner Bildverarbeitungsmethoden in Theorie, Hard- und Software, gibt [Jähne05]. Technische Bildverarbeitung wird heutzutage in vielen verschiedenen Bereichen wie Farbverbesserung oder Rauscheliminierung in digitaler Photographie, in industrieller Qualitätskontrolle und Fertigung, in Kartographie, in Umwelt-forschung, etc. sehr erfolgreich eingesetzt.

## I.2 Biologische Bildverarbeitung

Der im vorhergehenden Abschnitt geschilderten, relativ jungen technisch ausgelegten Bildverarbeitung steht eine Verarbeitung visueller Information gegenüber, die nicht so offensichtlich ist, jedoch schon wesentlich länger existiert und völlig ohne menschliche Intervention entstanden ist, die biologische Bildverarbeitung. Ihr Studium ist insbesondere interessant im Hinblick auf Bereiche der technischen Bildverarbeitung, bei denen entweder die manuelle (menschliche) Auswertung durch automatische Verfahren ersetzt werden soll (z.B. Qualitätskontrolle, Bildsortierung nach Inhalt, Objekterkennung, etc.), oder bei denen wie in der Robotik Bereiche der visuellen kognitiven Fähigkeiten von Lebewesen emuliert werden sollen.

Im Gegensatz zum wissenschaftlichen Ansatz der technischen Bildverarbeitung, folgte die Entstehung der visuellen Informationsverarbeitung in der Biologie nach Darwin dem einfachen Prinzip, die Überlebensfähigkeit zu verbessern [Darwin59, Norris07]. Viele Aspekte dieser Überlebensfähigkeit haben mit visueller Information zu tun, etwa die Partnerwahl, die Futtersuche, die Gefahrenabwehr, oder einfach die Navigation in einer komplexen Umwelt. Deshalb kommt auch in der Biologie der optischen Information über die Umgebung und ihrer Auswertung eine große Bedeutung zu. Die im vorhergehenden Abschnitt angesprochenen Robotersteuerungen versuchen etwa bzgl. der Navigation ähnliche Dinge zu leisten, können dies jedoch bis jetzt nur in kontrollierten, einfachen Umgebungen, wie z.B. Gebäuden oder industriellen Fertigungsumgebungen. Selbst ein halb-kontrolliertes System, wie z.B. eine Straße, mit einer endlichen Anzahl von Objekten, definierten Verhaltensregeln und insgesamt geringer Komplexität, überfordert heutige technische Ansätze bei weitem, so dass immer noch kein autonom navigierendes Straßenfahrzeug entwickelt wurde, welches außerhalb des Labors einsatzfähig wäre [Yoshida04].

Sobald also versucht wird, Eigenschaften der biologischen Bildverarbeitung mit Hilfe der (v.a. algorithmisch basierten) technischen Bildverarbeitung zu lösen, stößt diese an Grenzen. Aufgaben, die z.B. eine Biene mit Leichtigkeit erledigt, Orientierung in mehreren Quadratkilometern natürlicher Umgebung, Objektausweichen mit entsprechender Flugbahnsteuerung, Gefahrenerkennung, sicheres Wiederauffinden oder Erkennen von Futterpflanzen und heimischem Stock, können heute selbst bei Einsatz von hochleistender Rechentechnik nicht in solcher Präzision gelöst werden, obwohl diese gemessen an der Geschwindigkeit weit über der einer Biene liegt [Menzel01].

In den o.a. Anwendungsgebieten der technischen Bildverarbeitung kann es also von Vorteil sein, verschiedene Prinzipien biologischer Bildverarbeitung zu übernehmen, um durch einen Paradigmenwechsel bestimmte Funktionalitäten erst zu ermöglichen oder zumindest manche Anwendungen effizienter zu implementieren. Erschwert wird dies einerseits dadurch, dass die Entwicklung technischer Bildverarbeitung bereits eine sehr hohe Effizienz erreicht hat und auch an die verwendete Hardware, v.a. digitale Rechner, besser angepasst ist als die auf zellulärer elektrochemischer Basis laufende biologische Bildverarbeitung. Zum Anderen konnte bei der technischen Bildverarbeitung von einfachen Stufen ausgehend zunehmend komplexere Verarbeitung entwickelt werden, wohingegen man in der Biologie mit hochkomplexen Endergebnissen von Millionen Jahren Evolution konfrontiert ist. Bei diesen kann zudem manchmal nicht genau unterschieden werden, welche Teile eines Algorithmus notwendig zur Erfüllung einer bestimmten Funktion und damit interessant für eine technische Realisierung sind und welche Teile nur Unzulänglichkeiten der zugrunde liegenden biologischen Matrix kompensieren [Häusser03, Koch99]. Zusätzlich müssen auch die als relevant erkannten Teile eines solchen bildverarbeitenden Algorithmus für ihre technische Implementierung modifiziert werden, um auf konventioneller oder nur leicht angepasster Hardware lauffähig zu sein, wobei natürlich die relevanten informationsverarbeitenden Prinzipien erhalten bleiben sollen. Es gibt auch Ansätze zur stärkeren Anpassung der zugrunde liegenden Hardware an biologische Prozesse [Türel05], dies erhöht jedoch signifikant die Implementierungskosten und ist deshalb nur für Grundlagenforschung oder als Langzeitperspektive interessant, nicht jedoch als kurz- oder mittelfristig einsetzbare Alternative zur konventionellen technischen Bildverarbeitung. Es sei hier noch erwähnt, dass in der Grundlagenforschung auch das Verschalten biologischer neuronaler Netze mit technischen Schnittstellen untersucht wird, zum Einen, um als „Wetware“ Prothesen direkt mit Nervenzellen zu verbinden [Potter03], zum Anderen, um diese biologischen Netze direkt in technischen Anwendungen einzusetzen [Ruaro05].

# I.3 Das visuelle System des Säugetiers von Retina bis V1, nach Aufbau und Funktion

Da vom visuellen System von Säugetieren die meisten gesicherten Erkenntnisse vorliegen und es auch allgemein als repräsentativ für viele Aspekte biologischer Bildverarbeitung angesehen wird [Jones87a, Shepherd04 (Kapitel 12)], werden die dort postulierten oder nachgewiesenen Vorgänge i.d.R. als Grundlage für Modellierungen und technische Adaptionen verwandt. Im weiteren ist deshalb mit dem Begriff biologischer Bildverarbeitung die eines Säugetieres gemeint, solange keine weitere Erläuterung dazu gegeben wird.
Wie bereits erwähnt, sind für die technische Anwendbarkeit v.a. die ersten Stufen der biologischen Bildverarbeitung interessant, da deren Abläufe noch vergleichsweise leicht messtechnisch erfassbar sind [Hubel68, Shepherd04 (Kapitel 12)]. Dies gilt sowohl in phänomenologischer Sicht, also als Beschreibung der Übertragungsfunktion ohne Kenntnis der genauen Vorgänge, als auch in genauer auflösender Weise, d.h. wie durch die spezifische Vernetzungsstruktur und Aufbau und Funktion der einzelnen Elemente die Gesamtfunktionalität erreicht wird. Im folgenden wird als erstes die Funktion der Retina näher betrachtet, wobei hier als Vorgriff auf Kapitel II bereits auf die Vernetzungsstruktur der Retina eingegangen wird. Dies ist motiviert zum Einen durch die relativ leicht verständliche Grundstruktur der Retina, als auch durch ihre physische und funktionale Trennung von der weiteren Verarbeitung im visuellen Kortex.

## I.3.1 Bildaufnahme, -wandlung und -informationsverdichtung in der Retina

Durch die von Pupille und Augapfel gebildete ‚Kamera' wird ein Abbild des auf die Pupille einfallenden Bildes gespiegelt und fokussiert auf eine Zellschicht an der hinteren Innenwand des Augapfels, die Retina, projiziert (Abbildung I.2). Eine Beschreibung der optischen Eigenschaften und des generellen Aufbaus des Augapfels liefert [Kandel95 (Kapitel 22)], unter dem Aspekt der Signalverarbeitung soll hier die Betrachtung am Bildaufnehmer, einer Schicht der Retina, beginnen. Eine Schicht lichtempfindlicher Zellen in der Retina, die sogenannten Stäbchen und Zäpfchen detektieren dort das projizierte Bild:

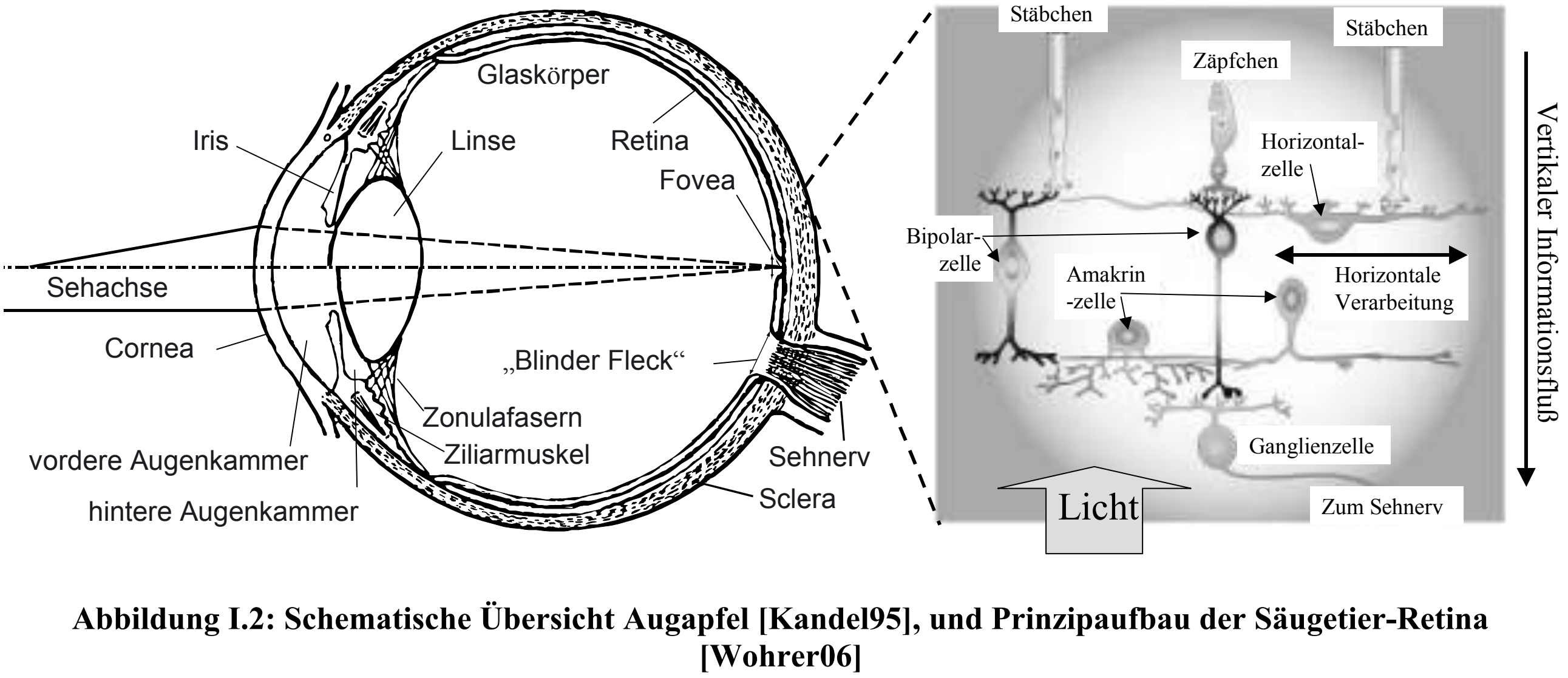

**Abbildung I.2: Schematische Übersicht Augapfel [Kandel95], und Prinzipaufbau der Säugetier-Retina [Wohrer06]**

Die Stäbchen nehmen Bilder nur in Helligkeitsstufen wahr und saturieren bei Tageslicht, die Zäpfchen existieren in drei Varianten mit unterschiedlicher spektraler Empfindlichkeit und stellen somit Farbkanäle für die Wahrnehmung bereit. Die Zäpfchen sind v.a. im mittleren Bereich der Retina angeordnet, mit einer Dichte $>1{,}5*10^5$, bei einem Dichteunterschied von $10^3$ zwischen Fovea und Peripherie. Die Stäbchen sind primär um diesen zentralen Bereich gruppiert, mit einer ähnlichen maximalen Dichte und einem zum Anstieg der Zäpfchendichte reziproken Abfall in der Mitte der Retina [Wohrer06]. Mithin stellen die Stäbchen einen Kanal zum Grauwert-, Dämmerungs- und Peripheriesehen bereit, während Zäpfchen für zentrales Tageslichtsehen und Farbsehen zuständig sind.

**Einschub:** Sowohl in der technischen als auch in der biologischen Bildverarbeitung kommt durch den höheren Gehalt an Information der Farbauflösung eines Bildes eine Sonderstellung zu. Jedoch soll im weiteren Verlauf dieser Arbeit nur ein von (verallgemeinerten) Rezeptorzellen bereitgestellter Grauwertkanal betrachtet werden. Dies lässt sich wie folgt begründen:

- Die Zäpfchen stellen über Sensorfusion der Farbkanäle vergleichbar den Stäbchen ebenfalls einen Grauwertkanal bereit. Dieser liegt flächendeckend an Neuronen im visuellen Kortex an, wohingegen zusätzliche Farbkanäle nur an ca. 20% der Neuronen gefunden wurden [Gegenfurtner03].
- Die für die Mustererkennung wichtige spatiale Vorverarbeitung der von Stäbchen und Zäpfchen gelieferten Information in der Retina kann durch die ähnliche Vernetzungsstruktur ebenfalls in einem verallgemeinerten Verarbeitungsmodell zusammengefasst werden [Wohrer06]. Experimentell belegt ist ausserdem, dass Objekt- bzw. Mustererkennung ähnlich gut mit Grauwertinformation funktioniert wie mit Farbkanälen [Mullen02].
- Die technische Bildverarbeitung findet primär auf Grundlage von Grauwertbildern statt, v.a. aufgrund der niedrigeren rechnerischen Komplexität, während Farbkanäle nur hinzugezogen werden, wenn ein Problem (z.B. Vordergrund/Hintergrundseparierung) nicht mit reiner Grauwertinformation lösbar ist [Jähne05]. In der biologischen Bildverarbeitung scheint die detaillierteste Verarbeitung ebenfalls auf Grauwerte aufzubauen, während die Farbinformation nur als unterstützender Hilfskanal Verwendung findet [Johnson01].
- Technische Anwendungen sollen zudem über einen Beleuchtungsbereich funktionsfähig sein, der in der Retina vom Dämmerungssehen (Stäbchen) bis Tageslicht (Zäpfchen) reicht, weswegen es auch unter technischen Aspekten sinnvoll erscheint ein verallgemeinertes Modell zu betrachten.
- Ebenso müssen Adaptionen biologischer Bildverarbeitung i.d.R. mit technisch erhältlichen optischen Sensoren arbeiten, die als Helligkeitssensoren ohne Farbkanalauflösung kostengünstiger herzustellen sind [Henker03].

Die Retina ist in einer Schichtstruktur aus mehreren verschiedenen Zelltypen aufgebaut, wobei die oben erwähnten Stäbchen und Zäpfchen (bzw. allgemein Rezeptorzellen) in der untersten Schicht sitzen. Darüber, d.h. weiter zum Inneren des Augapfels hin finden sich nacheinander die Horizontal-, Bipolar-, Amakrin- und Ganglienzellen, welche die optische Information vorverarbeiten und zur Übertragung auf dem Sehnerv codieren. Die in den verschiedenen Schichten stattfindende Verarbeitung lässt sich in guter Näherung über lineare Filteroperationen beschreiben [Meister99, Wohrer06].

Die Horizontalzellen bilden die erste Schicht der Verarbeitung, sie vernetzen Gruppen von Rezeptorzellen miteinander in Form einer diskret ausgeführten Glättungsmaske, in einem Vorgang ähnlich wie bei Diffusionsnetzwerken in elektrischen Schaltungen [Wohrer06, Mayr06b, Carmona02]. In der Bildverarbeitung entspricht dies einer Faltung beispielsweise mit einer Gaußglocke entsprechend Gleichung (I.1), wobei von den physiologischen Faktoren innerhalb der Horizontalzelle und ihrer Vernetzungsstruktur abhängt, mit welcher Form und Ausdehnung des Einflussbereichs geglättet wird.

$$gauss(x,y) = \frac{1}{2\pi\sigma^2} e^{-\frac{x^2+y^2}{2\sigma^2}} \qquad \textbf{(I.1)}$$

Der Glättungseffekt dieser Gaußschen Faltungsmaske mit verschiedenen $\sigma$, entsprechend der Ausdehnung der Horizontalzellen in der Retina, wird in Abbildung I.3 illustriert, wobei dies durch die mathematische Beschreibung natürlich nur eine Näherung des Ausgangsbildes der Horizontalzellen darstellt. In der Retina existieren diese verschiedenen Glättungsstufen parallel [Dacey00], so dass hier eine Art Bildpyramide erzeugt wird [Jähne05].

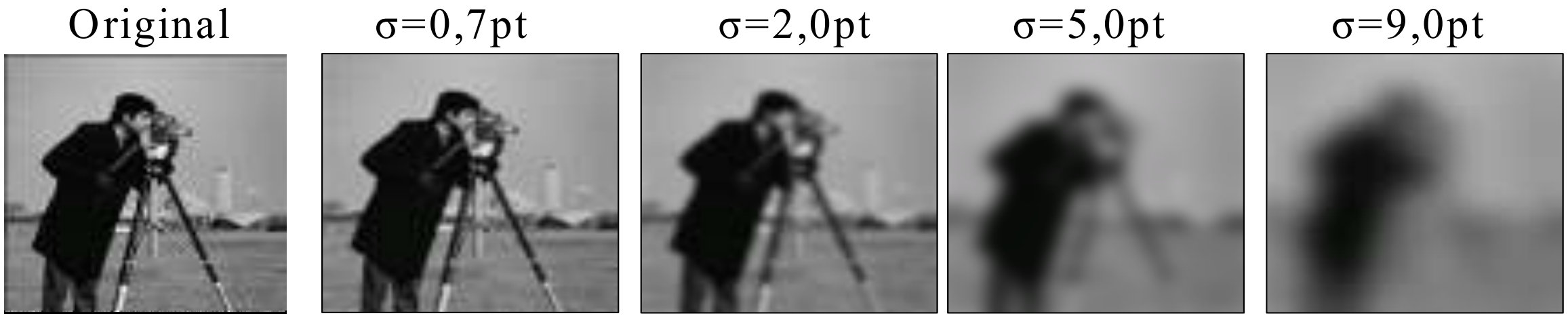

**Abbildung I.3: Gausssche Glättung, Faltungsmaske und Beispielbilder für verschiedene σ (in Bildpunkten – pt). Das Originalbild hat 256*256 pt Ausdehnung**

Die Bipolarzellen subtrahieren die geglätteten Bilder von zwei oder mehr Horizontalzellen mit unterschiedlichen Ausdehnungen voneinander, wodurch sich signaltheoretisch ein ‚Difference of Gaussian, (DoG)-Filter ergibt [Dacey00, Partzsch07a]:

$$d(x,y) = e^{-\frac{x^2+y^2}{2\sigma_1^2}} - \frac{1}{a^2} * e^{-\frac{x^2+y^2}{2a^2\sigma_1^2}} , \quad mit\ a = \frac{\sigma_2}{\sigma_1} \tag{I.2}$$

Da die Filterantwort bei der Faltung ohnehin auf die Summe der Maskenkoeffizienten normiert wird, kann die absolute Amplitude des DoG vernachlässigt werden. In der obigen Gleichung wurde deshalb der Faktor $2\pi\sigma_1^2$ im Nenner des ersten Summanden als 1 festgelegt, wodurch sich nach einer Multiplikation mit $2\pi\sigma_1^2$ die in Gleichung (I.2) dargestellte Form des DoG ergibt. Es entsteht ein Bandpassverhalten für Bildfrequenzen, begrenzt durch $\sigma_1$ und $\sigma_2$, mit einer spatialen Ausdehnung von $\sigma_1$ und einem Durchlassverhältnis von $a$. Dieses variiert bei vermessenen Ganglienzellen über einen Bereich von 1,1 bis 3, d.h. es existieren in der Retina sowohl schmalbandige Filter als auch solche mit breiterem Durchlassbereich. Die Filterantwort einer Bipolarzelle an den Koordinaten $(x_0,y_0)$ auf ihr zugehöriges DoG-Filter $d(x,y)$ bei einem Eingangsbild $b(x,y)$ ergibt sich dann zu:

$$D_{x_0,y_0} = \sum_{(x)} \sum_{(y)} d(x-x_0, y-y_0) * b(x-x_0, y-y_0) \tag{I.3}$$

Die Wirkung dieses Filters auf ein Bild lässt sich anhand eines ‚Laplacian of Gaussian' (LoG) beschreiben, den der DoG für ein $a$ von 1,4 approximiert [Dacey00]. Der LoG berechnet die zweite Ableitung des Helligkeitsverlaufes eines Bildes [Jähne05], d.h. es werden konstante Grauwertbereiche und Grauwertänderungen mit konstantem Anstieg verworfen, nur noch Änderungen im lokalen Grauwertverlauf ergeben von Null verschiedene Antworten. LoG-Filter werden deshalb in der technischen Bildverarbeitung als Kantenfilter eingesetzt [Tabbone95]. Abbildung I.4 gibt ein Beispiel einer solchen Faltungsmaske und ihren Effekt auf ein Beispielbild wieder.

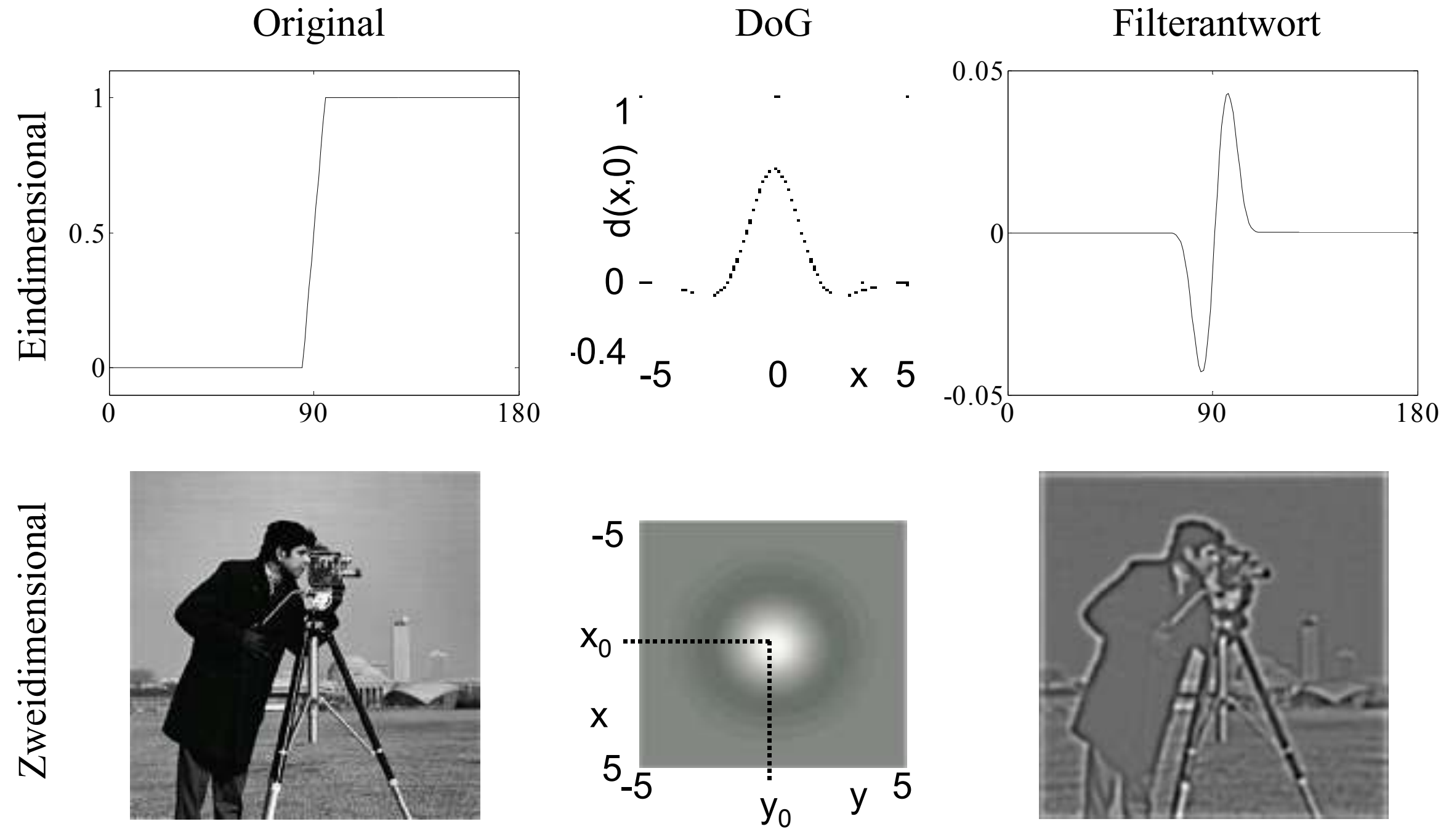

**Abbildung I.4: Faltungsmaske Retina, Ausgangsbild und gefiltertes Bild**

Die DoG-Maske liefert positive und negative Werte, die in der Darstellung der DoG-Antwort (Abbildung I.4 rechts unten) auf einen Wertebereich von 0 bis 255 Graustufen normiert wurden, d.h. die grauen Bereiche entsprechen einer minimalen (positiven oder negativen) Antwort des Filters, dunkle Bereiche sind stark negative DoG-Faltungsresultate, helle Bereiche entsprechend positive Antworten. Die DoG-Antwort wird von den Bipolarzellen an die Ganglienzellen weitergegeben, wobei diese in der zweiten Stufe horizontaler Verarbeitung durch die Amakrinzellen vernetzt sind. Die Amakrinzellen in Abbildung I.2 wirken inhibitorisch für eine enge Nachbarschaft von Ganglienzellen, d.h. wenn eine der an die Amakrinzellen angeschlossenen Ganglienzellen ein Ausgangssignal liefert, blockiert dieses die anderen Ganglienzellen. Somit selektieren die Amakrinzellen zwischen DoGs, deren Einzugsbereiche deutlich überlappen, die also ähnliche Bildinhalte codieren, so dass nur die (zeitlich) erste Antwort übertragen wird, wodurch Redundanz in den Antworten der Ganglienzellen vermieden wird. Diese durch die Vernetzung der Amakrinzellen spatial diskretisierte Natur der gegenseitige Hemmung der DoGs kann in guter Näherung durch eine kontinuierliche Inhibition ersetzt werden [Partzsch07a]. Die zeitdynamischen Vorgänge des gegenseitigen Pulsblockierens werden von einem statischen spatial kontinuierlichen Hemmungsmaß ebenfalls mit abgebildet, da die Pulshemmung ebenfalls mit zunehmender Entfernung durch ihre längere Laufzeit und Abschwächung an Wirkung verliert:

$$I_{AB} = \langle d_A, d_B \rangle = \int_{-\infty}^{\infty} \int_{-\infty}^{\infty} d_A(x,y) * d_B(x,y) dxdy \ , W_{AB}^{inh} = C * I_{AB} \qquad \textbf{(I.4)}$$

Ein Ähnlichkeitsmaß $I_{AB}$ für die DoGs von zwei Ganglienzellen wird über ihr Flächenintegral gewonnen, das Gewicht $W^{inh}{}_{AB}$ der inhibitorischen Verbindung zwischen beiden Zellen ergibt sich daraus über einen zusätzlichen Proportionalitätsfaktor C. Die Gesamtantwort einer Ganglienzelle $D^{inh}{}_{x0,y0}$ auf ein Eingangsbild ergibt sich somit aus der Antwort auf die DoG-Maske, vermindert um die mit dem Inhibitionsgewicht verrechneten Antworten der benachbarten Ganglienzellen:

$$D_{x_0,y_0}^{inh} = D_{x_0,y_0} - \sum_i W_{0i}^{inh} * D_{x_i,y_i}, \ bzw.\ am\ \ Ausgang:\ \ r_{x_0,y_0}(t) = f(D_{x_0,y_0}^{inh}) \qquad \textbf{(I.5)}$$

An den Ganglienzellen findet ein Wandel der Verarbeitungsmodalitäten statt, die bisherige Verarbeitung durch analoge Ströme zwischen den Neuronen wird wie oben angedeutet in eine Pulsfolge $r_{x0,y0}$ gewandelt. Dabei wird der in die Ganglienzelle fließende Strom zur Aufladung der Membrankapazität verwendet, die bei Erreichen einer Schaltschwelle einen Puls generiert [Wohrer06] (siehe auch Abschnitt II.1.1). Je nach Betrachtungsweise erfolgt somit eine Codierung der Retinaausgangssignale in ein Phasen- oder Frequenzsignal [Warland97].

Die Filtermasken der Ganglienzellen lassen sich in zwei Klassen einteilen, so genannte On- und Off-Zellen, wobei die Filtermaske einer On-Zelle der in Abbildung I.4 gezeigten entspricht, während eine Off-Ganglienzelle auf entgegengesetzten Kontrast reagiert, d.h. eine dunkle Mitte und helle Umgebung. Dies scheint der Tatsache geschuldet, dass sich in einer Pulsrate nur ein unipolares Signal codieren lässt, also an jeder Bildkoordinate Masken mit genau entgegengesetzten Charakteristiken nötig sind, um die volle Dynamik der Maskenantwort abzudecken. Messungen an Ganglienzellen, die an derselben Bildkoordinate die Antwort einer On- und Off-Maske codieren, ergeben keine Signalredundanz zwischen beiden [Warland97].

Von der Retina werden somit nur Beleuchtungskontraste an die höheren Stufen der visuellen Informationsverarbeitung weitergesendet, was zu einer starken Verdichtung des Informationsflusses führt, von ca. 100 Mbit/s auf 1Mbit/s [Meister99]. Eine weitere Informationsverdichtung findet insofern statt, als auch die Beleuchtungskontraste nur in komprimierter Form weitergegeben werden, d.h. ein absoluter Dynamikbereich der Kontraste von $10^9$ wird auf einen Dynamikbereich der Pulsraten auf dem Sehnerv von $10^2$ übertragen, wobei Änderungen des Kontrastes sehr viel detaillierter übertragen werden [Meister99, Smirnakis97]. Die beschriebene retinale Verarbeitung wird vereinzelt in VLSI nachempfunden, um eine ähnliche Informationsverdichtung zu erhalten oder Bildanalysen durchführen zu können [Carmona02, Erten99, Koch96, Mahowald89].

### I.3.2 Steuerungs- und Relaisstation nach dem Sehnerv: Der seitliche Kniehöcker

Die von den Ganglionzellen codierten DoG-Antworten werden über den Sehnerv zu einem Bereich des Thalamus geschickt, der als seitlicher Kniehöcker, Corpus geniculatum laterale bezeichnet wird (im Folgenden nach seinem englischen Namen ‚Lateral Geniculate Nucleus' als LGN abgekürzt) [Shepherd04 (Kapitel 8)]. Dieser agiert als Signalverstärker und -former für höhere Verarbeitungsstufen im Kortex, wobei hier eine nochmalige Selektion der vom Sehnerv eintreffenden Information v.a. hinsichtlich Redundanzeliminierung vorgenommen wird [Freeman02]. Der LGN ist zweifach ausgeführt, wobei der an einem Sehnerv angeschlossene LGN auch Informationen des anderen Sehnervs enthält, so dass hier bereits eine rudimentäre binokulare Verarbeitung stattfindet [Swiercinsky01]. Im folgenden ist der funktionale Aufbau des LGN mit seinen verschiedenen Signalpfaden wiedergegeben:

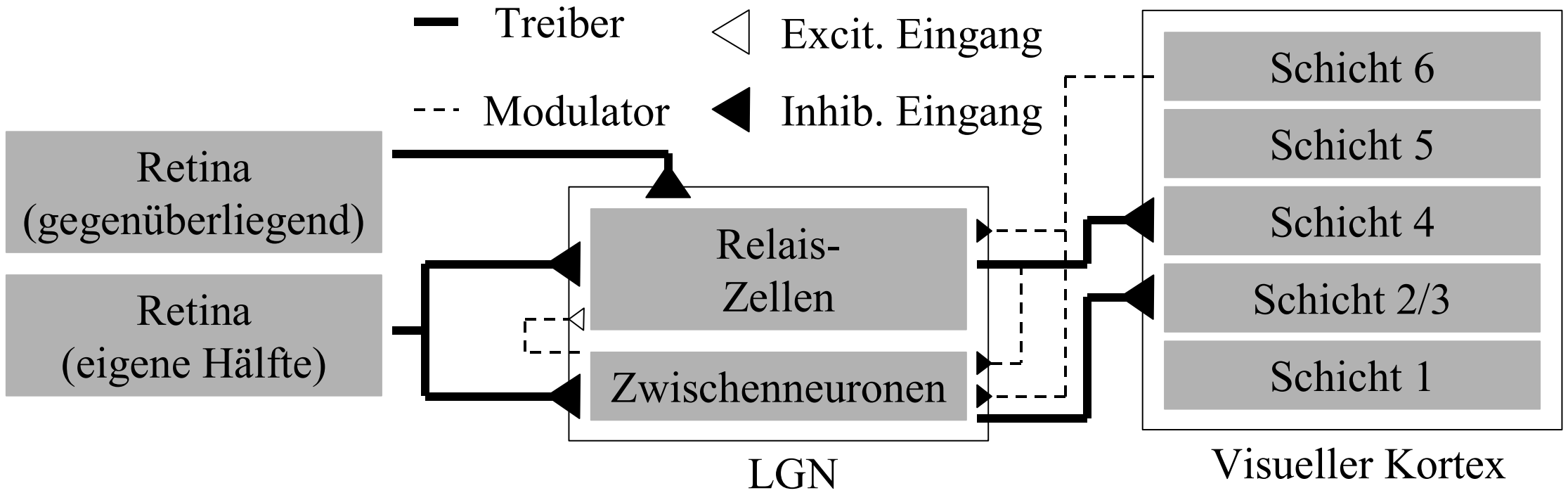

**Abbildung I.5: Überblick der Vernetzung von Retina und V1 durch den LGN mit wichtigsten Signal- und Steuerleitungen [Einevoll03, Shepherd04 (Kapitel 8)]**

Der LGN besteht aus Schichten von Relaisstellen, die verschiedene Aspekte der retinalen Information aus beiden Gesichtshälften repräsentieren. Die Aufteilung dieser Information beginnt bereits am Ausgang der Retina über verschiedene Ganglienzellen, die den sogenannten parvozellulären und magnozellulären Pfad bereitstellen, der aber in dieser Stufe nur leicht unterschiedliche Informationen beinhaltet [Kandel95]. Zwischen den Schichten im LGN, welche parvo- und magnozelluläre Informationen repräsentieren, sitzen Interneuronen, welche die Relaiszellen über laterale inhibitorische Verbindungen vernetzen und damit eine Art ‚Sensor fusion' zwischen den verschiedenen Informationen der Relaiszellenschichten durchführen. Die weiterverarbeiteten DoGs der Retina werden von den Relaiszellen an die Schicht vier des primären visuellen Cortex (V1) weitergegeben [Shepherd04 (Kapitel 8)]. Einen weiteren Informationskanal bilden die Interneuronen mit der Schicht 2/3 des V1 aus [Einevoll03]. Von Schicht 6 des visuellen Cortex existieren Rückkopplungsleitungen Richtung Relaiszellen, welche die Redundanzeliminierung im LGN steuern [Freeman02] und die intrinsische Kontrastadaption und Arbeitspunktnachführung des LGN unterstützen [Mukherjee95]. Im Rest der Arbeit werden die unterschiedlichen Informationspfade in Retina und LGN nicht separat betrachtet. Dies begründet sich zum Einen daraus, dass v.a. statische Bildverarbeitung betrachtet wird, die auf parvozelluläre Bahnen beschränkt ist [Kandel95]. Generell sind beide Bahnen außerdem wie oben angeführt sehr stark entlang ihres Wegs verkoppelt, so dass viele Informationen redundant übertragen werden. Zusätzlich soll in dieser Arbeit primär die Informationsverarbeitung im V1 untersucht werden, in dem an einer großen Mehrzahl von Neuronen sowohl parvo- als auch magnozelluläre Informationen ankommen [Vidyasagar02] und damit eine getrennte Behandlung nicht notwendig erscheint. Die Aufteilung in die unterschiedlichen Bahnen findet vornehmlich in höheren Stufen des visuellen Kortex statt (V3-V5), die beispielsweise mit Gesamtbildanalysen oder Bewegungsfilterung befasst sind [Kandel95].

### I.3.3 Komplexe Bildfilterung: Rezeptive Felder im V1 Bereich des visuellen Kortex

Der V1 Bereich des visuellen Kortex stellt die erste Stufe komplexer visueller Verarbeitung im Säugetiergehirn dar. In ihm erfolgt eine Filterung nach statischen und dynamischen orientierten Strukturen (z.B. bewegten oder unbewegten Kanten unter einem bestimmten Winkel). Eine wegweisende Charakterisierung dieser Filterung an einzelnen Neuronen des V1 erfolgte durch

Hubel und Wiesel bei Katzen und Primaten [Hubel68]. Die rezeptiven Felder (RF) der Neuronen wurden hierbei durch Präsentation von visuellen Stimuli im Sehfeld der Tiere und Messung der zugehörigen Feuerraten der Neuronen bestimmt, wobei diese Methode nur eine qualitative Charakterisierung erlaubt, d.h. Lage und Verteilung von hemmenden und verstärkenden Bildbereichen, nicht ihre Amplitude. Zusätzlich wird nicht zwischen statischen und dynamischen rezeptiven Feldern unterschieden. Was jedoch in der zitierten Arbeit deutlich wird, ist die weitaus höhere Komplexität dieser RFs gegenüber den DoGs der Retina und des LGN. Die Untersuchung der rezeptiven Felder im V1 wurde maßgeblich durch Jones und Palmer [Jones87a, Jones87b] weitergeführt, wobei hier durch eine Korrelationsanalyse zwischen Feuerraten und Stimulus und eine entsprechend rückgekoppelte Veränderung des Stimulus zum Einen statische und dynamische RFs unterschieden werden [Jones87a]. Zum Anderen erlaubt diese verbesserte Methode die quantitative Analyse der rezeptiven Felder, z.B. auf Ähnlichkeit mit bekannten Bildverarbeitungsoperatoren. In [Jones87b] wird ein Vergleich zwischen den so gefundenen RFs und Gabormasken [Jähne05] beschrieben:

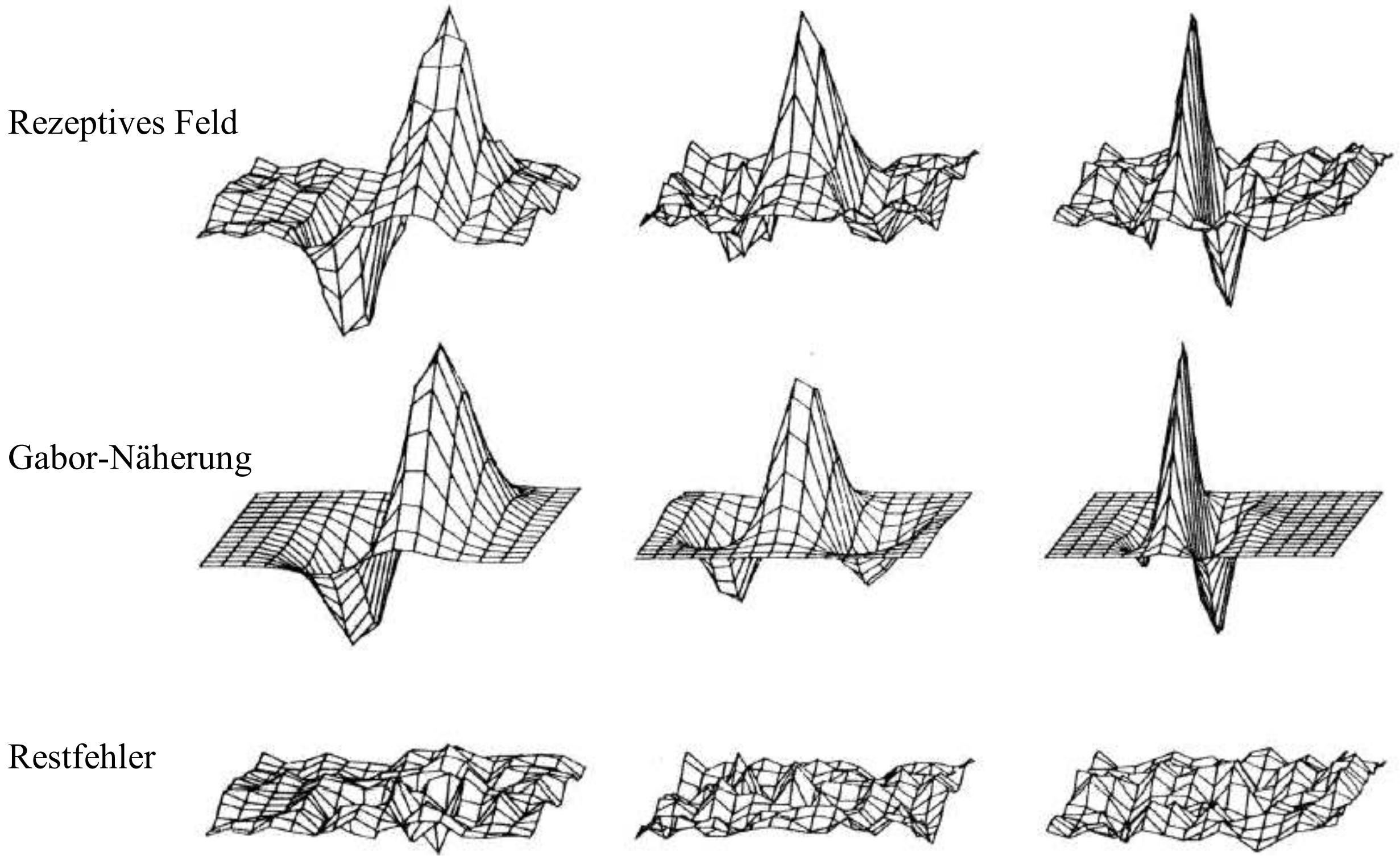

**Abbildung I.6: Jones et.al., Vergleich zwischen RFs und (angepassten) Gaborfiltern [Jones87b]**

Der entstehende Restfehler lässt keine systematische Abweichung erkennen, d.h. Gabormasken scheinen eine gute Näherung der rezeptiven Felder der Neuronen im V1 zu sein. Gaborfilter sind eine Untergruppe der Waveletfilter, die in der klassischen Bildverarbeitung seit Jahrzehnten sehr erfolgreich in Bereichen von Kantenfilterung über Robotik bis zu Bildklassifizierung verwendet werden [Jähne05, Loupias99]. Ein Gaborfilter besteht aus einer Kombination eines Bildfrequenzfilters und einer Gaußschen Glättung, welche die Frequenzfilterung räumlich begrenzt, d.h. es wird in einem Teilbereich des Bilds nach dem Auftreten eines wiederholten Hell/Dunkel-Wechsels in einer bestimmten Orientierung gesucht. Basierend auf der Gaboranpassung in [Jones87b] wurde in [Lee96, Partzsch07a] ein Gaborfilter hergeleitet, der sich an biologisch gemessenen rezeptiven Feldern orientiert und gleichzeitig für Randbedingungen der Filtertheorie optimiert ist:

$$g(x,y) = \frac{\omega_0}{\sqrt{\pi d} * k} * e^{-\frac{\omega_0^2}{2k^2}\left(x^2+\frac{y^2}{d^2}\right)} * \left( e^{j\omega_0 x} - e^{-\frac{k^2}{2}} \right) \qquad \textbf{(I.6)}$$

Grundsätzlich handelt es sich dabei um eine räumlich lokalisierte Filtermaske, ersichtlich aus der Multiplikation einer Gaussmaske in x- und y-Richtung (im Faktor vor der Klammer). Durch diese Gaussmaske wird der Wellenzahlfilter (innerhalb der Klammer) in seiner räumlichen Ausdehnung begrenzt. Die Orientierung dieses Filters ist 0°, d.h. er reagiert auf senkrechte Kontraste oder eine in der x-Achse verlaufende Wellenfront. Der Parameter $\omega_0$ gibt die Grundfrequenz oder auch Wellenzahl des Kontrastes an (in 2π/(Kontrastperiodizität in Pixeln)). Durch die mit $k/\omega_0$ definierte Standardabweichung der äußeren Glättungsmaske ist die Bandbreite der Frequenzen vorgegeben, auf die der Filter reagiert[1]. Das Verhältnis der Ausdehnung in x- und y-Richtung d.h. die Form des elliptischen Gaußfilters wird mit $d$ spezifiziert. Die folgende Illustration gibt Beispiele für die Filterung eines Bildes mit verschieden orientierten Gabormasken (Die Filter sind gerade, d.h. geben den Realteil von Gleichung (I.6) wieder):

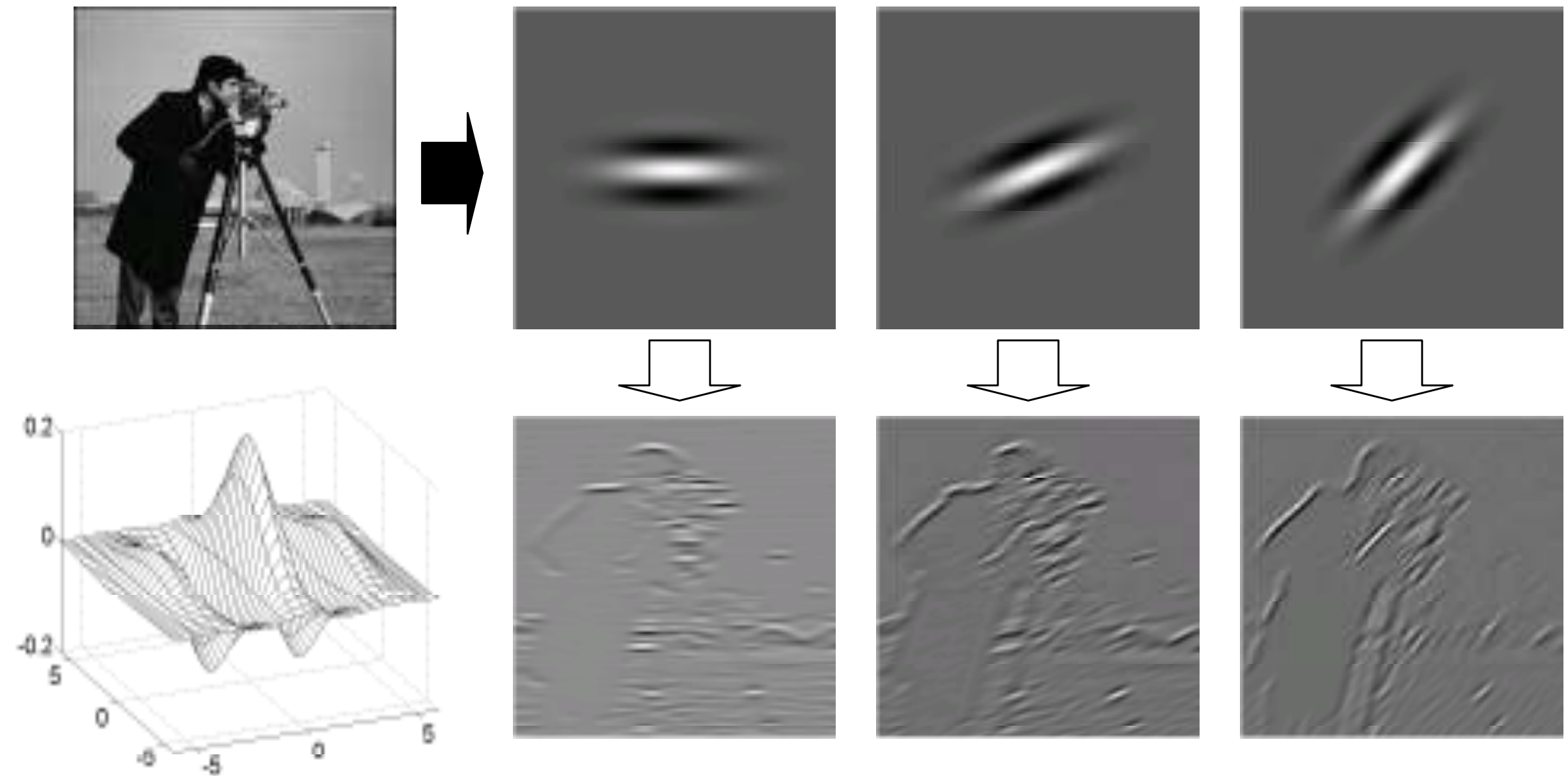

**Abbildung I.7.: Beispiele für Gaborfilterung, aus [Partzsch07a]**

Gabormasken oder rezeptive Felder werden im V1 hierarchisch aufgebaut, d.h. aus den retinalen DoGs werden einfache Filter mit z.B. kleiner Ausdehnung oder geringer Frequenzselektivität gebildet, aus denen in weiteren Stufen dann Filter mit zunehmender Komplexität gebildet werden [Partzsch07a, Riesenhuber99], siehe auch die folgende Darstellung:

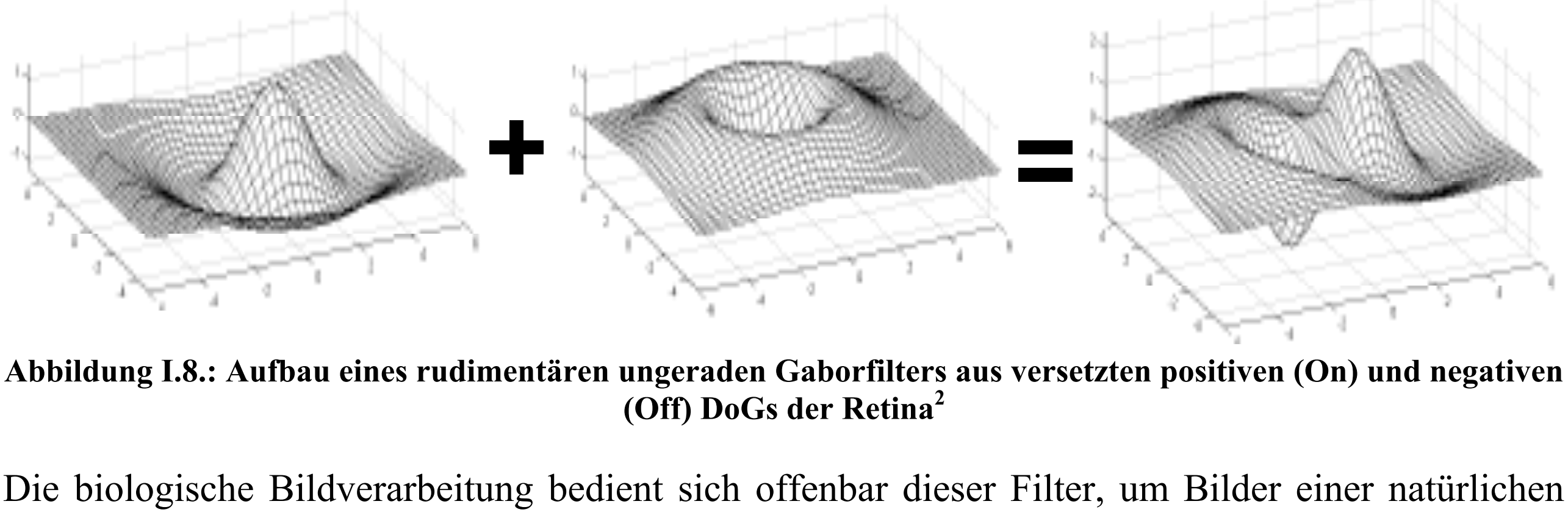

**Abbildung I.8.: Aufbau eines rudimentären ungeraden Gaborfilters aus versetzten positiven (On) und negativen (Off) DoGs der Retina[2]**

Die biologische Bildverarbeitung bedient sich offenbar dieser Filter, um Bilder einer natürlichen Umgebung möglichst effizient abzubilden. Bei einer statistischen Analyse von Naturbildern bzgl. der in ihnen wiederholt auftretenden Strukturen ergeben sich ähnliche Filtermasken, mit denen sich also diese Szenen gut beschreiben lassen [Olshausen02].

[1] Je größer k wird, desto weitere Ausdehnung im Ortsbereich hat die Gabormaske. Gleichzeitig verringert sich dadurch die Bandbreite der gefilterten spatialen Frequenzen.

[2] Algorithmische Aspekte von Gaborsynthese aus DoG-Filtercharakteristiken werden in Anhang B.2.1 behandelt.

# II Aufbau und Funktionalität von Neuronen und neuronalen Netzen

Um zu einem weitergehenden Verständnis der in biologischen neuronalen Netzen stattfindenden Informationsverarbeitung zu gelangen, muss ihr funktionaler Aufbau wesentlich feinmaschiger analysiert werden, als dies in den vorherigen Kapiteln geschehen ist, die eher an groben Baublöcken und Verhaltensbeschreibungen ausgerichtet war. Einer der ersten Forscher, die sich der detaillierten Untersuchung biologischer neuronaler Strukturen gewidmet haben, war Santiago Ramón y Cajal [Cajal09]. Von ihm wurde die physische Feinstruktur verschiedener Gehirnbereiche analysiert, wie in der folgenden Grafik wiedergegeben:

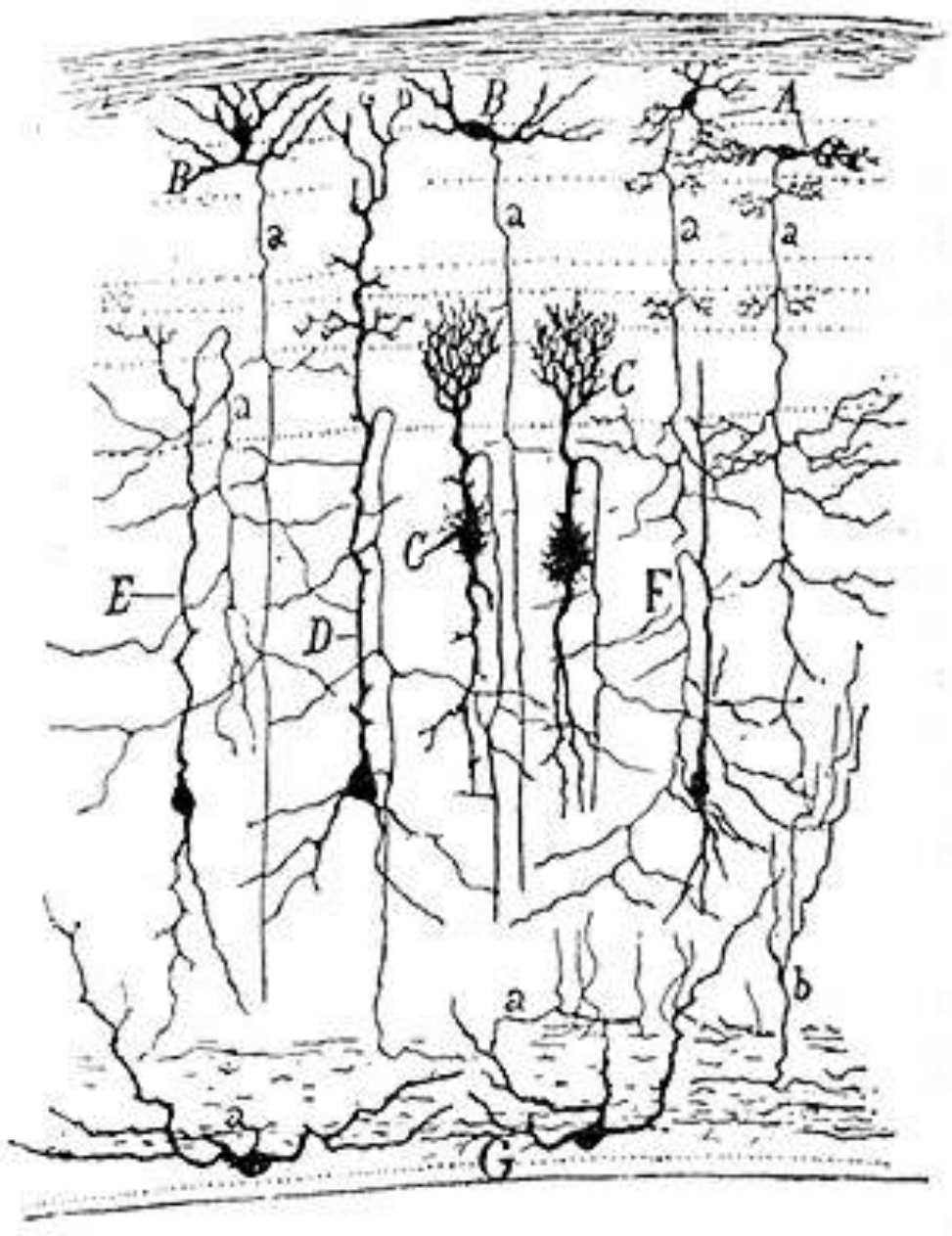

**Abbildung II.1: Entnommen aus [Cajal09], Schnitt durch das Tectum opticum beim Sperling**

Aus einem solchen topologischen Ausschnitt lässt sich bereits die Vielfältigkeit neuronaler Verarbeitung erahnen, jedoch reicht eine statische Beschreibung des Netzwerks nur teilweise aus, um ein Verständnis seiner Verarbeitungsweise abzuleiten. Viele der informationstheoretisch interessantesten Vorgänge sind zeitveränderlicher Natur, wobei im komplexen dynamischen Zusammenspiel von chemischen und elektrischen Größen des Netzwerks und seiner Einzelelemente Übertragungseigenschaften geändert werden, Verbindungen zu- oder abgeschaltet werden, oder z.B. Wachstumsvorgänge veränderte Topologien entstehen lassen [Koch99, Shepherd04]. Signale werden verschiedensten Transformationen unterworfen, wie z.B. Hoch- und Tiefpässen, Kompression und Expansion des Dynamikbereichs, Korrelationsberechnung, Summation, Integration, Differentiation, Skalierung, quasi-digitaler Interaktion (AND/NOT), etc. [Gerstner02, Blum72, Ohzawa82, Yu05, Shepherd04 (Tabelle 1.2)].
Im folgenden werden deshalb zuerst die physischen Grundbestandteile von biologischen neuronalen Netzen und ihre Verschaltung beschrieben. Danach wird auf die in diesem (biologischen) Substrat ablaufenden statischen (z.B. topologiebasierten) und dynamischen Verarbeitungsvorgänge eingegangen. Im weiteren wird ein kurzer Überblick über Simulations- und Hardwaremodelle gegeben, die versuchen, verschiedene Aspekte neuronalen Verhaltens nachzubilden. Den Abschluss bildet eine Motivation zum Einsatz neuronaler Elemente, Prinzipien und Schaltungen zur Lösung technischer Aufgabenstellungen.

# II.1 Baublöcke

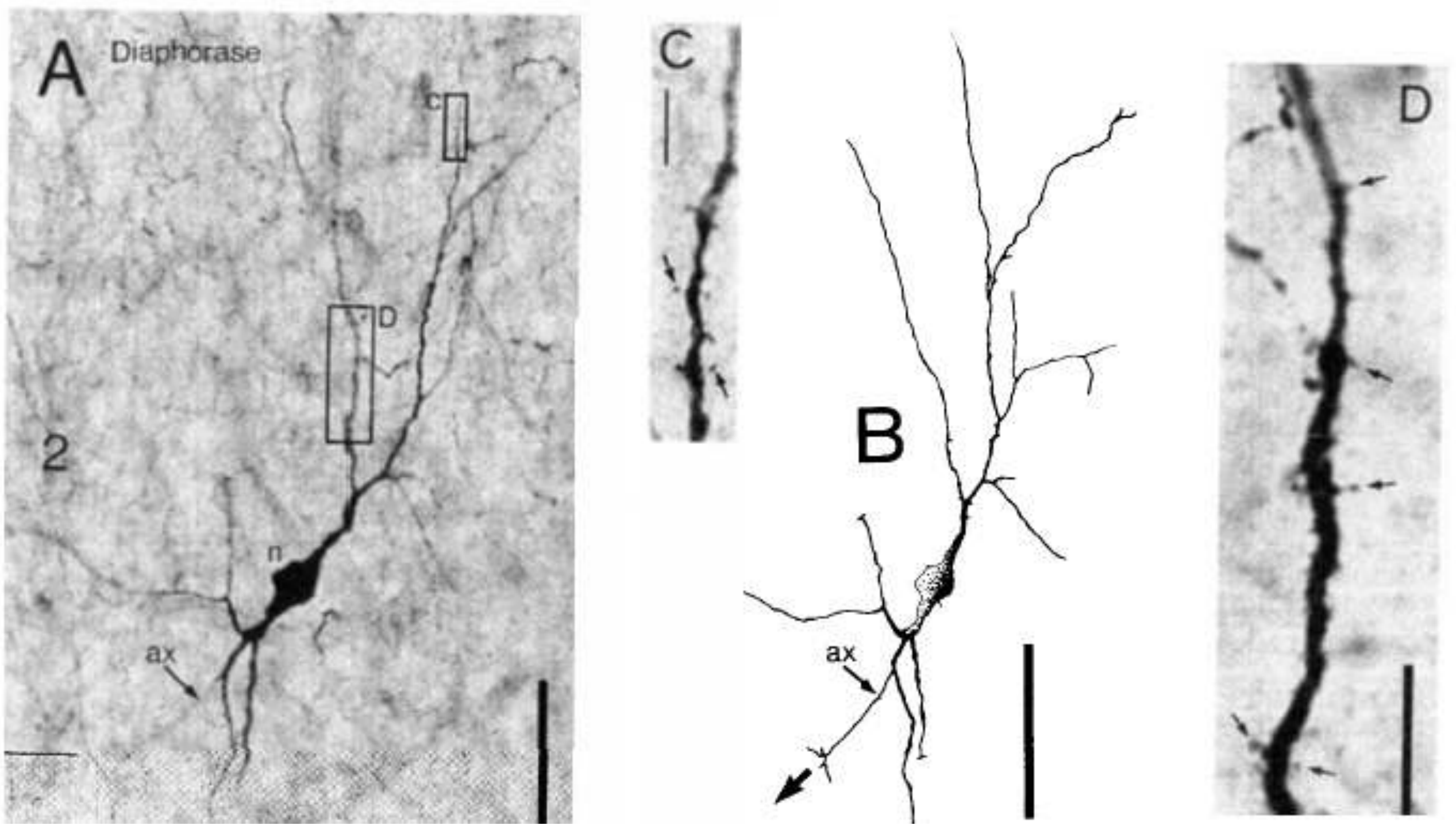

**Abbildung II.2.: Inhibitorisches bipolares Interneuron im mittleren präfrontalen Kortex des Primaten [Gabbott96], A Neuron in Mikroskopaufnahme (Axon und Neuron gekennzeichnet), B schematische Darstellung, C und D Dendritenabschnitte mit markierten dendritischen Dornen**

Neuronen haben im Vergleich zu anderen Zellformen keinen glatten Zellkörper, sondern bilden Ausläufer, so genannte Axone und Dendriten. In Abbildung II.2 ist ein inhibitorisches Zwischenneuron im visuellen Kortex des Primaten wiedergegeben, bei dem dieser Aufbau gut ersichtlich ist. Axone und Dendriten bilden an ihren Enden Ausstülpungen, die so genannten Synapsen, diese stellen das Kommunikationsrückgrat von Nerven dar, d.h. mit ihnen werden Verbindungen zu anderen Nervenzellen oder z.B. bei motorischen Neuronen zu Muskelzellen ausgebildet [Shepherd04 (Kapitel 3)], und in ihnen wird klassisch der Hauptanteil der Verarbeitungsfunktion angesiedelt. Dendriten sind sehr stark verzweigt, sie dienen dazu, einkommende Signale zu sammeln und zum Soma, d.h. der Membran des Zellkörpers weiterzuleiten. Selektiv werden dann diese Signale wieder über das Axon verteilt, welches in der Regel größere Distanzen überbrückt, jedoch weniger stark verzweigt ist als die Dendriten [Shepherd04 (Kapitel 1)]. Entlang des Signalpfades aus Dendriten, Soma und Axon wird ein stereotypischer Spannungsimpuls übertragen, das Aktionspotential (AP). Dieses stellt nach gegenwärtigem Stand der Wissenschaft die hauptsächliche Form der Signalübertragung sowohl innerhalb eines einzelnen Neurons als auch in neuronalen Netzen dar.

### II.1.1 Zentral: Zellkörper und Membran

Die Membran des Neuronenkörpers repräsentiert durch ihre Isolierung gegenüber der umgebenden Intrazellulärflüssigkeit elektrisch gesehen eine Kapazität. In die Membran eingebettet befinden sich mehrere so genannte Ionenkanäle, die ständig für einen aktiven, gesteuerten Transport bestimmter Ionen sorgen, abhängig vom Ionentyp entweder ins Zellinnere oder in der Gegenrichtung. Im Wechselspiel dazu ist die Membran durch den Diffusionsdruck (d.h. unterschiedliche Ionenkonzentrationen außerhalb und innerhalb der Membran) passiv durchlässig für dieselben Ionen. Die Zellmembran ist nicht für alle Ionen gleichermaßen durchlässig, für die verschiedenen Ionenarten liegen damit auch unterschiedliche Konzentrationen im Zellinneren vor. Die Membran ist für $K^+$-Ionen stärker durchlässig als für $Na^+$-Ionen, für $Cl^-$ Ionen dagegen fast gar nicht durchlässig. Durch aktiven Transport werden entgegen dem Konzentrationsgefälle laufend $Na^+$ aus der Zelle und $K^+$ in die Zelle befördert [Kandel95 (Kapitel 8)]. Auf diese Weise stellt sich als dynamisches Gleichgewicht ein Ruhepotential des Cytoplasma gegenüber der Umgebung ein, das an Neuronen relativ einheitlich zu ca. -70 mV gemessen werden kann. Für jeden Ionentyp lässt sich über die Ionenkonzentration im Intrazellulärraum $N_I$ und im Cytoplasma $N_C$ sowie der Ionenladung $q$ eine spezifische Potentialdifferenz, die sogenannte Nernstspannung, definieren (mit der Boltzmannkonstanten $k$ und der Temperatur $T$):

$$U_N = \frac{kT}{q} \ln \frac{N_I}{N_C} \qquad \textbf{(II.1)}$$

Von Hodgkin und Huxley wurden in einer einflussreichen Arbeit [Hodgkin52, Koch99 (Kapitel 6)] anhand von Messungen an einem motorischen Axon des Tintenfischs die qualitativen und quantitativen Vorgänge dieses Ionentransports charakterisiert und folgendes elektrische Verhaltensmodell aufgestellt:

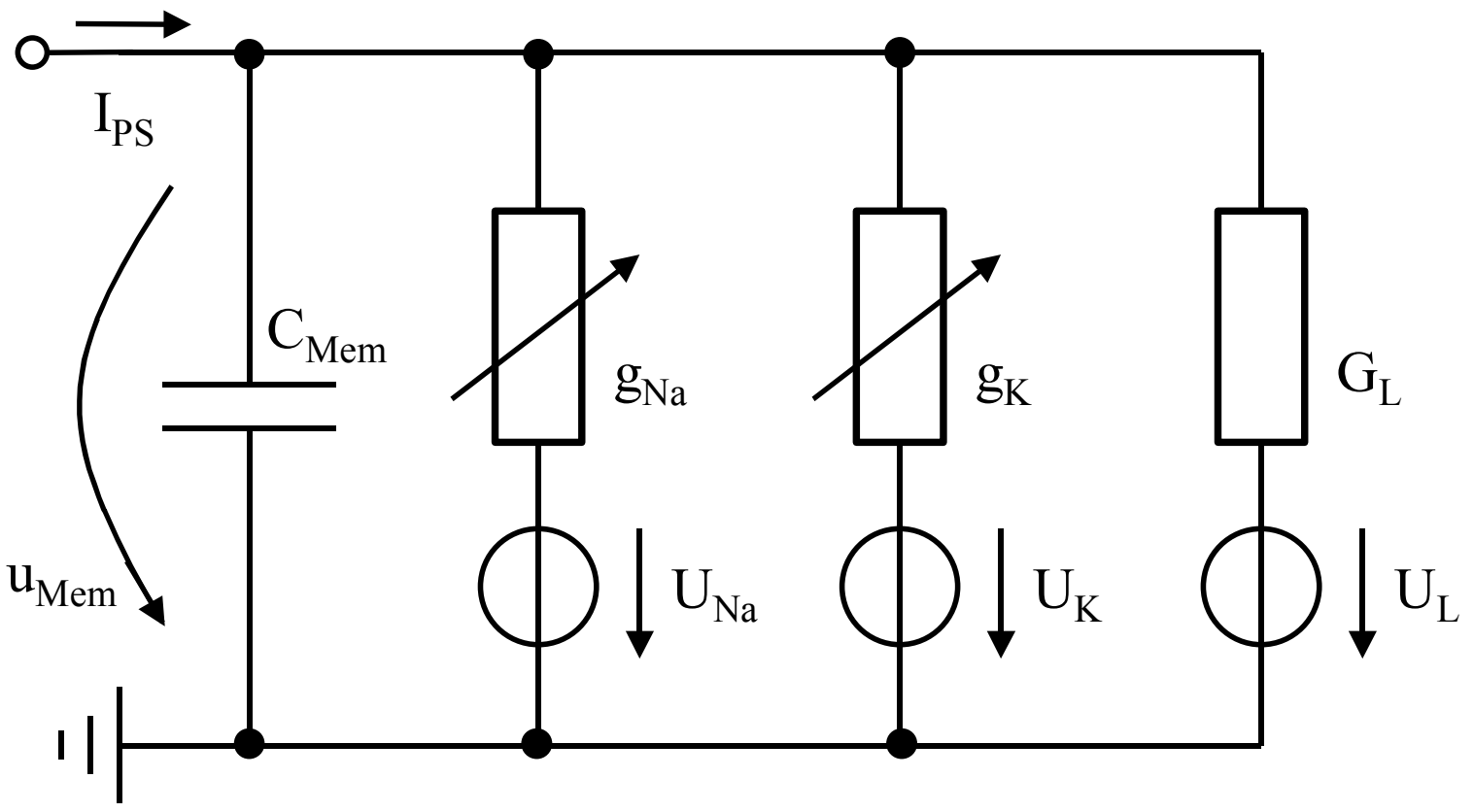

**Abbildung II.3.: Schaltung des Hodgkin-Huxley-Modells**

Hierbei stellt $C_{Mem}$ die Membrankapazität dar und $g_{Na}$ und $g_K$ die Leitwerte der betrachteten gesteuerten Ionenkanäle mit ihren zugehörigen Nernstspannungen. Der letzte Zweig beinhaltet die anderen Ionenkanäle (z.B. $Mg^+$, $Cl^-$, etc.) als passiven Kanal zusammengefasst.
Für die im HH-Modell verwendeten Ionen ergeben sich die Nernstspannungen und maximale Leitfähigkeiten der Ionenkanäle zu[3]:

| Ion | $U_N$ in [mV] | $\hat{G}$ in [mS*cm$^{-2}$] |
|---|---|---|
| $Na^+$ | 50 | 120 |
| $K^+$ | -77 | 36 |
| Rest | -75,6 | 0,36 |

**Tabelle II-1.: Nernstspannungen und maximale Leitfähigkeiten der Ionenkanäle im Hodgkin-Huxley-Modell**

Die Differentialgleichung des zeitlichen Verlaufs der Membranspannung lässt sich wie folgt aus dem elektrischen Verhaltensmodell ableiten:

$$C_{Mem}\dot{u}_{Mem} = I_{PS} + \hat{G}_{Na} m^3 h (u_{Mem} - U_{Na}) + \hat{G}_K n^4 (u_{Mem} - U_K) + G_L (u_{Mem} - U_L) \qquad \textbf{(II.2)}$$

Einkommende Aktionspotentiale erzeugen nach ihrer Übermittlung in der Synapse einen postsynaptischen Strom $I_{PS}$, welcher entlang des Dendriten als externe Erregung/Aufladung an das Soma weitergegeben wird. Die beiden mittleren Terme geben den Zusammenhang zwischen der

[3] Die einzelnen Ströme, Leitwerte und Membrankapazitäten im HH-Modell sind flächenbezogene Werte bzw. über den Durchmesser des Axons an die Länge des betrachteten Axonabschnitts gekoppelt. Für die Berechnung z.B. des Spannungsverlaufs im Aktionspotential spielt dies keine Rolle, da die Flächenbezogenheit wegfällt.

Membranspannung $U_{Mem}$ und dem jeweiligen Zustand der Ionenkanäle wieder ($m^3h$ bzw. $n^3$), in Abhängigkeit vom maximalen Leitwert des Kanals und der Spannungsdifferenz zwischen der Membranspannung und der Nernstspannung des Ions[4]. Der letzte Term fasst die nicht einzeln betrachteten Ionenströme als passiven Zweipol zusammen. Hodgkin und Huxley postulierten für den Zustand der Kaliumkanäle ein Aktivierungspartikel *n*, dieses stellt im biophysikalischen Modell die Anzahl der geöffneten Kanäle dar, als einheitenlose Größe zwischen 0 und 1, bezogen auf den maximalen Leitwert bei vollkommen geöffneten Ionenkanälen, von dem in vierter Potenz der Strom durch den Kaliumkanal abhängig ist. In ähnlicher Weise wurde für den Natriumkanal ein Aktivierungspartikel *m* eingeführt, jedoch war hier ein zusätzliches Inaktivierungspartikel *h* nötig, um die gemessene Kinetik des zeitlichen Verlaufs des Membranstroms abzubilden. Die zeitliche Entwicklung dieser Zustände lässt sich in Abhängigkeit ihres Ist-Wertes und sogenannter Übergangsraten α und β wie folgt beschreiben:

$$\begin{aligned} \dot{m} &= \alpha_m(u_{Mem}) * (1-m) - \beta_m(u_{Mem}) * m \\ \dot{n} &= \alpha_n(u_{Mem}) * (1-n) - \beta_n(u_{Mem}) * n \\ \dot{h} &= \alpha_h(u_{Mem}) * (1-h) - \beta_h(u_{Mem}) * h \end{aligned} \qquad \textbf{(II.3)}$$

Hierbei stellen die Übergangsraten die Geschwindigkeit des Übergangs zwischen den beiden Zuständen des jeweiligen Partikels dar (z.B. für *m*):

$$\begin{aligned} m &\xleftarrow{\alpha_m} 1-m \\ m &\xrightarrow{\beta_m} 1-m \end{aligned} \qquad \textbf{(II.4)}$$

Für die Abhängigkeit der Übergangsraten von der Membranspannung ergibt sich ein offsetbehafteter exponentieller Zusammenhang, wobei der Öffnungszustand der Ionenkanäle mit steigendem Membranpotential zunimmt, d.h. für die Aktivierungspartikel *m* und *n* nimmt *α* mit steigender $U_{Mem}$ zu und *β* ab [Koch99, Abschnitt 6.2.1]:

$$\alpha_m(u_{Mem}) = \frac{-4 - \frac{u_{Mem}}{10mV}}{e^{\left(-4-\frac{u_{Mem}}{10mV}\right)} - 1} \qquad \beta_m(u_{Mem}) = 4e^{\left(-3{,}6-\frac{u_{Mem}}{18mV}\right)} \qquad \textbf{(II.5)}$$

$$\alpha_n(u_{Mem}) = \frac{-0{,}55 - \frac{u_{Mem}}{100mV}}{e^{\left(-5{,}5-\frac{u_{Mem}}{10mV}\right)} - 1} \qquad \beta_n(u_{Mem}) = 0{,}125e^{\left(-0{,}8-\frac{u_{Mem}}{80mV}\right)} \qquad \textbf{(II.6)}$$

Für das Inaktivierungspartikel *h* dagegen verhält sich die Spannungsabhängigkeit genau umgekehrt:

$$\alpha_h(u_{Mem}) = 0{,}07e^{\left(-3{,}3-\frac{u_{Mem}}{20mV}\right)} \qquad \beta_h = \frac{1}{e^{\left(-3{,}5-\frac{u_{Mem}}{10mV}\right)} + 1} \qquad \textbf{(II.7)}$$

Wenn ein eingehender postsynaptischer Strom für eine Depolarisation der Membran, d.h. für eine Anhebung der Membranspannung sorgt, werden $Na^+$-Kanäle aktiviert (Gleichung (II.5)), damit steigt die $Na^+$-Leitfähigkeit. Bei geringer Auslenkung vom Ruhepotential klingt diese Depolarisation durch sich öffnende Kaliumkanäle schnell wieder ab. Oberhalb einer bestimmten Schwellspannung ergibt sich jedoch eine positive Rückkopplung zwischen sich verstärkender Depolarisation und der weitergehenden Öffnung der Natriumkanäle. Es kommt zu einem Zusammenbruch des Membranpotentials wobei Werte um +30mV erreicht werden. Die Übergangsraten für den Kaliumkanal und die Inaktivierung des Natriumkanals folgen hingegen einem verzögertem Zeitverlauf (Gleichung (II.6) und (II.7)), d.h. die Konzentration der $K^+$-Ionen steigt erst nach dem Erreichen der vollen Leitfähigkeit der Natriumkanäle an. Sie sorgt dann für eine Absenkung des Membranpotentials auf den Wert seiner Nernstspannung, d.h. ein Absinken der Membranspannung unterhalb des Ruhepotentials, die so genannte Hyperpolarisation. Mit dem darauf folgenden Schließen der Kaliumkanäle stellt sich wieder das Ruhepotential ein.
Dieses Modell besitzt trotz seines relativ hohen Abstraktionsgrades die Fähigkeit, die Bildung eines Aktionspotentials auf der Membran des Neurons in Abhängigkeit des eingehenden Stromes in sehr guter Übereinstimmung mit neurobiologischen Messungen zu modellieren, wie Abbildung II.4 verdeutlicht.

---

[4] Wie im Text und in Gleichung (II.3) angeführt, sind m, n und h zeitveränderliche Größen. In der Notation in Gleichung (II.2) wird dies der Einfachheit halber nicht angeführt, d.h. *n* statt *n*(*t*), etc.

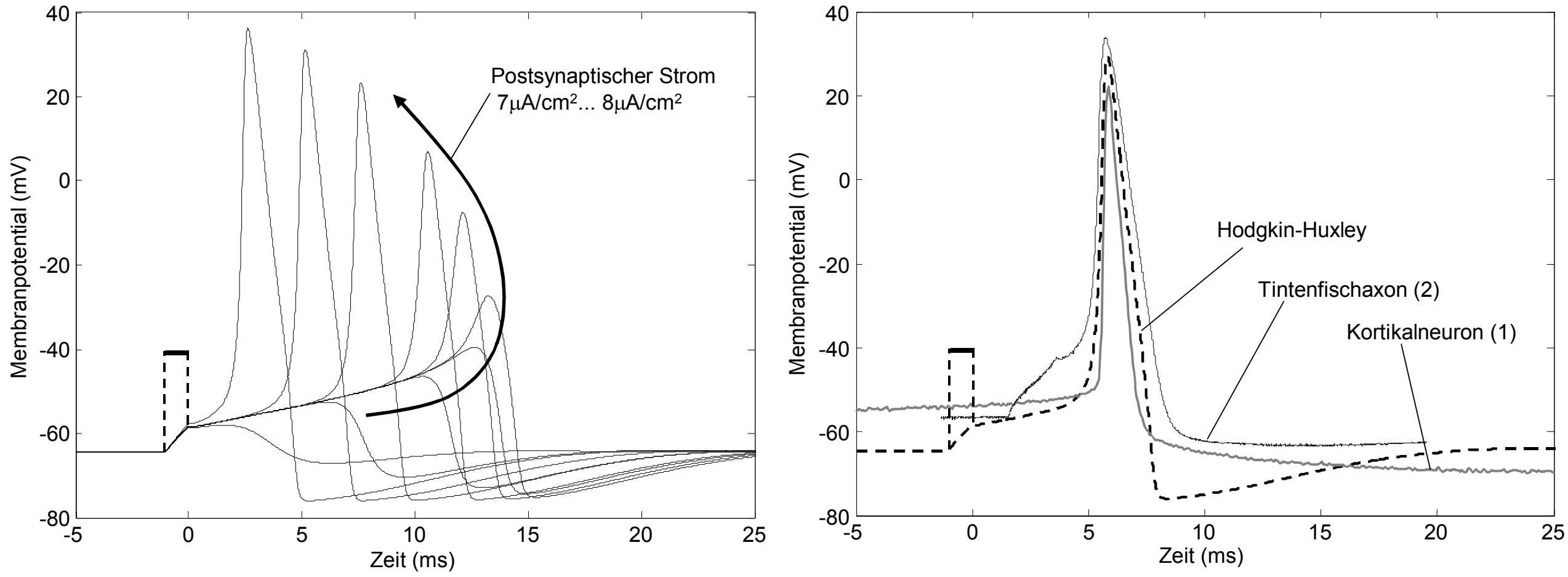

**Abbildung II.4.: Dynamisches Verhalten des Hodgkin-Huxley-Modells für verschiedene postsynaptische Ströme (links) und Vergleich mit Messdaten (rechts) typischer Aktionspotentiale von (1) ‚regular spiking' Neuronen in einer Scheibenpräparation aus dem okzipitalen Kortex eines Meerschweinchens [Piwkowska07] sowie (2) eines in vitro Abschnitts des Tintenfisch-Riesenaxons [Clay07] (beide unter Konstantstromanregung)**

Im linken Teil der obigen Abbildung ist der Zusammenhang zwischen postsynaptischem Strom und Membranspannungsverlauf im HH-Modell dargestellt. Der postsynaptische Strom wird von t=-1ms bis 0 angelegt (Balken und gestrichelte Linien), womit sich in einem Stromdichtebereich von 7µA/cm$^2$ bis 8µA/cm$^2$ die Spannungsverläufe entlang des Pfeils ergeben, von einer schnell abklingenden Auslenkung bis zu einem über 30mV hohen Aktionspotential.
Die Übereinstimmung zwischen am Kortikalneuronen eines Meerschweinchens und am Tintenfisch-Riesenaxon gemessenen Verläufen eines Aktionspotentials und dem HH-Modell ist in der rechten Hälfte von Abbildung II.4 dargestellt. Durch die Hyperpolarisation der Neuronenmembran ergibt sich im HH-Modell eine Zeitspanne von etwa 2-3 ms, in der das Neuron auf weitere eingehende depolarisierende Einflüsse nicht reagiert, die absolute Refraktärzeit. Daran anschließend folgt die relative Refraktärzeit, in der depolarisierende Ströme nur abgeschwächt zu einer neuerlichen Anhebung des Membranpotentials beitragen. Da die Kaliumkanäle im Säugetierkortex wesentlich längere Zeitkonstanten aufweisen [Koch99], finden sowohl die anfängliche Aufladung als auch die Rückkehr aus der Hyperpolarisation sehr viel langsamer statt als im Tintenfischaxon. Dies ist in der linken Hälfte der folgenden Abbildung verdeutlicht, mit ihren verschiedenen Zeitachsen für gleichlaufendes Subschwellwert-Membranverhalten:

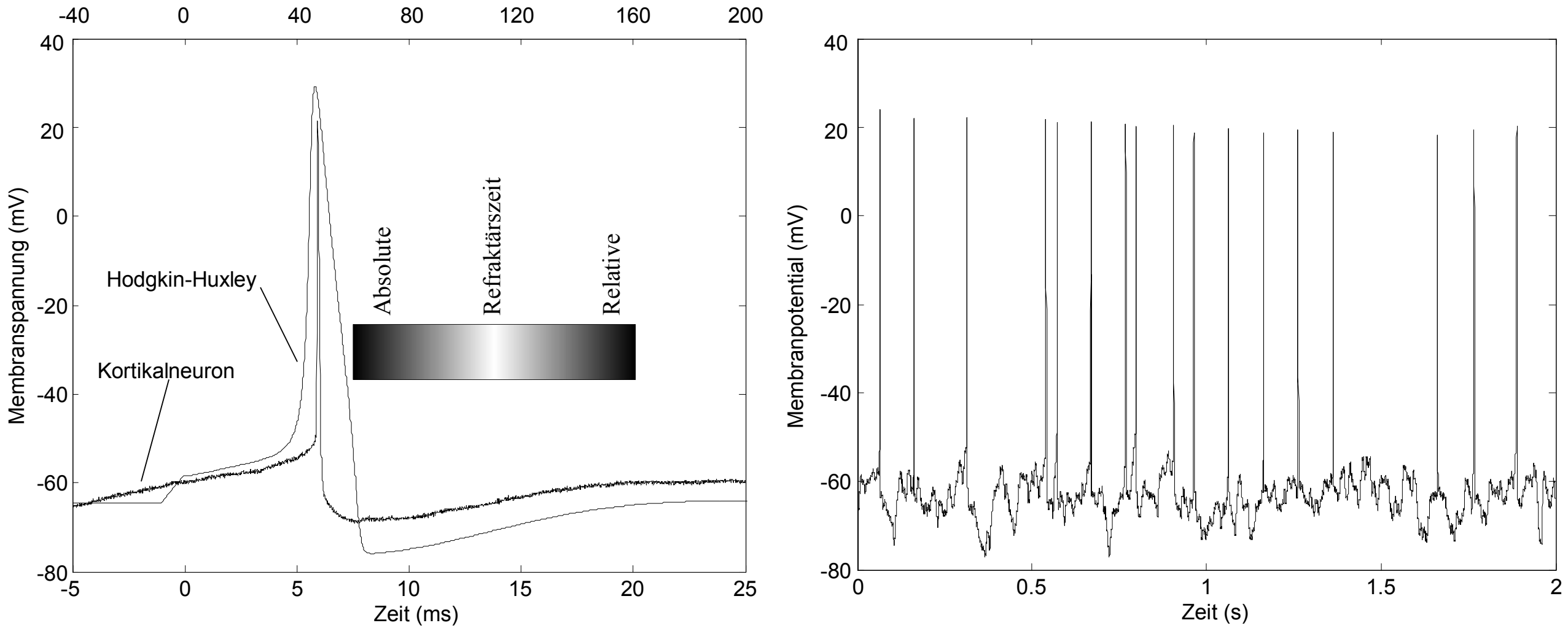

**Abbildung II.5, links: HH- und biologisches Aktionspotential [Piwkowska07] wie in Abbildung II.4 (rechts), ersteres bezogen auf die untere Zeitachse, letzteres bezogen auf die obere Zeitachse. Rechts: Pulsfolge von Neuron aus Messdaten zu linker Bildhälfte, jedoch Stimulus stochastische Aktionspotentiale statt Konstantstrom**

In der rechten Hälfte von Abbildung II.5 ist der typische Verlauf des Membranpotentials wiedergegeben, wenn es mit aus der Biologie abgeleiteten stochastisch verteilten Aktionspotentialen statt des Konstantstroms angeregt wird[5].

## II.1.2 Interaktion und Impulspropagierung: Synapsen, Dendriten und Axone

Am Anfang der Signalkette im Neuron liegt wie oben ausgeführt die Synapse, in der die Initialisierung eines postsynaptischen Aktionspotential erfolgt. Die dendritische Seite der Synapse empfängt hierbei Neurotransmitter, die durch ein eingehendes präsynaptisches Aktionspotential auf der axonalen Seite der Synapse ausgeschüttet werden und durch den synaptischen Spalt wandern (siehe Abbildung II.6). Der Empfang von Neurotransmittern verändert die Durchlässigkeit der Rezeptorstellen für Ionen und löst damit einen postsynaptischen Strom aus, der je nach Art des in der Synapse dominanten Transmitterstoffes eine erregende (exzitatorische, depolarisierende) oder eine hemmende (inhibitorische, polarisierende) Wirkung hat[6]. Inhibitorische Ströme können zu einer Hyperpolarisation des Dendriten führen, während ein exzitatorischer Strom bei genügender Amplitude in dem auf die Synapse folgenden dendritischen Abschnitt nach den oben angeführten HH-Formalismen ein Aktionspotential auslöst.
Eine wichtige Ausnahme dieser AP-basierten Informationsweitergabe stellt die Retina dar, deren Informationsverarbeitung auf direktem Stromaustausch basiert [Wohrer06], wie eingangs erwähnt stellen dort erst die Ganglienzellen einen Hodgkin-Huxley-mäßigen Integrations-/AP-Erzeugungs-mechanismus bereit, mit dem die Retinaausgangssignale als Aktionspotentiale codiert über den Sehnerv zum Gehirn geschickt werden.

[5]im folgenden werden die Begriffe ‚Puls' und ‚Aktionspotential' in austauschbarer Weise verwendet. Um welche Art von Puls es sich handelt (biologischer Messwert, simulierter analoger Pulsverlauf, mathematische Modellierung als Diracstoß oder eine zweiwertige, digitale Repräsentation), ergibt sich aus dem jeweiligen Kontext.
[6] Excitatory bzw. Inhibitory Postsynaptic Current (EPSC/IPSC)

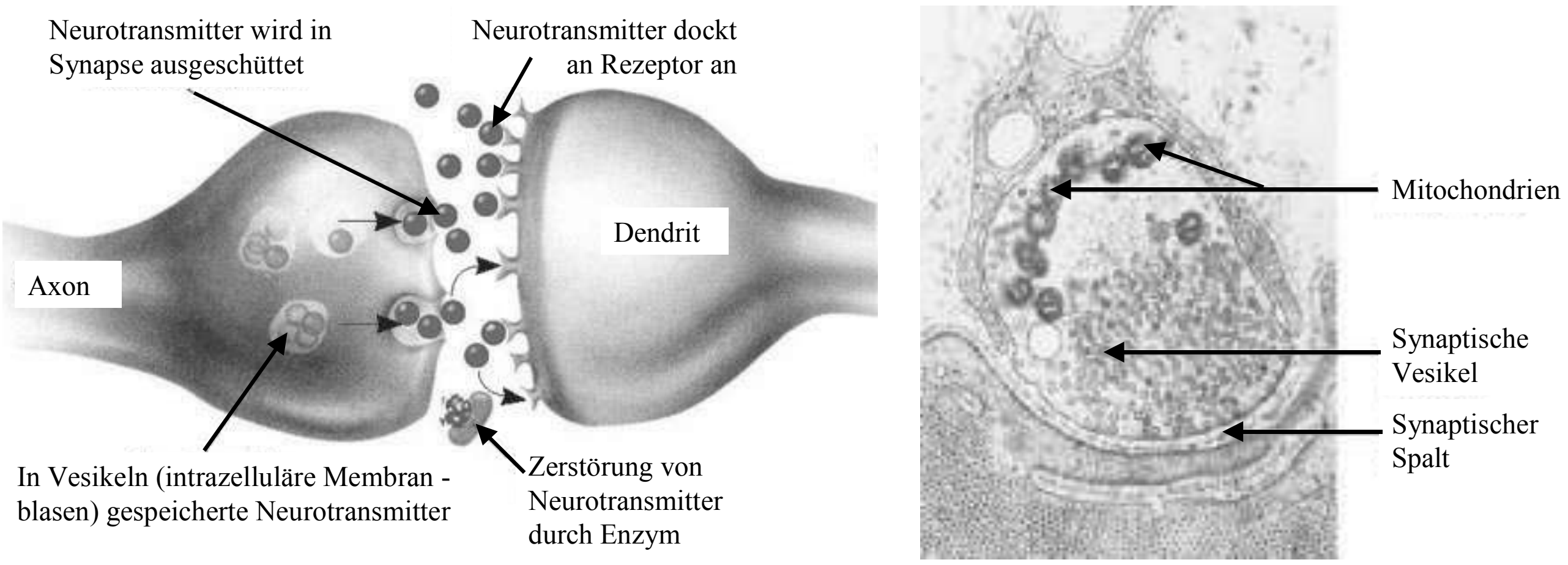

**Abbildung II.6.: Prinzipdarstellung und mikroskopische Aufnahme einer Synapse [Johnson07]**

Die Wirkung eines präsynaptischen Aktionspotentials auf den postsynaptischen Strom wird im allgemeinen über die sogenannte Quantenausschüttung oder Übertragungseffizienz $R$ modelliert:

$$R = n * p * q \qquad \textbf{(II.8)}$$

Diese Übertragungseffizienz wird auch oft vereinfacht als Gewichtswert $W$ angesehen, mit dem ein Aktionspotential beim Passieren der Synapse multipliziert wird. Die drei Größen, die in die Übertragungseffizienz eingehen, sind zum Einen die Anzahl der synaptischen Übertragungsstellen $n$, d.h. die Menge der Bereiche, in denen Neurotransmitter in den postsynaptischen Dendriten aufgenommen werden können. Einfluss hat auch die Wahrscheinlichkeit $p$, mit der ein präsynaptisches AP einen postsynaptischen Strom auslöst, sowie die Menge (‚Quanten') der ausgeschütteten Neurotransmitter q, d.h. die Größe dieses Stroms [Koch99 (Kapitel 13)]. Alle drei synaptischen Größen unterliegen aktivitätsabhängigen Veränderungen, d.h. Lernvorgängen auf unterschiedlichen Zeitskalen. Die schnellste Modifikation findet in der Ausschüttungswahrscheinlichkeit p statt, im Zeitbereich von 10-100 ms [Koch99 (Abschnitt 13.2.2)]. Im Sekunden- bis Minutenbereich liegen Modifikationen der Ausschüttungsmenge q [Markram98], zeitlich darüber finden Wachstumsvorgänge statt, welche die Anzahl der Synapsen zwischen Neuronen und damit n verändern [Song01, Yao05, Shepherd04 (Tabelle 1.2)]. Beim Sprachgebrauch wird im allgemeinen zwischen Adaption für Kurzzeitvorgänge und Plastizität für Langzeitlernen unterschieden. Da diese Begriffe aber nicht genau definiert sind und auch widersprüchlich verwendet werden, werden im Rest dieser Arbeit die Terme Lernen, Adaption und Plastizität austauschbar verwendet.

Nachdem ein Aktionspotential im postsynaptischen Dendriten erzeugt wurde, pflanzt sich dieses entlang des Dendriten, Zellkörper und Axon bis zur nächsten Synapse fort. Die Weitergabe des Aktionspotentials lässt sich ebenfalls über das HH-Modell erklären, da Dendrit und Axon als eine Reihenschaltung von gleichartigen Abschnitten modellierbar sind, bei denen jeder einzelne die o.a. Ionenkanäle und entsprechendes elektrisches und biophysikalisches Verhalten aufweist (Abbildung II.7).

Physiologisch gesehen geschieht hierbei räumlich entlang des Axons/Dendriten ein ähnlicher Vorgang wie der o.a. zeitliche Vorgang im HH-Modell, d.h. das ankommende Nervensignal bringt ein elektrisches Feld mit sich, welches die Natriumkanäle öffnet. Im weiteren Verlauf dieser Welle strömen durch die offenen Kanäle weitere Natriumionen in das Innere der dendritischen/axonalen Abschnitte und erhöhen wie oben angeführt das Membranpotential. Auf dem Höhepunkt des Aktionspotential aktivieren dann die Kaliumkanäle und das Aktionspotential klingt an dieser Stelle des Axons/Dendriten wieder ab.

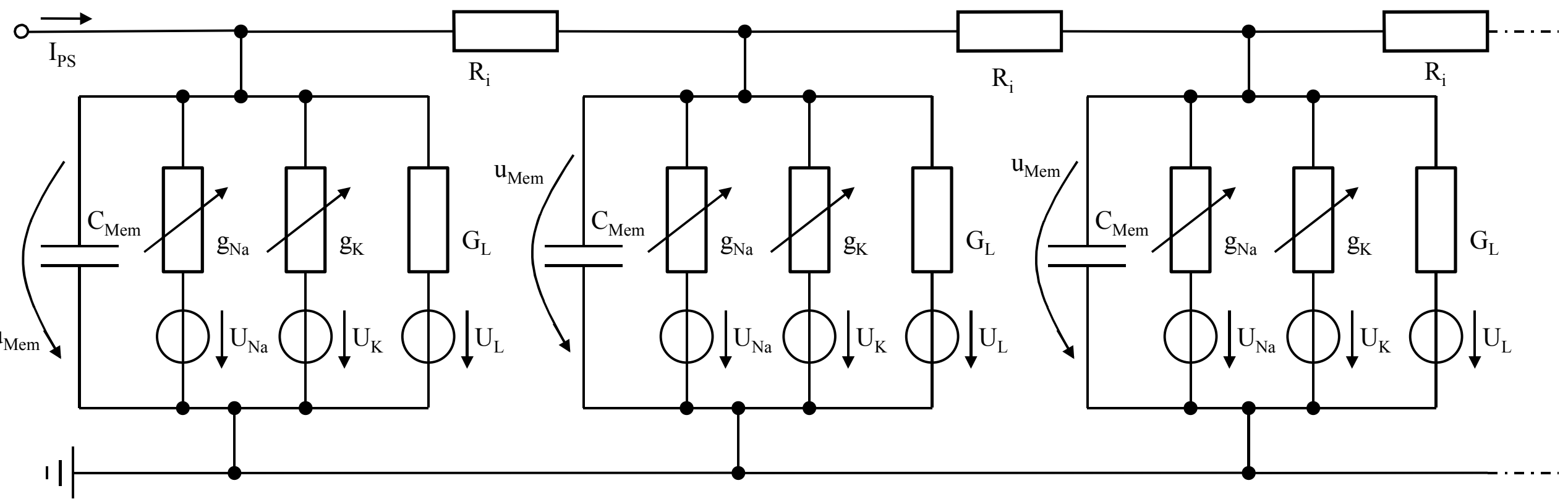

**Abbildung II.7.: Elektrisches Modell des dendritischen/axonalen Baums aus Einzelabschnitten**[7]

Für einen Leiter mit einem spezifischem elektrischen längenbezogenen Widerstand $R_i$ und einem Durchmesser von $d$ lässt sich der Spannungs- und Stromverlauf entlang des Leiters durch die folgende partielle DGL wiedergeben [Hodgkin52, Koch99 (Abschnitt 6.5)]:

$$I_m = \frac{d}{4R_i}\frac{\partial^2 u_{Mem}}{\partial x^2} \qquad \textbf{(II.9)}$$

Wenn man $I_m$ als postsynaptischen Strom in die Stromgleichung der Ionenkanäle (II.2) einsetzt, erhält man einen Ausdruck für die Membranspannung in Abhängigkeit vom Ort entlang des Dendriten/Axons $x$ und der Zeit $t$:

$$\frac{d}{4R_i}\frac{\partial^2 u_{Mem}}{\partial x^2} = C_m\frac{\partial u_{Mem}}{\partial t} + \hat{G}_{Na}m^3h(u_{Mem}-U_{Na}) + \hat{G}_K n^3(u_{Mem}-U_K) + G_L(u_{Mem}-U_L) \qquad \textbf{(II.10)}$$

Diese DGL-Form, mit ihrer zweiten partiellen Ableitung der Spannung nach dem Ort und ersten partiellen Ableitung nach der Zeit wird als Diffusionsgleichung bezeichnet. Für bestimmte Werte der Parameter existiert für diese DGL eine periodische Lösung [Bronstein87 (Abschnitt 3.3.2.3)]. Da im Dendriten/Axon von Hodgkin und Huxley eine wellenförmigen Ausbreitung des Aktionspotential beobachtet wurde, postulierten sie für die obige Gleichung folgende partikuläre Lösung [Hodgkin52]:

$$u_{Mem}(x,t) = u_{Mem}(x - vt) \qquad \textbf{(II.11)}$$

Bei zweimaliger partieller Ableitung nach Ort und Zeit ergibt sich für Gleichung (II.11) nach der Kettenregel entsprechend:

$$\frac{\partial^2 u_{Mem}}{\partial x^2} = \frac{1}{v^2}\frac{\partial^2 u_{Mem}}{\partial t^2} \qquad \textbf{(II.12)}$$

Die rechte Seite der obigen Gleichung lässt sich für die zweite partielle Ableitung nach dem Ort in der linken Hälfte von Gleichung (II.10) einsetzen, wodurch die folgende reguläre DGL zweiter Ordnung entsteht:

$$\frac{1}{K}\frac{d^2 u_{Mem}}{dt^2} = \frac{du_{Mem}}{dt} + \frac{1}{C_m}\left[\hat{G}_{Na}m^3h(u_{Mem}-U_{Na}) + \hat{G}_K n^3(u_{Mem}-U_K) + G_L(u_{Mem}-U_L)\right] \qquad \textbf{(II.13)}$$

$$mit \quad K = \frac{4R_i v^2 C_m}{d}$$

[7] Die diskrete Darstellung des Axons als einzelne Untereinheiten mit jeweils eigener Teilschaltung stellt nur ein Denkmodell dar, da die Ionenkanäle und der elektrische Widerstand entlang des Axons sehr fein unterteilt sind und deshalb als kontinuierlich angesehen werden können. Für eine Unterteilung des Axons in Abschnitte mit fester Länge ließen sich die absoluten Werte der diskreten Bauelemente aus den flächenbezogenen (bzw. bei $R_i$ längenbezogenen) Angaben aus diesem und letztem Abschnitt ermitteln.

Diese DGL lässt sich für $u_{Mem}(t)$ an einem bestimmten Ort des Axons unter Verwendung einer festen Fortpflanzungsgeschwindigkeit $v$ lösen. Hodgkin und Huxley fanden in einem iterativen Prozess eine Wellenlösung der obigen Gleichung [Hodgkin52]:

> "This is an ordinary differential equation and can be solved numerically, but the procedure is still complicated by the fact that $u_{Mem}(t)$ is not known in advance. It is necessary to guess a value of $v$, insert it in equation (II.13) and carry out the numerical solution starting from the resting state at the foot of the action potential. It is then found that $u_{Mem}(t)$ goes off towards either $+\infty$ or $-\infty$, according as the guessed $v$ was too small or too large. A new value of $v$ is then chosen and the procedure repeated, and so on. The correct value brings $u_{Mem}(t)$ back to zero (the resting condition) when the action potential is over."[8]

Über diese Methode wurde eine Ausbreitungsgeschwindigkeit von $v$=18,8 m/s ermittelt, bei einem spezifischen längenbezogenen Widerstand des Tintenfischaxons von $R_i$=35,4 Ωcm und einem Durchmesser des Axons von $d$=0,476 mm. Dieser Wert liegt sehr nahe am gemessenen Wert von 21,2 m/s [Einevoll03]. Für die axonale/dendritische Ausbreitungsgeschwindigkeit lässt sich grob eine Abhängigkeit der Ausbreitungsgeschwindigkeit von der Wurzel des Axondurchmessers angeben [Koch99 (Abschnitt 6.5.1)].
Elektrisch gesehen sind derartige Axone ineffizient, da bei ihnen eine Geschwindigkeitserhöhung mit einer quadratischen Erhöhung des Verluststroms über die entsprechend vergrößerte Außenwand des Axons einhergeht. Schnell leitende Axone bei Wirbeltieren bilden deshalb eine zusätzliche Isolierung aus, die sogenannte Myelisierung, welche den Ableitwiderstand und die Leitungskapazität zum umgebenden Gewebe verringert und damit die Impulsweiterleitung vereinfacht. Entlang des Axons wird die Myelinhülle durch s.g. Ranviersche Schnürringe unterbrochen [Kandel95]. In den Schnürringen findet ein Natriumionenaustausch statt, wodurch sich ein elektrisches Feld bildet. Bei nicht myelisierten Axonen entstehen kleine Stromschleifen, da die Isolierung fehlt und der Spannungsreiz durch die Ionenpumpen und spannungsgesteuerte Kanäle weitergeleitet werden muß. Bei myelisierten (d.h. abschnittsweise isolierten) Axonen entstehen große Stromschleifen, denn der Stromkreis kann erst am nächsten Schnürring geschlossen werden. Elektrisch gesehen findet in den Schnürringen eine Signalaufbereitung statt, die das Aktionspotential rekonstruiert und über die nächste myelisierte Teilstrecke weitersendet [Koch99 (Abschnitt 6.6)].

### II.1.3 Verschaltung: Netzwerkstrukturen

Ein erster Eindruck von der Komplexität der dreidimensionalen Verschaltung der Neuronen im Gehirn wird von Abbildung II.1 gegeben. Die Elemente dieser Verschaltung wurden im letzten Abschnitt eingeführt, d.h. Dendriten und Axone bilden ein vielschichtiges Netzwerk aus, an dessen Verbindungsstellen (Synapsen) Aktionspotentiale ausgetauscht werden. Wie in Topologien von elektrischen Schaltungen besteht ein klarer Zusammenhang zwischen Aufbau und Funktionalität dieser Netzwerke [Blinder05], wobei das Reengineering von Hirnstrukturen aufgrund der hohen Packungsdichte und dreidimensionalen Verbindungsstruktur eines der großen Probleme der Neuro-Biologie und -Informatik darstellt [Chklovskii04].

[8] Zitat entnommen aus [Hodgkin52], Referenzen auf Gleichungsnummern und Variablen wurden entsprechend angepasst.

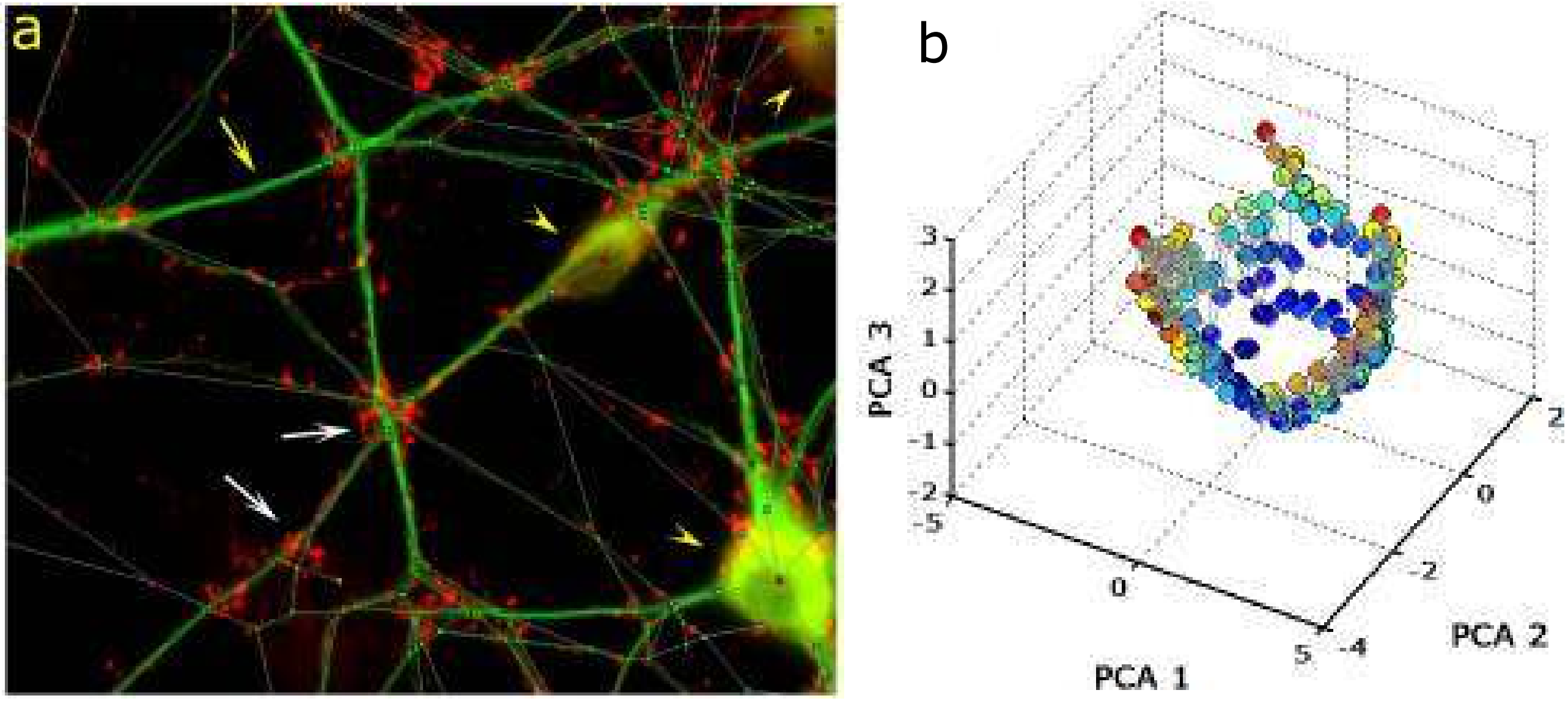

**Abbildung II.8.: Spontaner Netzaufbau von Neuronen in vitro und informationstechnische Erfassung des Netzwerks (aus [Blinder05], gelbe Pfeile Dendriten, gelbe Pfeilköpfe Zellkörper, weisse Pfeile Synapsen)**

Eine Übersicht der verschiedenen Strukturen von Axonen und Synapsen findet sich in [Koch99 (Abbildung 3.1)]. Im allgemeinen wird das Axon als lange unverzweigte Signalleitung angesehen, die sich erst im Zielgebiet auffächert, wie aus Abbildung II.2 ansatzweise ersichtlich. Der Dendrit ist deutlich stärker verzweigt, er sammelt Signale aus verschiedenen Arealen in der Nachbarschaft des Neurons zur weiteren Verarbeitung [Häusser03]. Je nach Hirnbereich ergeben sich stark unterschiedliche Ein- und Ausfächerungen des Neurons, so haben etwa Neuronen in motorischen Pfaden oder in bestimmten Teilen des auditiven Kortex nur wenige, stark gerichtete Verbindungen mit anderen Neuronen [Shepherd04 (Kapitel 3&4)], während z.B. ein Neuron im V1-Bereich des visuellen Kortex zwischen 2000 und 10.000 Synapsen besitzt und damit Informationen von 1000 bis 10.000 vorgeschalteten Neuronen empfängt [Binzegger04].

Die Struktur von Dendriten und Axon ist stereotypisch in der jeweiligen Neuronenart verankert, wobei individuelle Ausprägungen von Verbindungen durch wachstumslenkende Lernvorgänge stattfinden [Song01, Warren97]. Strukturierte Netzwerke finden sich im Kortex und anderen Hirnarealen auf allen Granularitätsebenen. Auf der untersten Ebene neuronaler Organisation stehen sogenannte dendritische Mikroschaltungen, bei denen 2-40 Neuronen eine quasi-digitale Grundfunktionalität für zeitlich korrelierte Pulse aufbauen [Blum72, Shepherd04 (Kapitel 1)], in dem z.B. eine entlang des Dendriten sitzende Synapse durch ein eintreffendes Aktionspotential die Ionenkanäle erschöpft und damit einen entlang des Dendriten laufenden Puls blockiert. Zusammenschaltungen von ca. 100 bis 200 Neuronen z.B. in den stereotypen Minikolumnen des visuellen Kortex stellen die nächste Organisationsstufe dar [Shepherd04 (Kapitel 4)], in der bereits komplexe Wahrnehmungsaufgaben wie etwa Richtungsfilterung ausgeführt werden [Hubel68]. Wiederkehrende Netzwerkstrukturen sind hierbei z.B. ein horizontal geschichteter Aufbau, die vertikale Integration der einzelnen Minikolumnen, eine gerichtete Feedforward-Architektur und horizontale Verschränkung der Neuronen in den jeweiligen Schichten. Auf einer Ebene von ca. 10000 Neuronen sind die Minikolumnen parallel zu Makrokolumnen zusammengefasst, die etwa einen bestimmten Ausschnitt des Sehfelds mit Gabormasken verschiedenster Orientierung abdecken [Hubel68, Riesenhuber99]. Eine ähnliche Dimension hat der in Abschnitt I.3.2 erwähnte seitliche Kniehöcker [Sherman96]. In dieser Größenordnung der Neuronenanzahl gibt es simulative Nachbildungen mit an die Biologie angenäherten strukturierten Netzen, mit denen der Umfang neuronaler Verarbeitung untersucht werden soll [Häusler07] oder unter Zuhilfenahme von Lernvorgängen einzelne Verarbeitungsfunktionen nachgebildet werden [Vogels05].

Die oberste Stufe der Analyse und Modellbildung findet auf einer ähnlichen Abstraktionsebene statt wie die Schilderung des Pfads der visuellen Informationsverarbeitung im Säugetier im Abschnitt I.3. Es werden komplette Hirnareale betrachtet, bei denen globale statistische

Verbindungsdichten [Binzegger04] zum Aufbau von funktionellen Repräsentationen der makroskopischen Signalpfade verwendet werden [Riesenhuber99, Swiercinsky01].

## II.2 Informationsrepräsentation, -propagierung und -verarbeitung

In diesem Kapitel werden die in der Überschrift angeführten drei unterschiedlichen Komponenten des neuronalen Verarbeitungsprozesses gemeinsam behandelt, da sich diese im jeweiligen Kontext gegenseitig bedingen, z.B. kann eine bestimmte Art der Informationspropagierung bereits eine Verarbeitungsfunktion darstellen, indem Information nur selektiv weitergegeben wird [Mukherjee95], oder eine Repräsentation der Information kann so gewählt sein, dass durch ihre Weitergabe inhärent z.B. eine komplexe Bearbeitung des Frequenzspektrums des Signals stattfindet [Gerstner99, Marienborg02, Spiridon99].
Eines der Hauptprobleme ist hierbei, aus den zugrunde liegenden biologischen Messungen die relevanten Mechanismen herauszufiltern, d.h. welche Teile sind für die jeweilige Verarbeitung (z.B. Bildanalyse) notwendig, und was findet nur aufgrund der Rahmenbedingungen der zugrunde liegenden biologischen Matrix in dieser Weise statt [Häusser03, Kass05, VanRullen05, Stiber05]. Eine der wichtigsten Fragen hierbei ist die Wahl des neuronalen Codes, welcher der Verarbeitung zugrunde liegt, da dieser wie oben angeführt starken Einfluss auf die Analyse der Verarbeitung hat. Biologisch realistische Codes sollten ein oder mehrere der folgenden Eigenschaften haben:

- Sie sollen eine schnelle, evtl. parallele Informationsverarbeitung ermöglichen. Dies ergibt sich aus der biologisch gemessenen Verarbeitungsgeschwindigkeit, bei der komplexe Aufgaben wie z.B. Bilderkennung in Zeiträumen stattfinden, in denen einzelne Neuronen nur wenige Aktionspotentiale abgegeben haben können [Guyonneau05, VanRullen01].
- Eine ‚intrinsische' Decodierung soll möglich sein, da empfangende Neuronen nur eingehende Aktionspotentiale sehen, sie jedoch die Codierung der vorhergehenden Stufe nicht kennen, und dennoch die Information zurückgewinnen müssen [Koch99]
- Einhergehend mit dem letztem Punkt ist eine ‚intrinsische' Plastizität, d.h. Lernvorgänge innerhalb eines Neurons, die auf diesem Code basieren, dürfen nur auf Zustandsvariablen zurückgreifen, die dem Neuron (oder bei Neurohormonen zumindest der lokalen Population [Izhikevich07]) vorliegen, eine (externe) Lernsteuerung findet in der Regel nicht statt. [Hopfield04]
- Idealerweise sollte diese Plastizität auf biologisch realistischen chemischen und elektrischen Mechanismen beruhen, die so im Neuron bereits gefunden wurden oder zumindest in Bezug auf die Struktur des Neurons sinnvoll erscheinen. [Markram98, Saudargiene04]
- Der Code sollte mit biologischen Messdaten übereinstimmen, beispielsweise hinsichtlich der statistischen Kenngrößen [Kass05, Shadlen98] oder I/O-Relationen von Aktions-potentialen an Neuronen [Aronov03, Steveninck97].
- Im Sinne der Verwendung im Rahmen dieser Arbeit sollte der Code einfach implementierbar sein und im Rahmen von technischen Anwendungen interessante Verarbeitungsmöglichkeiten eröffnen.

Thesen für relevante Codes werden aus biologischen Messdaten gewonnen, bei denen meist die Antwort bestimmter Neuronen auf einen externen Stimulus aufgezeichnet wird. Diese Antwort besteht aus Pulsfolgen, die durch Schwellwertbildung aus Messungen der Membranspannung (vgl. Abbildung II.5 rechts) gewonnen werden. Die folgende Darstellung verdeutlicht dies anhand der Pulsfolgen, die ein wiederholt (64 mal) präsentiertes bewegtes Gittermuster an einem „simple neuron" im V1 hervorruft:

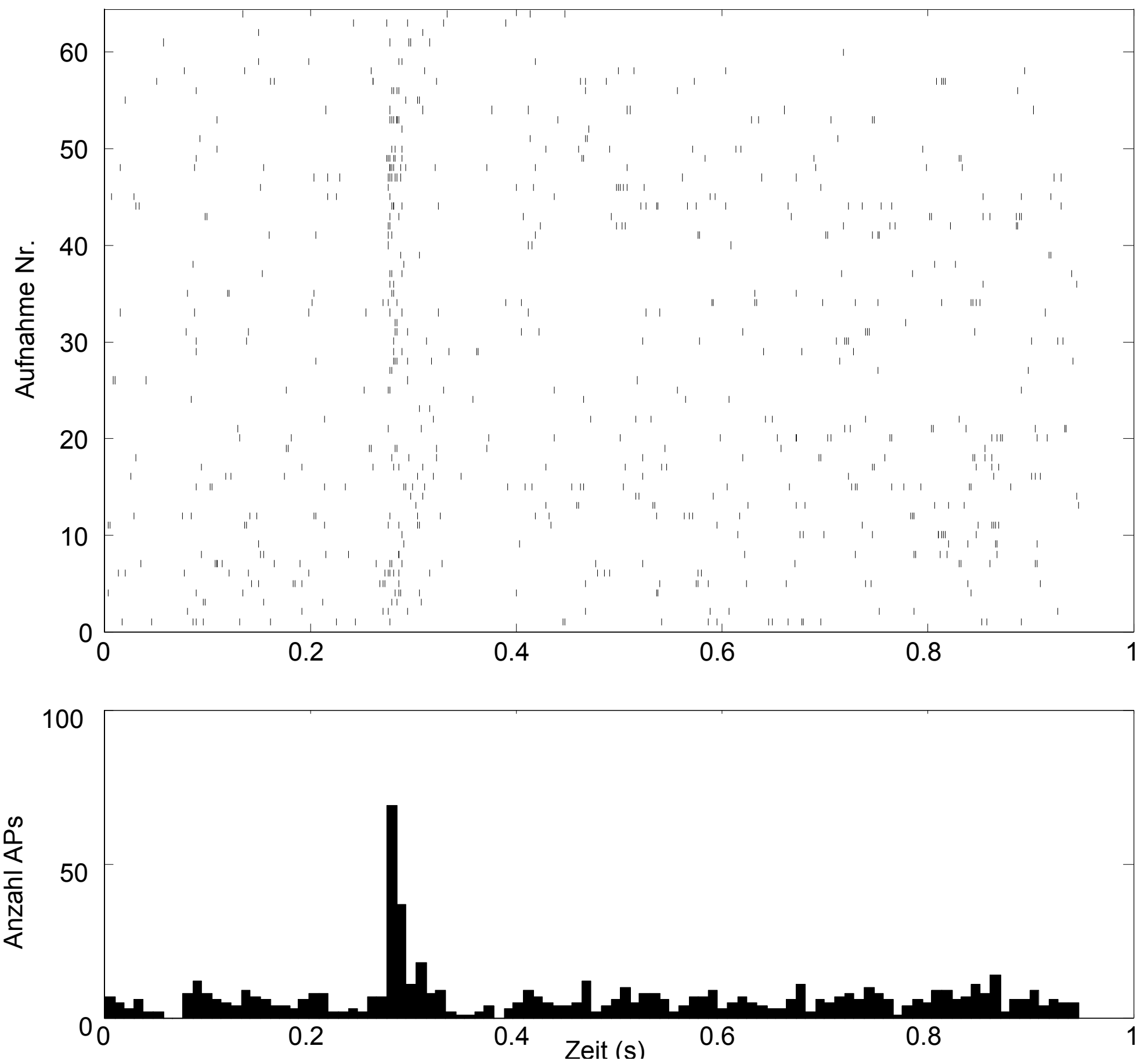

**Abbildung II.9.: Aktionspotentiale und Peristimulus Time Histogram (PSTH) aus 64 Einzelversuchen zur Pulsantwort eines V1-Neurons auf ein Gittermuster (aus [Aronov03], Rohdaten aus den zugehörigen Makaken-Experimenten)**

Der untere Teil der obigen Abbildung stellt ein sogenanntes Peristimulus Time Histogram (PSTH) dar, d.h. ein Histogramm der Pulshäufigkeiten in Abhängigkeit der Zeit nach Experimentbeginn, über viele Experimente aufsummiert. Am PSTH lässt sich die Korrelation zwischen aufeinanderfolgenden Experimenten visuell beurteilen, also welche Teile der Pulsfolge stochastischer Natur sind und welche sich annähernd reproduzieren lassen [Kass05, Koch99 (Kapitel 15)]. Ein weiteres wichtiges Mittel zur Beurteilung neuronaler Informationsverarbeitung ist die Analyse der zeitlichen Abstände zwischen aufeinander folgenden Pulsen, der Interspike Intervals (ISI), etwa als Histogramm aufgetragen:

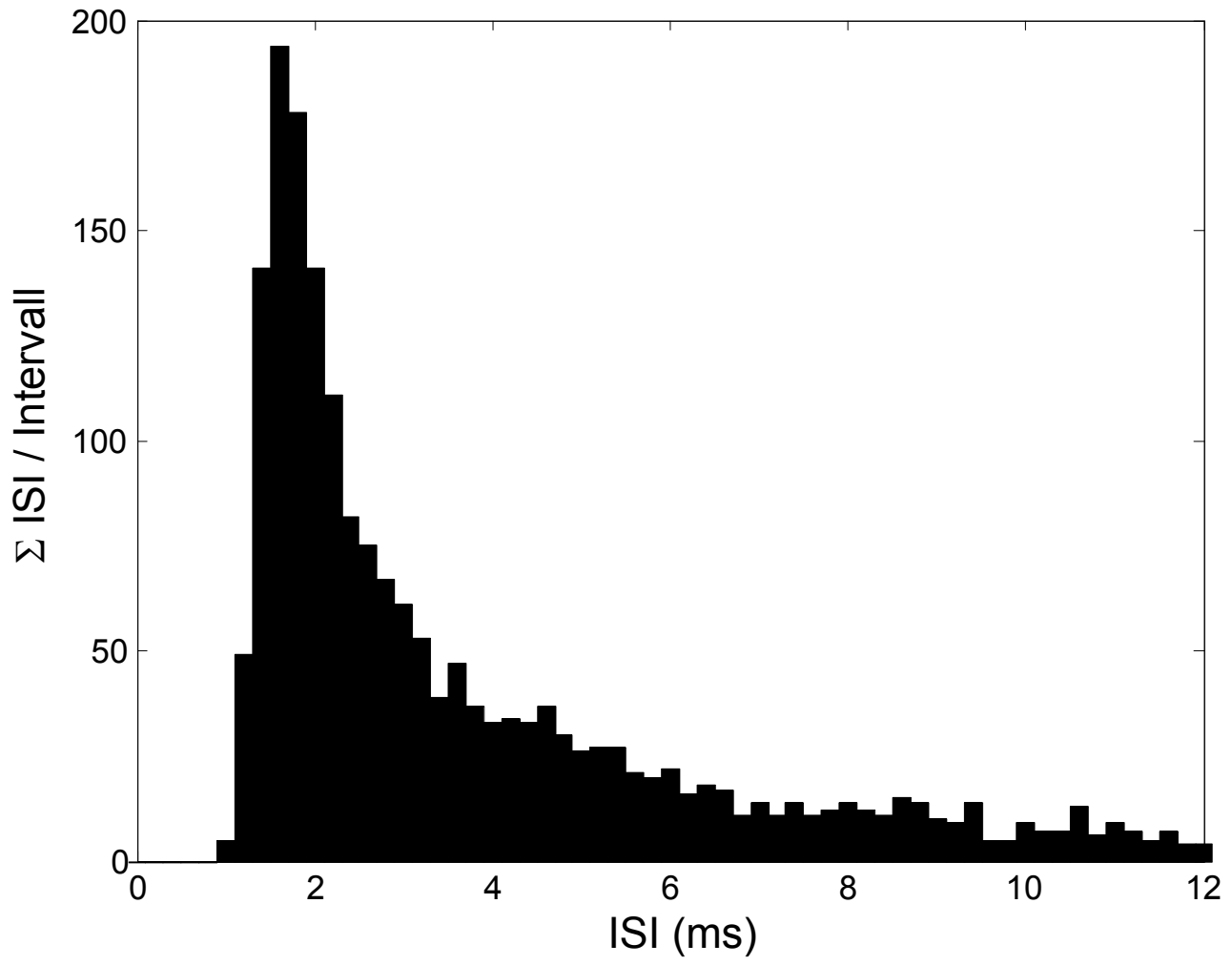

**Abbildung II.10.: ISI-Plot der Pulsfolgen aus Abbildung II.9 und weiterer Rohdaten aus [Aronov03]**

An einem ISI-Histogramm lässt sich die Verteilung der aufeinander folgenden Pulse ablesen, womit z.B. verschiedene Arbeitsmodi der Neuronen unterschieden werden können [Kass05]. Anhand der obigen Abbildung lässt sich auch die Auswirkung der absoluten und relativen Refraktärzeit sehr gut erkennen, beispielsweise existiert kein ISI, das kleiner als ca. 0,9ms wäre, bis zu diesem Zeitpunkt nach der Generierung eines Aktionspotentials ist die Membran durch ihre Hyperpolarisierung gesperrt. Im Anschluss daran nimmt die Wahrscheinlichkeit der ISIs langsam zu, ist aber während der relativen Refraktärzeit bis ca. 1,6ms immer noch durch das vorhergehende Aktionspotential gehemmt.
Die statistische Analyse von ISIs liefert u.a. Anhaltspunkte für die Variabilität einer Pulsfolge, was meist durch die auf den Erwartungswert normierte Standardabweichung ausgedrückt wird:

$$CV = \frac{\sqrt{V(ISI)}}{E(ISI)} \qquad \textbf{(II.14)}$$

Die normierte Standardabweichung der ISIs wird als Coefficient of Variation (*CV*) bezeichnet. Eine sehr regelmäßige Pulsfolge mit einem $CV \ll 1$ kann ein Anzeichen für eine deterministische Verarbeitung sein, jedoch kann in einer so gearteten Pulsfolge nur wenig Information übertragen werden [Koch99, Shannon49] (siehe auch die folgenden Unterabschnitte), während eine variable Pulsfolge mit einem $CV \geq 1$ evtl. einen hohen stochastischen Rauschhintergrund kennzeichnet [Shadlen98], jedoch auch eine hohe Informations-dichte erreichen kann [Steveninck97] (siehe auch folgende Abschnitte).
Eine weitere wichtige statistische Kenngröße von neuronalen Pulsfolgen ist ihre Rate $\lambda$. Die einfachste Definition dieser Rate ist die Anzahl der Pulse in einem Intervall $T$ [Koch99 (Kapitel 14)]:

$$\lambda_T(t) = \frac{1}{T} \int_t^{t+T} \sum_{i=1}^{N_i} \delta(t' - t_i)\, dt' \qquad \textbf{(II.15)}$$

Alle Ausgangspulse $N_i$ eines Neurons im Beobachtungszeitraum werden als Dirac-Impulse aufsummiert, und das entstehende Signal über ein Zeitfenster $T$ integriert und normiert. Als Grenzwert dieser Definition existiert die instantane Rate $\lambda(t)$ für ein gegen Null gehendes Beobachtungsintervall $T$. Andere Ratendefinitionen summieren die Aktionspotentiale einer Population, ebenfalls instantan $\lambda[\Sigma(N),t]$, oder intervallbasiert $\lambda[\Sigma(N),t,T]$. Im folgenden werden verschiedene auf den o.a. biologischen Messungen basierende Code- und Verarbeitungsmodelle diskutiert, angefangen mit Ratencodes.

### II.2.1 Ratencode

Modelle von Ratencodes beruhen auf der Annahme, das die in ISI-Histogrammen erscheinende Variabilität von Pulsfolgen nur stochastischen Hintergrund hat, mithin nur aus Rauschen ohne informationstragende Eigenschaften besteht [Shadlen98]. Einzig interessanter Parameter solcher Codes ist ihre mittlere Rate $\lambda$. Um statistische Betrachtung von z.B. informationsverarbeitenden Eigenschaften eines solchen Codes anstellen zu können, ist eine Modellierung der entsprechenden ISI-Dichtefunktion nötig:

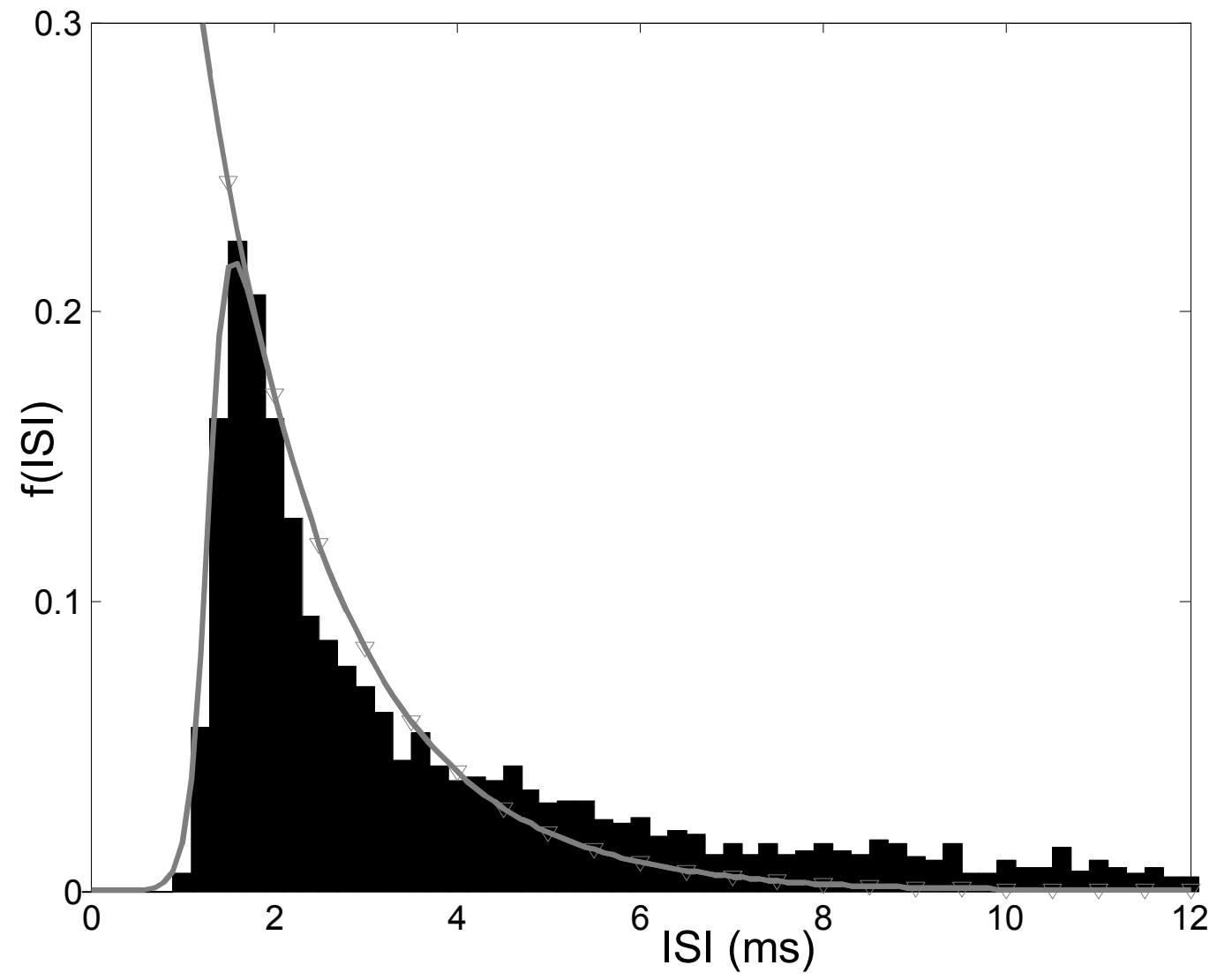

**Abbildung II.11.: Relative Häufigkeiten der ISIs aus [Aronov03], mit angepasster Poissonverteilung (Dreiecke) und Refraktärverteilung (Wahrscheinlichkeitsdichtefunktionen)**

In erster Näherung folgt die ISI-Verteilung dem charakteristischen Verlauf eines Erneuerungsprozesses, d.h. die Wahrscheinlichkeit eines Aktionspotentials in jedem beliebigen Intervall ist konstant und unabhängig von der Vergangenheit der Pulsfolge. Die einfachste Formulierung dieses Prozesses ist die Poisson-Verteilung [Koch99 (Kapitel 15)]:

$$f(T_{ISI}) = \lambda * e^{-\lambda T_{ISI}} \tag{II.16}$$

Der einzige Parameter dieser Verteilung ist die mittlere Rate $\lambda$. Diese wurde für die in Abbildung II.11 dargestellten Messwerte zu $714 s^{-1}$ bestimmt (obere Kurve). Die Annahme der Unabhängigkeit von Pulsereignissen von der Vergangenheit der Pulsfolge wird jedoch erst ab ca. 1,7ms erreicht, da die Hyperpolarisation in Form der absoluten und relativen Refraktärzeit ein Gedächtnis bereitstellt [Koch99 (Abschnitt 15.1)]. Dies muss in einer detaillierteren Beschreibung der ISI-Verteilung entsprechend berücksichtigt werden:

$$f(T_{ISI}) = \frac{1}{1 + e^{-\frac{T_{ISI} - T_{abs}}{T_{rel}}}} * \lambda * e^{-\lambda T_{ISI}} \tag{II.17}$$

Die Poissondichtefunktion wird für kleine ISIs zusätzlich mit einem inversen Cosinus Hyperbolicus gewichtet, der durch einen Offset um $T_{abs}$ und eine Skalierung um $T_{rel}$ entsprechend absolute und relative Refraktärzeit vorgibt. In Abbildung II.11 ist der Verlauf der verbesserten ISI-Dichtefunktion für ein $T_{abs}$ von 1,3ms und ein $T_{rel}$ von 0,1ms dargestellt (glatte, unmarkierte Kurve). Diese Näherung liefert v.a. für Analysen eines einzelnen Neurons genauere Ergebnisse. Wenn wie in Abbildung II.9 über mehrere Experimente am selben Neuron zusammengefasst wird oder eine Population betrachtet wird, tendieren die ISIs dazu, der Poissonverteilung aus Gleichung (II.16) zu folgen [Kass05]. Dies lässt sich dadurch erklären, dass dann aufeinander folgende Aktionspotentiale i.d.R. nicht mehr vom selben Experiment (bei wiederholter Durchführung) oder vom selben Neuron (bei Summierung über Population) stammen, d.h. die ISIs nicht mehr von Refraktärzeiten beeinflusst werden.
Eines der ersten Ratenmodelle wurde anhand der bei biologischen Neuronen gemessenen I/O Relation zwischen Eingangs- und Ausgangspulsen aufgestellt [Hopfield84]:

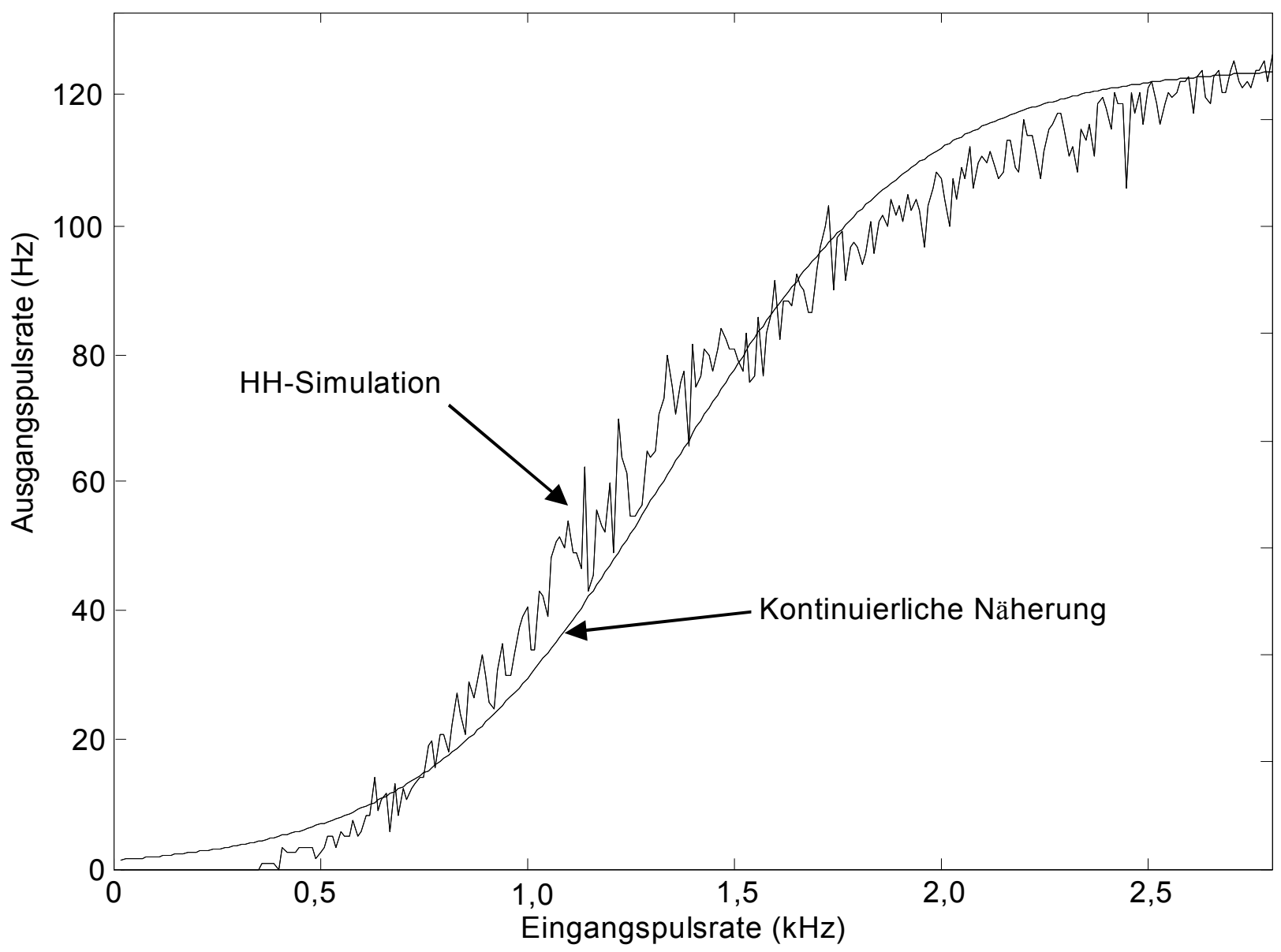

**Abbildung II.12.: Übertragungsverhalten des Hodgkin-Huxley-Neurons und sigmoide Näherung (jedes Neuron $\lambda$=10s$^{-1}$, Gesamteingangsrate wird erreicht über eine entsprechende Skalierung der Populationsgröße, einzelne EPSCs befinden sich um den Faktor 16 unter Schwellwert)**

Wenn ein biologisches Neuron oder wie in der obigen Abbildung dessen HH-Modell mit Poisson-Pulsfolgen stimuliert wird, ergibt sich ein charakteristisches Übertragungsverhalten, mit einem langsamen Anstieg, einem Übergangsbereich mit angenähert linearem Verlauf und Sättigungsverhalten für hohe Eingangsfrequenzen.

Im unteren Bereich agiert der Leckstrom als Tiefpass, d.h. der Einfluss niedrigfrequenter präsynaptische Pulse ist unterproportional, da das Membranpotential in den Pausen wieder zum Ruhepotential abklingt. Durch die stochastische Verteilung der Pulse ergeben sich jedoch Zeitpunkte, in denen genügend Pulse in einem entsprechenden Intervall auftreten, um das Membranpotential über den Schwellwert zu heben. Regelmäßige (d.h. nicht-Poisson) Pulsfolgen bewirken demgegenüber ein abruptes Einsetzen der Ausgangspulse mit einem starken Anstieg an der Einsetzstelle [Koch99 (Abschnitt 6.4)].

Mit zunehmender Eingangsfrequenz der Pulse erhöht sich die Wahrscheinlichkeit eines pulsauslösenden Zusammentreffens von Aktionspotentialen am Eingang, damit steigt die Übertragungskurve an, bis zu einem mittleren Bereich, in dem Eingangspulse mit größter Effizienz einen Ausgangspuls auslösen. Im Bereich hoher Eingangspulsfrequenzen wird über die Tiefpasscharakteristik der Integration sowie absolute und relative Refraktärzeit eine Sättigung im Übertragungsverhalten verursacht.

Das geschilderte Übertragungsverhalten wird klassisch über eine ‚Sigmoid'-Funktion angenähert:

$$N(\lambda_{in}) = C * \frac{1}{1 + e^{-\frac{\lambda_{in} - \lambda_{mitte}}{\lambda_{anstieg}}}} \qquad \textbf{(II.18)}$$

Entsprechende Neuronen werden als Sigmoid-Neuronen bezeichnet. Eine weitere Vereinfachung sind die s.g. Perzeptronneuronen, bei denen die sigmoid-Charakteristik auf eine Schrittfunktion reduziert wird, d.h. diese kennen nur zwei Ausgabewerte in Abhängigkeit eines Schwellwerts. Bei so gearteter Modellierung wird das zeitdynamische und individuelle Verhalten der Neuronen zugunsten eines Stereotyps vernachlässigt. Synapsen finden Eingang in dieses Modell als Gewichtungswert, mit dem der entsprechende Ausgabewert eines Neurons am Eingang des Ziel-Neurons skaliert wird.

Netzwerke aus diesen Neuronen finden v.a. in der Klassifikation und Mustererkennung weit reichenden Einsatz [Goerick94, König02, Mayr01, Zhang00]. Lernregeln für ihre Synapsen und die Netztopologien gehorchen hierbei meist einer empirisch ermittelten anwendungsspezifischen Schablone [Zhang00]. Dieses vereinfachte, abstrahierte Modell neuronaler Verarbeitung ist momentan das einzige mit breitem Einsatz in technischen Anwendungen. Durch die wesentlich höhere Komplexität von zeitdynamischen neuronalen Netzen gibt es noch keinen vergleichbaren Einsatz von PCNNs in Klassifikationsaufgaben, obwohl deren (theoretische) Rechenmächtigkeit zumindest vergleichbar ist [Maass99]. PCNNs werden vereinzelt in Vorstufen der Klassifizierung zur Projektion des Merkmalsraumes eingesetzt [Atmer03, Verstraeten05].

Etwas biologienäher als die obige abstrakte Repräsentation von Pulsraten als Zustandsvariablen und Übertragungsfunktionen ist ein ‚echter' Ratencode, bei dem die Summe der Aktionspotentiale über einen bestimmten Zeitraum als Informationsträger angesehen wird. Basierend auf einem Ratencode mit zwei unterscheidbaren Bereichen, d.h. niedriger und hoher Rate, wird beispielsweise in [Vogels05] ein Netz aus modifizierten IAF-Neuronen darauf trainiert, anhand der Raten der Eingangssignale Logikfunktionen (AND, XOR) auszuführen, wobei die Antworten der Logikfunktionen wieder als hohe/niedrige Rate an den Ausgang gegeben werden. Für eine detailliertere Übertragung/Verarbeitung wäre eine stärkere Differenzierung der unterscheidbaren Ratenbereiche nötig, diese soll im folgenden hergeleitet werden:

Wenn eine Poissonverteilung zugrunde gelegt wird, lässt sich die Wahrscheinlichkeit von $N$ Pulsereignissen im Zeitraum $T$ in Abhängigkeit des Erwartungswertes $\mu=\lambda*T$ (bei konstanter Rate $\lambda$) wie folgt formulieren [Kass05]:

$$p(N|\mu) = e^{-\mu}\frac{\mu^N}{N!} \qquad \textbf{(II.19)}$$

Für eine Rückgewinnung der Rateninformation aus der Zahl der Pulsereignisse in einem Zeitraum wird eine Ratenbandbreite definiert, d.h. ein Bereich mit einer unteren Schranke $\Delta_{US}$ und einer oberen Schranke $\Delta_{OS}$ um den Erwartungswert $\mu$, bei dem eine Anzahl von Pulsereignissen als Anzeiger für die Übertragung dieser Rate angesehen wird. In der obigen Gleichung wird entsprechend $N=\mu+\Delta$ substituiert und für die Fakultät von $N$ eine Näherung über die Stirlingsche Formel vorgenommen [Bronstein87]:

$$N! = \left(\frac{N}{e}\right)^N * \sqrt{2\pi N} \qquad \textbf{(II.20)}$$

Dies führt zu folgender modifizierter Wahrscheinlichkeitsdichtefunktion für Pulse mit Poisson-Verteilung:

$$p(\mu+\Delta|\mu) = \left(\frac{1}{1+\frac{\Delta}{\mu}}\right)^{\mu+\Delta} * \frac{e^{\Delta}}{\sqrt{2\pi(\mu+\Delta)}} \quad , \quad \Delta,\mu \in \mathrm{N} \qquad \textbf{(II.21)}$$

Die Wahrscheinlichkeit, mit der bei einem festen Erwartungswert $\mu$ die tatsächliche Zahl der Pulse in den von den $\Delta s$ definierten Bereich fällt, ergibt sich aus der entsprechenden Aufsummierung:

$$P(\mu+\Delta_{US} \leq N \leq \mu+\Delta_{OS}) = \sum_{\Delta=\Delta_{US}}^{\Delta_{OS}} p(\mu+\Delta|\mu) \qquad \textbf{(II.22)}$$

Gleichung (II.22) stellt die Trefferquote des Codes dar, d.h. mit welcher Wahrscheinlichkeit ein durch seine Bandbreite und Erwartungswert definiertes Signal richtig erkannt wird. Für eine Rate $\lambda$=(1...40)Hz, und einen Beobachtungszeitraum $T$=2s, ergibt sich ein Erwartungswert an Ereignissen $\mu=\lambda*t$ von (2...80). Wenn eine Trefferwahrscheinlichkeit $P$ von ca. 0,95 erreicht werden soll, ergibt sich die folgende Einteilung für $\Delta_{US}$, $\Delta_{OS}$ und $\mu$:

| μ | 2 | 11 | 28 | 53 | 80 |
|---|---|---|---|---|---|
| $\Delta_{US}$ | 0 | 5 | 18 | 39 | 67 |
| $\Delta_{OS}$ | 4 | 17 | 38 | 66 | ∞ |
| Σ p | 0,947 | 0,953 | 0,954 | 0,947 | 0,939 |

**Tabelle II-2.: Zahlenbeispiel für die Unterscheidbarkeit von Poisson-Ratencodes in einem vordefinierten Pulsereignisbereich**

Mit den gegebenen Parametern lassen sich demnach 5 Signale im angegebenen Zeitraum unterscheiden. Da die Standardabweichung der Poissonverteilung aus Gleichung (II.19) gleich $\sqrt{\mu}$ ist, und sich die in der obigen Tabelle angegebenen Bereiche $[\Delta_{US};\Delta_{OS}]$ in guter Näherung als $(\mu\text{-}2\sqrt{\mu}\ldots\ \mu+2\sqrt{\mu})$ beschreiben lassen, ergibt eine z.B. um den Faktor 2 erweiterte Beobachtungszeit nur ca. 1,5 weitere unterscheidbare Pulsbereiche. Die Erweiterung der Beobachtungszeit erzeugt zwar gegenüber der obigen Tabelle mehr Spielraum im Bereich höherer Erwartungswerte, diese ergeben jedoch nur eine unterproportional mitwachsende Anzahl neuer unterscheidbarer Pulsraten, da die zu ihrer Unterscheidung nötige Pulsraten-Bandbreite in der Wurzel der Pulsrate mitwächst.
Mögliche technische Implementierungen von Ratencodes jittern i.d.R. deutlich weniger (z.B. Anhang C.1), aber selbst für jitterfreien Code, der z.B. in einem 1s-Intervall zwischen 1 und 100 Pulse überträgt, muss das komplette Intervall abgewartet werden, um zu entscheiden, welches Wort übertragen wurde, d.h. die Datenrate beträgt 6,64 Bit/s.

### II.2.2 Zeitfolgencodes

Eine deutlich höhere Informationsdichte lässt sich erreichen, wenn von einem geringeren Rauschhintergrund ausgegangen wird, also die Zeitpunkte der einzelnen Pulse oder ihre ISIs Informationen enthalten [Gutkin03, VanRullen05]. Generell scheinen einzelne Neuronen eher geringe intrinsische Rauschquellen aufzuweisen [Kretzberg01], Variabilität, d.h. augenscheinliche Stochastik wird offenbar durch verschiedene extrinsische Mechanismen hervorgerufen, die jedoch nicht notwendigerweise wirklich statistische Schwankungen verursachen. Zum Einen scheint die Reaktion auf externe Stimuli eine Rolle zu spielen, d.h. konstante Stimuli werden ‚wegadaptiert' und die Neuronen pulsen zu zufälligen Zeitpunkten in Ermangelung einer zu übertragenden Information, während variable Stimuli für einen permanenten Informationsfluss sorgen und damit zwar variable, aber sehr präzise reproduzierbare Pulszeitpunkte verursachen [Gutkin03]. Ein weiterer Grund für die Variabilität einzelner Pulsfolgen kann die gegenseitige Inhibition zwischen gleichartig verarbeitenden Neuronen sein, wodurch ihre Pulsfolgen dekorreliert werden [Mar99]. Dadurch können sich z.B. die Pulsfolgen von einzelnen Neuronen partiell vertauschen, d.h. Pulse tauchen zwar zum selben Zeitpunkt auf wie im letzten Experiment, aber an einem nicht gemessenen Neuron. Dies kann erst durch moderne Parallelmessungen an vielen Neuronen analysiert werden [Zeitler06]. Durch diese Inhibition können einzelne Spikes von Experiment zu Experiment bis mehrere 10ms verschoben sein, wodurch auch die PSTHs sehr viel ‚falsche' statistische Schwankung enthalten, während bei einer Berücksichtigung dieses möglichen temporalen Offsets die ISIs von Experiment zu Experiment bei einem einzelnen Neuron ebenso wie die relativen ISIs von verschiedenen Neuronen zueinander wieder wesentlich reproduzierbarer sind [Aronov03]. Ein weiterer Grund für Variabilität zwischen Experimenten können externe Einflüsse sein, z.B. Augenbewegungen [Gur97] oder andere Veränderungen entlang des Eingangspfades, etwa Neurohormon-Ausschüttungen durch Erschöpfung/Aufregung des Versuchstieres.
Im folgenden wird ein hypothetischer neuronaler ISI-Code aus 40 Symbolen untersucht, der auf Grundlage der ISIs von 1 bis 40ms aus Abbildung II.10 und einer intrinsischen zeitlichen Präzision von 1ms [Kretzberg01] postuliert wird. Abbildung II.13 zeigt ein entsprechendes Histogramm mit 1ms breiten Klassenbereichen:

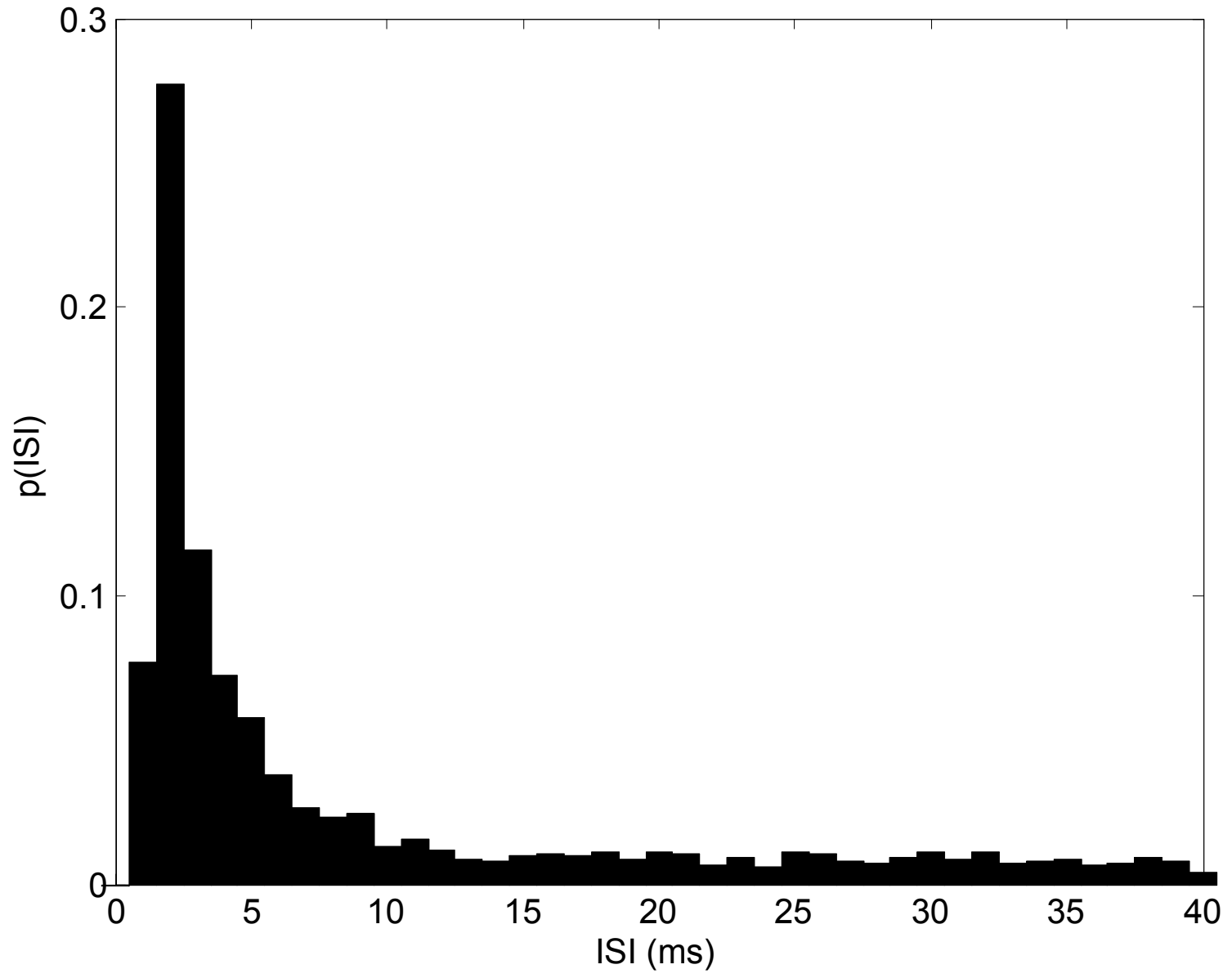

**Abbildung II.13.: ISI-Häufigkeiten wie in Abbildung II.11, aber in einem Zeitfenster von 0 bis 40 ms, Bingröße 1ms als Ausdruck Jitter der einzelnen Pulszeitpunkte (Präzision 1ms)**

Der theoretische Informationsgehalt eines Codes lässt sich mittels der Wahrscheinlichkeit der einzelnen Symbole über die Shannon-Entropie berechnen [Shannon49]:

$$H = \sum_{i=1}^{40} p(i) * ld\left(\frac{1}{p(i)}\right) \qquad \textbf{(II.23)}$$

Die maximal mögliche Entropie bei einem Code aus 40 unterschiedlichen ISIs ist 5,32 Bit/ISI für den Fall, dass alle Symbole/ISIs gleichwahrscheinlich sind. Die reale Verteilung der ISIs soll anhand der relativen Häufigkeiten aus der obigen Abbildung berücksichtigt werden, damit ergibt sich eine Entropie von 4,17 Bit/ISI. Vergleichbare Werte finden sich in [Warland97], mit einem maximalen Informationsgehalt/Entropie von Ganglienzellen aus ihrer ISI-Statistik von 6,6 Bit pro ISI/Spike[9]. Warland et. al. führen eine Rekonstruktion des von den Ganglienzellen übertragenen Signals durch, wodurch sich die von den Neuronen real verwendete Entropie zu 1,9 Bit/ISI bestimmen lässt. Ähnliche Daten ergibt die Auswertung der Informationsübertragung von sensorischen Neuronen in [Gabbiani99]. Neuronen scheinen demnach zwar nicht den vollen Umfang eines ISI-Codes auszunutzen, jedoch ist die Datenrate weit über einer Ratencodierung, mithin werden also zumindest Teile der temporalen Feinstruktur von Pulsfolgen in biologischen Neuronen ausgewertet. Möglicherweise wird diese Redundanz in Neuronen zur Fehlerkorrektur verwendet d.h. es wird mit einer höheren zeitlichen Präzision als nötig übertragen, entsprechend einem Überangebot an Symbolen bei einer Codierung wie vorab geschildert, um z.B. Jitter oder Pulsverluste ausgleichen zu können [Stiber05].

Gedanken zu möglichen Mechanismen in Neuronen, mit denen temporale Strukturen in Pulsfolgen eines einzelnen Neurons ausgewertet werden kann, liefern [Delorme01, Delorme03a, Guyonneau05]. Technische Anwendung finden derartige Stimulus-codierende Pulsfolgen beispielsweise als Eingangssignal für Liquid-Computing-Netzwerke [Schrauwen03].

Das Gedankenexperiment zum technischen Ratencode aus dem letzten Abschnitt lässt sich auch auf einen ISI-Code erweitern. Wenn der technische Code mit einer ähnlichen zeitlichen Präzision implementiert ist wie der oben postulierte biologische, d.h. einen Jitter unterhalb 1ms besitzt, kann jedes Symbol bereits nach einer ISI decodiert werden. Wenn alle *N* Codes gleichwahrscheinlich sind, ergibt sich als durchschnittliches Intervall:

[9] In [Warland97] werden größere maximale ISIs zugelassen, dadurch ergibt sich eine erhöhte Anzahl Symbole und damit erhöhte Entropie.

$$\overline{T_{ISI}} = \frac{1}{N}\sum_{i=1}^{N} T_{ISI,i} \qquad \textbf{(II.24)}$$

In Zahlen wäre das ein durchschnittliches ISI für einen 1-100Hz Code von 51,9 ms, d.h. im Mittel können 6,64Bit*1/51,9ms übertragen werden, oder 128,1 Bit/s.
Alternativ lässt sich die Pulsfolge eines Neurons auch als Frequenzspektrum darstellen, bei dem dann die frequenzabhängige Kanalkapazität aus dem Rauschpegel bestimmt werden kann [Gabbiani99, Mar99]. Für die der Abbildung II.10 zugrunde liegenden Pulsfolgen ergibt sich das folgende Leistungsdichtespektrum:

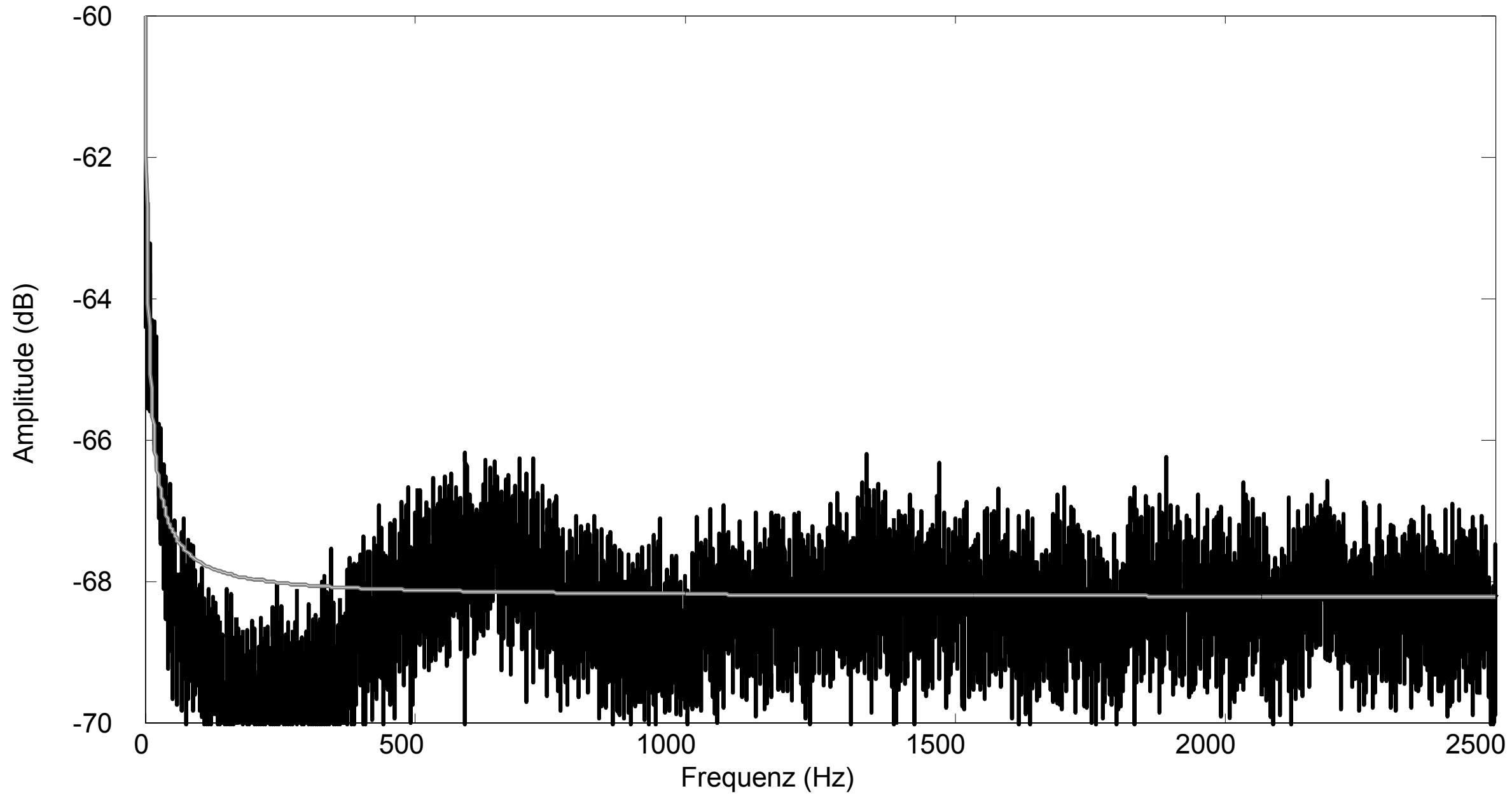

**Abbildung II.14.: Amplitudenfrequenzspektrum[10] der ISIs aus den vorherigen Abbildungen und analytische Näherung aus Poisson-Dichtefunktion**

Der Verlauf des Spektrums lässt sich aus der ISI-Verteilung herleiten, eine Poisson-Verteilung etwa ergibt ein flaches Spektrum, da bei gleichverteilten Pulszeiten jede Frequenz mit derselben Häufigkeit vertreten ist. Im allgemeinen wird dabei der Gleichanteil der Pulsfolgen als einzelner Dirac bei 0 Hz modelliert [Mar99, Spiridon99]. Aus Abbildung II.14 lässt sich entnehmen, dass dies nur eine unzureichende Wiedergabe des Spektrums im Bereich niedriger Frequenzen zulässt. In Anhang A.1 wurde anhand einer Poisson-verteilten Pulsfolge eine exakte Berechnung des Amplitudenspektrums ausgeführt. Eine dB-Darstellung von Gleichung (A.18) für eine Kreisfrequenz $\omega$ größer 1Hz lautet:

$$|R(\omega)|_{dB} = 10\log_{10}\left[|R(\omega)|^2\right] = 10\log_{10}\left(\lambda + \frac{2\lambda^2}{\omega}\right) \quad , \quad \omega \geq 1 \qquad \textbf{(II.25)}$$

Das mittlere Interspike Intervall aus den obigen biologischen Daten beträgt 65,2 ms, d.h. eine Rate $\lambda$ von 15,3 Hz (aus 4062 ISIs gemittelt). Basierend auf dieser Rate und Gleichung (II.25) ergibt sich die durchgezogenen Linie in Abbildung II.14, wobei diese um 80 dB nach unten verschoben werden muss. Dies ergibt sich aus den unterschiedlichen Integralflächendefinitionen der Pulse in der analytischen Beschreibung und der Rekonstruktion der biologischen Daten, erstere werden als

[10] Alle Frequenzspektren dieser Arbeit wurden erzeugt mit der Matlab-FFT-Funktion für beliebige Vektorlängen, Amplitude aus 20*$\log_{10}$|komplexe FFT-Antwort|. Soweit nicht anders angegeben ist die Samplefrequenz 10kHz, der Messdatenvektor wird mit einem Hann-Fenster der Vektorlänge gefiltert. Vektorlänge in Abbildung II.14 ist 2648435 Samples, verkettet aus den kompletten Messdaten zu [Aronov03]. Samplevektorlängen in Simulationen der folgenden Kapitel ergeben sich, soweit nicht anders angegeben, aus der Simulationszeit multipliziert mit der o.a. Samplefrequenz. Meist erfolgt keine komplette Darstellung des Spektrums bis zur Hälfte der Samplefrequenz, da die neuronal interessanten Effekte i.d.R. auf den Frequenzbereich bis ca. 1000 Hz begrenzt sind.

Diracimpuls mit Fläche 1 angenommen, letztere als Rechteckimpuls mit Höhe 1 und Länge 100µs. Der Unterschied beträgt damit $1*10^4$ oder 80 dB. Für niedrige Frequenzen zeigt die analytische Lösung denselben Abfall der Amplitude bis zum endgültigen Grenzwert $\lambda$ bei hohen Frequenzen. Der Bereich mittlerer Frequenzen (ca. 100-1200 Hz) weicht von der analytischen Kurve ab, da die angenommene Poisson-Verteilung im Gegensatz zu den biologischen Daten keine Refraktärzeit enthält. Die aus Abbildung II.10 ersichtliche relative Refraktärzeit sorgt für eine leichte Verringerung des Rauschpegels bei niedrigen Frequenzen [Koch99, Mar99] (im Bereich von ca. 100 bis 500Hz in obiger Abbildung). Durch die absolute Refraktärzeit von ca. 1,6 ms wird eine zusätzliche additive Komponente ins Frequenzspektrum eingebracht, resultierend in ein Maximum bei der ihrem Kehrwert entsprechenden Frequenz (Frequenzbereich 600-700 Hz) und eine weitere leichte Oberwelle bei ca. 1400 Hz.
Die in Abbildung II.14 (Kurve der Messdaten) sowie in [Mar99] gewonnene Erkenntnis, dass eine relative Refraktärzeit die Rauschamplitude bei niedrigen Frequenzen verringert, scheint kontraintuitiv zu sein, da diese kurze ISIs und damit hohe Frequenzen unterdrückt. Bei einer genaueren Betrachtung ergibt sich jedoch, dass die Hochfrequenzkomponenten im Spektrum hauptsächlich durch minimale Unterschiede zwischen aufeinanderfolgenden ISIs hervorgerufen werden, während Niederfrequenzkomponenten direkt durch ISIs entstehen, wodurch die Reduktion bei niedrigen Frequenzen erklärbar ist.

### II.2.3 Populationscodes

In Populationscodes lässt sich durch die größere Anzahl an beitragenden Neuronen und zugehörigen Aktionspotentialen deutlich schneller Information übertragen als in Codes, die auf Einzelpulsfolgen beruhen. Selbst gegenüber einem der zuletzt geschilderten ISI-Codes ergeben sich hier Vorteile, da dort auf jeden Fall für die Abschätzung des ISI zwei Pulse desselben Neurons abgewartet werden müssen, während in Populationscodes nur z.B. das ISI zwischen zwei Aktionspotentialen von verschiedenen Neuronen relevant ist, welches entsprechend schneller geschätzt werden kann (vgl. hierzu Gleichung (II.28), ISI-Verteilung für Neuronen-Population relativ zu Einzelneuron). Populationscodes lassen sich anhand ihrer Komplexität und temporalen Strukturen in drei große Kategorien aufteilen:

---

Die einfachste Variante ist ein Ratencode, bei dem alle Neuronen der Population statistisch unabhängig voneinander versuchen, denselben Stimulus zu übertragen. Eine Signalrekonstruktion kann dann entweder als Summe der Aktivität der Einzelneuronen erfolgen [Shadlen98], oder es werden ähnliche statistische Betrachtungen wie in Gleichungen (II.19) bis (II.22) für Populationen angestellt. Die Bandbreite für den Erwartungswert $\mu$ und damit die Anzahl übertragbarer Ereignisse kann entweder wie in Abschnitt II.2.1 über den Zeitraum oder die Größe der Population beeinflusst werden. Mithin kann eine längere Beobachtungsdauer an einem Neuron bei gleich bleibender Präzision durch eine Beobachtung über einen kürzeren Zeitraum an einer Neuronenpopulation ausgetauscht werden, deren Neuronen versuchen, mit unabhängigen Poisson-Prozessen dieselbe mittlere Rate $\lambda$ als Information zu übertragen.
Detailreichere Signalübertragung kann mit den beiden anderen Varianten von Populationscodes erreicht werden. Hierbei wird zum Einen ähnlich wie in Abschnitt II.2.2 angenommen, dass die zeitliche Feinstruktur von Pulsfolgen Informationen enthält, und dass diese Pulsfolgen zumindest teilweise korreliert sind, d.h. dass die relative Phasenabfolge von Aktionspotentialen unterschiedlicher Neuronen zueinander der Informationscodierung dient.

---

Der als nächstes betrachtete Korrelationscode trifft dabei keine Unterscheidung einzelner Pulse nach aussendendem Neuron. Globales Merkmal dieses Codes ist seine Kopplung an bestimmte Hirnareale, die ein dichtes Netz lateraler inhibitorischer Verbindungen unterhalten [Zeitler06, Buchs02, Guyonneau05]. Die dadurch entstehende gegenseitige inhibitorische Verschränkung einer Neuronenpopulation tritt in vielen Bereichen des Kortex und der sensorischen Pfade auf [Shepherd04]. Eine Modellvorstellung der in diesen Bereichen stattfindenden Codierung lässt sich wie folgt entwickeln:
Eine moderate Population von Neuronen (50-500) ist meist für die Übertragung einer kleinen Bandbreite an Stimuli zuständig, z.B. haben benachbarte Neuronen im V1 sehr ähnliche rezeptive Felder [Aronov03, Zeitler06, Hubel68]. Über die inhibitorische Kopplung teilen sich die Aktionspotentiale einer Population die Signalübertragung auf, das am besten auf den Stimulus abgestimmte Neuron reagiert und hemmt die anderen Neuronen der Population [Buchs02]. Dieser Vorgang ist dynamisch, wenn beispielsweise dieses Neuron einen synaptischen Ausfall hat, d.h. kein Aktionspotential weitergibt, reagiert das nächste Neuron und hemmt dann seinerseits den Rest der Population [Guyonneau05]. Diese inkrementelle Organisation der Stimulusreaktion dient zum Einen der Maximierung der über das Neuron übertragenen Information [Delorme03a, VanRullen01], da dann zur Repräsentation eines Stimulus nur ein Neuron mit sehr gut passendem rezeptiven Feld reagiert, statt einiger Neuronen mit schlechter auf den Stimulus abgestimmten Feldern [Olshausen02]. Zum Anderen führt diese Organisation zu einer optimalen Mischung aus gesteuerter Fehlerkorrektur und Energieeffizienz, da wie oben angeführt durch die teilweise redundanten Antwortcharakteristiken im Fehlerfall sofort andere Neuronen einspringen können, aber zur Repräsentation nur die minimal mögliche Anzahl Aktionspotentiale benötigt wird [Laughlin03]. Redundanz und Verteilung der Signalübertragung in einem Korrelationscode lassen sich anschaulich aus Abbildung II.15 erschließen:

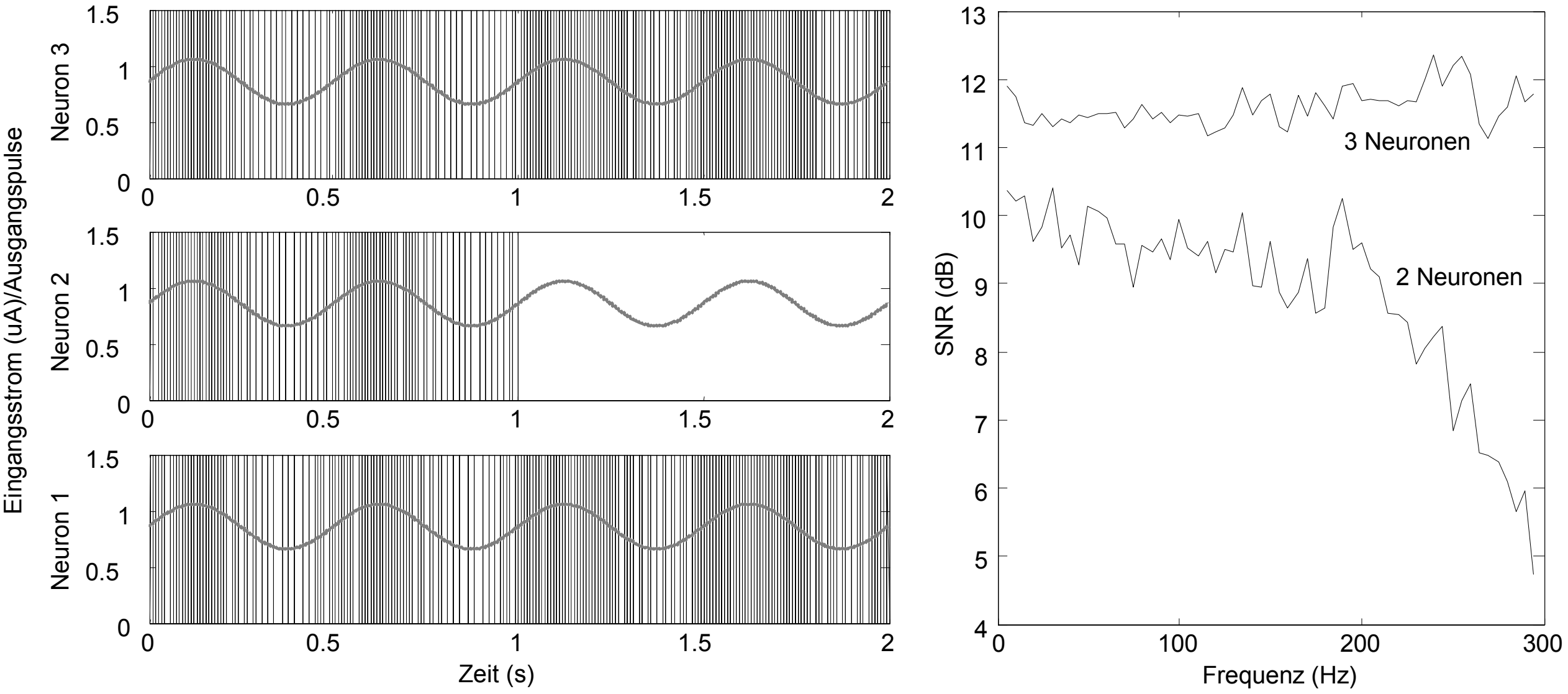

**Abbildung II.15.: Signalübertragung in einer Population aus drei inhibitorisch gekoppelten Neuronen, Pulsfolgen für ein 5Hz-Signal (links) und SNR im 0...300Hz Frequenzband als Funktion der Signalfrequenz für reguläre 3-Neuronen-Population und Ausfall von Neuron 2**

In einer Neuronenpopulation aus drei gegenseitig inhibitorisch gekoppelten Neuronen nach [Mar99] wird der Ausgang eines Neurons selektiv ab der Hälfte der Simulationszeit abgeschaltet. Wie in der linken Hälfte der obigen Abbildung ersichtlich, erhöhen dann die übrigen Neuronen durch die fehlende Inhibition von Neuron 2 ihre mittleren Pulsfrequenzen und übernehmen damit einen Teil der durch Neuron 2 ausgefallenen Übertragungskapazität. In der rechten Hälfte von Abbildung II.15 ist dies als SNR-Verlauf über der Signalfrequenz dargestellt, jeweils für die erste Hälfte der Simulationsdauer (mit regulärer Population) und für die zweite Hälfte der Simulation mit reduzierter Population. Für niedrige Signalfrequenzen ist die o.a. Verlagerung der Signalübertragung in der Lage, beinahe die originale Übertragungsqualität wieder herzustellen. Mit steigender Signalfrequenz verringert sich jedoch die Redundanz zwischen den einzelnen Kanälen, d.h. die Pulse jedes Neurons enthalten zunehmend unterschiedliche Details des Signals, und die reduzierte Population ist nicht mehr in der Lage, dies in vollem Umfang zu übertragen. Korrelationscodes sind also auch in der Lage, v.a. hochfrequente Stimuli durch Aufteilung der Signalübertragung in deutlich höherer Güte zu übertragen, als dies über ein Einzelneuron möglich wäre [Spiridon99]. Diese Fähigkeit reicht bis hin zur Übertragung von Signalen, die über der durch die Membrankonstante und Refraktärzeiten vorgegebenen Maximalfrequenz eines Neurons liegen [Mar99, Mayr05b]. Abgesehen von der generell höheren Datenrate liegt hierin einer der hauptsächlichen Vorteile von Korrelationscodes gegenüber Pulsratencodierungen, da bei letzteren die intrinsische Grenzfrequenz der Neuronen nicht überschritten werden kann [Shadlen98]. Ähnlich geartete Prinzipien wie die geschilderte inhibitorische Kopplung/Verteilung der Signalübertragung auf mehrere Einzelelemente finden auch in technischen Anwendungen Einsatz [Poorfard97].

---

Die dritte Variante eines Populationscodes ist das so genannte Rank Order Coding (ROC), bei dem jedes Neuron eindeutig für einen bestimmten Aspekt der Signalübertragung zuständig ist, ohne Austauschbarkeit zwischen den Neuronen wie im obigen Code. Informationstragendes Merkmal eines ROC ist die Reihenfolge, in der Neuronen relativ zueinander Aktionspulse generieren. In [VanRullen01] wird ein solcher ROC für die Übertragung der retinalen DoG-Information implementiert, mit dem Ergebnis, dass ein ROC in der absoluten Genauigkeit und Übertragungsgeschwindigkeit anderen Codierungen deutlich überlegen ist. Ein biologisch plausibles Modell für die Decodierung von Populationscodes findet sich in [Shamir04], es werden über mehrere Neuronen verteilte Winkelinformation decodiert. Jedes Neuron überträgt einen bestimmten, dem Neuron eigenen Aspekt der Winkelinformation, das Auslesen erfolgt über eine synaptisch realisierte gewichtete Summe der informationstragenden Neuronen. Ein Neuron kann die entsprechende Gewichtsverteilung zur Auswertung der Eingangsinformation beispielsweise über eine STDP-Adaption lernen [Delorme01]. Die genannte Adaptionsregel kann die dem ROC zugrunde liegende zeitliche Struktur der Aktionspotential direkt in entsprechende synaptische Gewichtswerte umsetzen [Bi98].

Für die Betrachtungen der Informationsdichte eines Rank Order Coding in Abhängigkeit des zeitlichen Rauschens der einzelnen Pulse wird eine Population von $n$ Neuronen angenommen, die homogen mit einer mittleren Rate $\lambda$ pulsen. Für die Analyse der Codierung wird ein externes Masterneuron postuliert, von dessen Pulsen aus die zeitliche Reihenfolge der Pulse der Population festgestellt wird. Diese verschiedenen Reihenfolgen repräsentieren die Symbole des Codes. Der Erwartungswert an Pulsen, die sich zwischen zwei Pulsen des Masterneurons in der Population ereignen, ist trivialerweise $n$, da der Erwartungswert jedes einzelnen Neurons der Population für die Anzahl Pulse im Intervall $1/\lambda$ gleich eins ist. Die Anzahl möglicher Codeworte von $n$ unterscheidbaren Pulse in diesem Intervall ist $n!$, oder als Informationsgehalt in Bit $ld(n!)$. Rauschen auf den Zeitpunkten der Aktionspotentiale wird diese Symbolanzahl in zweifacher Weise beeinflussen. Zum Einen wird ein gewisser Prozentsatz der Pulse der Population so dicht nach dem Puls des Masterneurons liegen, dass aufgrund des Rauschens nicht mehr sicher davon ausgegangen werden kann, dass diese Pulse wirklich Codierung in der aktuellen Periode des Masterneurons darstellen. Diese Pulse werden deshalb komplett verworfen. Für die Ermittlung dieses Prozentsatzes wird die Verteilung der Intervalle zwischen dem Puls des Masterneurons und jeweils einem der Neuronen der Population benötigt. Eine Poissonverteilung in Form von Gleichung (II.16) kann auf dieses Problem angewendet werden, wenn angenommen wird, dass zwischen unterschiedlichen Neuronen keine Refraktärzeit o.ä. auftritt, d.h. die Populationsneuronen müssen unkorrelierte Aspekte des Stimulus codieren [Kass05, Warland97]. Der Anteil der Pulse $k_1$, die verworfen werden, ergibt sich durch Integration von Gleichung (II.16) bis zum maximalen zeitlichen Rauschen:

$$k_1 = \int_0^{\tau_{\max}} \lambda e^{-\lambda T} dT = 1 - e^{-\lambda \tau_{\max}} \qquad \textbf{(II.26)}$$

damit ergibt sich die Anzahl an Pulsen, die noch für eine Codierung verwendet werden können, zu:

$$n_{res} = n * (1 - k_1) \qquad \textbf{(II.27)}$$

Die zweite Ursache, welche die Anzahl für die Codierung verwendbarer Pulse reduziert, ist der Jitter von Pulsen der Neuronenpopulation untereinander. Dies wird im Weiteren für den vereinfachten Fall betrachtet, dass immer zwei Pulse sich zeitlich so nahe kommen, dass in Rahmen von $\tau_{\max}$ nicht mehr entschieden werden kann, welcher Puls zuerst aufgetreten ist. Wenn dies ($k_2$*$n$)-mal auftritt, reduziert sich die Anzahl der möglichen unterscheidbaren Symbole um den Faktor $2^{k_2 * n}$. Um den Prozentsatz $k_2$ der verbliebenen $n_{res}$ Pulse zu bestimmen, für den das ISI entsprechend klein ist, wird ähnlich vorgegangen wie bei Gleichung (II.26), wobei für $\lambda$ die Populationsrate $n*\lambda$ eingesetzt wird:

$$k_2 = \int_0^{\tau_{\max}} n\lambda e^{-n\lambda T} dT = 1 - e^{-n\lambda\tau_{\max}} \qquad \textbf{(II.28)}$$

Aus diesen Vorüberlegungen ergibt sich die Gesamtanzahl an Symbolen $N_{Sym}$ in Abhängigkeit des zeitlichen Rauschens:

$$N_{Sym} = \frac{[(1-k_1) * n]!}{2^{k_2 * n * (1-k_1)}} \qquad \textbf{(II.29)}$$

Oder in anderen Worten der Fakultät der Gesamtanzahl Pulse, vermindert um den Prozentsatz, der zu nah am Puls des Masterneurons liegt, dividiert durch die Möglichkeiten, die durch ein zu enges Auftreffen von Pulsen der Population untereinander nicht mehr unterscheidbar sind. Folgendes Zahlenbeispiel verdeutlicht die Auswirkungen dieser Betrachtung: Angenommen wird eine Population von $n$=10 Neuronen (plus Masterneuron), ein zeitliches Rauschen $\tau_{max}$ von 2ms und eine Einzelneuronenrate von $\lambda$=50s$^{-1}$. Dies resultiert in einen Prozentsatz $k_1$=9,52% nicht auswertbarer Pulse nach dem Puls des Masterneurons, d.h. einer der erwarteten 10 Pulse kann nicht verwendet werden, weil er zu dicht nach dem Aktionspotential des Masterneurons auftritt. Ein zu nahes Zusammentreffen von Pulsen der Population untereinander findet mit $k_2$=63,2% statt, mithin hat jeder Puls mindestens einen zu nahen Nachbarn. Unter Verwendung von Gleichung (II.20) ergibt sich für den Zähler des Bruchs aus Gleichung (II.29) ein Wert von 400,98*10$^3$ (mögliche Symbolanzahl nach Berücksichtigung Jitter am Anfang der Beobachtungszeit). Diese wird um den Faktor 52,7 reduziert (Nenner von Gleichung (II.29)). Die Anzahl nutzbarer Symbole $N_{Sym}$ reduziert sich damit auf ca. 7609. Die Reduktion durch zu enges Aufeinandertreffen von Pulsen der Population wird durch die obige Formel tendenziell unterschätzt, für das Zahlenbeispiel ergeben sich auch Triplets von Pulsen ($k_2$>50%), deren Berücksichtigung für einen Teil der Pulse einen Faktor von 6 (=3!) statt 2 (=2!) in der Basis des Ausdrucks im Nenner von Gleichung (II.29) benötigen würde. Trotz dieser starken Einschränkungen, die bei einem ROC-Ansatz v.a. durch das zeitlich nicht mehr genau auftrennbare Eintreffen von vielen Pulsen in kurzem Zeitraum zustande kommt, ist diese Codierungsmöglichkeit selbst bei starkem zeitlichen Rauschen den beiden anderen diskutierten Möglichkeiten deutlich überlegen [VanRullen01]. Abträglich für eine technische Verwendung ist seine geringe Fehlertoleranz, da durch die nicht vorhandene oder zumindest geringe Redundanz zwischen den einzelnen Neuronen/Kanälen fehlende Pulse individuelle Auswirkungen auf die Informationsübertragung haben. Somit ist eine ‚graceful degradation' unabhängig von dem Ort des Pulsverlustes nicht mehr gegeben. Biologische Codes scheinen zumindest im V1 eine Mischung aus ROC und den zuletzt diskutierten Korrelationscodes zu verwenden, insofern als ein visueller Stimulus zwar wie im ROC in einzelne Bestandteile zerlegt und individuelle übertragen wird, jedoch diese Übertragung wieder über mehrere Neuronenunterpopulationen/Kanäle parallel-redundant stattfindet, wodurch ähnlich wie im Korrelationscode die Nachbarn desselben Kanals für Pulsverluste eines einzelnen Neurons einspringen können [Guyonneau05, Olshausen02, Zeitler06].

### II.2.4 Verarbeitung durch Membran-Übertragungsfunktion

Viele Analysen von PCNNs beschäftigen sich in erster Linie mit den Auswirkungen der membranbasierten Integrationsfunktion auf die Gesamtverarbeitung des Netzwerks, Synapsen werden als zeitkonstant angenommen. Dies ist zum Einen der Tatsache geschuldet, dass durch diese Vereinfachung eine aussagekräftige mathematische Beschreibung möglich ist. Bei Berücksichtigung von weiteren Nichtlinearitäten oder beispielsweise synaptischer Adaption ist dies entweder prinzipiell nicht mehr möglich oder lässt zumindest keine analytischen Aussagen über das Netzwerkverhalten mehr zu. Zum Anderen können viele interessante Anwendungen aber auch bereits mit diesem reduzierten Modell realisiert werden. Da beispielsweise ein mit Leckstrom behaftetes IAF-Neuron (Leaky Integrate and Fire Neuron, LIAF) für entsprechend zeitlich getaktete Eingangspulse einen Koinzidenzdetektor darstellt, eignet es sich zum Aufbau pulsbasierter Logikgatter [Maass99]. Über die Verbindungsgewichte lassen sich gewichtete Summen aufbauen [Shadlen98] oder z.B. einzelne Pulse um einen bestimmten Betrag verzögern, so dass komplexe zeitliche Zusammenhänge ausgewertet werden können oder PCNNs wie statische Klassifizierernetze als universelle Funktionsapproximierer über ihre Eingänge fungieren [Maass99]. Wenn ein Ratencode angenommen wird, können IAF-Neuronen für bestimmte Arbeitsbereiche eine Ratenmultiplikation ausführen [Koch99 (Kapitel 17), Maass99]:

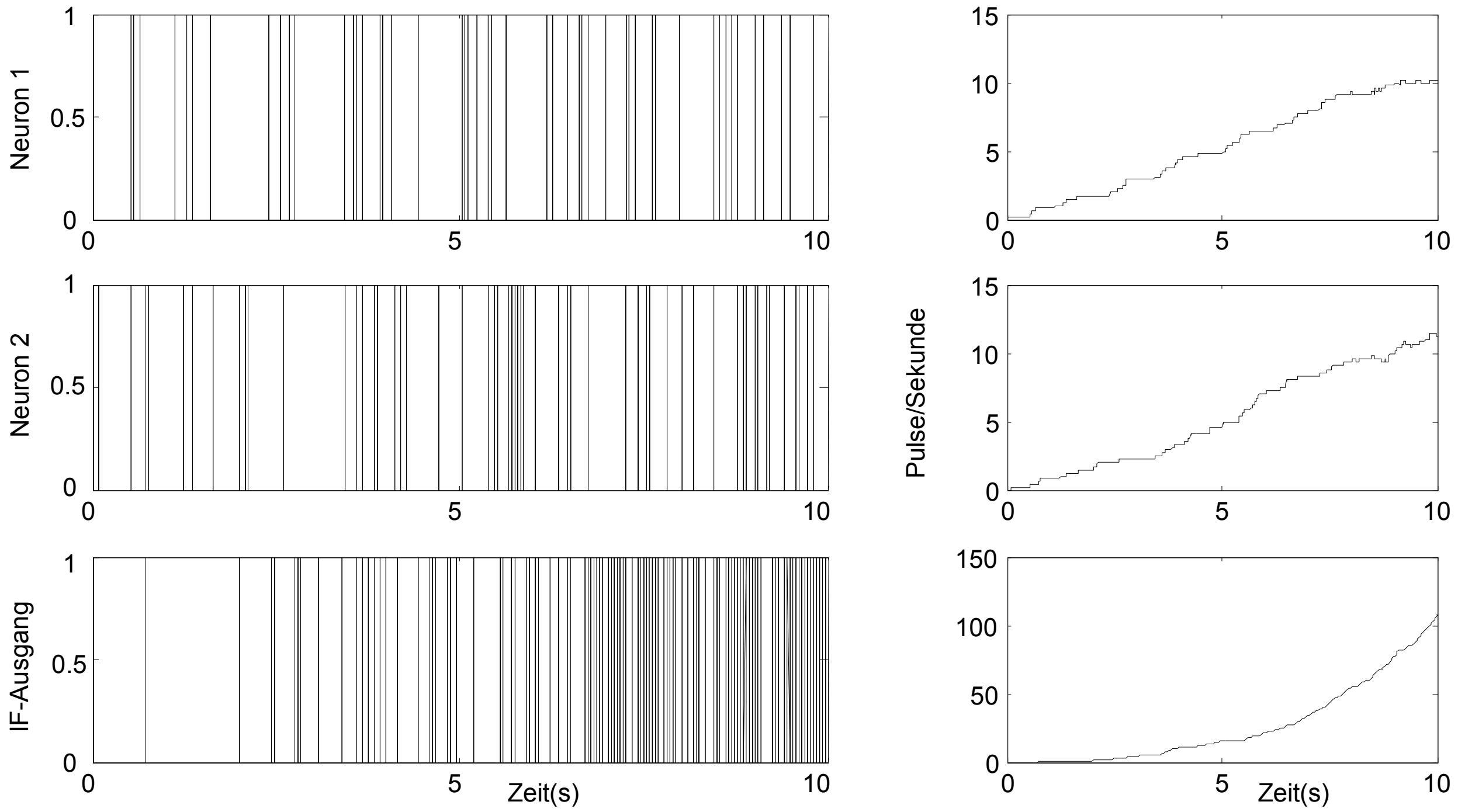

**Abbildung II.16.: Pulsratenmultiplikation bei einem IAF-Neuron; Poisson-Eingang aus Neuron 1 und 2 und Ausgang IAF-Neuron, Pulsfolgen mit stereotyper Amplitude (links) und Glättung mit 3s Zeitfenster (rechts)**

Weitere interessante Netzwerkdynamiken ergeben sich bei Rückkopplung der Pulsausgänge auf den Eingang, und zusätzlichem Eingangssignal. In Abhängigkeit des Eingangssignals entstehen dann über die zeitlichen Dynamiken der Integration verschiedene intrinsische Zustände, welche in eine verbesserte Klasseneinteilung der Eingangssignale resultieren können [Atmer03, Maass02]. Diese extern abrufbaren intrinsischen Zustände können als mögliche Modellierung des Gedächtnisses dienen, eine allgemeine Betrachtung findet sich in [Maass06]. Vogels et. al. [Vogels05] geben ein konkretes Beispiel, bei dem ein Netzwerk aus IAF-Neuronen in Abhängigkeit des Eingangssignals logische Pegel am Ausgang abrufen kann.

Derart rückgekoppelte Netze sind auch von Relevanz in der Signalübertragung, da in ihnen durch die Integration, (Puls-)Quantisierung des Signals und Rückkopplung des quantisierten Wertes/Pulses eine ähnliche Verarbeitung stattfindet wie in technischen Delta-Sigma-Modulatoren (siehe Anhang C.2). Durch diesen Mechanismus wird das im Abschnitt II.2.3 als Populationscode eingeführte neuronale Noise Shaping ermöglicht [Mayr05b, Norsworthy96, Spiridon99].

### II.2.5 Direkte Pulsinteraktion in Neuronenperipherie

Wie in Abschnitt II.1.3 und in [Häusser03, Poirazi01] ausgeführt, besitzt der Dendrit eine komplexe räumliche Struktur mit vielen Unterabschnitten mit variablen elektrischen und physiologischen Eigenschaften, mit verschiedensten Verschaltungen mit lokalen Axonen, etc. Eine Vereinfachung dieser Struktur wie in den vorigen Abschnitten auf ein einzelnes Soma, in dem alle eingehenden Pulse gleichwertig und mit derselben dynamischen Charakteristik behandelt werden, vernachlässigt einen Großteil der dendritischen Verarbeitungsmöglichkeiten [Poirazi01]. Bei genauerer theoretischer und biologischer Betrachtung des Dendriten ergibt sich eine große Zahl von Möglichkeiten der zeitlichen und spatialen Interaktion von Aktionspotentialen auf Dendritenabschnitten, bei der einzelne Pulse analoge [Poirazi01] oder digitale [Koch99 (Abschnitt 19.3.2)] Verarbeitung untereinander ausführen können.
Aktionspotentiale können auf den Dendriten in ähnlicher Weise analog verarbeitet werden, wie dies in [Maass99] und im vorherigen Abschnitt für eine zusammengefasste Verarbeitung auf dem Soma beschrieben wird: Zeitliche Korrelationen zwischen eingehenden Pulsen in Verbindung mit synaptischen und Membran-Zeitkonstanten können Pulse selektiv addieren, verzögern, subtrahieren, multiplizieren, mit Schwellwerten versehen, etc. Durch eine Erweiterung und Differenzierung dieser Verarbeitungsmöglichkeiten in einzelne Abschnitte des dendritischen Baumes potenziert sich die Anzahl der implementierbaren Interaktionen, hier fehlt jedoch noch die biologische Bestätigung [Poirazi01]. Manche Messdaten deuten sogar darauf hin, dass die unterschiedlichen physiologischen Randbedingungen entlang des Dendriten eher dafür geschaffen sind, etwaige Unterschiede in der synaptischen Einspeisung zu korrigieren, z.B. weiter von der Soma entfernte Aktionspotentiale zu verstärken und zu beschleunigen, um sie in möglichst gleicher Weise auf das Soma wirken zu lassen [Häusser03].
Es gibt jedoch starke Hinweise aus der Biologie, dass dendritische Verschaltungen zumindest dafür verwendet werden, über entsprechende lokale Mikroschaltungen quasi-logische Interaktionen hervorzurufen, d.h. Pulse selektiv zu blockieren oder zu übertragen, oder logische Grundfunktionalitäten wie AND- und OR-Gatter zu realisieren [Blum72], indem z.B. lokale Ionenkanäle durch einen postsynaptischen Puls für kurze Zeit so erschöpft werden, dass keine Übertragung eines weiteren Pulses entlang des Dendriten möglich ist [Koch99 (Abschnitt 19.3.2)], oder exzitatorische Pulse durch nachgeschaltete inhibitorische Synapsen blockiert werden:

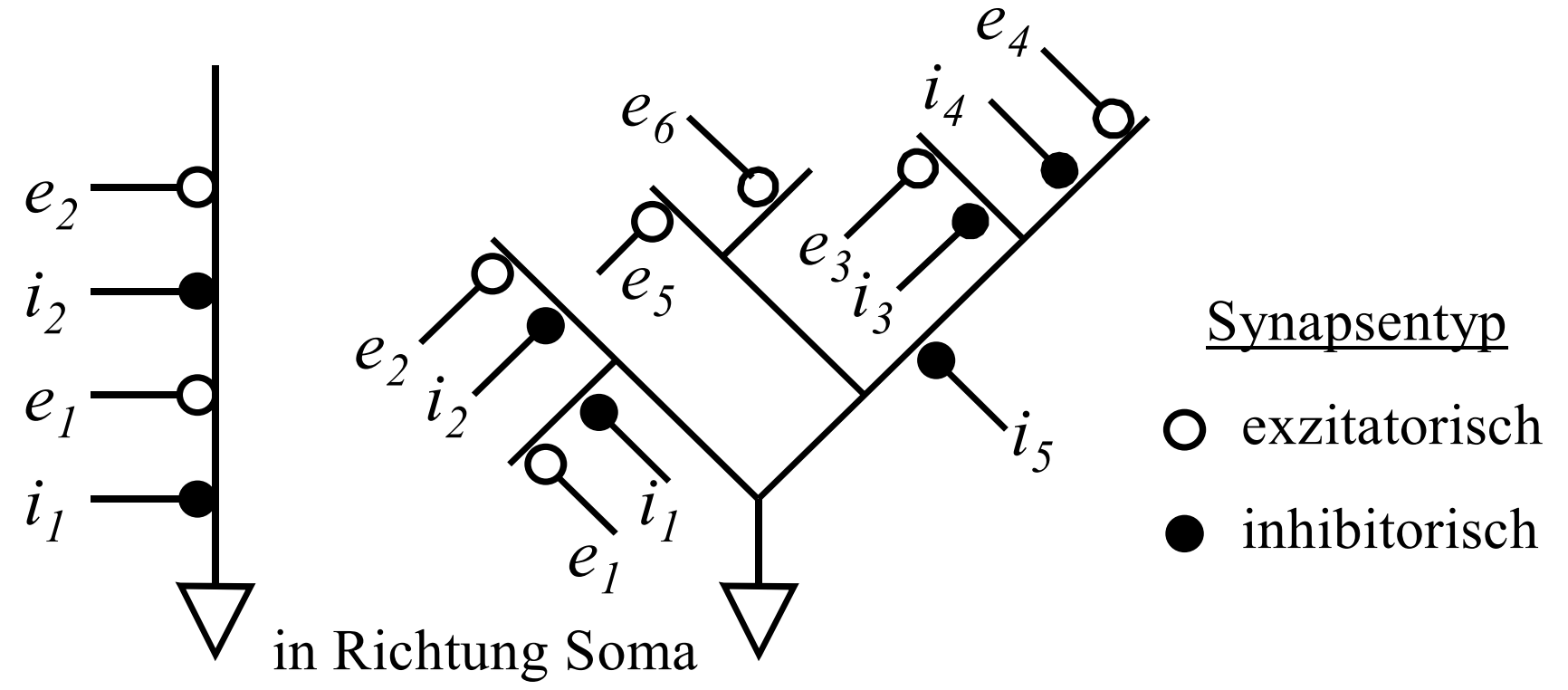

**Abbildung II.17.: Dendritischer Baum mit möglicher Verschaltung, nach [Shepherd04(Kapitel 1)]**

In der linken Hälfte von Abbildung II.17 wird vom Soma aus gesehen beispielsweise die Operation $(e_2\cap(\overline{i_1}\cup\overline{i_2}))\cup(e_1\cap\overline{i_1})$ ausgeführt, mithin blockiert ein entweder auf $i_1$ oder $i_2$ eintreffendes Aktionspotential einen zeitnah stattfindenden Puls von $e_2$, ein Puls von $e_1$ kann von $i_1$ ausgeblendet werden. In einem verzweigten dendritischen Baum sind vielfältige weitere Verschaltungen möglich, etwa $e_5\cup e_6$ oder $[(e_3\cap\overline{i_3})\cup(e_4\cap\overline{i_4})]\cap\overline{i_5}$. Bei Nutzung dieser Möglichkeiten durch die Biologie in entsprechenden makroskopischen Zusammenhängen ergibt sich auch hier ein großes Feld möglicher neuronaler Funktionalität [Häusser03, Shepherd04 (Kapitel 1)].

### II.2.6 Topologiebasierte Verarbeitung

Wie in Abschnitt II.1.3 dargestellt, wird im Kortex aus den im letzten Absatz geschilderten fundamentalen Verschaltungen von Dendriten eine große Bandbreite an Netzwerkstrukturen aufgebaut. Für verschiedenste Netzwerkgrößen und Bereiche des Kortex wurde durch Messungen und Simulationen ein klarer Zusammenhang zwischen Struktur und Funktion etabliert, z.B. für die Relais- und Gain-Control-Funktion des LGN [Sherman96]. Die Struktur der lokalen Verschaltungen im MT-Bereich des visuellen Kortex lässt sich über die in diesem Bereich nachgewiesenen Segmentierungseigenschaften und Objektvervollständigung erklären [Eckhorn99]. Im V1-Bereich des visuellen Kortex lässt sich der hierarchische Aufbau der rezeptiven Felder über die vertikale Struktur der Kolumnen erklären, ebenso die horizontale Verteilung der Charakteristiken der rezeptiven Felder über die parallele Anordnung der Kolumnen und ihre gegenseitige laterale Verschränkung durch inhibitorische Verbindungen [Shepherd04 (Kapitel 12), Delorme03a]:

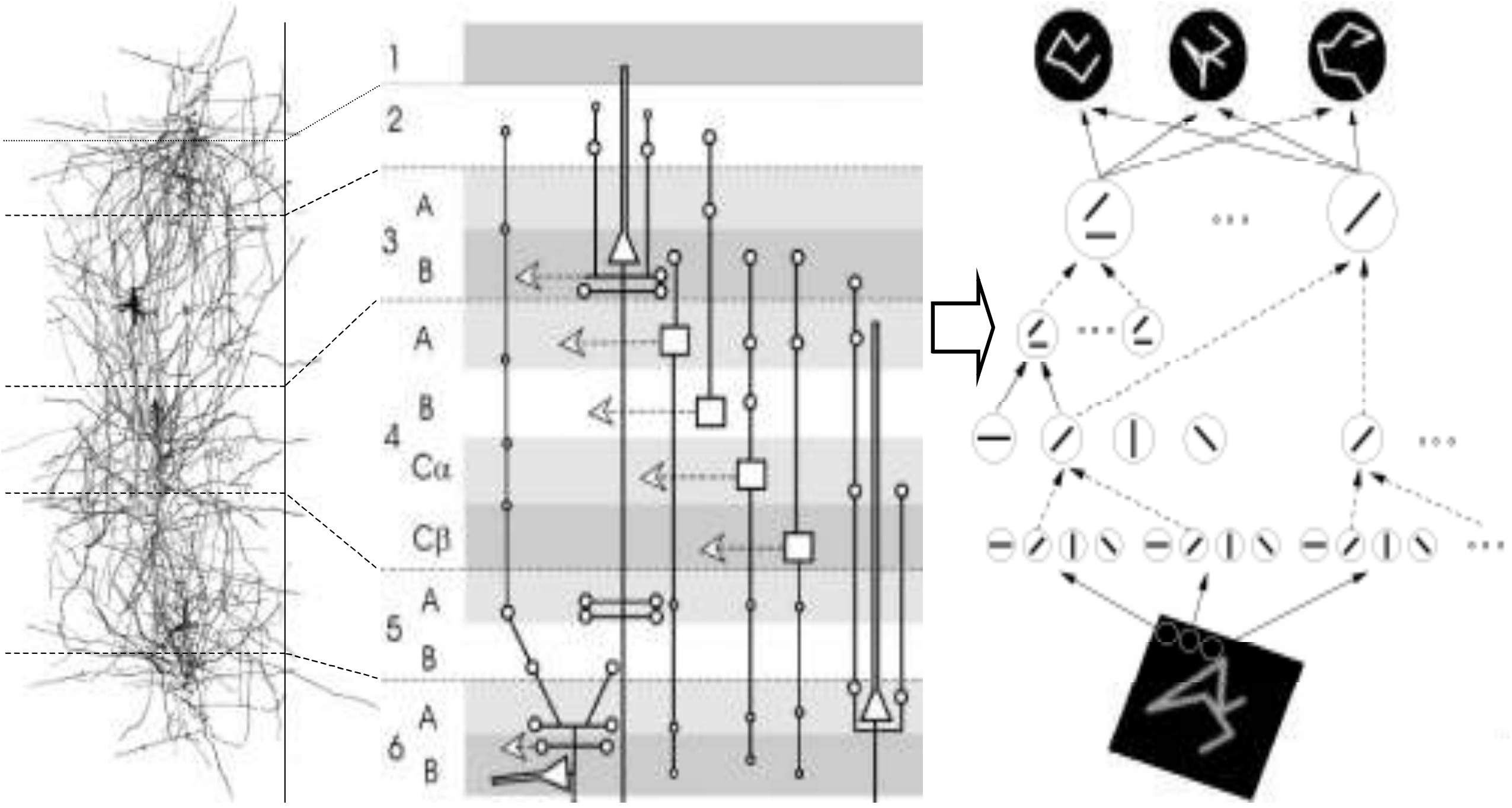

**Abbildung II.18.: Eine Minikolumne des V1, (von links nach rechts): ausgewählte Neuronen der Minikolumne, Vernetzung/Abhängigkeiten und Makromodell der Verarbeitung [Riesenhuber99, Thomson03]**

Diverse mathematische/algorithmische Ansätze versuchen, diesen Zusammenhang zwischen Struktur und Funktion zu analysieren, indem beispielsweise die Graphen der Netzwerke verschiedenen Projektionen unterworfen werden, welche Verbindungen, Gewichte oder interne Zustände berücksichtigen, um versteckte Abhängigkeiten sichtbar zu machen (siehe Abbildung II.8 und [Blinder05]). Andere Ansätze versuchen, anhand einer vorgegebenen Funktion und den biologischen Randbedingungen die spezifische dreidimensionale Netzwerkstruktur nachzubilden [Chklovskii04].

In technischen Schaltungsadaptionen neuronaler Verarbeitung lässt sich ebenfalls ein direkter Zusammenhang zwischen Vernetzungsstruktur und Funktionalität herstellen [Erten99], wobei auch hier die Komplexität dieser Strukturen einen weiten Bereich überspannt, angefangen bei Mikroschaltungen aus wenigen Neuronen [Heittmann04]. Die nächste Stufe wäre eine Organisationskomplexität ähnlich der Retina, mit mehr Neuronen, aber einer einfachen, sich wiederholenden lokalen Vernetzung in einer [Schreiter04, Mayr05d] oder mehreren Schichten [Morie01]. Das obere Ende technischer neuronaler Implementierungen besteht aus Modellen, die in biologie-ähnlicher Weise und Komplexität [Riesenhuber99] substanzielle Teile des Bild-verarbeitungspfades nachempfinden [Mayr07c].

Im weiteren Sinne enthält eine Topologie nicht nur binär gewichtete Netzwerkverbindungen, d.h. eine Verbindung existiert oder existiert nicht, sondern analoge Gewichte, die Informationen wesentlich differenzierter übertragen können. Damit sind selbst in Netzen mit zufällig generierten, strukturlosen Verbindungen über eine entsprechende Wahl der Verbindungsgewichte Strukturen implementierbar, die z.B. eine gerichtete Signalübertragung oder pulsbasierte Logikschaltungen ermöglichen [Vogels05].

### II.2.7 Synaptische Plastizität als Verarbeitungsfunktion

Wie im vorigen Abschnitt geschildert, stellen synaptische Gewichte ein wichtiges Mittel bereit, Funktionalität in neuronalen Schaltungen zu erreichen und zu verankern. Wissenschaftliches Interesse weckt hierbei die Art und Weise der Topologieentstehung, d.h. die Veränderung der synaptischen Gewichte $W$ als Antwort auf neuronale Lernvorgänge. Diese so genannte synaptische Plastizität, wird seit über fünfzig Jahren sehr aktiv erforscht, angefangen mit dem klassischen Postulat von Donald Hebb [Hebb49]:

> "When an axon of cell *A* is near enough to excite a cell *B* and repeatedly or persistently takes part in firing it, some growth process or metabolic change takes place in one or both cells such that *A*'s efficiency, as one of the cells firing *B*, is increased"

Diese These geht von einer Verstärkung des Synapsengewichts aus in Abhängigkeit einer gewissen Korrelation zwischen der Aktivität des prä- und postsynaptischen Neurons. Es wurden seit dieser Arbeit viele Lernregeln mit biologischem und/oder informationstheoretischem Hintergrund aufgestellt, die anhand einer Variante des Hebbschen Postulats versuchen, eine bestimmte Netzwerkfunktionalität zu erreichen [Koch99 (Kapitel 13)]. Hierbei kommen Lernregeln zum Einsatz, die auf den verschiedenen oben diskutierten Aspekten neuronaler Codes aufbauen, z.B. Lernregeln, die auf Raten basieren [Bienenstock82], oder auf relative Phasen [Bi98] oder Reihenfolgen [Heittmann04]. Beispiele für den Zusammenhang zwischen Netzstrukturen, Funktionalität und einer STDP-Lernregel, sind in [Song01, Yao05] illustriert, v.a. für das Lernen von rezeptiven Feldern im visuellen Kortex anhand von natürlichen Stimuli.
Plastizität, d.h. Lernvorgänge, welche ein bestimmtes Verhalten/Funktionalität über einen längeren Zeitraum erwerben und auch über längere Zeit verankern, arbeiten in erster Linie an den Parametern $n$ und $q$ von Gleichung (II.8), wobei sich die Ausschüttungsmenge von Neurotransmittern $q$ über einen Zeitraum von einigen Stunden verändern lässt, während die Anzahl an Synapsen $n$ zwischen einem bestimmten Dendriten und Axon in einem Zeitraum von Tagen oder Wochen verändert wird [Koch99 (Kapitel 13)]. Plastizitätsvorgänge, welche $n$ verändern, beeinflussen nicht nur das synaptische ‚Gewicht', sondern sie können auch Verbindungen komplett neu entstehen lassen. Hierbei steuert der Lernvorgang synaptische Wachstumsprozesse, indem Synapsen sich spontan selbst bilden, und dann entsprechend ihrer ‚sinnhaften' Verwendung wieder abgebaut oder verankert werden [Poirazi01].
Veränderungen an $n$ können auch das Verhalten einer Verbindung zwischen zwei Neuronen für den jeweiligen Verwendungszweck beeinflussen. Mit einer Ausschüttungswahrscheinlichkeit $p$ von 0,1 bis 0,3 für eine einzelne Synapse je nach neuronalem Gewebe sind einzelne Aktionspotentiale mit einer starken Unsicherheit in der Übertragung behaftet. Abhängig von der Anzahl der Ausschüttungsstellen (=Synapsen zwischen zwei bestimmten Neuronen) kann die Gesamt-wahrscheinlichkeit für den Verlust eines Aktionspotentials jedoch moduliert werden:

$$p_{verlust} = (1-p)^n \qquad \textbf{(II.30)}$$

Somit kann entweder für ein großes $n$ eine sehr sichere Übertragung eines APs gewährleistet werden, etwa bei sensorischen Neuronen, oder die Wahrscheinlichkeit als informationsverarbeitendes Prinzip verwendet werden [Koch99 (Kapitel 13)]. In [Senn02] wird postuliert, dass die Ausfälle von APs ein Mittel sein können, gelernte Synapsen intermittierend auszuschalten, um zu testen, ob diese noch sinnvoll sind. Eine weitere mögliche Verwendung dieser Wahrscheinlichkeit besteht in Kurzzeitadaptionen, da *p* mit Zeitverläufen im 1-100 ms Bereich verändert werden kann [Koch99 (Kapitel 13)]. In [Markram98] werden biologische Messungen vorgestellt, die ein differenziertes Übertragungsverhalten einer Synapse in Abhängigkeit der kurz zuvor stattgefundenen Aktivität belegen. In einem mittleren präsynaptischen Frequenzbereich werden Pulse linear übertragen, d.h. der durchschnittliche postsynaptische Strom steigt linear mit der präsynaptischen Pulsfrequenz. In einen niedrigfrequenten Eingangssignal von 1 Hz werden hingegen die ersten Pulse unterproportional weitergegeben, während nach den ersten 2-5 Pulsen eine integrative Adaption stattfindet, die diese Pulse proportional weitermeldet. In höheren Frequenzbereichen existiert eine differenzierende Adaption, d.h. nur die ersten Pulse eines hochfrequenten Eingangssignals werden mit einem linearen Zusammenhang zwischen Frequenz und postsynaptischem Strom übertragen, bei fortdauerndem hochfrequentem Signal findet eine unterproportionale Weitergabe statt (siehe Abbildung A.1).
Ebenfalls in diesen Bereich der Kurzzeitadaption fällt die Fähigkeit des V1, für rezeptive Felder eine ähnliche Kontrastadaptierung zu implementieren wie Retina und LGN für die DoG-Charakteristiken [Meister99, Smirnakis97], mit der absolute Filtermaskenantworten stark in ihrem Dynamikbereich eingeschränkt werden, aber Änderungen in der Maskenantwort überproportional weitergemeldet werden [Ohzawa82].

### II.2.8 Schlussfolgerung

Für die biologische Verwendung von jedem der in den vorherigen Abschnitten diskutierten neuronalen Codes/Lernverfahren existieren messtechnische Indizien, deshalb besteht eine starke Kontroverse zwischen den o.a. Theorien. Verschiedene Autoren [Izhikevich03, Kass05, Koch99] postulieren jedoch, dass diese Theorien nur unterschiedliche Aspekte derselben Verarbeitung darstellen, z.B. lässt sich die Pulswandlung in den Ganglienzellen der Retina zugleich als Raten- und als Phasencode betrachten [Meister99, VanRullen01]. Der Konsens scheint zu sein, dass sich das Gehirn auf mehreren Ebenen statistischer Methoden bedient, um die Information aus Pulsen zu extrahieren, verarbeiten und weitersenden zu können. Rapide, aber unzuverlässige und grobe Signalverarbeitung findet innerhalb kurzer Zeit mit wenigen Neuronen statt, sie ermöglicht wichtige, einfache Entscheidungen. Hingegen ermöglichen Statistiken über größere zeitliche oder spatiale Zusammenhänge eine detailliertere Verarbeitung und z.B. das Ausführen von Lernvorgängen, welche die entsprechende Verarbeitung für zukünftige Verwendung optimieren. Interessanterweise scheint neuronale Fehlerkorrektur, z.B. Jitterverbesserung, ebenfalls auf verschiedenen zeitlichen Auflösungen stattzufinden [Stiber05].
Von einer direkten Verwendung der o.a. neuronalen Codemodelle wird bei technischen Anwendungen oft abgewichen, um eine a priori festgelegte Funktionalität zu erreichen, für die z.B. kein biologisches Äquivalent existiert. Dies trifft v.a. auf ratenbasierte Netze mit statischen Schrittfunktionsneuronen zu, die in ihrer Standardanwendung als Klassifizierer breite Anwendung gefunden haben, jedoch sowohl in Informationsrepräsentation als auch -verarbeitung sehr starke Abstraktionen der Biologie darstellen. In [Shepherd04 (Kapitel 12)] wird ein diesbezüglicher Abstraktions-Vergleich von Hopfield-Netz mit biologischen Netzstrukturen des Kortex angestellt, wobei wenig biologische Relevanz dieser künstlich geschaffenen Netze gefunden wird.
In den letzten Jahren werden jedoch auch verstärkt näher mit der Biologie verwandte Neuromodelle in technischen Anwendungen eingesetzt, die teilweise biologische Lernregeln verwenden und nicht mehr auf raten- sondern auf pulsbasierten Modellen aufbauen [Koickal06]. Die implementierten Verarbeitungsschemata versuchen hierbei meistens, ein oder mehrere Aspekte der oben diskutierten Modelle aufzugreifen, also z.B. die Information über Populationen aufzuteilen [Marienborg02] oder etwas auf mehreren zeitlichen Auflösungsstufen zu operieren [Mayr06d]. Im folgenden soll eine mehrdimensionale Taxonomie entsprechender simulativer and technischer neuronaler Modelle versucht werden.

# II.3 Neuronale Nachbildungen und Simulationen

## II.3.1 Einteilung nach Art der Verhaltensbeschreibung

Modelle, die das Verhalten biologischer neuronaler Strukturen nachbilden, lassen sich zum Einen anhand der Zielsetzung ihrer Verhaltensbeschreibung klassifizieren:

| | |
|---|---|
| Anwendungsorientiert | Netzstrukturen und neuronale Elemente werden nur ansatzweise und i.d.R. stark vereinfacht vom biologischen Vorbild übernommen, technische oder informationstheoretische Anwendungen stehen im Vordergrund. |
| Phänomenologisch | Strukturen und Elemente werden aus der Biologie in abstrakte, verhaltensorientierte und verallgemeinerte Modelle überführt |
| Biophysikalisch | Die Modellierung findet anhand detaillierter biologischer Zustandsgrößen statt. |

**Tabelle II-3.: Einteilung von Neuro-Modellen nach Verhaltensbeschreibung**

Beispiele für die erste Kategorie finden sich in [Atmer03, Schreiter04]. Meist handelt es sich hierbei um Anwendungen in der Bild- oder Datenanalyse, bei der das Prinzip gekoppelter Oszillatoren für Segmentierung oder Klassentrennung verwendet wird [Schreiter04]. PCNN-Pulswandlungen von analogen Eingangsgrößen werden auch zur Transformation von Merkmalsräumen verwendet, d.h. bestimmte Eigenschaften des von den Eingangsgrößen repräsentierten Datenobjekts lassen sich in pulsgewandelter Form leichter erkennen [Atmer03, Maass02, Verstraeten05].
Wie in Abschnitt II.2.1 eingeführt, lässt sich in einer weiteren Vereinfachung neuronaler Dynamik für biologische Neuronen bzw. deren HH-Modell für Pulsfolgen mit Poisson-Verteilung eine statische Übertragungskennlinie definieren. Netze, die mit so gearteten Neuronen aufgebaut sind, finden sich in vielen technischen Anwendungen zur Klassifikation und Rekonstruktion von Signalen/Mustern [Hopfield84, Zhang00]. Diese Netze vernachlässigen in der Regel die zeitdynamische Struktur und den in Zustandsgrößen implizit vorhandenen Speicher, ihr „Gedächtnis" besteht nur aus den in den Gewichten abgespeicherten Mustern.
Eine phänomenologische Beschreibung versucht, ein in biologischen Neuronen beobachtetes Verhalten nur anhand seiner Auswirkungen zu beschreiben, ohne dabei auf die zugrunde liegenden biologischen Vorgänge einzugehen. Meist wird hierbei eine große Anzahl Neuronen in vielen Versuchen mit stereotypen Eingangsmustern erregt und statistisch signifikante Vorgänge herausgefiltert. Ein Beispiel hierfür ist das „Spike-Timing-Dependent-Plasticity"-Modell (STDP), das Langzeitlernvorgänge an Synapsen mittels des Aufeinandertreffen von prä- und postsynaptischen Pulsen beschreibt, wobei das Lernverhalten meist in einer einzigen Gleichung zusammengefasst wird [Abbott00, Kepecs02]. Teilweise werden auch biologische detailliert bekannte Vorgänge wie z.B. das dynamische Verhalten der Ionenkanäle im HH-Modell in einfacherer Weise zusammengefasst, etwa als Leitwertmodell [Destexhe97] oder als „Integrate-and-Fire" (IAF) Neuron [Gerstner02]. Diese Vereinfachungen werden zum Einen vorgenommen, um Rechenzeit in der Simulation neuronaler Netze zu sparen, aber auch wegen ihrer Übersichtlichkeit, die detaillierte simulative und mathematische Analysen etwa der Signalübertragungseigenschaften, oder des generellen Netzverhaltens ermöglicht. Vereinfachende phänomenologische Verhaltensmodelle werden auch benötigt, um technisch handhabbare VLSI-Schaltungsnachbildung von neuronalen Netzen zu ermöglichen, da dort in der Regel nicht der zugrunde liegende biologische Vorgang mit allen Zustandsvariablen emuliert werden kann, sondern eine an schaltungstechnische Möglichkeiten angepasste Variante [Schemmel04].
Vor allem für Neuronen existieren sehr viele Modelle technischer Implementierungen, die versuchen, verschiedene Bereiche des Neuronenverhaltens nachzubilden [Izhikevich04b], zum Beispiel beide oben angeführten Modelle, oder das Quadratic-IAF-Modell, das das Burst-Verhalten genauer nachstellt als das Standard-IAF-Modell. IAF-Modelle werden auch um absolute und relative Refraktärzeiten erweitert, um näher an das zeitdynamische Verhalten echter Neuronen bzw. des HH-Modells zu gelangen [Indiveri03].
In einer biophysikalischen Beschreibung neuronaler Bauteile werden biologische Zustandsgrößen ohne nennenswerte Vereinfachung oder Zusammenfassung direkt als Variablen wiedergegeben, um eine möglichst detaillierte Modellierung aller in Neuron, Synapse, Dendriten und/oder Axon stattfindenden Vorgänge zu erhalten. Generell versuchen diese Modelle eine Brücke zu schlagen zwischen verschiedenen Arten der biologischen Messung, zum Einen der meist anhand von Pulsfolgen charakterisierten reinen Input/Output-Zusammenhänge, und den detailliert an einzelnen Zellen gemessenen Zeitverläufen von biologischen Zustandsgrößen, z.B. Ionen- oder Neurotransmitterkonzentrationen.
Hierbei zu nennen wären erweiterte, detaillierte HH-Formalismen als Neuronenmodell, z.B. zur Untersuchung des Frequenzverhaltens von LGN-Relaiszellen [Mukherjee95]. Für synaptische Lernvorgänge wie das o.a. STDP-Verhalten existieren ebenfalls auf biologischen Messdaten basierende biophysikalische Modelle [Saudargiene04]. Unabhängig von der speziellen Lernregel, die die Synapse ausführt, lässt sich an dieser Stelle ebenfalls noch die Realisierung der Synapse anführen, wie in Abschnitt II.1.2 beschrieben. Diese reicht von einem verhaltensorientierten einzelnen Synapsengewicht, wie es meistens in schaltungstechnischen Synapsen verwendet wird, bis zu biologisch realistischen Einzelmodellen für n, p und q (Abschnitt II.1.2). Beispielsweise existiert wie bereits erwähnt von Maass et. al. eine auf Messungen beruhende Biologie-basierte Modellierung der $p$-basierten Kurzzeitadaption einer Synapse [Markram98].

### II.3.2 Einteilung nach Granularität

Eine weitere Einteilung lässt sich bei Betrachtung der Granularität neuronaler Modelle treffen, d.h. mit welcher Detailtreue werden die einzelnen Komponenten modelliert:

| | |
|---|---|
| Nur Neuron | Nur das Neuron als nichtlinearer Schwellwertschalter wird betrachtet, die Koppelgewichte zwischen Neuronen werden nach a priori festgelegtem Schema statisch vergeben. |
| Neuron und Synapse | Beide Grundelemente neuronaler Verarbeitung werden modelliert, wobei die Synapse in der Regel relativ eng auf einen bestimmten Typ der Adaption/Verarbeitung beschränkt ist. |
| Multi-Kompartment | Die in den Abschnitten II.1.1 und II.1.2 beschriebenen Axone, Dendriten, Synapsen und Neuronen werden in eine variable Zahl von Unterabschnitten aufgeteilt, die jeweils mit eigenen Modellen bzw. Gleichungen beschrieben werden. |

**Tabelle II-4.: Einteilung von Neuro-Modellen nach Detailtreue**

Netzwerke der ersten Kategorie werden beispielsweise in ‚Liquid Computing'-Anwendungen verwendet, um Merkmalsräume anhand einer IAF-Verarbeitungsfunktion und der daraus entstehenden Netzwerkdynamik in leichter zu klassifizierende (Ausgangs-)Merkmalsräume zu projizieren [Verstraeten05]. Mathematische Vergleiche zwischen Klassifizierernetzwerken und Netzen aus dynamischen Neuronen werden i.d.R. auch anhand der ersten Kategorie getroffen, da diese aufgrund ihrer auf eine Funktion eingeschränkten zeitlichen Dynamik noch mathematisch erschließbar sind [Maass99].
Eine weitere Anwendung besteht wie eingangs erwähnt in verschiedenen Formen der unüberwachten Datensortierung und Klassifikation, bei der die durch Rückkopplung entstehenden Oszillatorstrukturen zur Synchronisation von Teilen des Netzwerks führen. Hierbei beeinflussen die statistischen Kenngrößen der zusätzlich an den (Neuronen-)Oszillatoren anliegenden Eingangsdaten die Größe, Zusammensetzung und Anzahl solcher Synchronisationsanhäufungen, d.h. Klassengebiete [Rhouma01]. Einige dieser Anwendungen verwenden auch bereits gesteuerte Anpassungen der Kopplungsgewichte zwischen den Neuronen, d.h. sie modellieren Synapsen, die beispielsweise das Klassifizierungsverhalten des Netzwerks beeinflussen sollen, etwa die Anzahl der entstehenden Klassen über eine mehr oder weniger starke wechselseitige Mitkopplung zwischen Gebieten mit unterschiedlicher Synchronisation [Nowotny03, Vogels05]. In den o.a. Anwendungen sind die synaptischen Lernregeln meist auf die Anwendung zugeschnittene synthetische Konstrukte [Schreiter04]. Aber auch in Nachbildungen biologischen Verhaltens wird häufig nur die Synapse als hauptsächliches verarbeitendes Element modelliert, wobei verschiedene Aspekte ihrer Plastizität nachempfunden werden, um eine bestimmte Informationsverarbeitung zu erhalten [Gerstner02, Gerstner99]. Der nächste Granularisierungs-Schritt wäre eine zusätzliche Einbeziehung einfacher Axone und Dendriten als Kanalmodelle [Häusser03], wobei diese jeweils als einzelnes HH-Modell o.ä. angesehen werden. Synapsen sind dann jeweils lokal an diese Membranabschnitte gekoppelt:

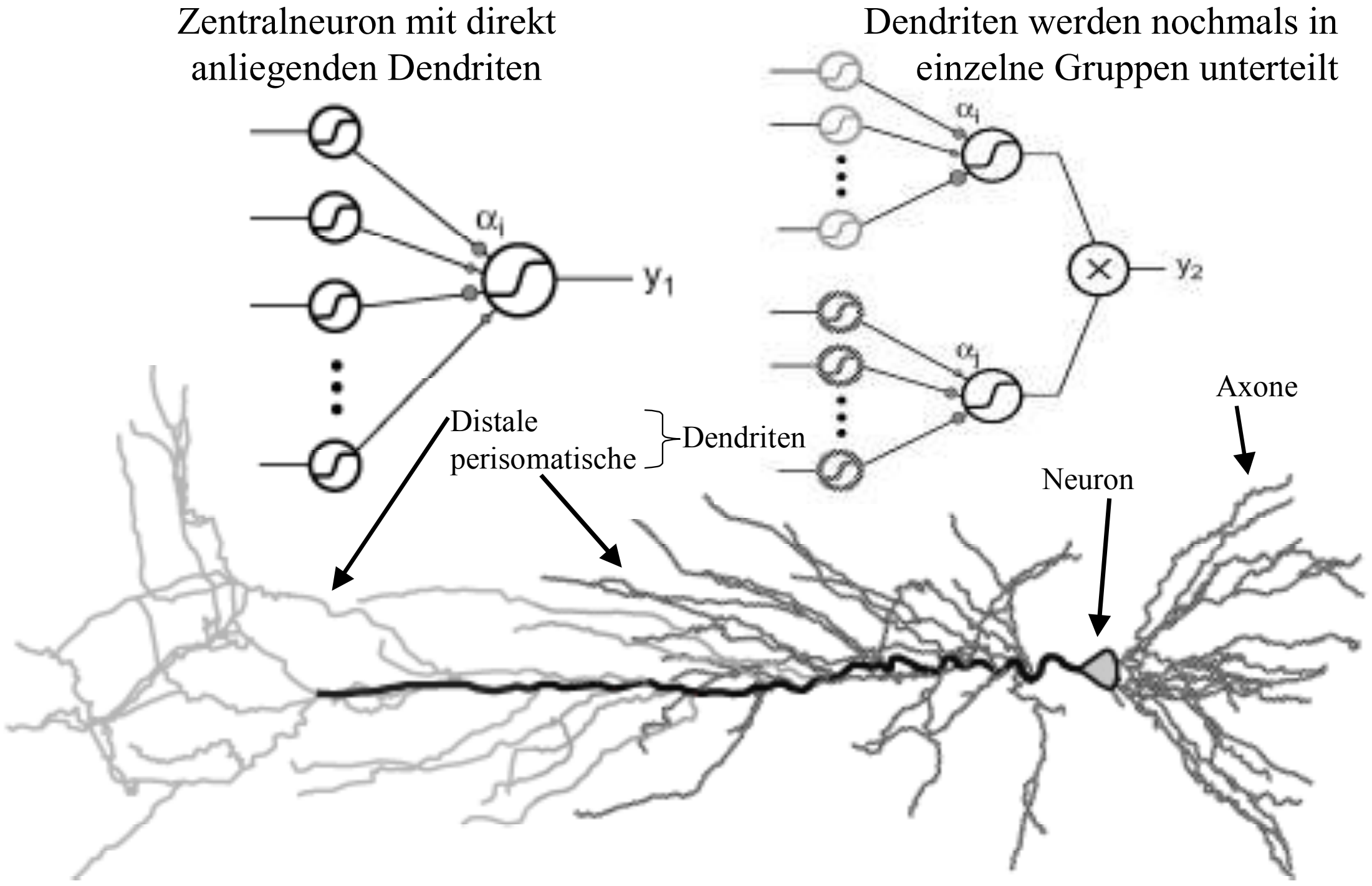

**Abbildung II.19.: Multi-Kompartment-Modell mit unterschiedlicher Granularisierung, adaptiert aus [Häusser03]**

Am oberen Ende dieser Skala wären dann Modelle, bei denen die Kette Synapse-Dendrit-Neuron-Axon in viele verschiedene Abteilungen (=Kompartmente) zerlegt wird [Koch99 (Abschnitt 3.3.2)], von denen jede mit Systeme von Differentialgleichungen beschrieben wird, die verschiedene biologische Zustandsvariablen und Vorgänge (z.B. Molekül- oder Ionenkonzentrationen) repräsentieren. Synapsen werden dann u.U. sogar einzeln modelliert, wobei ihre postsynaptischen Ströme (EPSCs und IPSCs) dann an den entsprechenden Stellen in die Kanalabschnitte eingebunden werden und durch die Kompartmente auf dem Weg zur Soma vielfältig modifiziert werden [Markram06] und selbst wieder lokale Veränderungen hervorrufen [Saudargiene04]. Häufig liegt dieser detaillierten Unterteilung auch ein entsprechender 3D-Aufbau zugrunde, bei dem dann die lokalen Synapsen nur bei einem räumlichen Zusammentreffen von Axon und Dendrit ausgebildet werden [Chklovskii04, Markram06, Poirazi01].

### II.3.3 Einteilung nach Kommunikation

Die dritte Einteilung neuronaler Modellierung lässt sich über die Kommunikation treffen, d.h. welche Bestandteile der sich entlang von Axonen und Dendriten fortpflanzenden biochemischen und elektrischen Vorgänge werden als relevant für die neuronale Informationspropagierung angesehen und deshalb in das Modell eingebaut:

| | |
|---|---|
| Pulsbasiert | Das Pulsereignis wird als eigentlicher Informationsträger angesehen, sein genauer Zeit-/Ortsverlauf wird als stereotyp angenommen. Somit wird nur Zeitpunkt und Adresse (pseudo-)digital zur Netzwerkkommunikation übertragen. |
| Analoges Aktionspotential | Die genaue Form des Aktionspotential wird als wichtig für Lernvorgänge, Verarbeitung, etc. angesehen. Evtl. werden auch weitere ortsveränderliche Zustandsgrößen als Informationsgrößen miteinbezogen. |

**Tabelle II-5.: Einteilung von Neuro-Modellen nach Detailtreue**

Meistens ist die entsprechende Modellierung eng verwandt mit der in der Tabelle aus Abschnitt II.3.2 geschilderten, d.h. wenn Kompartmente simuliert werden, wird auch ein analoges Aktionspotential angenommen, dagegen werden bei auf Synapsen bezogenen Modellen gerne nur Pulszeitpunkt und Adresse in Form einer Adress-Event-Representation (AER) weitergegeben, also nur das Pulsereignis per se als Information angesehen [Koch99 (Kapitel 14)] (siehe auch Abschnitt III.3.2). Diese Modelle haben in der Simulation den Vorteil, Netzwerkelemente analog zu entkoppeln, so dass deren DGLs zwischen ihren Feuerzeitpunkten unabhängig voneinander berechnet werden können [Delorme03b]. In Hardware ergibt sich der Vorteil pseudo-digitaler Signalübertragung, so können z.B. Standard-Digitaltreiber für die Signalleitungen verwendet werden. Es existieren auch Mischmodelle, bei denen z.B. zwar ein digitales (=AER) Signal das Aktionspotential signalisiert, der zeitliche Verlauf des Aktionspotential jedoch im empfangenden Dendriten oder Synapse wieder rekonstruiert wird [Schemmel04].
Argumente gegen eine Beschränkung auf Pulszeitpunkte finden sich z.B. in [Saudargiene04], wo das STDP-Lernverhalten vom zeitlichem Verlauf der Membranspannung (d.h. Aktionspotential) abhängig ist, weswegen sich entlang des Dendriten biologisch belegt unterschiedliches Plastizitätsverhalten ergibt.
Nennenswert sind in diesem Zusammenhang auch die Makromodelle, anhand derer die diskutierten Modellierungen von neuronalen Einzelelementen makroskopisch verschalten werden, um bestimmte Verarbeitungsfunktionen oder Signalpfade im Kortex oder anderen Hirnregionen nachzuempfinden. Eine Diskussion verschiedener Möglichkeiten, den Pfad der visuellen Informationsverarbeitung zu modellieren, findet sich z.B. in [Einevoll03, Riesenhuber99].

# II.4 Bedeutung für technische Bild- und Informationsverarbeitung

Als Abschluss der allgemeinen Betrachtungen zu neuronalen Verarbeitungsprinzipien soll hier nochmals ihre eingangs angeschnittene Relevanz für die Implementierung von technischer Bildverarbeitung und Informationsverarbeitung diskutiert werden.

## II.4.1 Parallelität

Für die technische Anwendbarkeit ist v.a. interessant die extrem hohe Parallelität, z.B. sind an einer Wandlung des kompletten visuellen Feldes in die in Abschnitt I.3.3 vorgestellten rezeptiven Felder im V1-Bereich des Kortex ungefähr $3*10^6$ Neuronen beteiligt, die in ca. 2-5 Verarbeitungsschritten ihre komplette Verarbeitung ausführen [VanRullen05], also das vom Sehnerv eintreffende Bild gleichzeitig mit etwa $1*10^6$ Neuronen verarbeiten [Shepherd04]. Ähnliche Zahlen treffen auch für die im selben Abschnitt diskutierten Einzelbausteine (Zelltypen) der Retina zu, die mit extrem hoher Parallelität eine Kontrastfilterung, Informationsverdichtung und Analog-zu-Phasenverzögerung (vergleichbar mit einem Analogwert-PWM-Wandler) des auf die Retina einfallenden Bildes durchführen [Meister99]. Ansätze zur Parallelisierung finden sich auch in der konventionellen Rechnertechnik [Dmitruk01], da aus Kostengründen oder technischen Grenzen die Beschleunigung der seriellen Datenverarbeitung mehr und mehr in den Hintergrund tritt. Die Parallelisierung reicht hier von den bei Dmitruk et al. angeführten Rechnerclustern bis zu

Mehrprozessorrechnern und Parallelverarbeitung im einzelnen Prozessor, wobei dies verglichen mit neuronaler Verarbeitung immer noch eine sehr grobkörnige Parallelität darstellt. Diese Grobkörnigkeit wird unter anderem durch die Tatsache vorgegeben, dass die Komplexität der Netze exponentiell mit dem Grad der Parallelisierung zunimmt, so dass sehr hohe Parallelitäten nicht mehr technisch beherrschbar sind [Herbert02].
In diesem Zusammenhang ist auch die Aufgabenverteilung auf einzelne Verarbeitungselemente zu erwähnen, die natürlich einen wesentlichen Anteil daran hat, die gegebene Parallelität auch ausnutzen zu können. Bei konventioneller Rechentechnik muss fast immer eine externe, gesteuerte Lastaufteilung erfolgen, die vielfach im Quellcode verankert ist, während neuronale Netze ihre Lastaufteilung autonom organisieren und dabei sogar in der Lage sind, die Aufteilung dynamisch zu ändern, wenn Verarbeitungsaufgaben an Rechenaufwand zunehmen [Laughlin03].
Manche Prozesse, wie z.B. (zeit-)lineare Rechenaufgaben, die vom Menschen parallel ausgeführt werden, können auf Rechnern konventioneller Architektur nur seriell ausgeführt werden, was eng mit der entsprechenden Repräsentation der Daten zusammenhängt (siehe auch Abschnitt 2.2.). Eine Zahl, die ähnlich einer Kohonenkarte als paralleles Muster vorliegt[11] [Hopfield84, Shepherd04 (Kapitel 12)], schneidet unter dem Aspekt der Speicherdichte schlecht gegenüber einer technischen Bitrepräsentation ab, kann jedoch z.B. bei einer Multiplikation parallel verarbeitet werden, während eine als MSB...LSB vorliegende Zahl eine Anzahl serieller Verarbeitungsschritte benötigt, die der Bitauflösung der Zahlen entspricht:

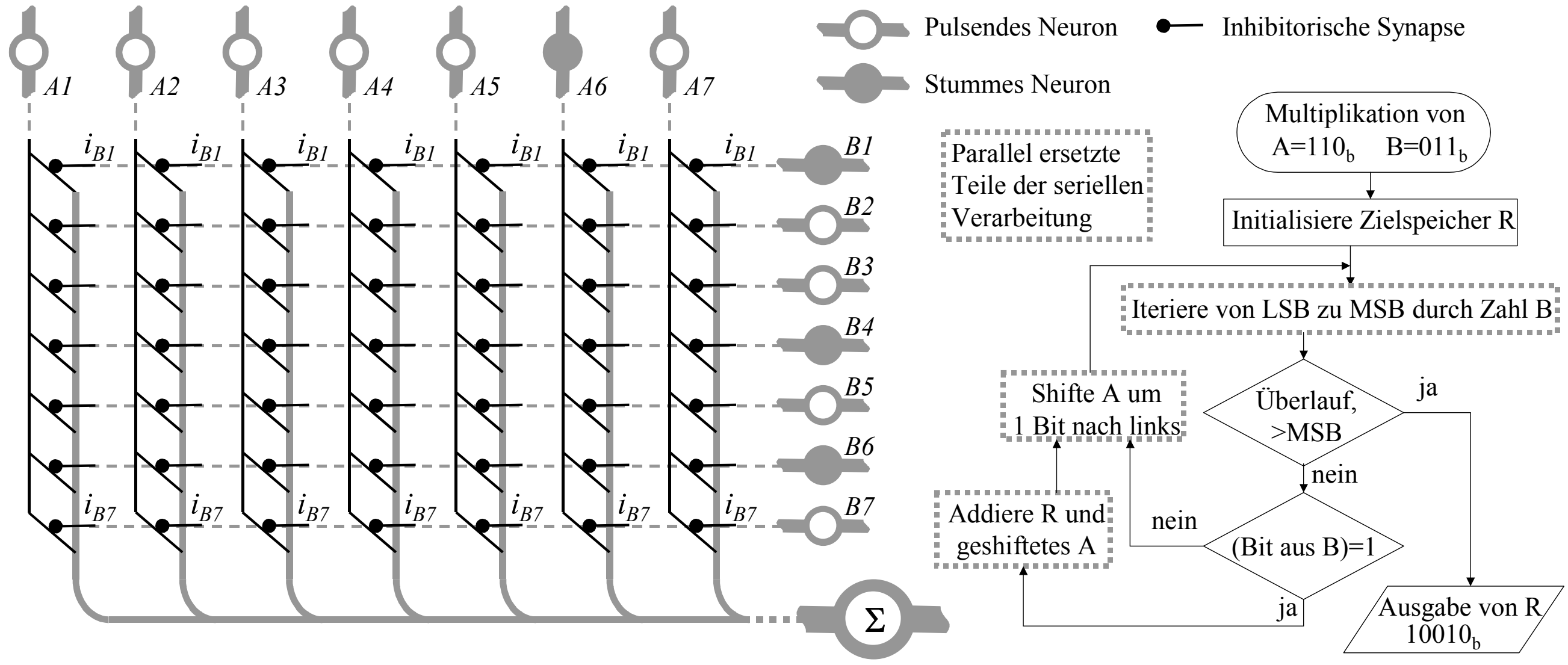

**Abbildung II.20.: Multiplikation im neuronalen und technischen Kontext**

In der obigen Abbildung sind eine (hypothetische) neuronale Multiplikationsarchitektur und ihr konventionelles Gegenstück dargestellt. Die unterschiedlichen Randbedingungen von Computer und Hirnstruktur sind deutlich zu sehen, d.h. eine hohe mögliche Vernetzungsdichte und viele Einzelelemente bei niedriger Verarbeitungsgeschwindigkeit gegenüber einer Architektur mit wenigen Elementen und hoher Taktrate. Verarbeitungsschritte, die in der konventionellen Version seriell ausgeführt werden, lassen sich in einer entsprechenden Neuronen/Dendriten-Realisierung parallel ausführen, beispielsweise wird die Iteration über B durch die Decodierung zum Thermometercode ersetzt, das damit zusammenhängende Shiften von A (entsprechend der Wertigkeit des betrachteten Bits von B) entfällt aufgrund der parallelen Struktur der von B auf A wirkenden inhibitorischen Verbindungen. Die serielle Summation von R und A in der konventionellen Architektur wird durch die parallele analoge Stromsummierung auf dem Zielneuron abgelöst. Dem Geschwindigkeitsgewinn durch die parallele Verarbeitung steht natürlich ein entsprechend erhöhter Schaltungsaufwand gegenüber, für die parallele neuronale Variante wächst der Aufwand quadratisch mit A und B, bei der seriellen, schleifenbasierten Version nur linear.
Durch die auf der linken Seite von Abbildung II.20 illustrierte Repräsentation der Information und ihrer Verarbeitung wird ein hohes Maß an Redundanz und Fehlertoleranz erreicht, was einen der Gründe darstellt, warum neuronale Netze trotz ihrer Komplexität robust arbeiten. Im oben angeführten Beispiel hätte in der konventionellen Architektur ein Fehler im MSB katastrophale Konsequenzen im Vergleich zu einem Fehler im LSB, beide treten jedoch mit gleicher Häufigkeit auf. Demgegenüber verursacht in einer parallelen Zahlenspeicherung und -verarbeitung der Fehler jeder einzelnen Synapse nur einen Fehler im LSB.
Somit kann von Neuronen trotz ihrer individuellen Fehleranfälligkeit bei entsprechender Netzstruktur und Informationsrepräsentation eine robuste Signalverarbeitung ausgeführt werden, die der technischen Zielsetzung des ‚graceful degradation' entspricht, d.h. Fehler führen nicht zu einem völligen Ausfall des Systems, sondern nur zu entsprechend verminderter Leistungsfähigkeit, und weitere Fehler haben nur einen additiven, überschaubaren Effekt, wohingegen bei modernen, komplexen Informationsverarbeitungssysteme die Interaktion verschiedener Fehler u.U. zu unvorhersehbaren, sich gegenseitig steigernden Konsequenzen führen kann [Herbert02].

[11] Die technische Analogie wäre ein Thermometercode

### II.4.2 Asynchronität

Ein weiterer Aspekt neuronaler Netze mit Relevanz für technische Informationsverarbeitung stellt ihre Asynchronität dar. Zum Einen können in zunehmendem Maße in sich synchrone Bausteine wie Rechner, Mikrokontroller, etc. nicht mehr als getrennte Einheiten betrachtet werden, weil z.B. über verschiedene Taktdomänen hinweg miteinander kommuniziert werden muss [Xia02]. Zum Anderen kann auch in einem räumlich kohärenten System ein globaler Systemtakt zu einem erhöhten Leistungsverbrauch führen, weil z.B. Speicher nur sporadisch abgefragt werden, jedoch mit dem Systemtakt erneuert werden, um jederzeit verfügbar zu sein. Die nächste Entsprechung zu einem Takt im digitalen System wäre in biologischen neuronalen Netzen eine Synchronisation der Pulszeitpunkte, die jedoch nur in jeweils eng begrenzten Bereichen auftritt [Gerstner99] und dort selbst zur Informationsverarbeitung beiträgt, d.h. die Ereignisse sind selbst der „Takt" [Kretzberg01], womit eine effizientere Nutzung gegeben ist als für die separate Takt- und Datenübertragung in konventionellen Systemen. Diese Synchronisation ist auch stark verarbeitungsabhängig, in den meisten Fällen werden z.B. in den Synapsengewichten gespeicherte Informationen nur abgerufen, wenn entsprechende präsynaptische Pulse einen postsynaptischen Puls verursachen [Gutkin03]. Da der Ruheenergieverbrauch von Neuronen klein gegenüber der Pulsenergie ist, wird signifikante Leistung deshalb nur im Fall der Speicherabfrage verbraucht [Laughlin03]. Für sichere, effiziente Datenübertragung findet sich in biologischen neuronalen Netzen eine Rückkopplungsstruktur [Maass06], die permanent die Fortpflanzung der Signale in Vorwärtsrichtung korrigiert und damit eine Art zeitkontinuierliches ‚Handshake' herstellt. Dies scheint ähnlich einem Handshake im Technischen der Realisierung einer zuverlässigen asynchronen Übermittlung zu dienen. Ein klassisches Beispiel hierfür ist die Rückkopplung aus dem visuellen Kortex auf den Thalamus, welche selektiv die Weitergabe sensorischer Information sicherstellt [Masson02]. Für die Informationsverarbeitung bedeutet dies, dass Gebiete mit sehr unterschiedlicher Aktivität damit koexistieren und interagieren können ohne dass, wie in technischen Systemen die Regel, alle Systemkomponenten mit der höchsten Einzelfrequenz getaktet werden müssen [Breakspear03].
Leider kann diese Asynchronität aufgrund von z.B. Limitationen der VLSI-Entwurfssoftware nur begrenzt bei technischen Implementierungen neuronaler Netze eingesetzt werden. Dies führt dazu, dass man i.d.R. den Hauptanteil des Energieverbrauchs solcher Netze in den getakteten Digitalschaltungen findet, mit denen global alle Pulse kommuniziert werden, während die Pulserzeugung nur unwesentlichen Anteil daran hat (siehe [Schemmel04] oder auch Abschnitt III.4). Somit ist meistens die VLSI-Realisierung einer Bildverarbeitung in Form eines neuronalen Netzes nicht konkurrenzfähig mit einer konventionellen Variante, die wesentlich besser an die physikalischen und technischen Gegebenheiten des Entwurfs und Herstellungsprozesses angepasst ist. Eine Ausnahme stellen hier z.B. die in [Morie01] dargelegten Arbeiten oder der in Abschnitt IV.2 geschilderte Bildoperator dar, bei denen durch handentworfene Puls- und Digitalschaltungen und bewussten Einsatz von asynchroner Verarbeitung wesentliche Vorteile vor allem beim Energieverbrauch gegenüber synchronen Varianten erreicht werden.

In diesem Zusammenhang soll auch die bereits eingangs und v.a. in Abschnitt II.2 umrissene Art der Darstellung analoger Größen in neuronalen Netzen erwähnt werden. Die Wandlung einer wert- und zeitkontinuierlichen analogen Größe in einen Raten- oder Phasencode resultiert in beiden Fällen in einer deutlich höheren Robustheit gegenüber Störungen im Signalübertragungspfad. Wie in Abschnitt II.2 ausgeführt, lässt sich, normiert auf die mittlere Pulsrate, in einem Phasencode wesentlich mehr Informationen übertragen, wobei der Ratencode demgegenüber deutlich robuster ist. Beide Prinzipien wurden bereits in technischen Realisierungen zur Verringerung der Störungsempfindlichkeit verwendet, siehe z.B. Kapitel III oder [Marienborg02, Schreiter04] für Ratencode, und Abschnitt IV.2. und [Cameron05] für Phasencode, wobei i.d.R. Ratencodes mit höheren mittleren Pulsraten arbeiten, um ähnliche Übertragungsgeschwindigkeiten zu erreichen.

# III Komplexe optische Verarbeitung in VLSI mittels Pulse-Routing

Eines der hauptsächlichen Probleme bei Implementierungen von neuronalen Netzen als integrierte Schaltkreise (VLSI) betrifft die Vernetzung der Neuronen über nachempfundene Dendriten, Synapsen und Axone. Bei fester, leitungsbasierter Verdrahtung kann nur eine festgelegte Funktionalität erreicht werden, zusätzlich erhöht sich die Anzahl der Verbindungen und damit der Verdrahtungsaufwand mit $N^2$, wobei N die Zahl der miteinander verbundenen Neuronen darstellt. Meistens sind neuronale Netze mit statischer Vernetzung daher auf Nächster-Nachbar-Kopplung beschränkt, wodurch nur einfache Bildverarbeitung wie z.B. Grauwertsegmentierung über Pulssynchronisation [Schreiter04] realisiert werden kann, oder die Neuronen übernehmen nur eine Transformationsaufgabe, während die weitere Verarbeitung in nachfolgenden Softwarestufen erfolgt [Atmer03].
Um komplexere Verarbeitungsschritte direkt mit pulsgekoppelten neuronalen Netzen ausführen zu können, sind, wie in Abschnitt II.1.3 und II.2 dargelegt, wesentlich aufwändiger strukturierte Netze nötig. Um die Komplexität der Verdrahtung gering zu halten und das Netzwerk bzgl. seiner Verarbeitungsaufgaben umkonfigurieren zu können, ist eine symbolische Übertragung der Pulse nötig (Pulse-Routing), bei dem Pulse als Paketinformation mit Zeitpunkt und Zielort ohne feste Leitung übertragen werden. Es soll eine VLSI-Implementierung eines neuronalen Netzes dargestellt werden, das ausgehend von pulscodierter (Bild-)Grauwertinformation mit einer auf neuronalen Adaptionsregeln beruhenden Mikroschaltung und symbolischem Pulse-Routing mehrschichtige, komplexe Bildverarbeitungsfunktionen bis hin zu Gabortransformationen ausführen kann. Dieser Schaltkreis ist zusätzlich für eine 3D-Integrationstechnologie vorbereitet, so dass Pulse auch vertikal übermittelt werden können [Mayr07a].

## III.1 Adaptionsregeln und neuronale Mikroschaltung

### III.1.1 Mikroschaltung

Von verschiedenen Autoren [Maass02, Shepherd04] werden in der Biologie vorkommende stereotypisierte neuronale Mikroschaltungen beschrieben, die eine bestimmte, durch ihre Struktur festgelegte Funktion erfüllen. Ein bestimmendes Merkmal dieser Mikroschaltungen ist der Kontrast zwischen ihrer individuellen Simplizität, und den komplexen Verarbeitungsfunktionen, welche aus ihnen konstruierte Netzwerke ausführen können. Die Mikroschaltungen, die in [Maass02] behandelt werden, enthalten inhibitorische und exzitatorische Synapsen, jedoch wurden rein inhibitorische Mikroschaltungen ebenfalls in Säugetieren nachgewiesen [Moore04]. Sowohl von Maass et. al. als auch von Eckhorn [Eckhorn99] wird postuliert, dass die Extrahierung von Korrelationen zwischen Eingangspulsfolgen eine der Hauptverarbeitungsfunktionen dieser Mikroschaltungen darstellt, z.B. in der Merkmalsverknüpfung im V4 Bereich der Säugetier-Bildverarbeitung. Hierbei nimmt Eckhorn an, dass die für die Korrelationsfindung nötige synchrone Aktivität zwischen Neuronen nur mit inhibitorischen Synapsen erreicht werden kann. Dies liegt im Widerspruch zu den in [Schreiter04] dokumentierten Arbeiten, bei denen nur mit exzitatorischen Synapsen gekoppelte Neuronen ebenfalls synchrone Aktivität für korrelierte Eingänge aufweisen. Es gibt also keine grundsätzlichen Hindernisse auf dem Weg zu rein exzitatorischen Mikroschaltungen, auch wenn diese in der Biologie noch nicht explizit nachgewiesen wurden. Ein Vorteil einer solchen Mikroschaltung wäre die Ausnutzung schneller Signalpropagierung entlang exzitatorischer Synapsen [VanRullen01, Gerstner99], wenn sie mit einer entsprechend gerichteten Struktur entworfen wird. In diesem Kontext wurde die folgende Schaltung entworfen [Heittmann04]:

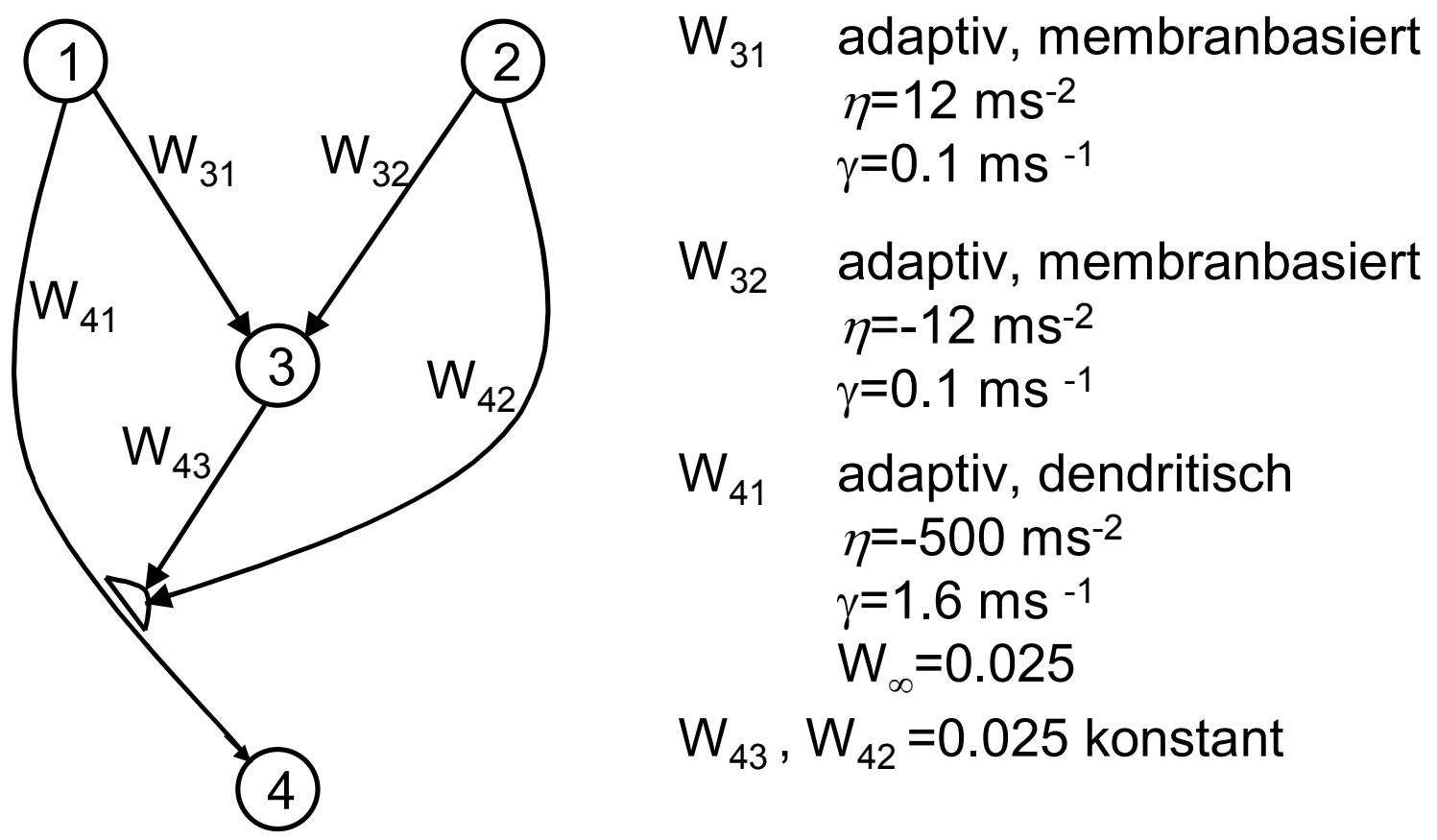

**Abbildung III.1:Die im weiteren betrachtete neuronale Mikroschaltung**

Diese technische Realisierung einer (hypothetischen) neuronalen Mikroschaltung besteht aus Integrate-and-Fire (IAF) Neuronen, die mit zwei Arten adaptiver Synapsen verbunden sind. Das Verhalten des IAF-Neurons ergibt sich aus der Multiplikation der entlang der Synapsen laufenden Pulse mit den Synapsengewichten und ihrer Integration auf der Membrankapazität. Wenn diese eine Feuer-Schwellspannung $\theta$ von (symbolisch) 1 erreicht, erzeugt das Neuron einen Ausgangspuls und setzt die Membrankapazität zurück auf 0 .Es wurde keine Refraktärzeit in das Modell einbezogen, d.h. das Neuron ist unmittelbar nach dem Rücksetzen wieder für neue Eingangspulse offen. Der o.a. Mechanismus, d.h. Korrelationsfindung durch wechselseitige Verstärkung oder Abschwächung von vorwärtsgerichteten Synapsen statt einer lateralen Inhibition, ist durchaus biologisch plausibel. In [Freeman02] werden biologische Messungen vorgestellt, die nahe legen, dass die im V1 Bereich des visuellen Kortex gefundene Interaktion zwischen korrelierten rezeptiven Feldern nicht durch laterale Inhibition, sondern durch Modulation der vertikalen exzitatorischen Synapsen erreicht wird (vergleiche Abbildung III.1).

### III.1.2 Adaptionsregeln

Die Adaptionsregel aus [Heittmann04] für die ersten beiden Synapsen ($W_{31}$ and $W_{32}$) ist in Gleichung (III.1) wiedergegeben, sie wird im Folgenden Membranadaption genannt. Die Indizes beziehen sich auf die Synapse von Neuron 1 auf Neuron 3, repräsentiert durch das Synapsengewicht $W_{31}$.

$$\frac{d}{dt}W_{31} = -\gamma \cdot W_{31} + \eta \cdot (a_3 - \tfrac{\theta}{2}) \cdot \chi(X_1) \qquad \textbf{(III.1)}$$

Es handelt sich dabei um eine Hebbsche Lernregel [Hebb49], die Pulse mit beinahe konstanter Phasendifferenz synchronisieren soll. Das Verhalten wird gesteuert durch die Parameter $\gamma$ als Abklingkonstante und Lernrate $\eta$. Der erste Term auf der rechten Seite ‚verlernt' das Gewicht, wenn es nicht durch Feuern des präsynaptischen Neurons gefestigt wird. Die Indikatorfunktion $\chi$ überwacht den Ausgang von Neuron 1, $X_1$, sie wird 1 bei einem Feuern von Neuron 1, sonst 0, und steuert entsprechend die Addition des zweiten Terms auf der rechten Seite, der Lernfunktion. Die Lernfunktion verhält sich im Hebbschen Sinne, d.h. eine zeitlich asymmetrische Korrelation zwischen Integratorzustand $a_3$ des postsynaptischen Neurons und dem Feuern $X_1$ des präsynaptischen Neurons. Wenn $a_3$ über die Hälfte ($\theta/2$) des Schwellwertes erreicht hat, wird das Gewicht verstärkt, ansonsten abgeschwächt. Je näher Neuron 3 sich an seiner Feuerungsschwelle befindet, wenn Neuron 1 feuert, desto mehr wird das Gewicht zwischen beiden Neuronen verstärkt, somit ergibt sich das Hebbsche Ziel, ein Gewicht zu verstärken, wenn das präsynaptische Neuron am Feuern des postsynaptischen Neurons beteiligt ist [Hebb49].

In der vorliegenden Anwendung in der neuronalen Mikroschaltung wird Gleichung (III.1) dazu verwendet, korrelierte Pulse der Neuronen 1 und 2 zu extrahieren. Zum besseren Verständnis der Korrelationsfunktion der Neuronen 1 bis 3 nehme man an, dass Neuron 3 gerade gefeuert hat und damit ein Membranpotential nahe Null besitzt. Wenn als nächstes Neuron 2 einen Puls abgibt, wird das zugehörige Gewicht $W_{32}$ verstärkt ($\eta$<0 und $a_3$-θ/2<0) und $a_3$ über den Schwellwert θ/2 geschoben. Falls Neuron 1 jetzt einen Puls abgibt, wird auch dessen zugehöriges Gewicht verstärkt ($\eta$>0 und $a_3$-θ/2>0) und Neuron 3 damit über seinen Feuerschwellwert aufgeladen. Nur diese spezielle Pulsabfolge, d.h. einem Puls von Neuron 2 folgt ein Puls von Neuron 1, resultiert in einen Ausgangspuls von Neuron 3, damit zeigt Neuron 3 die zwischen Neuron 1 und 2 korrelierten Pulse an.
Die zweite Adaptionsregeln, die hier dendritische Adaption genannt wird, wird durch die folgende Formel definiert:

$$\frac{d}{dt}W_{41} = -\gamma \cdot (W_{41} - W_\infty) + \eta \cdot (X_3 \cdot W_{43} + X_2 \cdot W_{42} - I_\theta) \cdot W_{41} \cdot \chi(X_1) \qquad \textbf{(III.2)}$$

Die Funktionsweise dieser Adaption ergibt sich wie folgt: Falls Neuron 1 keine Aktivität zeigt, zieht der erste Term Gewicht $W_{41}$ asymptotisch nach $W_\infty$, was Pulse von Neuron 1 zu Neuron 4 passieren lässt. Falls aber durch einen Ausgangspuls $\chi(X_1)$ von Neuron 1 der zweite Term addiert wird, und zeitgleich ein korrelierter Puls weiter oben im dendritischen Baum durch $X_2$ oder $X_3$ signalisiert wird, verkleinert sich das Gewicht $W_{41}$ rapide ($\eta$<<0) und verhindert damit die Weitergabe des Ausgangspulses von Neuron 1 an Neuron 4. Beide Zeitkonstanten, $\eta$ und γ sind betragsmäßig deutlich größer als bei der Adaption in Gleichung (III.1) (siehe Abbildung III.1). Damit wird deutlich, dass diese Adaption bei Zeitskalen von einzelnen Pulsen wirksam wird und damit eine quasi-digitale axonale Einzelpulsinteraktion darstellt wie in Abschnitt II.2.5 beschrieben. Bei entsprechender Parametrisierung von μ und dem Adaptionsschwellwert $I_\theta$ wäre diese Art des Pulsnegierens auch nur über den Term $X_3*W_{43}$ möglich, dies würde jedoch durch die Zeitverzögerung über die Integration auf Neuron 3 den Puls von Neuron 1 nicht mehr hinreichend blockieren, so dass $X_2*W_{42}$ zusätzlich in Gleichung (III.2) eingefügt wurde, um die Adaption ‚vorzuladen'.

### III.1.3 Verhalten/Simulationsergebnisse

Im folgenden wird die Korrelation zwischen den Ausgangspulsen der Neuronen 1 und 3 wie folgt definiert:

$$C_{13} = \frac{1}{T_1 - T_0} \int_{T_0}^{T_1} \chi(X_1(\tau)) \cdot \chi(X_3(\tau)) d\tau \qquad \textbf{(III.3)}$$

Mit der normalisierten Korrelation gegeben als $\overline{C_{13}}=C_{13}/C_{11}$, und der normalisierten Dekorrelation definiert als $\overline{D_{13}}=1-\overline{C_{13}}$, ergibt sich ein Gesamteffekt der o.a. Regeln in der Mikroschaltung wie in Abbildung III.2 dargestellt. Wenn unkorrelierte, d.h. überzählige Pulse von Neuron 1 empfangen werden, werden diese über die Mikroschaltung weitergegeben, korrelierte Pulse von Neuron 2 werden subtrahiert, während unkorrelierte Pulse von Neuron 2 ignoriert werden. Die Mikroschaltung agiert damit als Pulsratensubtrahierer, der auf den positiven Wertebereich begrenzt ist, d.h. es gibt auch bei einer höheren Aktivität von Neuron 2 gegenüber Neuron 1 kein negatives Ausgangssignal, Neuron 4 bleibt aktivitätslos. Hierin unterscheidet sich die diskutierte Mikroschaltung von biologischen Vorbildern, die eine spontane Aktivitätsrate kennen und bei subtrahierenden Eingangssignalen auch unter diese spontane Rate sinken können (siehe Abschnitte II.1.1 und II.1.3). Eine detaillierte Diskussion der Mikroschaltung findet sich in [Heittmann04].

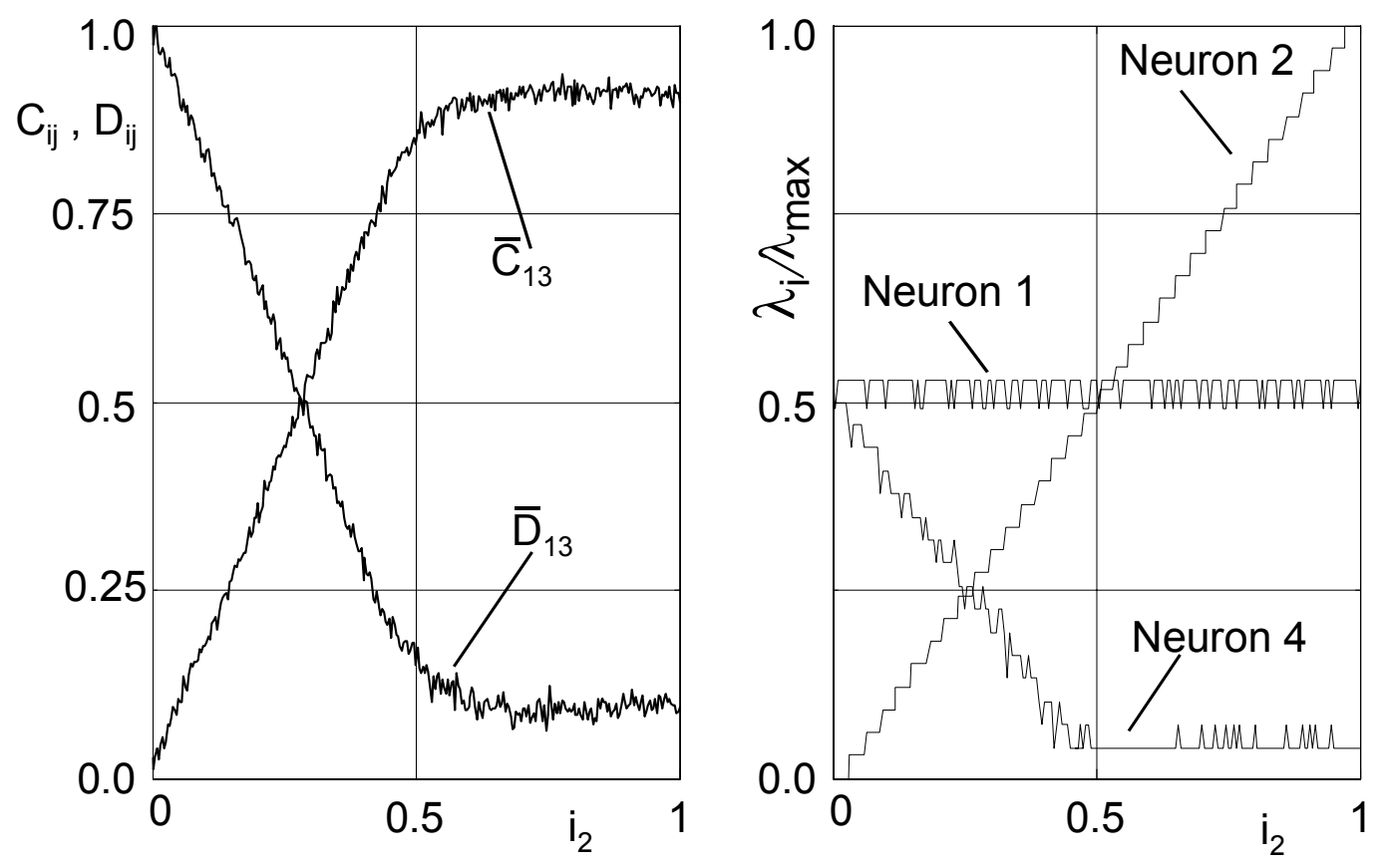

**Abbildung III.2: Verhalten der Mikroschaltung für variable Eingangspulsraten**

Durch die o.a. Mikroschaltung werden Grundfunktionalitäten von Neuronen wie Hebbsche Adaption, Korrelation und dendritische Verschaltung zu einer neuen Verarbeitungseinheit zusammengesetzt. Im weiteren wird aufgezeigt, welche neuen Verarbeitungsmöglichkeiten mit dieser Mikroschaltung als Grundbaustein aufgebaut werden können. Der Weg von Grundfunktionalitäten einzelner Neuronen über repetitive Mikroschaltungen hin zu komplexen globalen Verarbeitungsfunktionen erfolgt hierbei analog zum biologischen Vorbild [Shepherd04].

# III.2 Verarbeitungsmöglichkeiten mit kaskadierten, vernetzten Mikroschaltungen

## III.2.1 Kantendetektor

Eine Gruppe der neuronalen Mikroschaltungen kann z.B. zum Aufbau eines einfachen Kantendetektors verwendet werden, indem ihre Eingänge mit benachbarten Pixeln verbunden werden und sie gemäß Abbildung III.3 angeordnet werden [Mayr06b]. Ihre zugehörigen Eingangsneuronen 1(+) und 2(-) werden dann mit einer pulsgewandelten Version eines Eingangsbildes versorgt und die Ausgaben der Mikroschaltungen aufsummiert. Die gepulste Version des Eingangsbildes wird mittels einer AHDL-Beschreibung der in Abschnitt C.1.1 dokumentierten Pixelzelle erzeugt, die mit einer Stromrepräsentation des Eingangsgrauwertbildes versorgt wird. Im rechten Teil von Abbildung III.3 wird die summierte Ausgabe der Mikroschaltungen über der Translation einer Grauwertkante aufgezeigt.

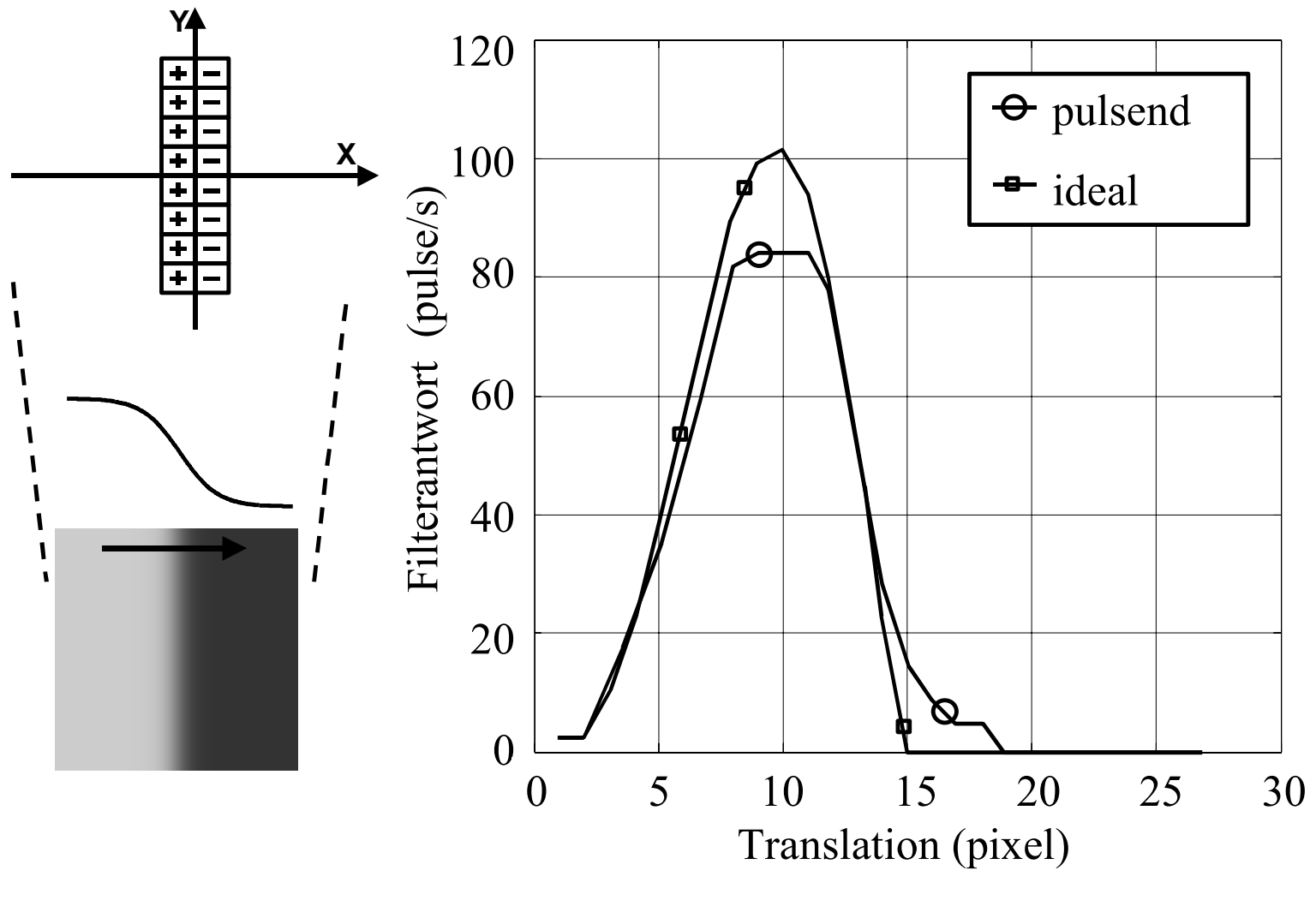

**Abbildung III.3: Antwort eines aus den neuronalen Mikroschaltungen aufgebauten Kantendetektors auf die Translation einer Kante in Detektorrichtung**

Wie aus der obigen Abbildung erkennbar, steigt die Pulsantwort an, wenn der Beginn der Kante in die Nähe des Detektors gelangt, wobei sich ein deutliches Maximum für die Mitte der Kante ergibt. Die ideale Kurve wurde mit einer wert- und zeitkontinuierlichen Version des Kantendetektors simuliert, wobei im Unterschied zwischen der idealen und der auf den Pulsantworten basierenden Kurve Fehler der Adaptionsregeln und die (Puls-)Diskretisierung des Netzwerks erkennbar sind. Der Kurvenverlauf der Grauwerte der Kante (angedeutet in der Kurve oberhalb der Grauwertkante) mit variablem Gradienten wurde eingeführt, um die Lernregeln über einen weiten Bereich an Pulsratendifferenzen zu testen. Es ergeben sich vor allem im Bereich kleiner Pulsratendifferenzen durch die Lernregeln leichte Phasenverschiebungen, die zu fehlerhaften Antworten der Mikroschaltung führen. Das Maximum ist niedriger als für die ideale Variante, da durch den Abklingterm der Membranregel in Gleichung (III.1) für hohe Pulseingangsaktivität verstärkt Pulse unterdrückt werden. Mit zunehmendem Gradienten ergeben sich entsprechend in der Ausdehnung begrenztere Verläufe der Pulsantwort. Beispielsweise ergibt eine nur einen Pixel breite Kante durch das exakte Übereinstimmen mit dem Detektorprofil eine Punktantwort an dieser spatialen Koordinate, die jedoch aus dem oben angeführten Grund nicht die volle Höhe erreichen wird, die aus einer Subtraktion der äquivalenten Pulsraten auf beiden Seiten der Kante zu erwarten wäre.

### III.2.2 Einstufige Gabortransformation

Eine Schar dieser Mikroschaltungen kann zum Aufbau eines einfachen Gaborfilters verwendet werden [Mayr06d, Heittmann04], indem sie wie in Abbildung III.4 angeordnet werden, wobei hellere Grauwerte als der Durchschnitt Zugriff von positiven Eingängen „1“ ≙ (+) der Mikroschaltungen anzeigen, dunklere Grauwerte hingegen negative Eingänge „2“ ≙ (-). Um die unterschiedlichen Koeffizienten für die verschiedenen Bereiche der Gabormaske zu erhalten, wird diese Maske diskretisiert, und die diskretisierten Werte werden in entsprechend mehrfachen Zugriff von Mikroschaltungen auf dieselben Koordinaten umgewandelt. Auf diese Weise wird in der pulsbasierten Faltungsmaske eine Gewichtung der einzelnen Pixel erreicht, die den originalen Gaborkoeffizienten entspricht. Die entsprechenden Eingänge der Mikroschaltungen 1 (+) und 2 (-) erhalten dann als Eingangssignal eine mit den pulsenden Pixelzellen (Abschnitt C.1.1) pulsgewandelte Repräsentation des Eingangsbildes. Die Ausgangssignale werden zur Gesamtantwort der Gabormaske aufsummiert. In der folgenden Abbildung ist diese Gesamtantwort als Funktion des Drehwinkels einer rotierenden Kante dargestellt:

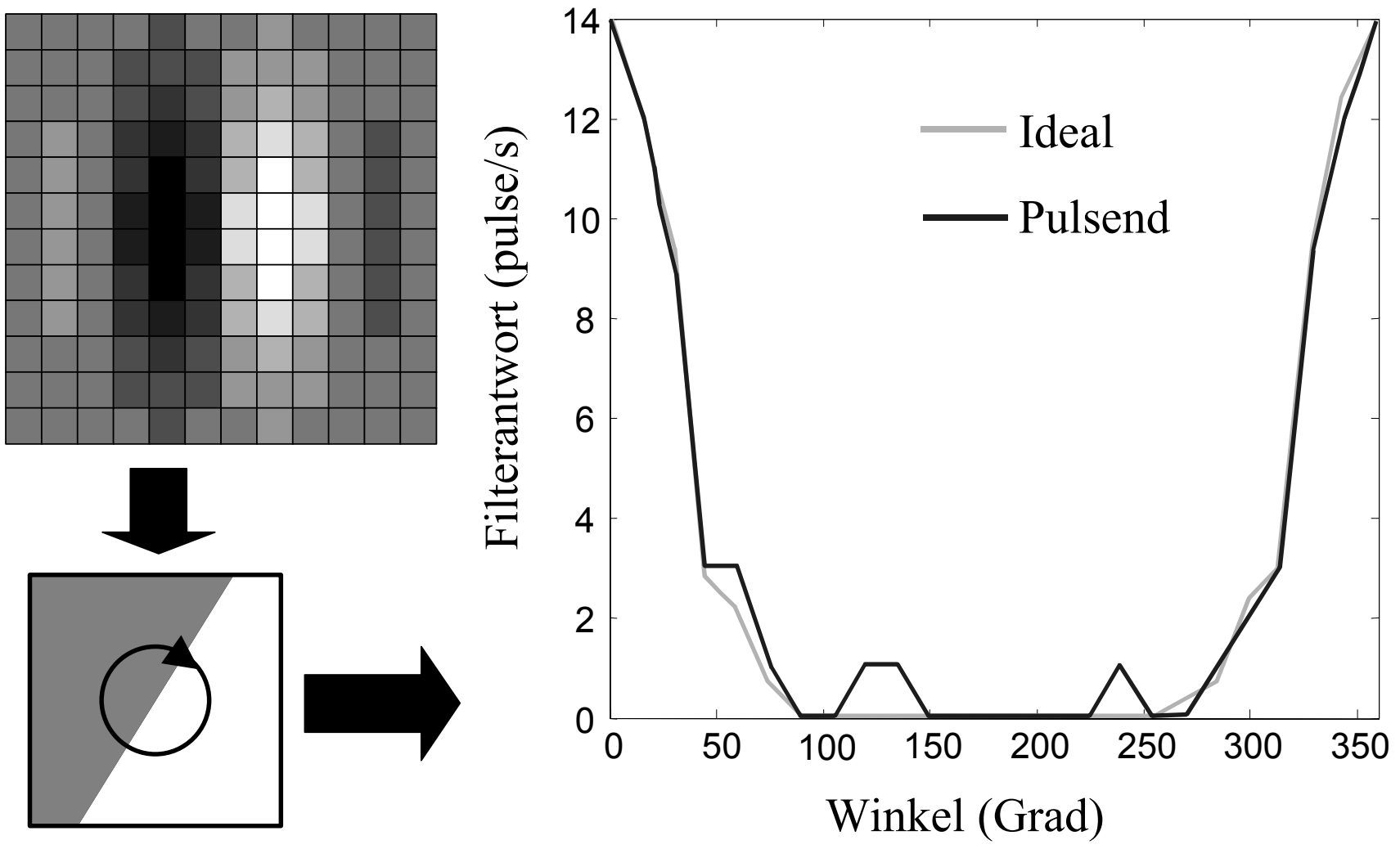

**Abbildung III.4: Antwort eines pulsbasierten einstufigen Gaborfilters auf eine rotierende Kante**

Die Abweichung zwischen idealer und pulsbasierter Gaborfaltung (z.B. zwischen 100° und 150°) wird von der Subtraktion im ersten Quadranten verursacht, d.h. die äußeren Bereiche der Gabormaske erzeugen falsche positive Antworten, die normalerweise von der negativen Antwort des inneren Hauptteils der Gabormaske ausgelöscht werden. Die einseitige Antwort der Mikroschaltungen eignet sich also nur bedingt zu einer echten Bildfaltung, da sich dabei sowohl die positiven als auch die negativen Anteile der Faltungsmaske in der Struktur des Ausgangsbildes wieder finden müssen.

### III.2.3 Mehrstufige Gabortransformation mit Gegenmaske

Ein erster Schritt zur Verbesserung der pulsbasierten Bildfaltung wäre die Unterdrückung der Fehler aus Abbildung III.4. Dies kann erreicht werden, indem die Maskenantwort mit einem Korrektursignal verrechnet wird, das aus der Antwort der entgegengesetzten Faltungsmaske an denselben Bildkoordinaten besteht [Mayr07c]. Zur Veranschaulichung dieses Prinzips wird eine eindimensionale Faltungsmaske von (1 –2 1) angenommen, die aus zwei Mikroschaltungen mit einem Pixelzugriff von (+ -- +) aufgebaut werden kann, bei dem die beiden negativen Eingänge auf denselben Bildpunkt zugreifen. Wenn eine nicht-pulsbasierte Version dieser Faltungsmaske ein Eingangsmuster von (3 2 1) sieht, ergibt sich als Maskenantwort 0. Demgegenüber resultiert die Verwendung von Mikroschaltungen zum Einen in 3-2=1 und zum Anderen in 1-2=0 (siehe Abbildung III.2). In Summe ergibt sich damit eine Antwort von 1. Wenn eine zweite, entgegengesetzte Maske von (- ++ -) dasselbe Eingangsmuster erhält, ergibt sich ebenfalls 1 als Antwort. Die richtige Antwort der originalen Faltungsmaske von 0 ergibt sich damit als Differenz zwischen dieser und ihrer Gegenmaske. Diese Korrekturmethode beeinflusst die Antwort der Maske auf ein ‚zu ihr passendes' Eingangsmuster nicht, da dann die Antwort der Gegenmaske 0 ist. Es werden nicht alle Fehler von dieser Korrektur beseitigt, z.B. würde ein Eingangsmuster von (2 2 1) für die ideale Maske eine Faltungsantwort von –1 ergeben, während die Verwendung von Mikroschaltungen selbst für die korrigierte Antwort natürlich in 0 resultiert. Grundsätzlich erzeugen alle Eingangsbilder einen Fehler, bei denen eine der Mikroschaltungen eine negative Ausgabe erzeugen müsste, um eine korrekte Gesamtantwort zu liefern.
Da in der Bildanalyse die Gaboramplitude am meisten Verwendung findet, wird im nächsten Verarbeitungsschritt der Betragswert der Gaboramplitude aus der korrigierten Maskenantwort $R_+ - R_-$ (Differenz aus Maske und Gegenmaske) berechnet. Der Betrag dieses Terms ist natürlich wie folgt definiert:

$$\left|R_+ - R_-\right| = \begin{cases} R_+ - R_- & für\ R_+ \geq R_- \\ R_- - R_+ & für\ R_+ < R_- \end{cases} \quad \textbf{(III.4)}$$

Wenn $R_+$ und $R_-$ jeweils einmal in beiden Richtungen in eine Mikroschaltung eingegeben werden und ihre Ausgaben dann summiert werden, ergibt sich aus der einseitigen Differenz der Mikroschaltungen dieselbe Rechnung (d.h. ein Betrag der Gaboramplitude):

$$\left(R_+ - R_-\right) + \left(R_- - R_+\right) = \begin{cases} R_+ - R_- + 0 & für\ R_+ \geq R_- \\ 0 + R_- - R_+ & für\ R_+ < R_- \end{cases} \quad \textbf{(III.5)}$$

Diese Rechnung korrigiert gleichzeitig die inhärenten Fehler der Gabormaskenbildung durch eine wechselseitige Subtraktion der Maske von der Gegenmaske, wie im Abschnitt über Gleichung (III.4) dargelegt. Abbildung III.5 belegt die Übereinstimmung dieser pulsbasierten Faltungsmethode mit konventioneller Bildfaltung, wobei die Faltungsmaske der in Abbildung III.9 dargestellten entspricht. Die in Abbildung III.5 (c) ersichtlichen Wellen ergeben sich aus der nicht-idealen Korrekturmethode, bei dem sich im gefalteten Bild die Grundfrequenz der im folgenden Abschnitt beschriebenen deterministischen Gabormaske wieder findet.

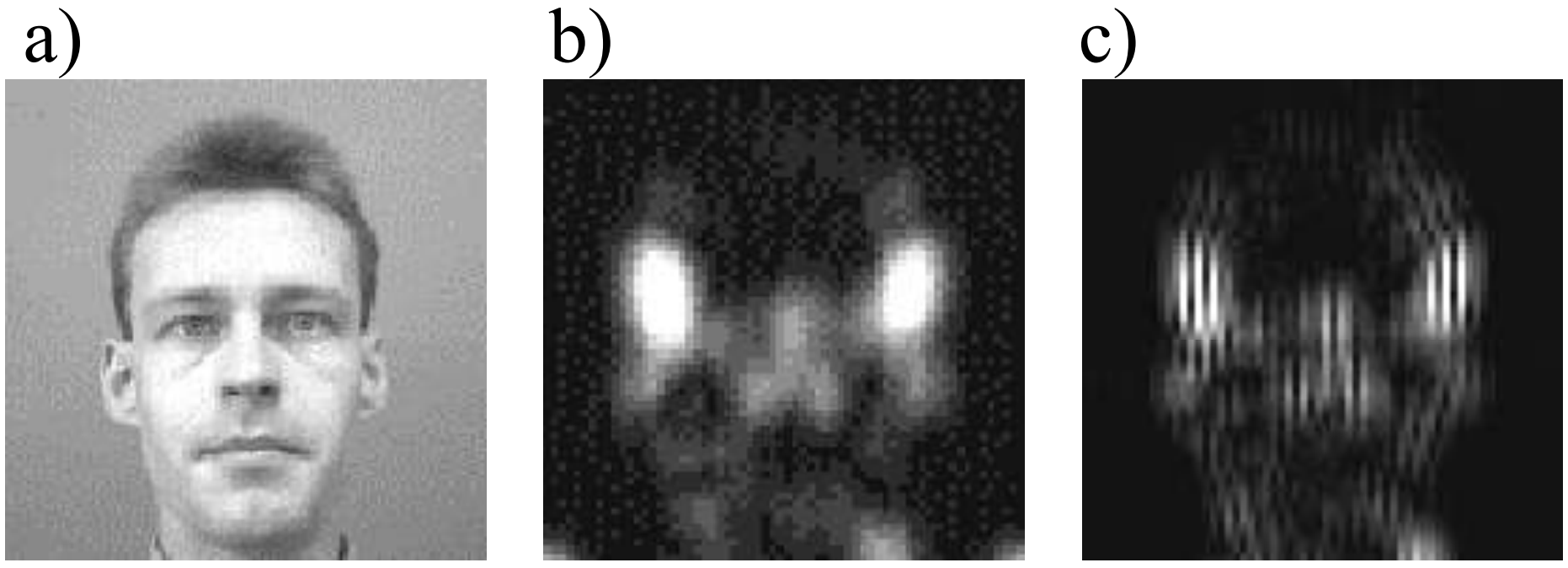

**Abbildung III.5: Vergleich zwischen Originalbild, wertkontinuierlicher und pulsbasierter Gaborfaltung**

Die gepulste Berechnung der Gabormaske weist zwei deutliche Vorteile gegenüber konventioneller Berechnung auf, zum Einen Robustheit gegenüber zeitlichem Rauschen durch die Integrationsvorgänge in den einzelnen Verarbeitungsstufen. Dies ist deutlich zu sehen in Abbildung III.6 b, die gegenüber Abbildung III.5 c mit 20% weißem Rauschen auf den Pixelströmen berechnet wurde. Zur Rauschsicherheit tragen auch der generelle hierarchische Aufbau und die Verrechnung mit der Gegenmaske bei, wodurch temporale lokale Störungen unterdrückt werden.
Zum Anderen agiert die pulsbasierte Gaborfaltung inhärent auf verschiedenen zeitlichen/quantitativen Auflösungen. Wie aus Abbildung III.6 a ersichtlich, die den Ausgang der Gaborfaltung nach 3 Ausgangspulsen des hellsten Pixels im Ausgangsbild zeigt, sind die wichtigsten Aspekte der späteren vollen Gaborantwort bereits zu einem sehr frühen Zeitpunkt ersichtlich. Die hier aufgebaute Verarbeitungsstrecke zeigt damit Analogien zu biologischen neuronalen Netzen, d.h. schnelle, grobkörnige und langsame, detaillierte Verarbeitung [VanRullen01].

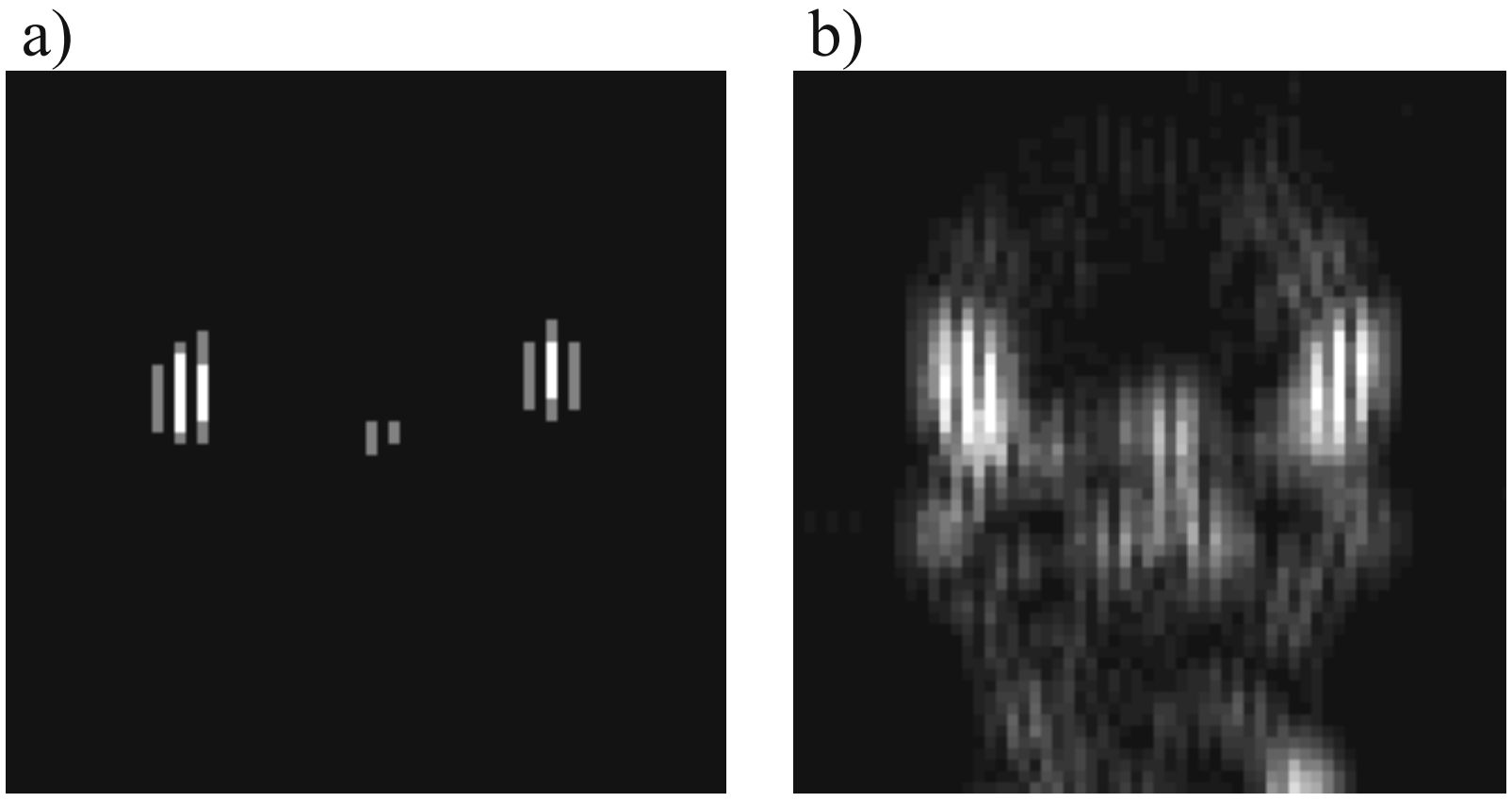

**Abbildung III.6.: Vergleich der Gaborantwort nach maximal vier Pulsen der Pixelzellen und komplette Antwort**

### III.2.4 Stochastischer vs. hierarchischer Aufbau von Gabormasken

Die in den vorigen Abschnitten beschriebenen Gabormasken lassen sich grundsätzlich auf zwei verschiedene Weisen aus den einzelnen Mikroschaltungen aufbauen. Eine Möglichkeit ist ein stochastischer Aufbau, bei dem die Koeffizienten der vorgegebenen Gabormaske als Wahrscheinlichkeiten eines Zugriffs der positiven und negativen Eingänge der Mikroschaltung gesehen werden. Hierbei werden die Mikroschaltungen disjunkt verwendet, d.h. für jede neue Gaborfaltung an anderer Stelle im Bild wird wieder die volle Anzahl Mikroschaltungen benötigt. Begonnen wird mit einer regulären Gaborfaltungsmaske (Abbildung III.7 links oben), in diesem Beispiel der gerade Anteil oder Realteil der komplexen Gabormaske aus Gleichung (I.6). Die zugehörigen Faltungskoeffizienten ergeben sich wie folgt:

$$g_{real}(x,y) = \frac{\omega_0}{\sqrt{\pi d} * k} * e^{-\frac{\omega_0^2}{2k^2}\left(x^2+\frac{y^2}{d^2}\right)} * \left(\cos(\omega_0 x) - e^{-\frac{k^2}{2}}\right) \qquad \textbf{(III.6)}$$

Die Parameter für die in Abbildung III.7 links oben dargestellte Gabormaske sind: Grundfrequenz $\omega_0$=1,57 (≅ Periode von 4 Pixeln), Exzentrizität $d$=1 (≅ runde Maske), Ausdehnung/Bandbreite k=3,5.

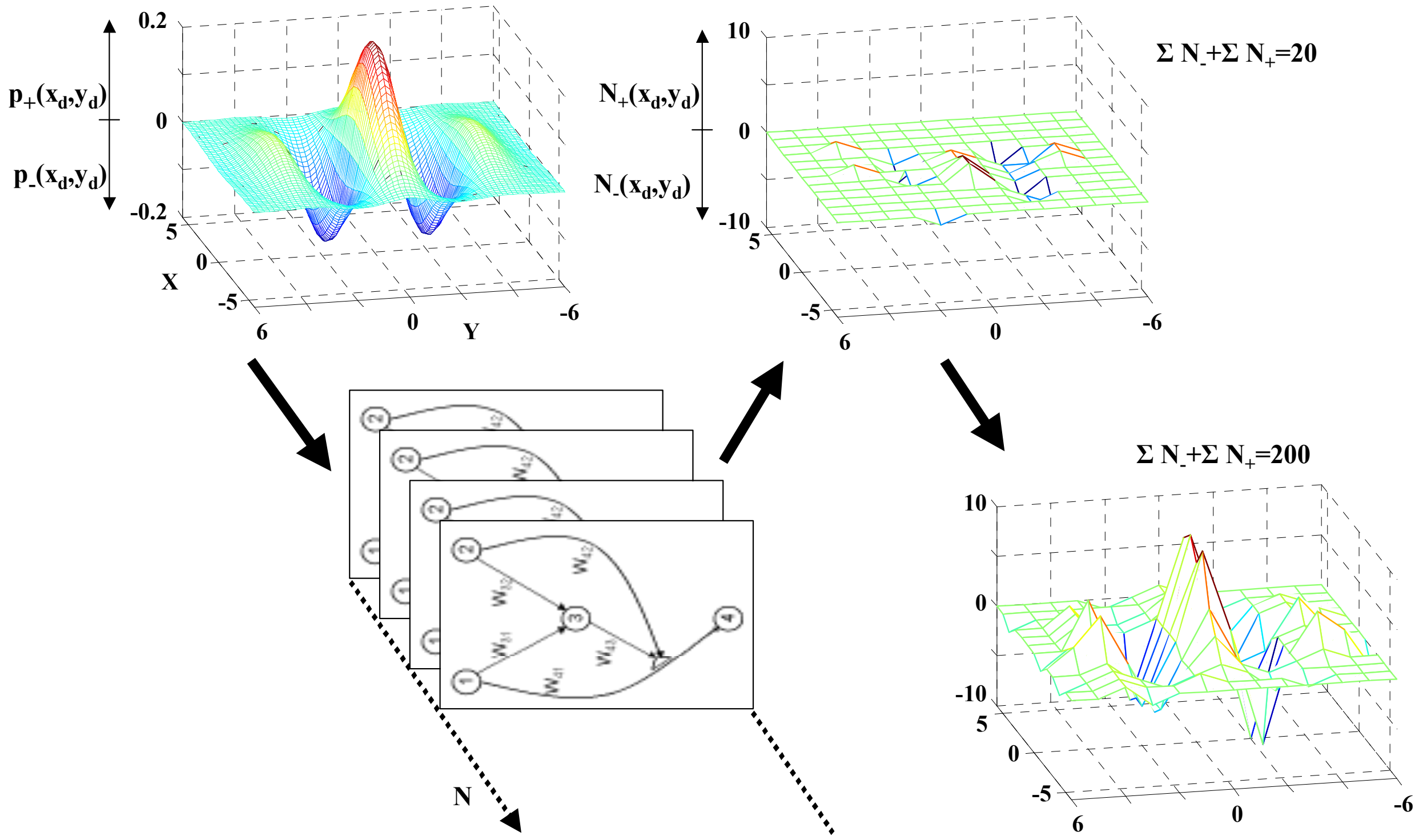

**Abbildung III.7.: Generierung einer stochastischen Gabormaske (Koeffizienten als Wahrscheinlichkeiten, disjunkter Aufbau)**

Im Weiteren werden die äußeren Bereiche der Maske mit vernachlässigbaren Koeffizienten abgeschnitten und die Berechnung auf das Pixelgitter beschränkt:

$$\langle x_d / y_d \rangle = \left\{ (x_d, y_d) \middle| x_d, y_d \in Z \;\cap\; -6 \le x_d, y_d \le 6 \right\} \qquad \textbf{(III.7)}$$

Ab hier wird nur noch die Zuweisung von Koordinaten an die positiven Eingänge der Mikroschaltungen behandelt, die negative Zuweisung erfolgt dann analog. Zuerst werden die Maskenkoeffizienten auf den positiven Wertebereich beschränkt:

$$p_+(x_d, y_d) = \begin{cases} g_{real}(x_d, y_d) \,|\, g_{real}(x_d, y_d) \ge 0 \\ 0 \,|\, g_{real}(x_d, y_d) < 0 \end{cases} \qquad \textbf{(III.8)}$$

Die Auswahl der Koordinaten für den Zugriff von positiven und negativen Eingängen der Mikroschaltungen erfolgt dann durch eine an das ‚Stochastic Universal Sampling' [Chipperfield96] angelehnte Methode. Die Faltungsmaske wird hierbei in zufälliger Koordinaten-Reihenfolge durchlaufen, und ihre zugehörigen Koeffizienten als zusammenhängender Wertebereich fortlaufend akkumuliert auf einem Zahlenstrahl angetragen (untere Schiene von Abbildung III.8). Es wird ein Satz Zeiger in regelmäßigen Abständen Δ generiert, die durch ihre Zugriffe auf den Koeffizientenstrahl festlegen, ob von der jeweiligen Mikroschaltung mit einem (positiven) Eingang auf die entsprechende Koordinate zugegriffen wird. Die Anzahl der Mikroschaltungen (=Σ $N_+$= Σ $N_-$) entspricht dabei der Anzahl der Zeiger, sie bestimmt die Güte der Reproduktion der Ausgangsmaske und wird a priori festgelegt (vgl. Maske mit 10 oder 100 Mikroschaltungen in Abbildung III.7 rechts). Der Abstand Δ der Zeiger berechnet sich wie folgt:

$$p_{sum} = \sum_{x_d=-6}^{6} \sum_{y_d=-6}^{6} p_+(x_d, y_d) \qquad \textbf{(III.9)}$$

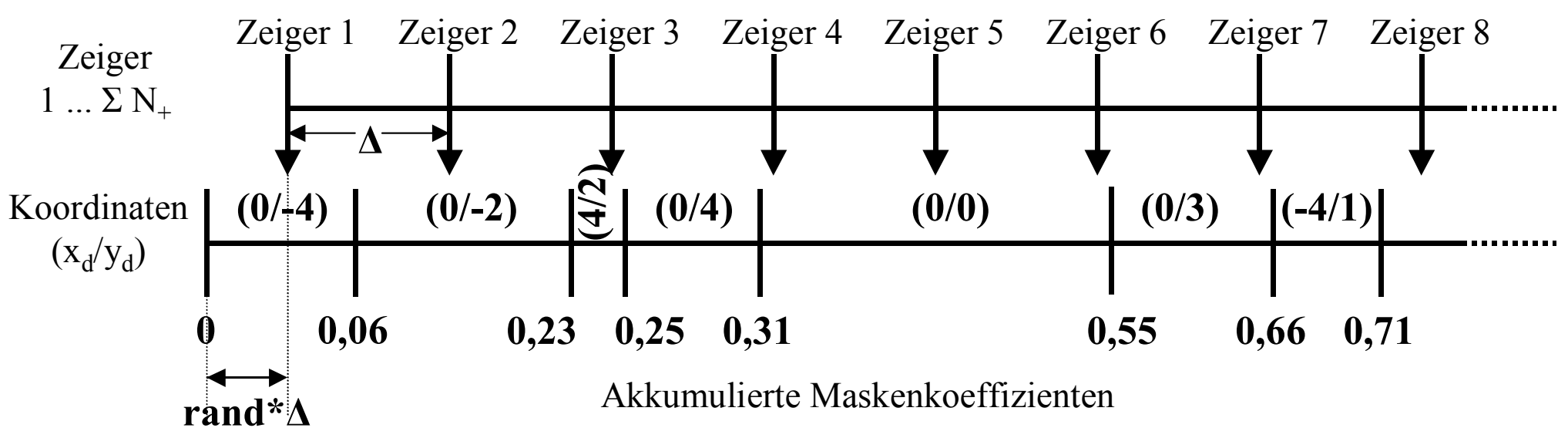

**Abbildung III.8.: Prinzipdarstellung Stochastic Universal Sampling: Anwendung auf Generierung von stochastischen Faltungsmasken**

Zuerst wird die Summe aller positiven Koeffizienten gebildet, d.h. die Länge des Zahlenstrahls. Das Δ ergibt sich dann aus der Division dieser Summe durch die Gesamtanzahl der Mikroschaltungen:

$$\Delta = \frac{p_{sum}}{\sum_{x_d=-6}^{6} \sum_{y_d=-6}^{6} N_+(x_d, y_d)} \qquad \textbf{(III.10)}$$

Der Versatz, mit dem der erste Zeiger relativ zum Beginn des Zahlenstrahls auf diesen zugreift, wird als *rand*∗Δ festgelegt, wobei *rand* eine gleichverteilte Zufallszahl im Bereich [0,1[ bezeichnen soll. Wie eingangs erwähnt, wird dieselbe Vorgehensweise dann auf die negativen Eingänge der Mikroschaltungen angewendet. Alle auf diese Weise zugewiesenen Mikroschaltungen werden im nächsten Verarbeitungsschritt über ein als Summationsglied fungierendes Einzelneuron zusammengefasst. Die Ausgangspulse dieses Summationsneurons bilden die Antwort des Gaborfilters auf das am Eingang der Mikroschaltungen präsentierte Bild.
Da dieser Maskenaufbau eine große Anzahl Mikroschaltungen benötigt, wird in einem Alternativansatz die o.a. Gabormaske in geeignete Unterbausteine zerlegt. Diese sind für mehrere Masken verwendbar, womit sich eine wesentlich dichtere Abdeckung eines Bildes mit weniger Mikroschaltungen erreichen lässt. Dies ist in Abbildung III.9 demonstriert:

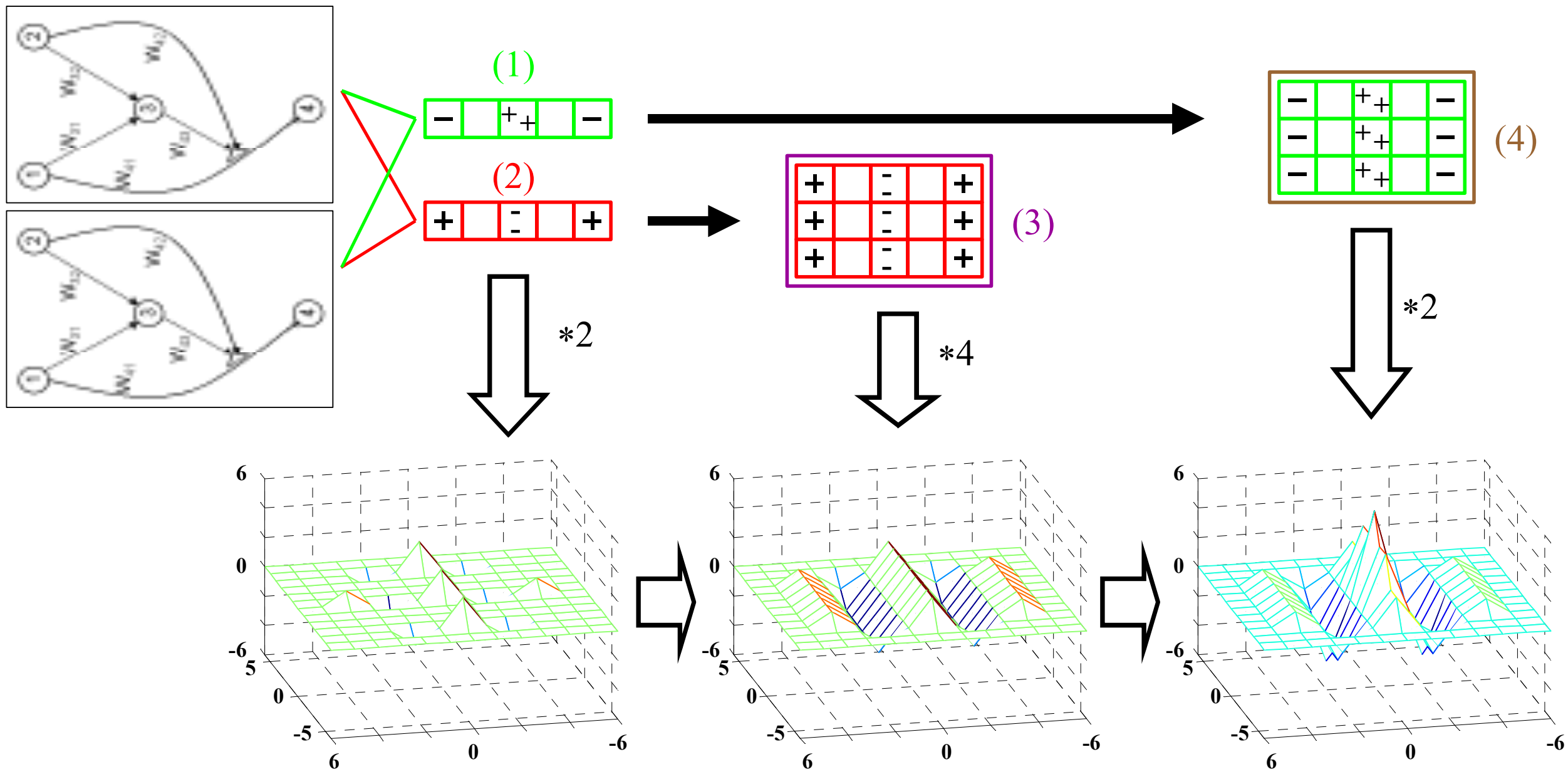

**Abbildung III.9.: Generierung einer deterministischen Gabormaske (hierarchischer Aufbau, Wiederverwendung von Untereinheiten), Achsen wie in Abbildung III.7 rechte Hälfte**

Die in Abbildung III.7 am Ende des stochastischen Verteilungsprozesses erreichte Gabormaske zeigt eine starke Ähnlichkeit mit den in [Jones87a] bei Säugetieren gefundenen rezeptiven Feldern (siehe Abbildung I.6). Trotzdem ist ein Aufbau wie der in Abbildung III.8 beschriebene nicht repräsentativ für die biologische Generierung im V1, da hier ein eher hierarchisches Prinzip zur Anwendung kommt, bei dem rezeptive Felder mit fortschreitender Komplexität aus Feldern geringerer Komplexität aus vorhergehenden Schichten aufgebaut werden [Riesenhuber99]. Da die für die obigen Masken verwendeten Mikroschaltungen außerdem disjunkt sind, stellt dies auch technisch eine ineffiziente Lösung dar. Der Vorteil der stochastischen Faltungsmasken gegenüber den hierarchisch aufgebauten aus Abbildung III.9 ist die universelle Verwendbarkeit, d.h. die oben beschriebene stochastische Maskengenerierung kann automatisiert auf beliebige Faltungsmasken angewendet werden. Wenn für eine bestimmte Bildzerlegung, etc., einzelne Gabormasken mit unterschiedlichen Charakteristiken an verschiedenen Bildkoordinaten benötigt werden, ist ebenfalls kein Vorteil gegenüber hierarchischen Masken erkennbar. Der hierarchische Aufbau hat dort Vorteile, wo eine flächenmäßige Abdeckung mit identischen Gabormasken nötig ist und sich der Aufwand lohnt, die Gabormaske von Hand in einzelne Unterelemente zu zerlegen. Der in diesem Fall erreichbare Vorteil lässt sich wie folgt beziffern: Mit einem Pool von 16384 Mikroschaltungen sind 150 stochastische Gabormasken (Parameter wie oben angeführt) oder 1024 hierarchische Masken realisierbar.

Beide Ansätze zeigen auf, was bei Verwendung von einfachen Mikroschaltungen an Bildverarbeitung realisiert werden kann. Wenn die Analogie zum biologischen Aufbau rezeptiver Felder weitergeführt werden soll, so kann dort natürlich nicht von einer solchen Strukturierung und Typisierung des hierarchischen Aufbaus ausgegangen werden, da im V1-Bereich zwar eine hierarchische Verarbeitung mittels (anders gearteter) Mikroschaltungen erfolgt [Häusler07], diese jedoch weitaus weniger regelmäßig verbunden sind [Buchs02, Yao05]. In der Natur wird also eher ein Mittelweg zwischen den beiden obigen Methoden beschritten.

# III.3 Einzelkomponenten des Router-Schaltkreises

Zu der in den vorangegangenen Abschnitten und in [Heittmann04] geschilderten Mikroschaltung und ihren topologiebasierten Verarbeitungsmöglichkeiten wurde eine VLSI-Realisierung erstellt. Designziele hierbei waren eine möglichst flexible Konfiguration der Vernetzung, genaue Abbildung des simulierten Einzel- und Ensemble-Verhaltens der Mikroschaltung, geringe Verlustleistung, Integration von Pixelzellen, um die Mikroschaltungen direkt mit Eingangssignalen versorgen zu können, und der wahlweise Einsatz des ASIC als direkt gebondeter Einzel-IC oder als Teil eines 3D Chipstapels.

Entsprechende Übertragungselemente zur Weitergabe der Pulsantworten der Mikroschaltungen vertikal im Chipstapel wurden in die einzelnen Neural Processing Units (NPU) integriert (siehe Abschnitt III.3.1). Informationsverarbeitung mit wertdiskreten Signalen, z.B. pulsweitenmoduliert oder neuronal/pulsbasiert, ist besonders für die Übertragung von analogen Werten über die 3D Kontakte geeignet, da deren Widerstand und Leckstrom über weite Bereiche variieren können [Benkart05], was eine direkte Übertragung analoger Ströme oder Spannungen erschwert. Zusätzlich kann eine Kompensation für fehlerhafte 3D-Kontakte (z.B. ein Multiplexen) wesentlich leichter für Signale mit nur zwei gültigen Zuständen durchgeführt werden als für wertkontinuierliche.

Eine der Adaptionsregeln der Mikroschaltung, die Membranadaption (Gleichung (III.1)), wurde bereits für andere Anwendungen in analoger Schaltungstechnik realisiert. Dabei wurden die simulativen Ergebnisse bestätigt, etwa im Rahmen einer Bildsegmentierung [Schreiter04] oder als Bestandteil eines Assoziativspeichers [Kaulmann05]. Für die letztlich durchgeführte IC-Implementierung der Mikroschaltung wurde jedoch aufgrund von Platzbedarf und reduzierter Entwurfszeit die Schaltung nicht aus einzelnen Neuronen und Synapsen aufgebaut, sondern eine phänomenologische Digitalschaltung entworfen, die das Verhalten der gesamten Mikroschaltung emuliert.

Die NPUs sind in einer regelmäßigen 128*128 Matrixstruktur über den ASIC angeordnet, um den Designaufwand gering zu halten und ansatzweise eine Art neuronale FPGA zu realisieren, bei der einfache Grundelemente je nach Konfiguration zu komplexen Gesamtverarbeitungsfunktionen verschaltet werden können[12]. Um die Realisierung der Topologie frei programmierbar zu halten, wurde ein zentraler, voll konfigurierbarer Routing-Baustein entworfen, der Pulse aus der gesamten NPU-Matrix empfängt und als Random Access wieder an ein oder mehrere NPUs verteilt (Abschnitt III.3.3). Für die Implementierung der Pulserfassung auf der NPU-Matrix wurde auf die IP eines AER (Abschnitt III.3.2) aus einem vorhergehenden Projekt zurückgegriffen, welches ursprünglich dafür entwickelt wurde, die Ausgangspulse eines lokal gekoppelten Netzwerks mit Hebbscher Adaption zur Bildsegmentierung zu analysieren [Schreiter04].
Während der Realisierungsphase des Router-ICs wurde auf der jeweils implementierten Abstraktionsstufe eine laufende Verifikation des Entwurfs vorgenommen, um den Entwurf gezielt auf Fehlertoleranz und Erfüllung der Spezifikation auszulegen. Hierbei kamen unter anderem VHDL-Simulatoren, Timingsimulationen, lehrstuhlinterne pulsbasierte symbolische Simulatoren und designspezifische ausführbare Systembeschreibungen in Matlab zum Einsatz.

### III.3.1 Neural Processing Unit

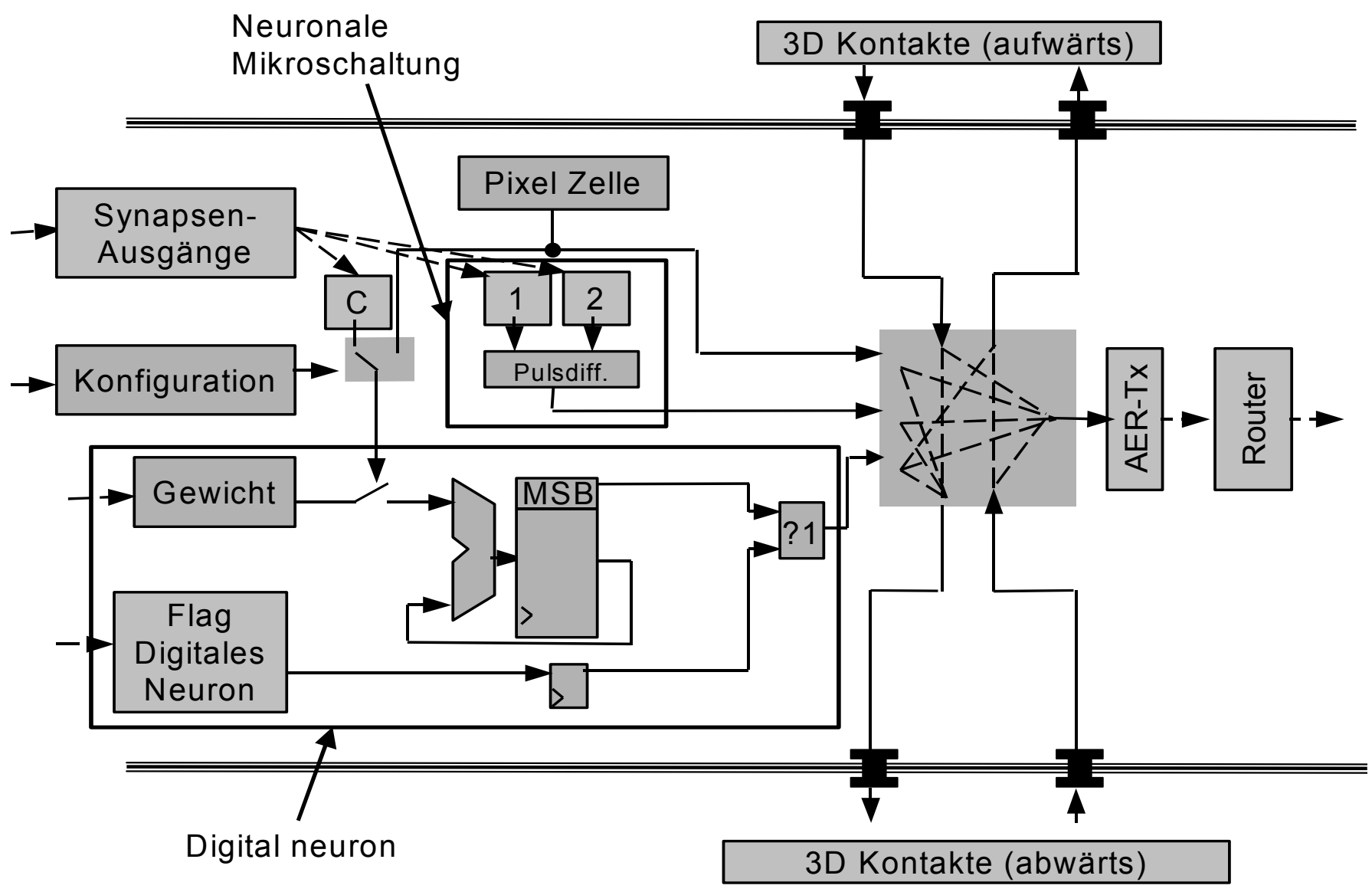

**Abbildung III.10.: Neuronale Verarbeitungseinheit (Mikroschaltung, Konfiguration, digitales Neuron, etc.)**

Die o.a. digital emulierte Mikroschaltung ist wie oben erwähnt Teil einer NPU, die weitere neuronale Funktionalität enthält, so z.B. ein digitales Neuron mit konfigurierbarem Eingangsgewicht zum Summieren der Maskenantwort, und eine pulsende Pixelzelle als Eingangssignal der Bildverarbeitung. Die Gewichtung der eingehenden Pulse des digitalen Neurons kann während der Laufzeit von außen gesteuert werden, so dass z.B. weitere Adaptionsregeln mittels einer externen FPGA realisiert werden können. Die NPU erhält Pulse vom Router, schickt diese an die entsprechenden Eingänge der Untereinheiten (Mikroschaltung/digitales Neuron), und die entstehenden Ausgangspulse (Neuron/Pixelzelle/Mikroschaltung) werden je nach Konfiguration über die oberen oder unteren 3D Kontakte an die entsprechenden ICs im Stapel weitergegeben oder an das AER auf dem eigenen IC gesandt. Weitere Verbindungsmöglichkeiten in der Schaltmatrix zwischen AER, 3D Kontakten und NPU sind z.B. das bidirektionale vertikale Durchreichen von Signalen, oder die horizontalen Weiterleitung der von oberen oder unteren 3D Kontakten kommenden Signale über das AER.

[12] Ein ähnliches Konzept mit vollständig digital emulierter Neuronenfunktionalität wird in [Eickhoff06] vorgestellt. Dieses offeriert mehr Konfigurierbarkeit/Flexibilität in den einzelnen Untereinheiten, enthält jedoch beispielsweise keine Pixelzellen. Die Rekonfigurierbarkeit im oben angeführten Konzept dient in erster Linie dazu, die Grundfunktionalität der Mikroschaltungen in möglichst effizienter Weise direkt für verschiedene Bildverarbeitungsaufgaben einzusetzen. Im Gegensatz dazu wird in [Eickhoff06] ein allgemeinerer Ansatz verfolgt, der jedoch für den vorliegenden Verwendungszweck überdimensioniert ist.

### III.3.2 Adress-Event-Representation

Eine Repräsentation von Pulsereignissen nach Ort (Adresse) und Zeit (Adress-Event-Representation, AER) wird oft in VLSI Realisierungen von pulsenden neuronalen Netzen angewandt, um die Kommunikation zwischen den einzelnen neuronalen Bausteinen kompakt paketbasiert abwickeln zu können. Die Motivation für eine solche Wiedergabe des neuronalen Informationsaustauschs beruht auf dem Postulat, dass nur der Zeitpunkt, nicht jedoch die genaue Pulsform für die Informationsverarbeitung von Belang ist [Indiveri06, Koch99 (Kapitel 14)].
Im Folgenden wird eine kurze Übersicht der in [Mayr06a] beschriebenen AER-Codierung gegeben. Das Codierungsschema ist kollisionsfrei, da die maximale Pulsfrequenz einzelner NPUs hardwareseitig auf 10kHz begrenzt ist, was der maximalen Verarbeitungsgeschwindigkeit des AER entspricht. Wenn Pulsereignisse in den selben Verarbeitungstakt des AER fallen, entscheidet ein Arbitrierer darüber, entsprechend eines der beiden Codewörter aus Abbildung III.11 für die Weitergabe multipler Ereignisse zu verwenden. Spitzen in der lokalen Aktivität werden damit parallel übertragen und später im Router wieder serialisiert. Die Pulscodierung und –kompression ist lokal auf 4*4 Neuronen große Blöcke orientiert, d.h. zeitlich nah beieinander liegende Pulse werden am effizientesten übertragen, wenn sie sich im selben Block ereignen. Die folgende Darstellung illustriert das Codierungsschema:

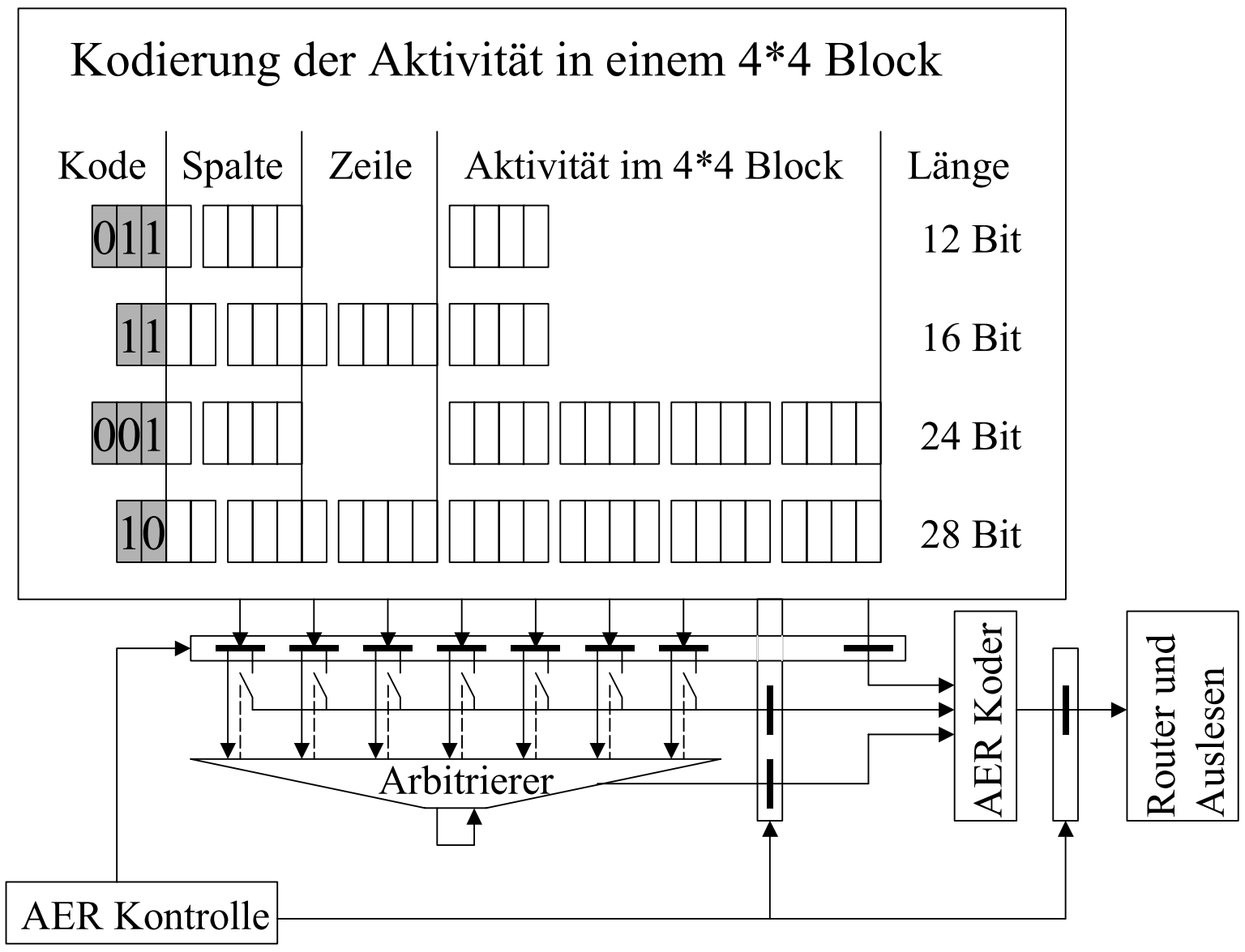

**Abbildung III.11.: Codierungs-/Kompressionsschema des im Router-IC verwendeten AER**

Die AER-Kontrolle taktet die zugehörigen AER-Schaltungsteile mit 120-200 MHz, was aufgrund der seriellen Architektur der Pulsrate entspricht, die vom AER abgefangen und weitergemeldet werden kann. Die mit der Taktfrequenz korrespondierende Genauigkeit der zeitlichen Auflösung beträgt demnach 5-8,3 ns, was bei einer mittleren Neuronenfrequenz von 10kHz ausreichend ist, um die in [Schreiter04] erwartete Wellenaktivität zu analysieren. Wie erwähnt, wird zwischen zeitlich zusammenfallenden Pulsen über den Arbitrierer entschieden, welcher dann die Pulse an den Codierungsbaustein weitermeldet. Mit dem dort implementierten Codierungsschema kann die Aktivität einer 128*128 Matrix von neuronalen Bausteinen beschrieben werden. Die ersten 2 oder 3 Bit stellen eine Kennung des Codeworts dar, dann folgen jeweils 5 Bit zur Codierung der Spalte/Zeile des 4*4 Unterblocks, in dem die Pulsaktivität stattgefunden hat, und die übrigen Bit codieren entweder ein Ereignis in einem einzelnen Element des 4*4 Blocks (12 und 16 Bit Codeworte). Oder es wird bei mehreren Ereignissen im selben Zeitfenster jedem neuronalen Element ein Bit zugewiesen, welches entsprechend 1 (Puls) oder 0 (kein Puls) gesetzt wird (24 und 28 Bit Codeworte). Wenn die Pulsaktivität in derselben Zeile stattfindet wie im vorhergehenden Takt, wird nur die neue Spalte übertragen (12 und 24 Bit Codeworte), ansonsten wird die komplette

Adresse des aktiven Blocks mitgesendet. In der weiteren Verarbeitungsfolge finden sich Router und Ausleseschaltung, die die Möglichkeit bieten, die Pulse zur weiteren Verarbeitung wieder in die Matrix zu verteilen und/oder zur Analyse über die Chipkante nach außen zu geben. Beide Komponenten sind als FIFO(First-In-First-Out)-Speicher realisiert, um kurze Burstaktivitäten abzupuffern. Bei längerer überdurchschnittlicher Aktivität werden die ältesten Pulse gelöscht, was v.a. für das Routing einen Kompromiss zwischen Pulsverlust (=Genauigkeitsverlust) und zeitlicher Relevanz der Pulse darstellt. Diesem Kompromiss liegt die Überlegung zugrunde, dass Pulse, die das Ende des FIFOs erreichen, zeitlich bereits soweit verzerrt sind, dass ihr Verlust weniger Konsequenzen hat als ihr Weiterleiten zu diesem späten Zeitpunkt, was für die hier diskutierte korrelierende Mikroschaltung oder phasenbasierte Codes [Bi98] zutrifft. Eine gezielte Ausnutzung der blockbasierten Natur des AER kann dadurch erfolgen, dass korrelierte Verarbeitungsfunktionen durch entsprechende Konfiguration im selben Block stattfinden.

### III.3.3 Pulse-Router

Der Aufbau des Routers ist in [Mayr06a] dokumentiert. Die Pulsverteilung findet in Form einer Zuordnungstabelle statt, d.h. wenn die durch $y_{empf}$ und $x_{empf}$ definierten Pulse die AER-Codierung und den FIFO-Speicher durchlaufen haben, werden die ihnen zugeordneten Adressen in der Tabelle abgefragt und die zugehörigen Ziel-Neuroelemente identifiziert. Für die Zieladresse wird zusätzlich noch der Eingang des Neuroelements abgefragt, mit dem der Puls verarbeitet werden soll, d.h. den positiven oder negativen Eingang der Mikroschaltung oder den Eingang des digitalen Neurons. Der Puls wird dann entsprechend des Inhalts der Zuordnungstabelle mit einem 1 aus 128 Decoder wieder auf der Matrix verteilt ($x_{send1}/y_{send1}\ldots y_{sendn}$). Der Decoder arbeitet mit derselben Taktfrequenz wie das AER, kann also dieselbe Anzahl Pulse ($120\text{-}200 \cdot 10^6 s^{-1}$) wieder über die Verarbeitungsmatrix verteilen.

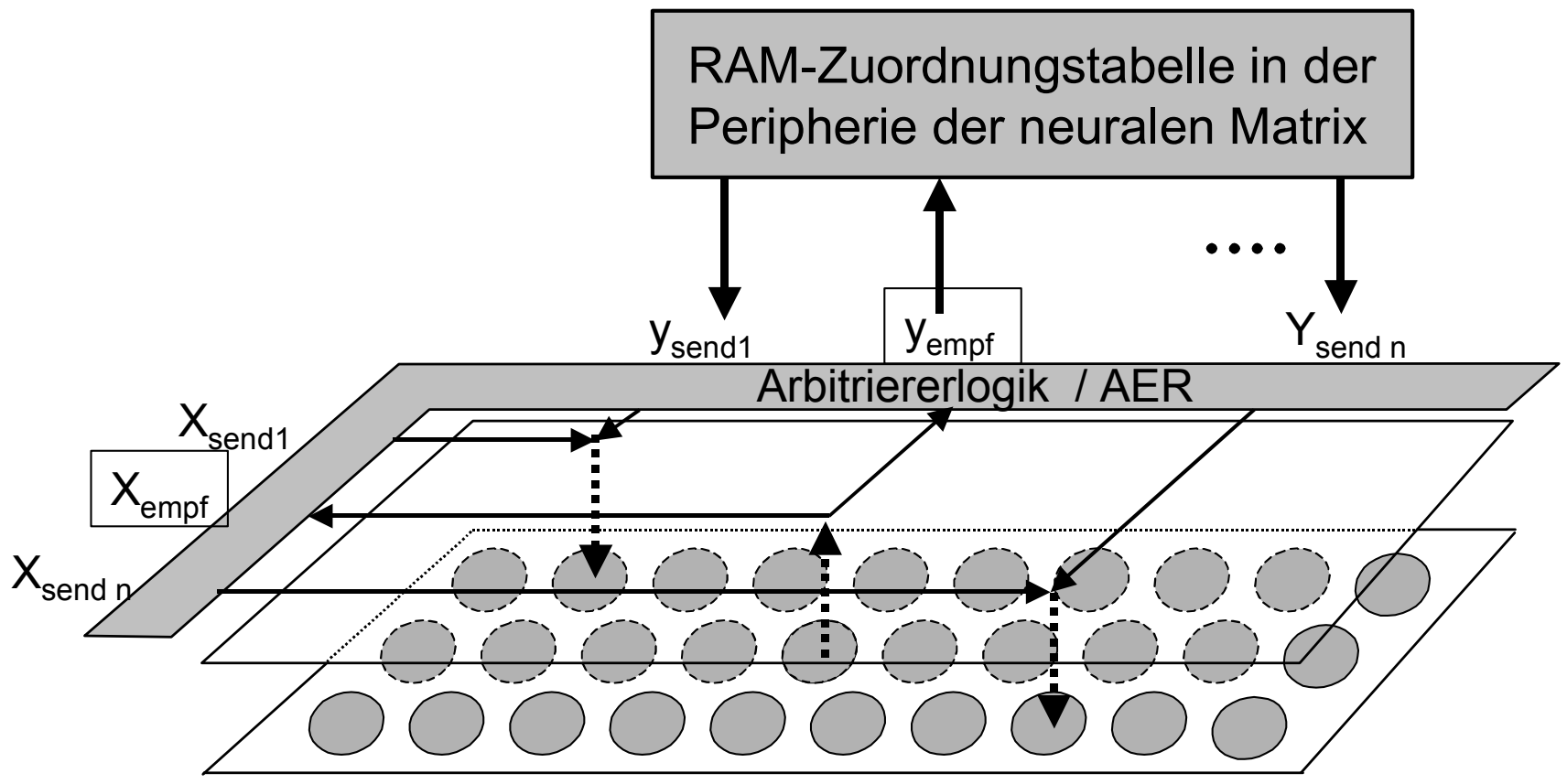

**Abbildung III.12.: Matrix-Organisation der Kombination aus AER und Router**

In Abbildung III.13 wird die Speicherorganisation der Zuordnungstabelle genauer dargestellt. Die Speicherorganisation ist zweigeteilt, wobei der erste Speicher, der fest den Neuroelementen zugeordnet ist, eine Adresse für den Zielspeicher enthält, in dem dann eine (variable) Anzahl an Zielen für den eingehenden Puls enthalten ist. Der Signalweg stellt sich damit wie folgt dar:
Den vom AER kommenden Pulsen wird in einer ersten Zuordnung, dem Lookup RAM, eine Adresse (‚baseaddr') und ein Offset im Target RAM zugewiesen. Diese Adresse gibt die Stelle im Target RAM an, ab der dort Ziele für den einkommenden Puls zu finden sind, während der zugehörige Adressbereich über den Offset definiert ist. Demnach kann ein einkommender Puls an bis zu 32 Zieladressen (5 Bit) weiterverteilt werden. Der Inhalt des Target RAM ist in zwei Hälften unterteilt, die zum Einen die Adresse des Ziel-NPU enthalten (2*7 Bit, in einem 128*128 Array), und zum Anderen 2 Bit, die den Eingang des NPU definieren, an den der Puls geleitet werden soll. Hierbei stellt A den positiven Eingang der Mikroschaltung dar, B den negativen, und C ist der Eingang des digitalen Neurons. Der Target RAM wird demnach je nach zu realisierender

Netztopologie variabel vergeben. Die ausgehenden Pulse werden entsprechend ihrer Adresse über Multiplexer auf der NPU-Matrix verteilt. Auf diese Weise realisiert die Kombination aus AER und Router virtuelle Axone/Dendriten, mit denen sich die NPUs in komplexen Netzwerken verbinden lassen.

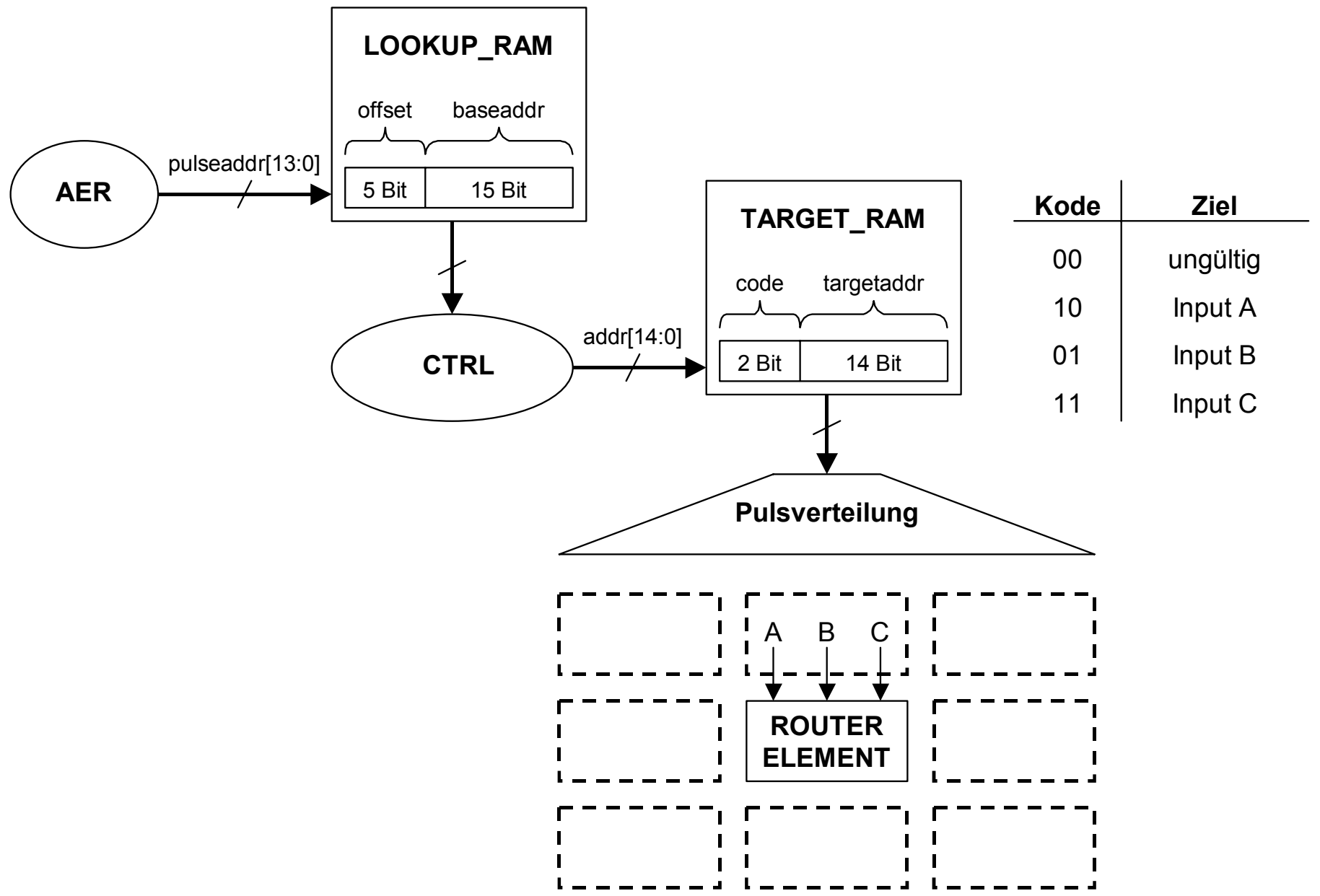

**Abbildung III.13.: Aufbau und Funktionsablauf des Routers**

Zusätzlich können in den Router parallel zum AER auch externe Pulse eingelesen werden, so dass sich für Testzwecke oder nicht-Pixelzellen-gebundene Verfahren Stimuli von außerhalb des Router-IC einspeisen lassen.

Diese Art von Routing ist konstant, d.h. Pulsereignisse werden bei einer feststehende Konfiguration immer zu denselben Zielen weitergegeben, unabhängig von der Aktivität auf den virtuellen Dendriten/Axonen. Jedoch können die o.a. Speicher zur Laufzeit umkonfiguriert werden, so dass z.B. Plastizitätsvorgänge basierend auf Pulsaktivität durch externe Analyse auf einer FPGA und entsprechende Umkonfiguration realisierbar wären.

# III.4 Gesamtkonzept und Simulationsergebnisse

## III.4.1 Implementierung des Gesamtkonzeptes

Der Router IC wurde in einer 130 nm Infineon Technologie als Full-Custom Digital-Design entworfen, mit automatisch generierten RAM-Makros und Mixed-Signal Inserts. Die NPUs wurden aus hochsprachlichen Beschreibungen synthetisiert und mit einem Place-and-Route-Werkzeug gesetzt, wobei die Layouts der handentworfenen pulsenden Pixelzellen in Platzhalterstellen in die NPUs eingesetzt wurden. Die folgende Abbildung gibt den Floorplan des Router-ICs wieder, aufgeteilt in die einzelnen Unterbaugruppen.

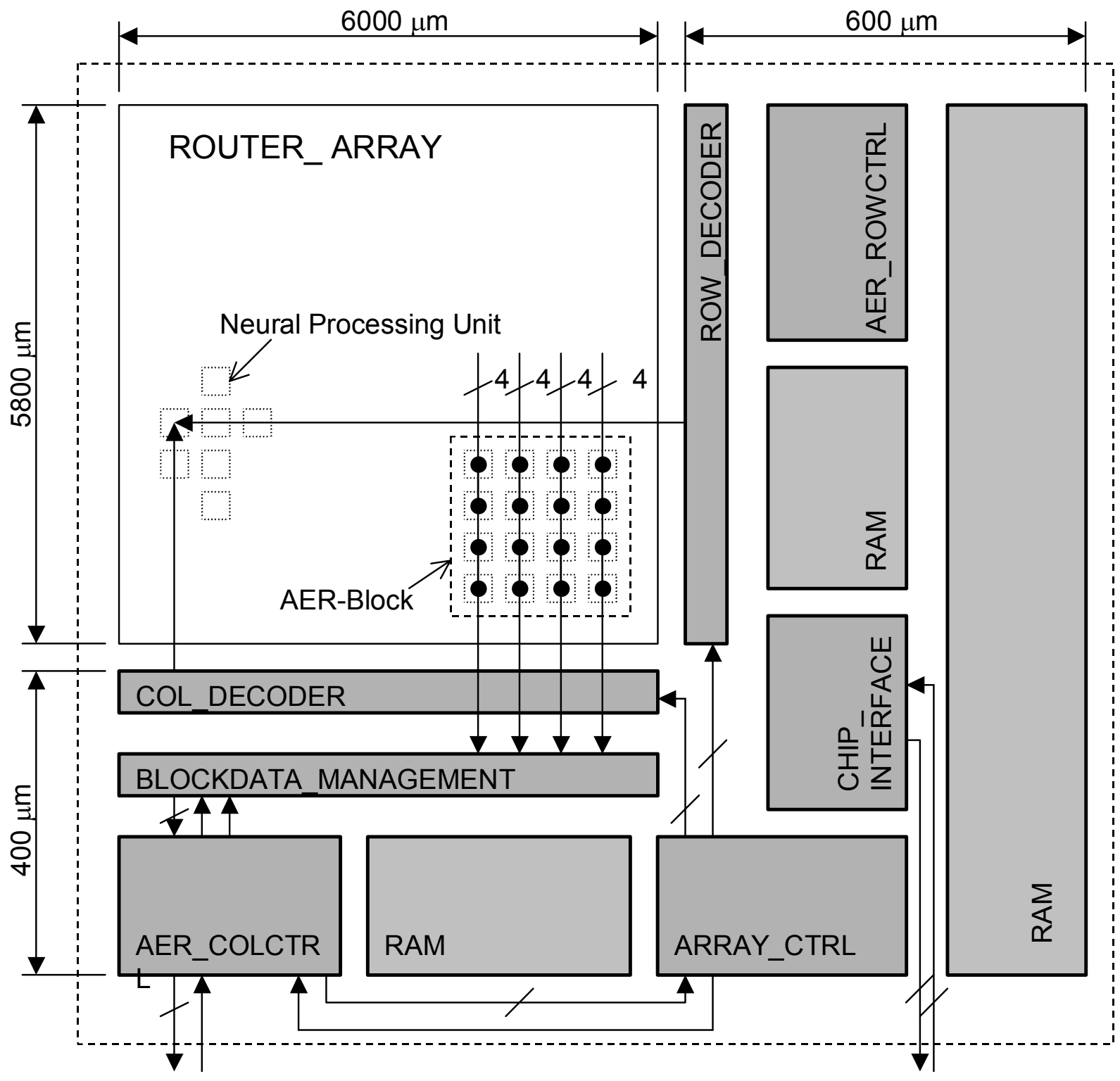

**Abbildung III.14.: Floorplan des Router-IC mit Einzelkomponenten**

Der Floorplan ist nicht maßstabsgerecht, einen Hinweis auf die echten Größenverhältnisse geben die Abmessungsangaben am Rand. Im Router Array wird angezeigt, wie ein Puls von Zeilen- und Spaltendecoder an ein Ziel-NPU verteilt wird. Ebenso verdeutlicht ist die Blockstruktur des AER. Die Konfiguration der NPUs und des Router-Speichers, d.h. der gesamten Verarbeitungsfunktion des ICs, wird über eine JTAG-Schnittstelle vorgenommen, die im Block CHIP_INTERFACE enthalten ist. Die simulierte Energieaufnahme bei einer mittleren Pulsverteilungsrate von $160*10^6$ Pulsen/s ist ca. 2 W für Router und Peripherie, die zusätzliche Leistungsaufnahme der NPUs hängt stark von ihrer Konfiguration ab (Pixelzellen, Mikroschaltung oder digitales Neuron aktiv), und variiert zwischen 100mW und 1,5 W.

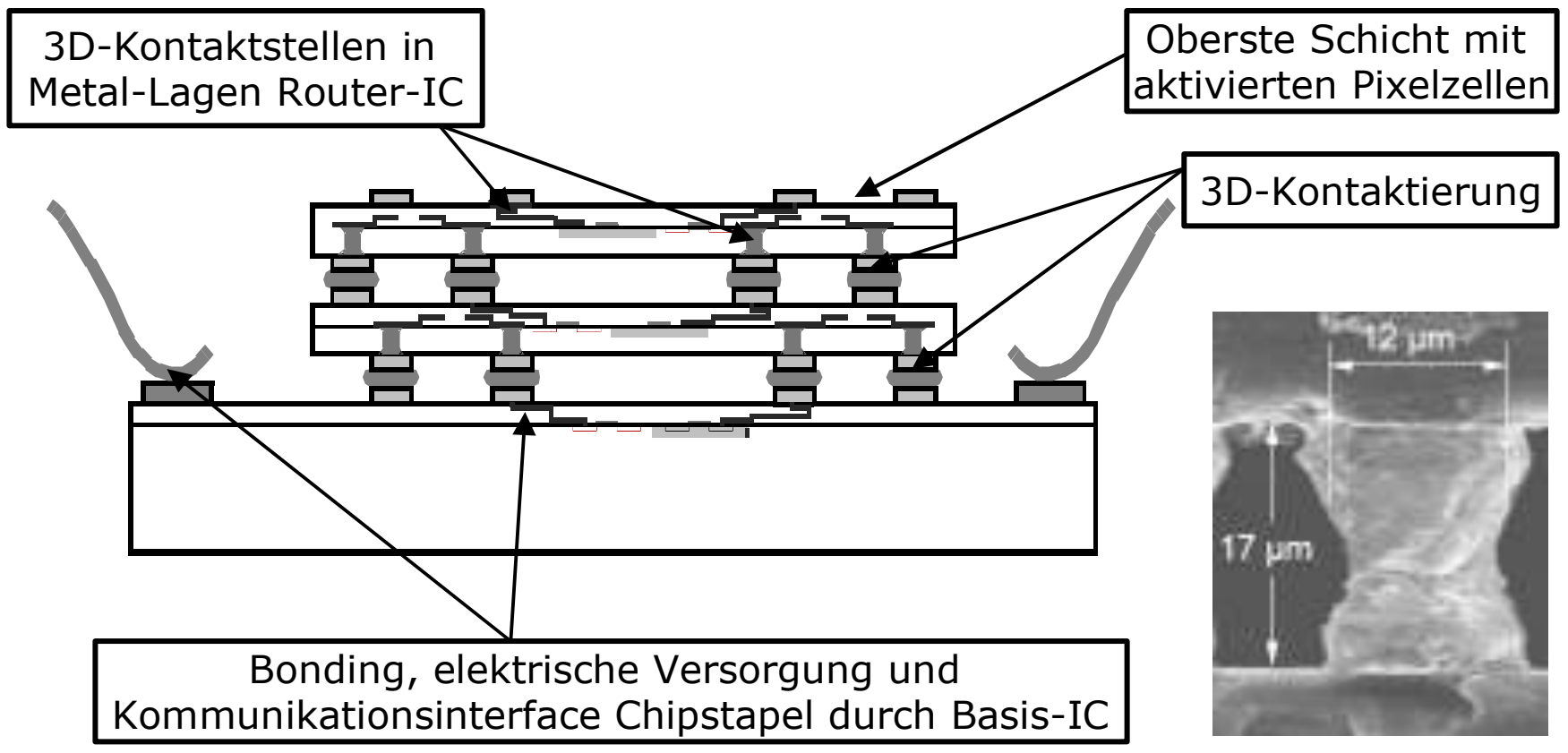

**Abbildung III.15.: 3D IC-Stapel und Bild einer 3D-Verbindung [Benkart05]**

Eine schematische Darstellung des zu fertigenden Chipstapels gibt Abbildung III.15 wieder. Auf einem Basis-IC sind mehrere Lagen des Router-ICs aufgebracht und durch 3D-Kontakte verbunden, die auf jeweils in der obersten und untersten Metalllage ausgeführte Kontaktflächen aufsetzen [Benkart05]. Der Basis-IC wird in einem regulären IC-Gehäuse gebondet, er enthält die entsprechenden analogen Biassignale, Spannungsversorgung und Digitalschnittstelle zum Router-IC-Stapel, um von der externen PCB-Platine Kontrolle und Kommunikation mit dem Chipstapel sicherzustellen. Statt des ursprünglich vorgesehenen ICs aus [Schreiter04] kann als Basis-IC des Chipstapels auch ein Assoziativspeicher verwendet werden, mit dem die berechneten Gaborcharakteristiken klassifiziert werden können [Kaulmann05].

### III.4.2 Simulationsergebnisse

Im folgenden wird der Router-IC für die in Abbildung III.4 vorgestellte Kantenfilterung konfiguriert. Der IC wird in 4*6 einzelne Bereiche aufgeteilt (Abbildung III.16 links), die jeweils einen Gaborfilter mit Orientierung 0° enthalten, wobei die Ausgänge der Mikroschaltung räumlich korreliert und jeweils rechts neben den Eingangsbildern versetzt so angeordnet wurden, dass sie für Analysezwecke leicht den Pixeln des Eingangsbildes zugeordnet werden können (Abbildung III.16 rechts oben). In den linken Hälften der Unterbereiche sind die Pixelzellen aktiviert und werden mit einer in diskreten Schritten drehenden Kante als Eingangsbild versorgt.

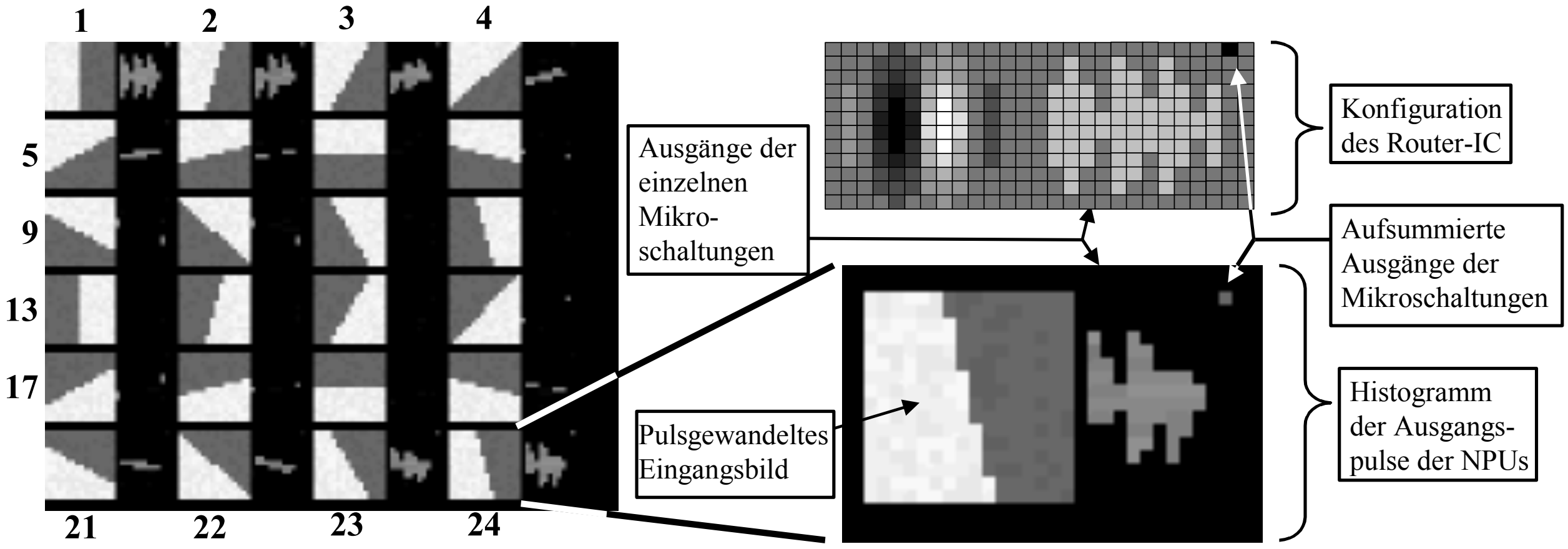

**Abbildung III.16: Konfiguration des Router-IC für Kantenfilterung mittels Gabormaske**

In Abbildung III.16 rechts unten (vergrößerter Teilbereich 24 der Matrix) ist anhand der Ausgänge der Mikroschaltungen deutlich zu sehen, wie im Vergleich zu Teilbereich 1 bereits ein Teil der Mikroschaltungen aufgrund der leichten Drehung der Kante nur auf Gebiete gleichen Grauwerts zugreift und damit kein Ausgangssignal liefert. Damit reduziert sich das Gesamtausgangssignal gegenüber Teilbereich 1. Das Ergebnis dieser Konfiguration als Pulsantwort über Drehwinkel aufgetragen stimmt mit Abbildung III.4 überein.

In Abbildung III.17 wird das Beispiel aus Abschnitt III.2.3/Abbildung III.5 für die Systemsimulation des Router-IC aufgegriffen. Die komplette Verarbeitungspyramide aus dem o.a. Abschnitt wurde hierbei über die NPUs auf einem Router-IC realisiert, die mittels der Pulsweiterleitung/Verteilung aus Abbildung III.13 entsprechend in die Hierarchie eingebunden sind. Die Abbildung gibt eine Grauwertrepräsentation des Puls-Histogramms aller NPUs des Router-IC an, so dass die Verarbeitung in den einzelnen Bereichen aus der relativen Häufigkeit ihrer Ausgangspulse deutlich wird.

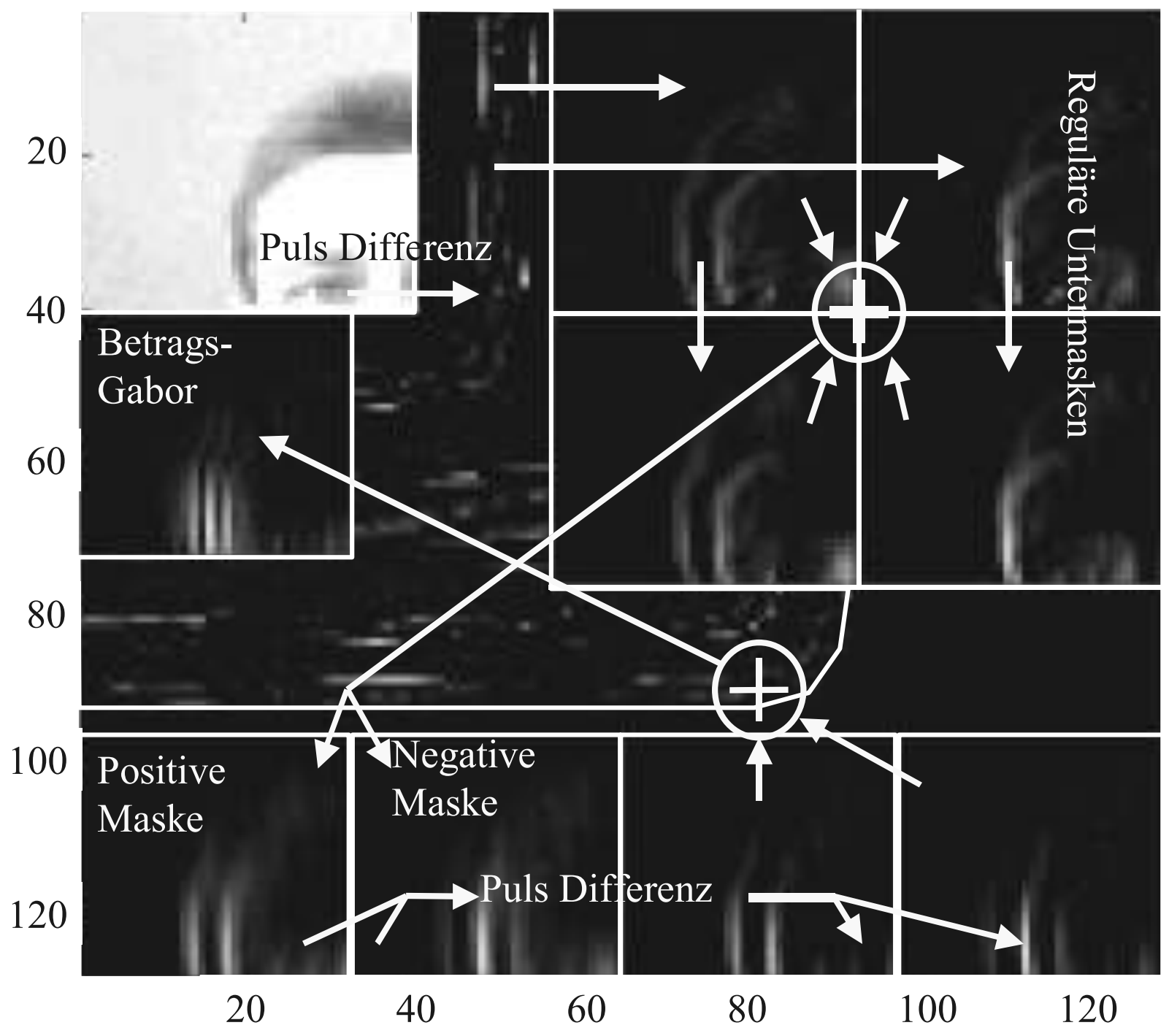

**Abbildung III.17.: Histogramm der Pulsdifferenzen für eine vollständige, hierarchische Gaborfaltung**

Die Maskenantwort wird hierbei aus Maske und Gegenmaske gewonnen, entsprechend Abschnitt III.2.3. Der Aufbau von Maske und Gegenmaske erfolgt mit der in Abschnitt III.2.4 vorgestellten hierarchischen Gabormaskenerzeugung. In der obigen Konfiguration des Router-ICs wurden diese einzelnen Schritte zur besseren Übersichtlichkeit räumlich in Untergruppen zusammengefasst.
16384 NPUs sind ausreichend, um ein Viertel des Originalbildes komplett mit der entsprechenden Gabormaske (Phase/Ausdehnung/Wellenzahl/Orientierung) abzudecken. Eine komplette Filterung des Bildes könnte z.B. in vier Schichten des Chipstapels erfolgen, bei der das Bild im obersten IC von den pulsenden Pixelsensoren aufgenommen wird und jeweils Teilbereiche des Bildes über die 3D-Kontakte in die darunterliegenden Schichten zur Weiterverarbeitung durchgeleitet werden.

# III.5 Schlussfolgerungen

Es wurde ein Bildverarbeitungskonzept dargestellt, das mittels neuronaler Mikroschaltungen in entsprechenden Netzwerktopologien in der Lage ist, pulsbasierte Faltungsoperation vorzunehmen. Die Mikroschaltungen basieren auf zwei Adaptionsregeln, die aus biologischen Messdaten postuliert wurden, zum Einen einer Hebbschen Pulskorrelation, zum Anderen einer quasi-digitalen Pulsinteraktion auf Dendriten, und führen als Gesamtverhalten eine Pulsdekorrelation zwischen zwei Eingangspulsfolgen durch (Abschnitt III.1.1). Die einzelnen Mikroschaltungen können als einfache Grundbausteine für Merkmalsverbindung (‚Feature Linking') gesehen werden, d.h. aus einzelnen, simplen Zusammenhängen (Rezeptor A höhere Feuerrate/Helligkeit als Rezeptor B) können komplexe Bildzusammenhänge aufgebaut werden. Im vorliegenden Beispiel wird gezeigt, wie die als Biologienäherung gefundenen Gabormasken [Jones87b] über eine entsprechende Verschaltung von Mikroschaltungen aufbaubar sind (Sektion III.2.4).
Die Faltung mit Gabormasken verwendet hierbei einige Informationsverarbeitungsprinzipien, die aus biologischen Messdaten postuliert werden, so etwa den Aufbau der Masken über mehrere Zwischenschritte mit zunehmend komplexerer Faltung in einer geschichteten, hierarchischen Struktur. Ebenso kann die Verwendung von Gegenmasken an denselben spatialen Koordinaten zur Abdeckung des gesamten Bild-Dynamikbereiches in den On/Off Zentren der Retina [Dacey00] wiedergefunden werden. Weitere Charakteristiken des biologischen Bildverarbeitungspfades, die ebenfalls von der vorliegenden pulsbasierten Verarbeitung aufgewiesen werden, sind Robustheit,

Parallelität, und die Eigenschaft, gleichzeitig auf verschiedenen zeitlichen Auflösungen zu operieren, d.h. sowohl schnelle, grobkörnige, als auch langsame, detaillierte Bildverarbeitung auszuführen.

Um die genannte Funktionalität in VLSI testen zu können, wurde ein Mixed-Signal Router-IC entwickelt, der alle nötigen Grundbestandteile, wie Mikroschaltungsrepräsentation, pulsende Pixelzelle, und einen Puls-Router enthält. Die pulsenden Pixelzellen haben eine mittlere Pixelfrequenz von 3 kHz, d.h. eine Gabormaskenantwort liegt nach 85 ms in einer Auflösung von 256 Graustufen vor. Für eine höhere Auflösung können für die Maskenantwort bis zu $10^3$ Einzelpulse aufsummiert werden, entsprechend einem Dynamikbereich von 60dB. Weitere Aufsummierung ist nicht sinnvoll, da die der Verarbeitung zugrunde liegenden Pixelzellen nur etwa 60dB SNR besitzen [Henker07] (siehe auch Abschnitt C.1).

Der Puls-Router wurde dafür ausgelegt, bis zu 200 Millionen Pulse pro Sekunde aus der NPU-Matrix zu sammeln und wieder zu verteilen, kann also eine 128*128 Matrix bei der maximalen (schaltungstechnisch begrenzten) Pulsfrequenz von 10kHz versorgen. Das Pulsverteilungskonzept ist universell anwendbar, d.h. es kann als Testumgebung für zukünftige pulsbasierte/digitale verteilte Verarbeitungsmechanismen dienen. Dies ist vor allem interessant in Hinblick auf die Verwendung von pulsbasierter Kommunikation in einem 3D-IC-Stapel, in dem Information redundant und robust verteilt sein muss (Kompensation fehlerhafter 3D-Kontakte) und die Informationsverarbeitung aus thermischen Gründen gleichmäßig verteilt sein soll. Es wurden Softwareroutinen entwickelt, die Gabormasken aus den Mikroschaltungen generieren, die Verarbeitungspyramide aufbauen, und eine Konfiguration für den Router-IC erzeugen, so dass dieser in entsprechend verteilter Weise die gewünschte Bildverarbeitung ausführt.

Neuroinspirierte VLSI-Realisierung von Gaborfilterung findet sich auch in der Literatur, z.B. [Morie01]. Das hier vorgestellte Konzept unterscheidet sich jedoch von diesen Arbeiten in seiner Vielseitigkeit, z.B. lassen sich die Gabormasken völlig frei parametrisieren, auch können komplett andere Faltungsmasken realisiert werden, oder die NPUs werden über externe Pulseinspeisung für nicht-bildgebundene Verarbeitung eingesetzt[13]. Zusätzlich wurde in den vorgestellten IC 3D-Funktionalität integriert, so dass dessen Verarbeitungsfunktionen auch im Rahmen eines Chipstapels einsetzbar sind.

Ein möglicher Nachteil des vorgestellten Konzeptes in seiner jetzigen Form ist die bedingte Skalierbarkeit, da die gesamte Pulsverteilung und damit die Vernetzung über den zentralen Router ausgeführt wird. Mithin müsste für größere Matrizen entweder die mittlere Pulsfrequenz gesenkt werden oder der Router beschleunigt, was jedoch durch die Zugriffsgeschwindigkeit des Router-RAMs nur bedingt möglich ist. Deshalb ist für Weiterentwicklungen angedacht, die Pulsvernetzung über ein verteiltes Netzwerk auszuführen, ähnlich dem Konzept in [Eickhoff06] bzw. der in Kapitel V geschilderten Layer1-Pulskommunikation.

[13] Ein Vergleich der in [Morie01, Nagata99] verwendeten Pixelzelle mit der in dieser Arbeit entworfenen findet sich in Abschnitt C.1.

# IV Verschiedene neuroinspirierte Informationsverarbeitungskonzepte

## IV.1 Signal und Rauschen in neuronalen Netzen

In Erweiterung des Abschnitts II.2 soll in Kapitel IV.1 vor allem die Rolle des in einem biologischen neuronalen Netz immer vorhandenen Hintergrundrauschens auf die dort stattfindende Signalübertragung untersucht werden. Hierzu werden Konzepte aus den Abschnitten II.2.3 und II.2.4 fortgeführt, zusätzlich wird die Betrachtung des Rauschens erweitert auf den Einfluss, den verschiedene neuronale Adaptionen auf die Signalübertragung über ein neuronales Netz haben. Wie im Weiteren ausgeführt, scheint ein maßgeblicher Mechanismus der Adaptionen zu sein, verschiedene Aspekte dieses Rauschen zu beeinflussen, beispielsweise das stochastische synaptische Übertragungsrauschen [Koch99 (Kapitel 13.2.2)] oder das Frequenzspektrum des pulsenden Hintergrundrauschens [Häusler07, Indiveri06].

### IV.1.1 Frequenz-/Spektrumsanalyse neuronaler Netze

Bei einer Betrachtung der Informationsübertragung in Neuronen fällt auf, dass diese durch viele frequenzabhängige Phänomene geprägt ist, beispielsweise auf dem Niveau von Einzelneuronen in der Übertragung von externen Stimuli über sensorische Nervenpfade [Gabbiani99]. Durch eine Analyse des Frequenzspektrums einer Pulsfolge können mithin frequenzabhängige Verarbeitungsmerkmale/Charakteristika sichtbar gemacht werden, bei denen beispielsweise reine statistische Auswertungen nicht greifen [Kass05]:

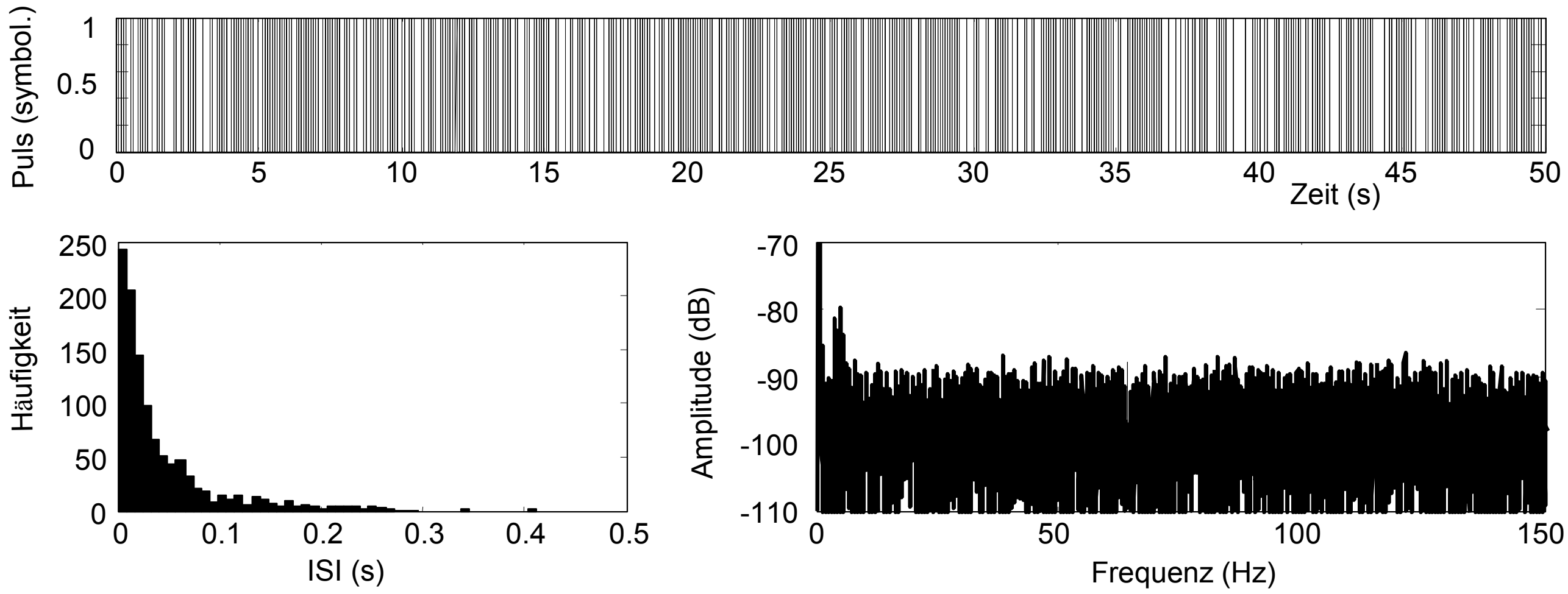

**Abbildung IV.1.: Burstende Pulsfolge mit niederfrequenten Grundelementen, von oben im Uhrzeigersinn: Darstellung der Pulsfolge, des zugehörigen Frequenzspektrums (50s*10kHz Datenpunkte) und ISI-Histogrammplot**

Eine visuelle Analyse der obigen Pulsfolge lässt abgesehen von einer leichten Tendenz zu Bursts[14] keine weiteren Strukturen erkennen. Auch bei der Betrachtung eines ISI-Histogramms ergeben sich keine signifikanten Unterschiede zu einer konventionellen Poisson-Verteilung [Kass05] (siehe auch Abbildung II.11). Eine Darstellung des Frequenzspektrums zeigt jedoch deutlich die Niederfrequenzkomponente der Burst-Aktivität bei ca. 5Hz.

Wie in der Einleitung ausgeführt, scheint auf Netzwerk-Ebene v.a. die Regulierung des Populationsrauschens interessant zu sein, um Signale nicht in der spontanen Hintergrundaktivität zu verlieren [Spiridon99]. In diesem Sinne ist eine Analyse des Rauschspektrums dahingehend relevant, dass die frequenzabhängige Amplitude des Rauschens festgestellt werden kann und damit die Frequenzbereiche, in denen Signalübertragung mehr oder weniger gut stattfinden kann. Ein typisches Frequenzspektrum der Ausgangspulse von schwach gekoppelten [Aronov03] Einzelneuronen wurde bereits in Abbildung II.14 dargestellt, in dem sich vor allem Auswirkungen der Refraktärcharakteristiken (d.h. von Einzelneuron-Effekten) erkennen lassen, jedoch keine der typischen Populationseffekte [Mar99 (Fig.3)]. Eine bessere Näherung für das Frequenzspektrum einer generischen Neuronenpopulation gibt die in derselben Abbildung dargestellte analytische Lösung. Diese Formel wurde gemäß Abschnitt II.2.2 für Poisson-Verteilungen hergeleitet und ist somit relevant für generische Populationen, da für deren Ausgangssignal in erster Näherung die Hypothese „Poisson-Verteilung“ zutrifft [Kass05].
Wie in Abbildung IV.1 dargestellt, kann jedoch auch ein auf den ersten Blick als Poisson-Verteilung identifiziertes ISI-Histogramm zeitliche/frequenzmäßige Feinstrukturen enthalten, die sich signifikant von dem weißem Rauschen unterscheiden, das eigentlich aus einer Poisson-Verteilung entsteht (Anhang A.1). Insbesondere für Populationen von Neuronen, die lateral inhibitorisch gekoppelt sind, ergibt sich eine deutliche Verminderung des Rauschpegels bei niedrigen Frequenzen [Mar99]. Dieses Verhalten kann über eine Analogie zu im Rahmen technischer AD-Wandlung eingesetzter Delta-Sigma-Modulatoren (DSM) erklärt werden (Anhang C.2). Die Grundstruktur eines DSM erster Ordnung, mit Integrator, 1-Bit Quantisierung und negativer Rückkopplung zeigt große Ähnlichkeit mit der Membranintegration und Pulsauslösung eines Neurons (siehe Abbildung C.5), wobei die gegenseitige laterale Inhibition die Rolle der direkten Rückkopplung übernimmt [Norsworthy96]. Aus der Tiefpasscharakteristik des Integrators ergibt sich die erwähnte Verschiebung des Puls/Quantisierungsrauschens hin zu höheren Frequenzen, ersichtlich aus der systemtheoretischen Analyse der Schleife in Abbildung C.6.
Im neuronalen Fall erstreckt sich diese Reduktion im Rauschpegel bis in Frequenzbereiche, die deutlich oberhalb der maximalen intrinsischen Pulsrate eines einzelnen Neurons liegen [Mayr04] (siehe auch Abbildung IV.5). Ein derartiges Netzwerk hat, wie in Abbildung II.15 angedeutet, die Fähigkeit, einen relativ hochfrequenten Stimulus auf mehrere Neuronen verteilt zu übertragen, wobei die Rauschreduktion für einen verbesserten Rauschabstand (SNR) sorgt. Mar et. al. [Mar99] sprechen von einer Dekorrelation der Neuronen über die inhibitorische Kopplung, d.h. es wird mit zunehmender Stimulusfrequenz immer weniger redundante Information übertragen. Die Relevanz dieser Informationsaufteilung und redundanzoptimierten Übertragung lässt sich am Absinken des SNR in der rechten Hälfte von Abbildung II.15 ablesen. Mit einer derartigen Netzstruktur wird dem Netzwerk ermöglicht, auf sehr schnelle, transiente Stimuli zu reagieren[15] [Gerstner99].
Eine andere, gleichlaufende Sichtweise orientiert sich an konventionellen, zeitversetzt betriebenen AD-Wandlern [Poorfard97]. Unter dieser Sicht wirkt die inhibitorische Kopplung wie ein dynamisches Time-interleaving, d.h. wenn ein Neuron im aktuellen Zeitverlauf das Signal aufintegriert hat und pulst und damit diesen Verlauf „quantisiert“ und überträgt, hält es durch die inhibitorische Kopplung die anderen Neuronen ab, ebenfalls diesen Abschnitt zu quantisieren. Die restlichen Neuronen werden (zumindest teilweise) zurückgesetzt und starten entsprechend die Signalquantisierung bis zum nächsten Ausgangspuls.
In den nächsten beiden Abschnitten werden eigene frühere Arbeiten zur Signalverarbeitung dieser speziellen kortikalen Struktur unter Verwendung der in den Abschnitten II.2.3 und II.2.4 gelegten Grundlagen erweitert. Im Besonderen wird die Auswirkung bestimmter Formen aktiver neuronaler Adaption auf das Frequenzspektrum und das generelle Übertragungsverhalten dieser inhärent passiven Struktur untersucht.

[14] Kurze Zeitbereiche, in denen im Ausgangspulssignal die Anzahl der Aktionspotentiale deutlich über der mittleren Rate liegt [Koch99].

### IV.1.2 Synaptische Kurzzeitadaption als selektive Regulierung des SNR

Wie in Abschnitt II.1.2 erwähnt, existieren verschiedene Arten der synaptischen Adaption, die auf Zeitskalen im Sekundenbereich agieren und damit unter signaltheoretischen Gesichtspunkten die

[15] Die Fähigkeit, als Population Stimuli zu übertragen, die über der mittleren Spikefrequenz eines einzelnen Neurons liegen, ist im Endeffekt gleichbedeutend mit schneller Signalpropagierung.

Übertragung eines Signals von seiner unmittelbaren Vorgeschichte abhängig machen. In früheren Arbeiten [Mayr05c] zum technischen Einsatz des neuronalen Noise Shapings wurde eine adaptive Nachbearbeitung des (akkumulierten) Netzwerkausgangssignals durchgeführt, dahingehend, dass für eine große Anzahl an Ausgangspulsen des Noise-Shaping-Netzwerks innerhalb eines bestimmten (kurzen) Zeitraumes die Amplitude der Pulse progressiv erhöht wurde. Diese Adaption hat starke Ähnlichkeit mit der sogenannten Posttetanic Potentation (PTP) [Koch99 (Abschnitt 13.2.2)], bei der die synaptische Übertragungswahrscheinlichkeit $p$ für jeden einzelnen Puls steigt, wenn viele Pulse zu einem Zeitpunkt eintreffen. Diese Modellierung der Wahrscheinlichkeits-modulation als Erhöhung der Amplitude scheint zulässig, wenn ein Populationssignal betrachtet wird, bei dem ähnlich wie in Abbildung II.15 ein homogenes Eingangssignal verteilt übertragen werden soll. Die Ausgangsaktivität der Neuronen ist damit korreliert, wodurch die einzelnen Synapsen des Netzwerks eine korrelierte Adaption von $p$ erfahren, was sich im Populationssignal als ein mit der Anzahl der Synapsen quantisierter adaptiver Skalierungsfaktor darstellt. Von Koch et al. wird postuliert, dass dies auf Einzelsynapsenniveau der sicheren Signalübertragung v.a. bei hochfrequenten, starken Stimuli dient, z.B. bei sensorischen Neuronen [Koch99 (Abschnitt 13.2.2)]. Auf Populationsebene scheint diese Art der synaptischen Kurzzeitadaption aber eher der Signalüberhöhung und damit der Erhöhung des SNR zu dienen [Mayr05c].
Synaptische Adaptionen beeinflussen neben $p$ auch die Neurotransmitterausschüttungsmenge $q$, ebenfalls auf Zeitskalen im Sekundenbereich [Markram98] (siehe auch Abschnitt II.1.2). Die Auswirkungen dieser Adaption lassen sich am ehesten als Transientenübertragung charakterisieren, d.h. Änderungen der präsynaptischen Pulsraten werden in vollem Dynamikumfang postsynaptisch weitergemeldet. Wenn diese Signaländerungen länger anhalten bzw. ein Konstantsignal anliegt, wird der Dynamikbereich der zugehörigen Antwort zunehmend begrenzt (siehe Abbildung A.1). Phänomenologisch scheint dieses Modell damit eine gegenteilige Funktion zu PTP auszuführen, d.h. es wird keine Expansion des Signals vorgenommen, das den Dynamikbereich am Ausgang erhöhen würde, sondern eine Kompression. Erklärbar ist dieser Widerspruch über die etwas größere Zeitkonstante dieser sogenannten ‚quantalen' Kurzzeitadaption und das neuronale Bestreben, eine Maximierung der übertragenen Information zu erreichen [Yu05]. Ein starker Stimulus wird damit im Zusammenspiel der beiden Adaptionen in der ersten Phase der Übertragung durch die $p$-Modulation überhöht dargestellt, während ein Fortbestand des Stimulus wenig neue Information beinhaltet und damit über die Anpassung der Neurotransmitter-ausschüttung ‚wegadaptiert' wird. Die so entstehende Kompression kann ebenfalls dazu dienen, sensorische Information mit einem sehr großen Dynamikbereich auf die dynamikbegrenzte Pulsantwort eines Neurons abzubilden [Ohzawa82].
Von Markram et al. wird in [Markram98] ein mathematisches Modell für die quantale Kurzzeit-adaption aus biologischen Messdaten hergeleitet. Es kennt zwei Parameter, die (momentan) verwendete synaptische Ausschüttungsmenge $u_n$ und die (noch) freie synaptische Ausschüttungsmenge $R_n$, jeweils als Bruchteil der maximalen Menge. Die iterativen Bildungsvorschriften für $u_{n+1}$ bzw. $R_{n+1}$ aus den vorhergehenden Gliedern werden wie folgt formuliert:

$$u_{n+1} = u_n e^{-\frac{\Delta t}{\tau_{facil}}} + U * \left(1 - u_n e^{-\frac{\Delta t}{\tau_{facil}}}\right) \quad \textbf{(IV.1)}$$

$$R_{n+1} = R_n \left(1 - u_{n+1}\right) e^{-\frac{\Delta t}{\tau_{rec}}} + 1 - e^{-\frac{\Delta t}{\tau_{rec}}} \quad \textbf{(IV.2)}$$

Die ersten Glieder der Folgen können aus dem Startwert $U$ (im Ruhezustand der Synapse) für die verwendete synaptische Ausschüttungsmenge berechnet werden, mit $u_1=U$ bzw. $R_1=1-U$ [Markram98]. Der durch einen präsynaptischen Puls hervorgerufene exzitatorische postsynaptische

Strom (EPSC) kann aus dem Produkt von $u_n$ und $R_n$ berechnet werden, gewichtet mit einem Umrechnungsfaktor $A$ von Ausschüttungsmenge auf im Mittel dadurch hervorgerufenen Strom[16]:

$$EPSC_n = A * R_n * u_n \qquad \textbf{(IV.3)}$$

Aus Gleichung (IV.1) kann durch Gleichsetzen von $u_n$ und $u_{n+1}$ das konvergente $u_k$ für eine feste Pulsrate λ hergeleitet werden:

$$u_k(\lambda) = \frac{U}{1-(1-U)*e^{-\frac{1}{\lambda\tau_{facil}}}} \qquad \textbf{(IV.4)}$$

Mit diesem $u_k$ und einer ähnlichen Gleichsetzung erhält man das konvergente $R_k$ für eine feste Pulsrate λ:

$$R_k(\lambda) = \frac{1-e^{-\frac{1}{\lambda\tau_{rec}}}}{1-(1-u_k(\lambda))*e^{-\frac{1}{\lambda\tau_{rec}}}} \qquad \textbf{(IV.5)}$$

Dieses Modell bewirkt, dass die EPSC-Antwort für niedrige Eingangspulsraten mit einer Zeitkonstanten $\tau_{facil}$ angehoben wird, während die Antwort auf hohe Pulsraten mit $\tau_{rec}$ abgeschwächt wird, so dass insgesamt der Dynamikbereich der Eingangssignale nach einer kurzen Adaptionsphase stark komprimiert wird [Markram98] (siehe auch Abbildung A.1).
Für veränderliche Pulsraten sollte jedoch die Komprimierung nicht so stark zu Tage treten, um nicht wichtige Information über den Stimulus zu verlieren [Steveninck97]. Im folgenden soll deshalb bei diesem Modell untersucht werden, was eine periodische Modulation der Pulsrate vergleichbar etwa mit einer Synfire-Chain [Durstewitz00] oder einem natürlichen Stimulus [Gabbiani99] für Auswirkungen auf den mittleren EPSC hat. In Anhang A.2 wurde eine Methode hergeleitet, mit der die in den Gleichungen (IV.1) und (IV.2) angegebenen iterativen Zeitfunktionen mit absoluten Zeitabhängigkeiten einer einfachen e-Funktion und den konvergierten Werten aus den Gleichungen (IV.5) und (IV.4) genähert werden können. Die analytische Untersuchung der Modulationsabhängigkeit des EPSC soll anhand dieser Näherung und einer periodischen Rechteckmodulation der präsynaptischen Pulsrate erfolgen:

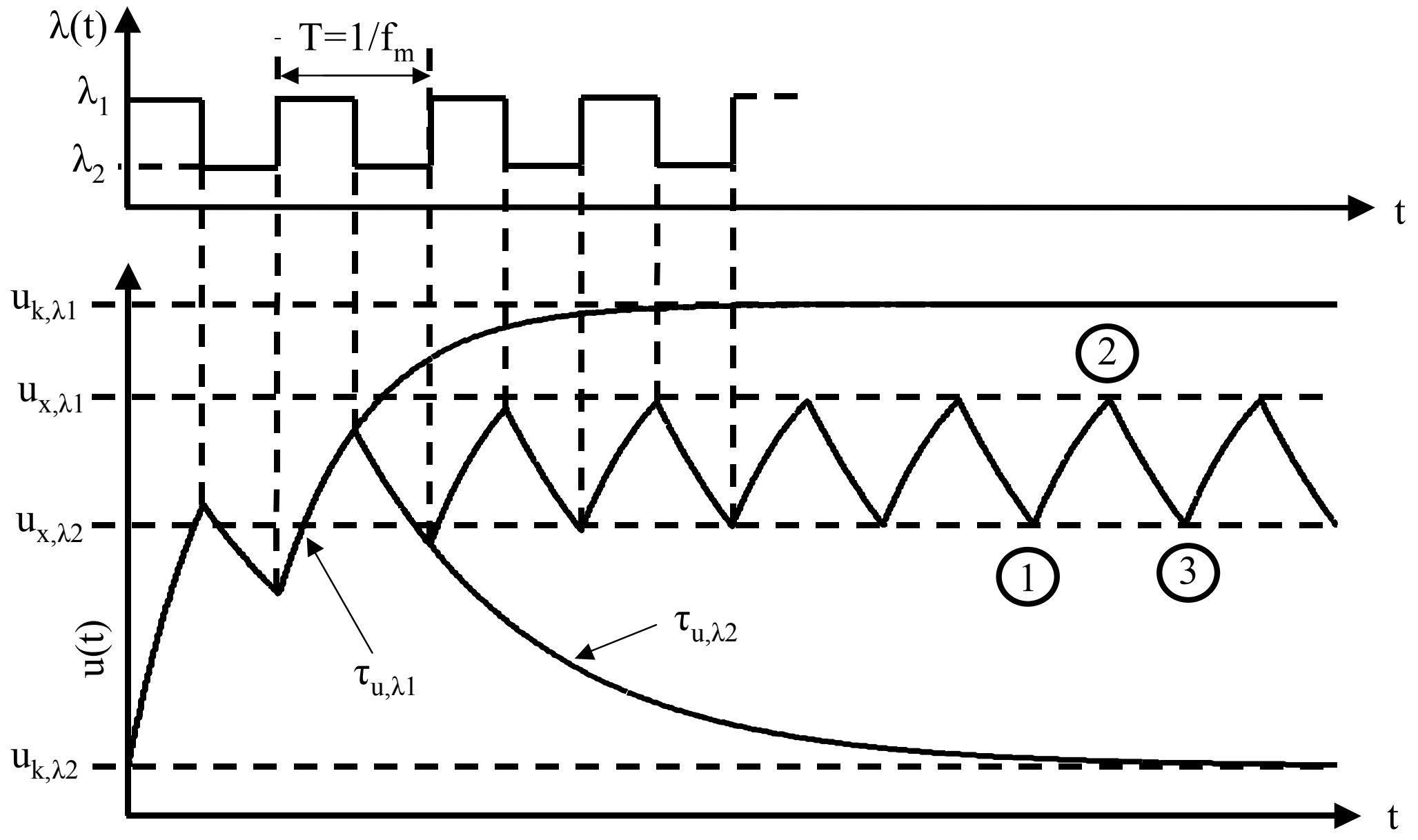

**Abbildung IV.2.: Zeitverlauf von u(t) in Abhängigkeit von der Modulationsfrequenz $f_m$ und der jeweiligen hohen und niedrigen Pulsrate**

[16] $EPSC_n$ gibt den mittleren Strom wieder, der bei einem einzelnen Puls entsteht. Die Umrechnung in einen effektiven EPSC unter Berücksichtigung der aktuellen zugrunde liegenden Pulsfolge wird in Gleichung (IV.12) hergeleitet.

Wie aus der obigen Darstellung exemplarisch für $u_n$ entnehmbar, erfolgt ein periodisches „Einschwingen" auf einen festen Amplitudenbereich in Abhängigkeit von der Modulationsfrequenz $f_m$, konvergiertem $u_k$ für niedrige und hohe Pulsrate, sowie den Zeitkonstanten $\tau_{u,\lambda 1}$ und $\tau_{u,\lambda 2}$ der Näherungsformeln[17].

Als Ausgangsbasis für die Herleitung der Modulationsabhängigkeit des EPSC wird zuerst Gleichung (A.26) in derselben Notation wie Gleichung (A.34) aufgeschrieben. Sie ist gültig für den aufsteigenden und abklingenden Ast, je nach dem Vorzeichen der Differenz $(u_0 - u_k)$:

$$u(t) = (u_0 - u_k)e^{-\frac{t}{\tau_{konv,u}}} + u_k \qquad \textbf{(IV.6)}$$

Zur Berechnung des Integrals über den mittleren EPSC müssen die Startwerte zu Beginn des Taktes berechnet werden; dies sind wie in Abbildung IV.2 dargestellt in der Regel nicht die jeweiligen konvergierten Werte, sondern Zwischenwerte. Anhand von $u_{x,\lambda 2}$ soll dies exemplarisch durchgeführt werden. Ausgangspunkt ist die Überlegung, dass der Weg von Punkt 1 bis Punkt 3 in Abbildung IV.2 über die beiden Annäherungsvorgänge an $u_{k,\lambda 1}$ und $u_{k,\lambda 2}$ mit ihren unterschiedlichen Zeitkonstanten wieder bei demselben Wert $u(t)$ ankommen muss. Von Punkt 1 ausgehend lautet die Gleichung für die zeitliche Entwicklung von $u(t)$ wie folgt (jeweils bezogen auf $t=0$ zum Zeitpunkt der Flanke von $\lambda(t)$):

$$u(t) = (u_{x,\lambda 2} - u_{k,\lambda 1})e^{-\frac{t}{\tau_{u,\lambda 1}}} + u_{k,\lambda 1} \quad bzw. \quad u\left(\frac{1}{2f_m}\right) = (u_{x,\lambda 2} - u_{k,\lambda 1})e^{-\frac{1}{2f_m\tau_{u,\lambda 1}}} + u_{k,\lambda 1} \qquad \textbf{(IV.7)}$$

Für den Weg von Punkt 2 zu Punkt 3 kann der oben berechnete Startwert dann in die neue Formel für $u(t)$ eingesetzt werden; wenn diese zum Zeitpunkt $1/(2f_m)$ ausgewertet wird, muss wieder der Wert $u_{x,\lambda 2}$ entstehen:

$$u_{x,\lambda 2} = \left[u\left(\frac{1}{2f_m}\right) - u_{k,\lambda 2}\right]e^{-\frac{1}{2f_m\tau_{u.\lambda 2}}} + u_{k,\lambda 2} \qquad \textbf{(IV.8)}$$

Die Auswertung der Gleichungen (IV.7) und (IV.8) ergibt folgenden Ausdruck für den unteren Wert $u_{x,\lambda 2}$ des periodischen eingeschwungenen Zustands von $u(t)$:

$$u_{x,\lambda 2} = \frac{u_{k,\lambda 1}e^{-\frac{1}{2f_m\tau_{u.\lambda 2}}}\left(1 - e^{-\frac{1}{2f_m\tau_{u.\lambda 1}}}\right) + u_{k,\lambda 2}\left(1 - e^{-\frac{1}{2f_m\tau_{u.\lambda 2}}}\right)}{1 - e^{-\frac{\tau_{u.\lambda 1}+\tau_{u.\lambda 2}}{2f_m*\tau_{u.\lambda 1}*\tau_{u.\lambda 2}}}} \qquad \textbf{(IV.9)}$$

In ähnlicher Weise lassen sich die entsprechenden Startwerte $u_{x,\lambda 1}$ sowie $R_{x,\lambda 1}$ und $R_{x,\lambda 2}$ herleiten. Für die Zeitdauer von Punkt 1 bis Punkt 2 berechnet sich die mittlere synaptische Ausschüttungsmenge $\overline{UR}_{12}$ als Integral über das Produkt der entsprechenden Zeitfunktionen $u(t)$ und $R(t)$ in diesem Abschnitt, normiert auf die betrachtete Integrationszeit:

$$\overline{UR}_{12} = 2f_m * \int_0^{\frac{1}{2f_m}} \left[(u_{x,\lambda 2} - u_{k,\lambda 1})e^{-\frac{t}{\tau_{u,\lambda 1}}} + u_{k,\lambda 1}\right] * \left[(R_{x,\lambda 2} - R_{k,\lambda 1})e^{-\frac{t}{\tau_{R,\lambda 1}}} + R_{k,\lambda 1}\right]dt \qquad \textbf{(IV.10)}$$

dieses Integral ergibt:

[17] $\tau_{u,\lambda 1}/\tau_{u,\lambda 2}$ bezeichnet hierbei die Zeitkonstante $\tau_{konv,u}(\lambda_1)$ bzw. $\tau_{konv,u}(\lambda_2)$ aus Gleichung (A.25), in gleicher Weise wird im Weiteren mit den Zeitkonstanten für konvergierte $R_k$ aus Gleichung (A.33) verfahren.

$$\overline{UR}_{12} = 2f_m * \left[ \left(u_{x,\lambda 2} - u_{k,\lambda 1}\right)\left(R_{x,\lambda 2} - R_{k,\lambda 1}\right) \frac{\tau_{u,\lambda 1} * \tau_{R,\lambda 1}}{\tau_{u,\lambda 1} + \tau_{R,\lambda 1}} \left(1 - e^{-\frac{\tau_{u,\lambda 1} + \tau_{R,\lambda 1}}{2f_m * \tau_{u,\lambda 1} * \tau_{R,\lambda 1}}}\right) + \right. \qquad \textbf{(IV.11)}$$
$$\left. \left(u_{x,\lambda 2} - u_{k,\lambda 1}\right)\tau_{u,\lambda 1} R_{k,\lambda 1} \left(1 - e^{-\frac{1}{2f_m \tau_{u,\lambda 1}}}\right) + \left(R_{x,\lambda 2} - R_{k,\lambda 1}\right)\tau_{R,\lambda 1} u_{k,\lambda 1} \left(1 - e^{-\frac{1}{2f_m \tau_{R,\lambda 1}}}\right) + \frac{u_{k,\lambda 1} R_{k,\lambda 1}}{2f_m} \right]$$

Auf gleich Weise lässt sich die Integration von Punkt 2 auf Punkt 3 durchführen. Wie bereits für Gleichung (IV.3) erwähnt, muss das oben berechnete Zeitmittel effektiver synaptischer Ausschüttung $\overline{UR_{xx}}$ noch auf die in der jeweiligen Wegstrecke erfolgten Eingangspulse bezogen werden. Dies kann erfolgen durch die Normierung der Pulsdauer auf eine Bezugszeit und Multiplikation mit den in dieser Bezugszeit aufgetretenen Pulsen:

$$\overline{EPSC}_{12} = A * \frac{T_{puls}}{T_{norm}} * N_{puls} * \overline{UR}_{12} = A * \frac{T_{puls}}{T_{norm}} * \lambda_1 * T_{norm} * \overline{UR}_{12} \qquad \textbf{(IV.12)}$$

Als Bezugszeit wird hier $T_{norm}=1/(2f_m)$ gesetzt, da in diesem Intervall die Pulsrate konstant ist und sich $N_{puls}$ als $\lambda_1 * T_{norm}$ (bzw. $\lambda_2 * T_{norm}$ für $\overline{EPSC_{23}}$) darstellen lässt. Im Weiteren wird als Näherung der biologischen Pulsdauern in Abbildung II.4 eine Pulsdauer $T_{puls}$ von 1,4ms Pulsdauer angenommen[18]. Der durchschnittliche EPSC über den kompletten Weg von Punkt 1 bis 3 in Abbildung IV.2 kann dann als $\lambda$-gewichteter Mittelwert der beiden mittleren postsynaptischen Ausschüttungen berechnet werden, multipliziert mit dem Äquivalenzfaktor $A$ für die Umrechnung Neurotransmittermenge→Strom.

$$\overline{EPSC} = A * \frac{1}{2} * T_{puls} * \left(\lambda_1 * \overline{UR}_{12} + \lambda_2 * \overline{UR}_{23}\right) \qquad \textbf{(IV.13)}$$

Aus den obigen Gleichungen ergibt sich ein EPSC-Übertragungsverhalten in Abhängigkeit von der Modulationsfrequenz. Analytisch ermittelte Datenpunkte aus den genäherten Gleichungen (IV.6), etc. zu diesem Übertragungsverhalten werden im Folgenden mit Simulationen der exakten iterativen Verhaltensgleichungen (IV.1) und (IV.2) verglichen. Diese Simulationen wurden mit dem Parametersatz von Fig. 4(D) aus [Markram98] durchgeführt, d.h. $A$=1540pA, $U$=0,03, $\tau_{rec}$=130ms, $\tau_{facil}$=530ms, sowie $\lambda_1$=130s$^{-1}$ und $\lambda_2$=6s$^{-1}$. Die Zeitkonstanten $\tau_{u,\lambda 1}$, etc. für die Näherungsformeln wurden verwendet wie aus den genannten Parametern und den Gleichungen in Anhang A.2 berechenbar; die asymptotischen Werte $u_{k,\lambda 1}$, etc. ebenfalls aus diesen Parametern gemäß den Gleichungen (IV.4) und (IV.5), sowie die eingeschwungenen Startwerte aus Gleichung (IV.9) und ihren Pendants für andere $\lambda$ und für $R$.

Im Unterschied zu der obigen analytischen Lösung diente den Simulationen ein sinus-modulierter Erneuerungsprozess[19] statt einer deterministischen Rechteckmodulation als Stimulus, um eine größere Biologienähe z.B. zu dem quasi-sinus-Signal aus [Gabbiani99] herzustellen. Der resultierende präsynaptische Pulsstimulus ist damit qualitativ ähnlich der Pulsfolge in Abbildung IV.4, jedoch sind die einzelnen ISIs deutlich variabler. Das Neuronenmodell für die postsynaptische Membran ist LIAF, mit einer Membrankapazität 10pF, einem Membranwiderstand 1GΩ, und einer resultierenden Zeitkonstanten 10ms, nach [Koch99 (Kapitel 1)].

Der Schwellwert wurde jedoch (biologisch unrealistisch) 80mV oberhalb des Ruhepotentials angenommen, um ein Pulsen zu verhindern, damit der mittlere EPSC aus dem Membranzustand wie folgt zurückgerechnet werden kann: In einer nichtpulsenden Situation sind die einzigen Einflüsse auf das Membranpotential der eingehende EPSC und die dynamische Entwicklung des

[18] Diese Pulsdauer wird in [Markram98] nicht explizit angegeben, scheint aber angesichts der in der dortigen Fig. 4(D) zugrunde gelegten Pulsraten, Parameter A und resultierendem EPSC in einem ähnlichen Bereich zu liegen (siehe auch die gute quantitative Übereinstimmung zwischen Abbildung A.1(obere Hälfte) und Fig. 4(D) aus [Markram98]).

[19] Gleichung (II.16), mit entsprechend sinusmoduliertem Erwartungswert λ, $\lambda_1$ und $\lambda_2$ werden als Maximum bzw. Minimum der modulierten Pulsrate gesetzt.

RC-Gliedes. Der mittlere Membranzustand kann aus der Simulation ermittelt werden und dafür ein äquivalenter Konstant-EPSC berechnet werden, der die Membran auf denselben Zustand heben würde. Die folgende Darstellung zeigt das entstehende EPSC-Übertragungsverhalten:

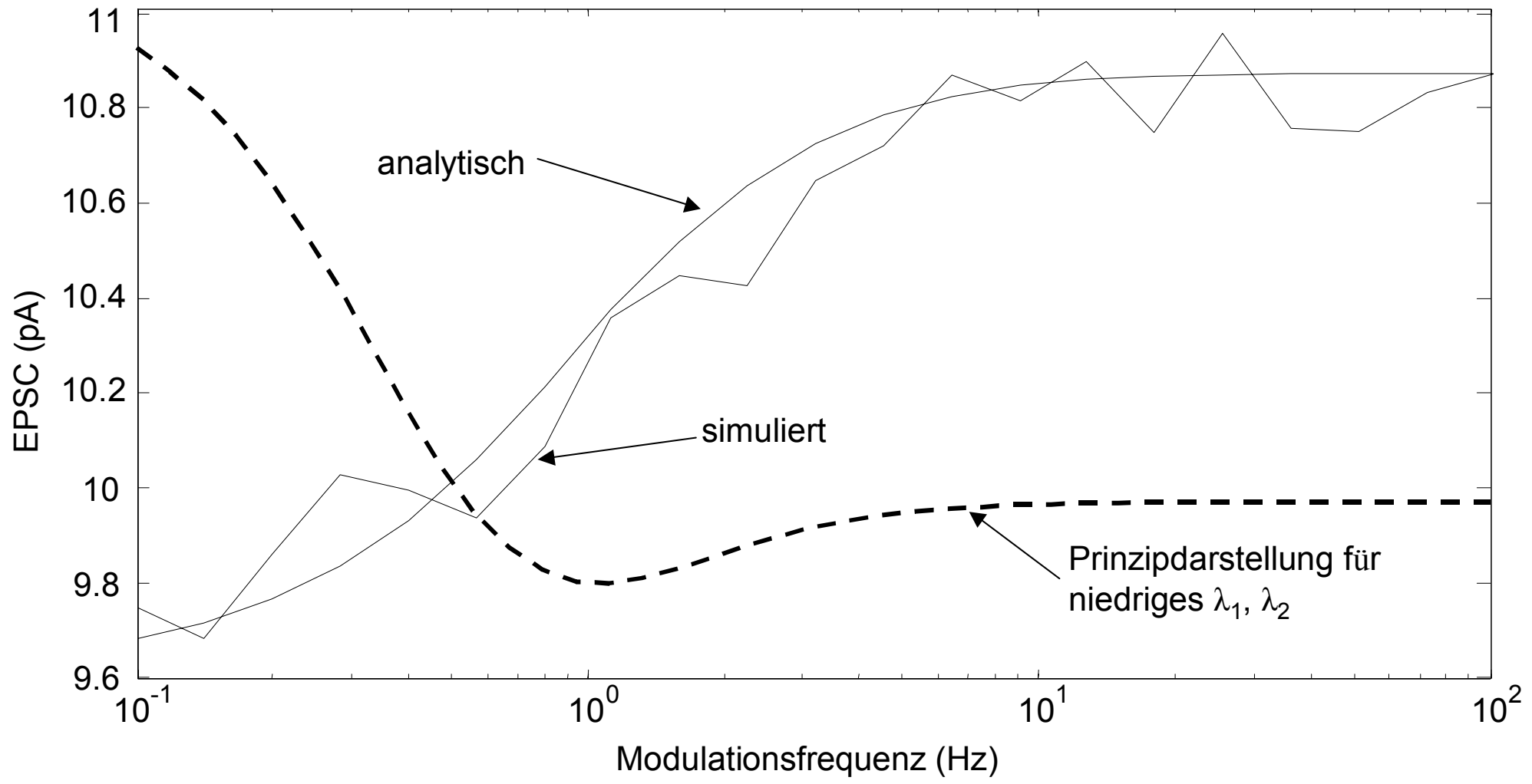

**Abbildung IV.3.: Analytisches und simuliertes EPSC (als Mittelwert über 30 Simulationen), dargestellt über der logarithmischen Modulationsfrequenz. Zusätzliche Prinzipdarstellung des Kurvenverlaufs bei $\lambda_1=70s^{-1}$ bzw. $\lambda_2=2s^{-1}$ (Darstellung bei angehobener Absolutamplitude)**

Für eine übersichtliche Darstellung des Kurvenverlaufs wurde eine logarithmische Einteilung der Modulationsfrequenz $f_m$ gewählt. Die analytische Kurve aus Gleichungen (IV.6) bis (IV.13) zeigt trotz ihrer Näherungen bzgl. Zeitverlauf von $u$ und $R$ sowie der Modulationswellenform gute Übereinstimmung mit einem Mittelwert aus 30 Simulationsdurchläufen mit stochastischem Sinus. Rechnung und Simulation weisen generell wenig Variation im postsynaptischen Strom auf, die mittlere EPSC-Antwort ist beinahe konstant über einen weiten Bereich der Modulationsfrequenz. Im Vergleich mit den konstanten $EPSC_{\lambda 1,\lambda 2}$ der hohen und niedrigen Pulsraten, 15,7pA bzw. 1,28pA zeigt sich, dass die mittlere EPSC-Antwort auf ein moduliertes Signal deutlich über dem Mittelwert zwischen den EPSCs von $\lambda_1$ und $\lambda_2$ liegt. Damit ist die postsynaptische ‚Effizienz', d.h. der durch eine bestimmte Anzahl Pulse in einer Zeitspanne ausgelöste Strom, deutlich stärker, wenn diese moduliert, d.h. gruppiert auftreten. Es findet somit für modulierte Signale in Übereinstimmung mit biologischen Messdaten [Gutkin03] eine bessere Übertragung statt als für ihre äquivalente Konstantpulsrate. Dieser Effekt greift in der Simulation selbst bei sehr hohen Modulationsfrequenzen, da auch dort noch kein Übergang zu einer äquivalenten mittleren Pulsrate stattfindet, sondern sich immer noch über die stochastische Modulation sehr kurze ISIs mit etwa doppelt so langen gruppiert abwechseln (für $f_m$ im Bereich von $\lambda_1$), wodurch die iterative Originalformeln nicht auf einen konvergierten Wert für die mittlere Pulsfrequenz einschwingen, sondern, wie in Abbildung IV.2 für die kontinuierlichen Näherungsformeln angedeutet, um einen Zwischenwert pendeln.

Der Kurvenverlauf über der Modulationsfrequenz ergibt sich daraus, dass in der dargestellten Simulation beide Einzelpulsraten relativ hoch sind, weswegen die Teilformel zur EPSC-Anhebung (IV.1) keine signifikanten Auswirkungen hat. Die Abschwächung des EPSC gemäß (IV.2) ist hier der dominierende Effekt, welcher bei niedriger Modulationsfrequenz voll zum Tragen kommt, da hier die Zeitkonstanten $\tau_{rec}$ in jeweils der hohen und niedrigen Modulationsphase $R$ sehr stark abklingen lassen kann. Bei hoher Modulationsfrequenz wirkt sich die Abschwächung in einer entsprechend kürzeren halben Periode der Modulationsfrequenz nicht so stark aus, deshalb steigt die EPSC-Antwort in Abbildung IV.3 zu hohen Modulationsfrequenzen hin an (analytische und simulierte Kurve).

Welcher Effekt bei niedrigerem $\lambda_1$ und $\lambda_2$ auftritt, ist in der Prinzipkurve in der obigen Darstellung wiedergegeben. Das „Großsignalverhalten", d.h. die oben angestellten Überlegungen zur

Übertragung modulierter Pulsraten bleiben gleich, das Delta während des Kurvenverlaufs ist ebenfalls klein gegenüber dem Absolutwert des EPSC. Der Verlauf unterscheidet sich jedoch insofern, als sich eine abnehmende EPSC-Antwort hin zu hohem $f_m$ beobachten lässt. Durch die niedrigeren Einzelpulsraten hat hier $\tau_{facil}$ ebenfalls einen Einfluss, wodurch sich folgendes Zusammenspiel der beiden Zeitkonstanten ergibt: Für sehr lange Modulationsdauern entsteht eine leicht erhöhte EPSC-Antwort daher, dass die Abschwächung über $\tau_{rec}$ zwar sehr schnell innerhalb einer Modulationsperiode voll einsetzt, jedoch die Verstärkung über $\tau_{facil}$ erst verzögert nachkommt und damit die Gesamtantwort wieder anhebt. Zu hohen Modulationsfrequenzen hin verringert sich der Einfluss von $\tau_{facil}$, weswegen der entsprechende EPSC abfällt. Dieser Effekt wird sowohl in der analytischen Lösung (Prinzipkurve) als auch in einer entsprechend parametrisierten Simulation beobachtet.

Eine ansatzweise Untersuchung der frequenzabhängigen Signalübertragung innerhalb entsprechender Netzwerke findet sich in [Häusler07], jedoch größtenteils beschränkt auf Diskriminierungsverhalten und Jitter. An dieser Stelle soll v.a. der Einsatz quantaler Adaption in inhibitorisch gekoppelten Netzen analog zu [Mar99,Spiridon99] untersucht werden. Hierzu wurde ein Netz aus zwanzig LIAF-Neuronen mit inhibitorischen Synapsen gekoppelt, die quantale Kurzzeitadaption aufweisen, diese sind vollständig vernetzt abzgl. direkter Rückkopplung. Die elektrische Parametrisierung wurde gegenüber den vorhergehenden Untersuchungen am Einzelneuron aus simulationstechnischen Gründen leicht angepasst. Die Membrankapazität wurde zu 3nF gewählt, der Membranleckwiderstand ist 3,3MΩ, damit ergibt sich wieder eine Zeitkonstante von 10ms, zusätzlich wurde der äquivalente postsynaptische Strom auf $A$= -1µA (inhibitorisch) angehoben, um mit einem membranbezogenen Pulsschwellwert von 1V biologisch realistisches Pulsverhalten zu erreichen. Abgesehen von $A$ wurden die Parameter der quantalen Adaption aus den vorhergehenden Simulationen übernommen, die Pulsdauer ist ebenfalls wieder 1,4ms. Die einzelnen Neuronen wurden mit zufälligen Integratorzuständen initialisiert und mit identischen sinusförmigen Stimulusströmen gemäß der folgenden Darstellung angeregt. Eine Kombination aus einem Neuron dieses Netzes mit entsprechend aus dem Stimulusstrom entstehender Pulsrate und quantaler Adaption am korrespondierenden postsynaptischen Neuron mit Zustandsvariablen $u$ bzw. $R$ sieht wie folgt aus:

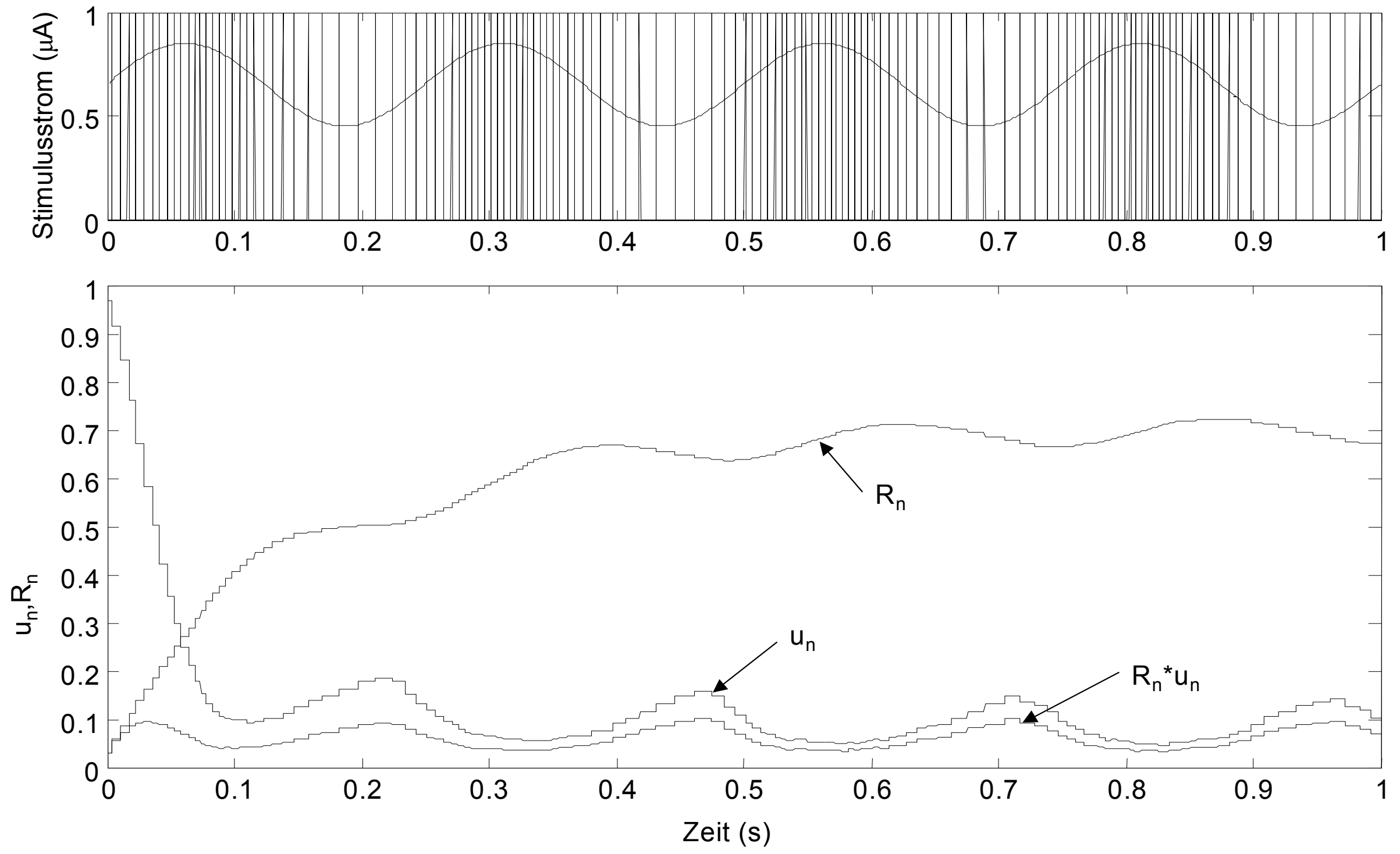

**Abbildung IV.4.: Einzelneuron des inhibitorischen Netzwerks, Stimulusstrom und resultierende Pulsrate eines präsynaptischen Neurons und zugehörige postsynaptische Zustandsvariablen der quantalen Adaption**

Die Zustandvariablen folgen mit entsprechend verzögerter Phase der Modulation der präsynaptischen Pulsfolge, wobei besonders bei $u_n$ eine Verzerrung der Flanken bemerkbar ist. Aus dieser Betrachtung der Adaption eines Einzelneurons lässt sich bereits für das Frequenzspektrum einer Neuronenpopulation herleiten, dass eine starke Oberwellenzunahme stattfindet, da der ursprüngliche Sinusstimulus asymmetrisch moduliert wird:

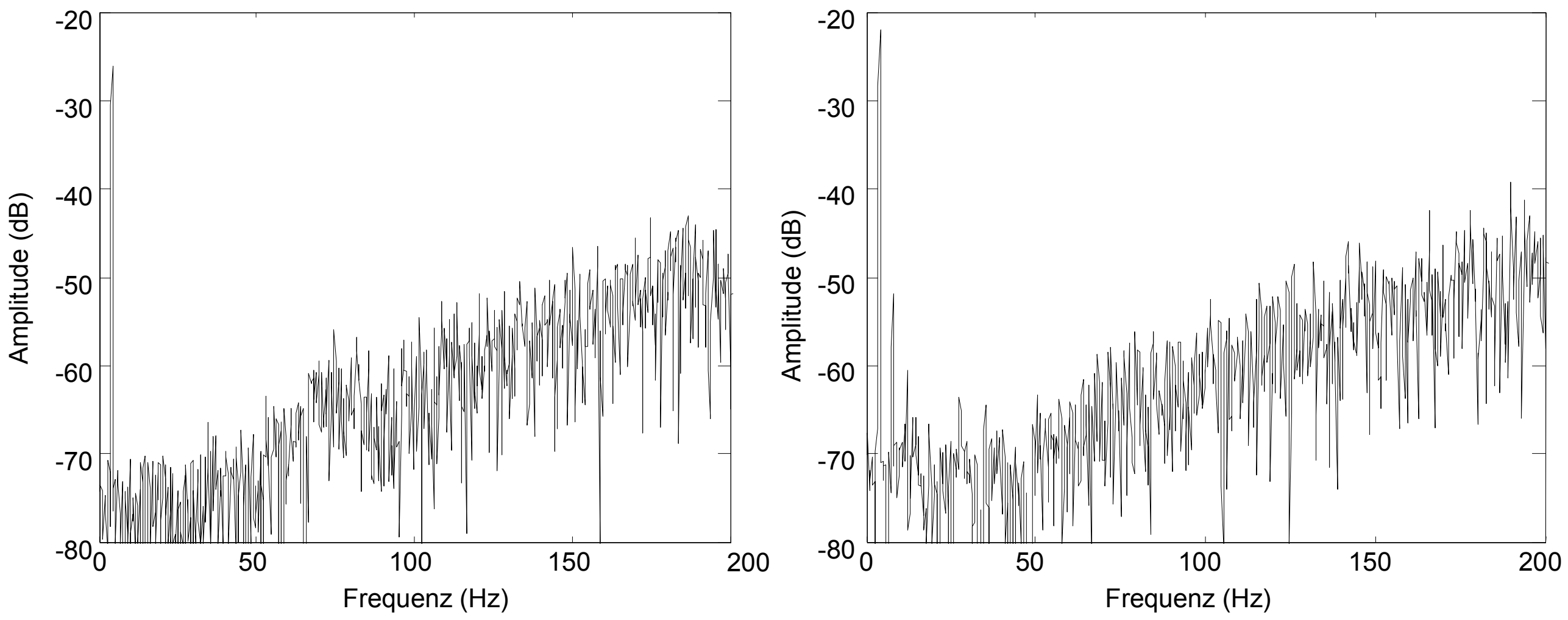

**Abbildung IV.5.: Amplitudenspektrum von konstanten synaptischen Gewichten (links) und quantaler Adaption (rechts) in einem inhibitorischen Netzwerk, Modulationsfrequenz 4Hz, Simulationsdauer 4s**

Der für das Noise Shaping charakteristische Anstieg des Rauschens mit der Frequenz ist in beiden Darstellungen gut erkennbar. Durch die quantale Adaption und die damit verbundene variable Inhibition innerhalb des Netzes ergibt sich zusätzlich eine Signalüberhöhung in der Stimulusübertragung. Im Zeitbereich kann dies über die entsprechende Abfolge der Adaptionsvorgänge erklärt werden (Abbildung IV.4). Nach einer ansteigenden Flanke des Stimulus wird mit leichter Verzögerung $u$ niedriger, wodurch sich die Inhibition verringert. In der Folge steigt zwar $R$ leicht an, hebt jedoch durch seine geringe relative Veränderung die Inhibition nicht signifikant an (siehe Kurve $u*R$ in Abbildung IV.4). Da dieser Effekt hauptsächlich über die Zeitkonstante $\tau_{facil}$ gesteuert wird, ist die Amplitudenanhebung mit etwa 5dB bei niedrigen Stimulifrequenzen wie oben am höchsten und nimmt hin zu höheren Frequenzen stetig ab, beispielsweise auf ca. 2dB bei 15Hz. Signifikant ist dabei, dass sich die Signalüberhöhung explizit aus dem Adaptionsverhalten ergibt, d.h. in einem Signalminimum muss sich $u$ wieder entsprechend erhöhen. Für konstant gehaltene $u$ und $R$, gleich welchen Absolutpegels, zeigt sich dieses Netzwerkverhalten nicht.

Ein weiterer durch die Adaption bedingter Effekt zeigt sich im Frequenzspektrum des Hintergrundrauschens bei niedrigen Frequenzen:

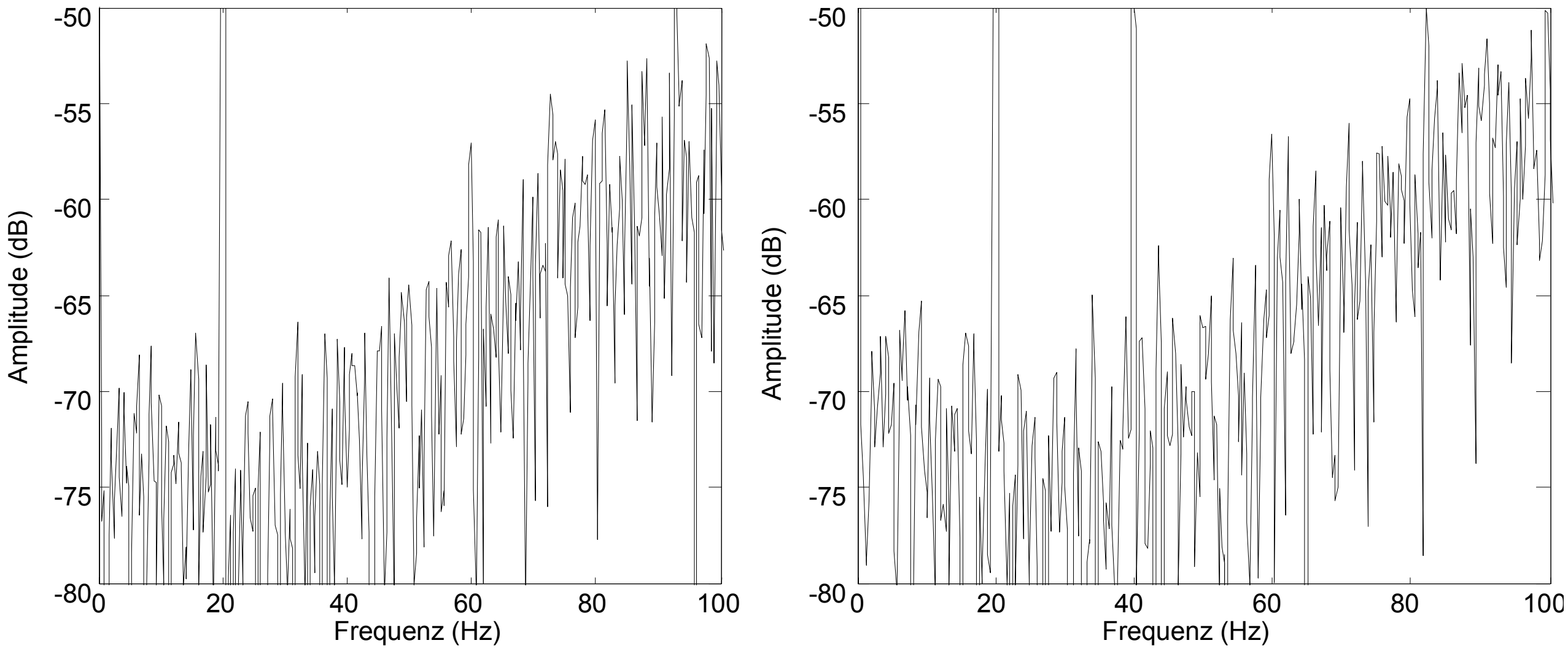

**Abbildung IV.6.: Amplitudenspektrum wie Abbildung IV.5, jedoch Modulationsfrequenz 20Hz, vergrößerte Darstellung des Hintergrundrauschens bei niedrigen Frequenzen**

Durch die Modulation der präsynaptischen Pulsfolgen des Netzwerks mit $u$ und $R$ erhöht sich der Rauschpegel im Frequenzband von 2Hz bis 10Hz, d.h. den Kehrwerten der Zeitkonstanten der quantalen Adaption $\tau_{rec}$ und $\tau_{facil}$. Da die Modulation durch das Produkt aus $u$ und $R$ sowie die unterschiedlichen Verzögerungen relativ zum Stimulus kontinuierlich stattfindet, verteilt sich das dabei entstehende zusätzliche Rauschen auf das oben erwähnte Frequenzband, statt einzelne Spektrallinien bei $\tau_{rec}^{-1}$ bzw. $\tau_{facil}^{-1}$ zu bilden.

### IV.1.3 Korrelationsadaption: Signalextraktion vor Rauschhintergrund

Wie das Signal in einem Netzwerk dynamisch auf mehrere Neuronen verteilt werden kann und sich das entstehende Pulsrauschen in hohe Frequenzbereiche verschiebt, wurde in Abschnitt II.2.3 (Abbildung II.15) kurz angesprochen. Die entsprechenden Dynamiken ergeben sich bereits durch eine inhibitorische Kopplung passiver Integratoren mittels statistisch verteilter konstanter Kopplungsgewichte (siehe Abschnitt II.2.4, letzter Absatz). Nachdem diese grundsätzliche Noise-Shaping-Charakteristik im letzten Abschnitt als gegeben vorausgesetzt wurde und nur zusätzliche Effekte wie Signalverzerrung und Spektralbänder mit erhöhtem Rauschen betrachtet wurden, soll in diesem Abschnitt die tief greifende Veränderung der Übertragungseigenschaften durch Langzeit-Lernvorgänge untersucht werden.

Langzeitadaptionen unterscheiden sich von den in Abschnitt IV.1.2 angesprochenen Kurzzeitadaptionen vor allem durch ihre ‚mehr zielgerichteten', spezifischeren Lernvorgänge, während letztere eher stereotype, von den Adaptionszeitkonstanten abhängige Änderungen der Übertragungseigenschaften aufweisen. Langzeitadaptionen sind beispielsweise in der Lage, die Übertragungseigenschaften einer Synapse in Abhängigkeit der Frequenzen von prä- und postsynaptischen Pulsen bezogen auf lange Zeiträumen zu verändern, um Signale bestimmter Frequenz gezielt zu übertragen, etwa bei der bekannten Bienenstock-Cooper-Munroe (BCM) Regel [Bienenstock82] (siehe auch Abschnitt V.2.2). Eine Analogie aus dem Bereich elektrischer Baugruppen wäre der Unterschied zwischen transientem Verhalten, das bei Eingangssignaländerungen immer in gleicher, abklingender Weise auftritt, und einer grundlegenden Veränderung der Baugruppeneigenschaften etwa durch Justierung von Übertragungskennlinien. Ein weiterer Unterschied zwischen dem o.a. BCM, aber auch sich ähnlich verhaltenden STDP-Regeln [Indiveri06] und der Kurzzeitadaption ist der zusätzliche Bezug auf die postsynaptische Pulsrate, welche wieder von der Gesamtaktivität an den eingehenden Synapsen des Neurons abhängig ist. Diese Abhängigkeit führt zu einer Beeinflussung des Netzwerkverhaltens und -rauschspektrums in Abhängigkeit von zum Einen des Eingangssignals, sowie der Pulscharakteristika der umgebenden neuronalen Strukturen und der Netzwerktopologie.

Ein Ansatz, bei dem mit Hilfe einer externen Adaption über genetische Algorithmen die Netzwerkgewichte zur Erlangung eines höheren SNR angepasst wurden, findet sich in [Mayr04]. Der Lernvorgang wurde hierbei jeweils nach einer bestimmten Simulationsdauer bei gestopptem Netzwerk ‚offline' durchgeführt, während einer Simulation wurden die Kopplungsgewichte auf den jeweils in der letzten Iteration optimierten Werten festgehalten. Kritikpunkt dieser Methode unter neuronalen Gesichtspunkten ist zum Einen die Wahl einer externen, gesteuerten Adaption, sowie die Wahl eines generischen Optimierungsalgorithmus ohne neurobiologische Motivation. Zusätzlich sind zeitliche Dynamiken des Netzwerkverhaltens, die sich beispielsweise durch das permanente Wirken einer inhärenten Adaption im Gegensatz zu der iterativen Optimierung ergeben, nicht möglich. Positiv kann hervorgehoben werden, dass grundsätzlich eine Verbesserung des Noise-Shaping-Verhaltens gegenüber zufälligen statistisch gleichverteilten Gewichten gezeigt wurde [Mayr04, Mayr05c].

In den folgenden Simulationen soll diese Optimierung der Kopplungsgewichte mit einer neurobiologischen Zielrichtung fortgeführt werden. Hierzu wird untersucht, was für Auswirkungen Spike Timing Dependent Plasticity (STDP), eine Lernregel aus aktueller Forschung [Bi98, Kepecs02], auf das Frequenzspektrum inhibitorisch gekoppelter Netzwerke hat. Es wurde eine Standardimplementierung von STDP gemäß Gleichung (V.1) gewählt, mit generischen Parametern $\eta$=0,02 und $\tau$=20ms aus der Literatur [Delorme01]. Motiviert wird die Verwendung dieser Lernregel dadurch, dass entsprechendes Hebbsches Verhalten zu einer Synchronisation bzw. Korrelation der Aktivität einzelner Neuronen führt [Nowotny03] und diese Korrelation wiederum dazu beiträgt, das SNR zu erhöhen [Zeitler06]. Die elektrischen Parameter des Netzwerks wurden aus der Auflistung oberhalb Abbildung IV.4 übernommen, wobei in den STDP-Simulationen das bei der Adaption entstehende Gewicht dem Produkt aus $u$ und $R$ entspricht, d.h. der während eines Pulses fließende postsynaptische Strom berechnet sich als Produkt aus $A$ und dem synaptischen Gewicht wie in der folgenden Abbildung dargestellt. Die Amplitude und der Offset des Stimulusstroms sind identisch mit denen aus Abbildung IV.4, Stimulusfrequenz ist 15 Hz, das Netzwerk besteht aus 20 vollvernetzten Neuronen. Im Zeitbereich ergibt sich folgendes typisches Adaptionsverhalten der Gewichte (bei zufälliger Initialisierung der Gewichte im Bereich 0 bis 1):

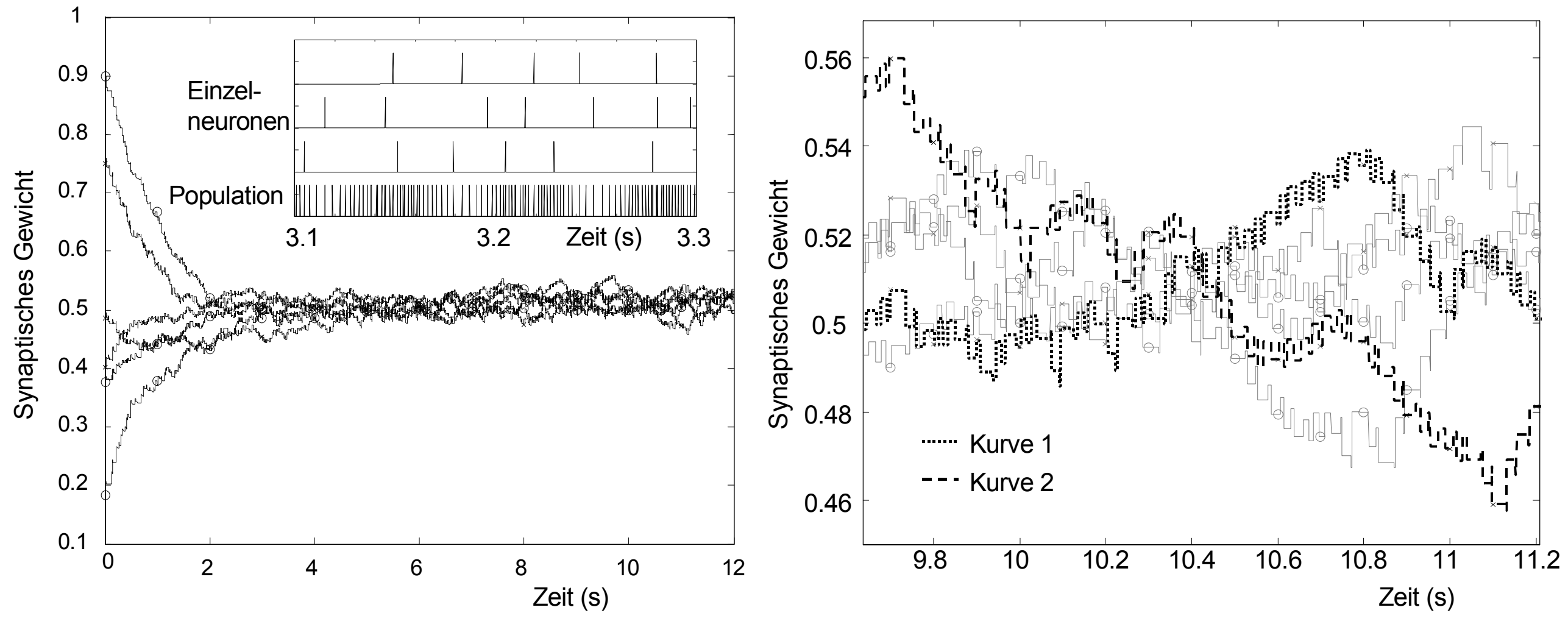

**Abbildung IV.7.: Entwicklung repräsentativer Gewichte in einem Netzwerk inhibitorisch gekoppelter Neuronen mit STDP-Adaption, global (inklusive einer exemplarischen Darstellung von zwei Einzelpulsfolgen und des Populationssignal) und erweiterte Betrachtung eines Zeitausschnittes**

Signifikant ist hier zunächst, dass alle Gewichte unabhängig vom Startwert in einen Bereich zwischen ca. 0,45 bis 0,55 streben. Dies scheint bedingt zu sein durch die stochastische Natur der Pulsabfolgen[20], welche dazu tendiert, häufig wiederkehrende Pulsabfolgen zu verhindern, so dass keine starken Lernvorgänge einsetzen [Izhikevich07]. Ein feingranulares Anpassen der Synapsengewichte im mittleren Bereich wird dadurch jedoch nicht ausgeschlossen. Wie die rechte Hälfte der obigen Abbildung zeigt, existiert im mittleren Bereich durchaus zielgerichtete Adaption, die nicht nur auf ein mittleres Synapsengewicht konvergiert, sondern Gewichte auch wieder aus der Mitte auslenkt. Beispielhaft angedeutet ist dies für die gepunktet markierte Kurve 1 zu Anfang und Ende des betrachteten Zeitausschnittes. Noch deutlicher sichtbar ist ein deterministisches Lernverhalten für die mit Strichlinien markierte Kurve 2. Lernvorgänge finden hierbei auf Zeitskalen im Sekundenbereich statt, und damit deutlich über der Periodendauer des 15Hz Stimulussignals. Was eine derartige Adaption der Gewichte in einem begrenzten Gewichtsbereich für die spektralen Eigenschaften des Netzwerks an Auswirkungen hat, ist in der folgenden Darstellung wiedergegeben:

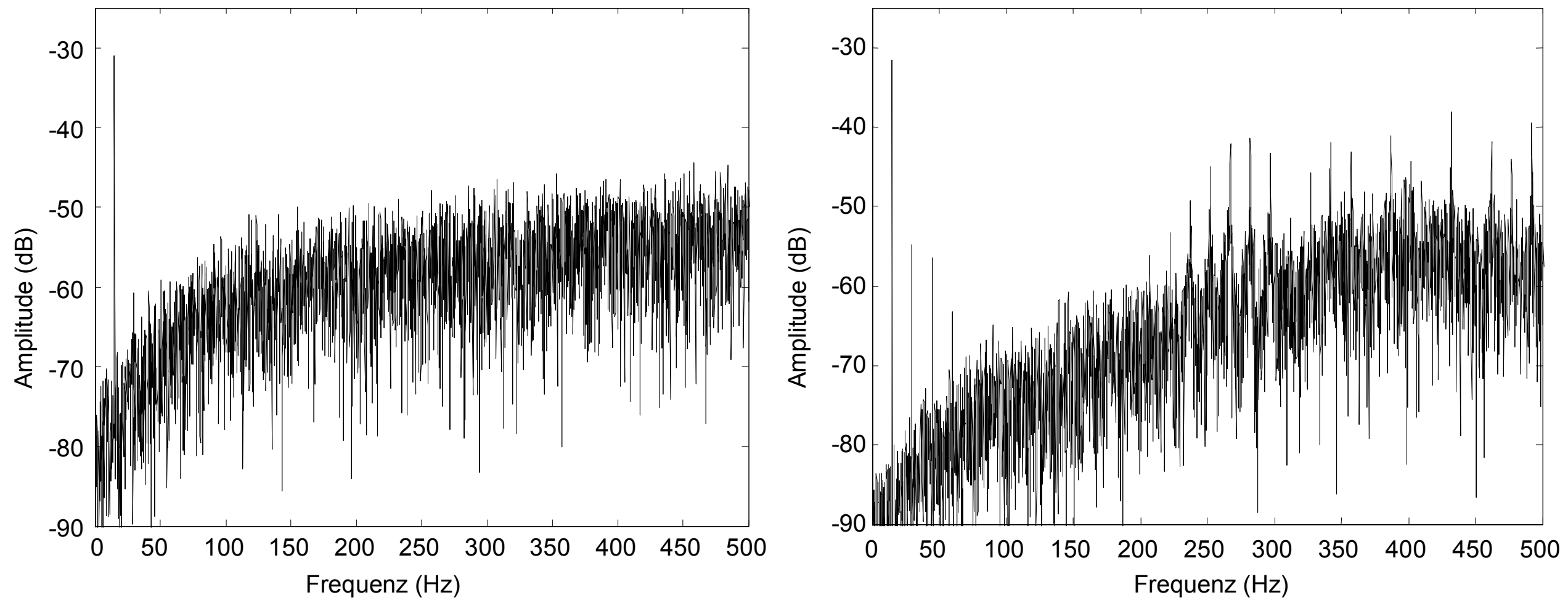

**Abbildung IV.8.: Frequenzspektrum des Populationssignals inhibitorischer gekoppelter 20-Neuronen-Netze, links konstante, zufällig verteilte Gewichte im Bereich 0,45 bis 0,55, rechts Gewichte wie in Abbildung IV.7**

Die beiden obigen Simulationen wurden jeweils für 12 Sekunden (biologische Zeit) simuliert, wobei nur die letzten 8 Sekunden mit der FFT ausgewertet wurden, da sich dann die STDP-Adaption laut Abbildung IV.7 eingeschwungen hat[21]. Für das Referenznetz mit konstanten Gewichten in der linken Hälfte der obigen Abbildung wurden gleichverteilte Gewichte im Intervall 0,45 bis 0,55 gewählt, um eine qualitative und quantitative Vergleichbarkeit der beiden Simulationen zu gewährleisten. Es ist deutlich zu sehen, dass die Feinstruktur der Gewichtsverteilung und dynamische Vorgänge im STDP-Fall ein signifikant anderes Verhalten schaffen, als wenn nur nominal der Bereich der Gewichtswerte stimmt. Der Haupteffekt ist hierbei eine Verringerung des Rauschpegels bei niedrigen Frequenzen um ca. 10dB und eine Erweiterung des Noise Shaping bis auf ca. 300Hz im Gegensatz zu dem Netz mit konstanten Gewichten, bei dem die Rauschminderung im Wesentlichen bereits bei 100 bis 150Hz abgeklungen ist. Weitere Simulationen haben ergeben, dass der Effekt nahezu unabhängig ist von der Stimulusfrequenz, für dieses Netzwerkverhalten scheint nur wichtig zu sein, dass überhaupt ein periodischer Stimulus am Netzwerk anliegt. Eine STDP-Adaption trägt damit generell dazu bei, die Signalübertragungseigenschaften zu verbessern, wobei sich v.a. bei hohen Stimulusfrequenzen eine deutliche Verbesserung des SNR ergibt. Der wesentliche Unterschied zu der eher modulationsähnlichen quantalen Kurzzeitadaption ist, dass sich durch die STDP-Adaption dauerhaft die Übertragungseigenschaften des Netzwerks ändern.
Als Nebeneffekt entsteht ein Oberwellenspektrum der Modulationsfrequenz, das einer ähnlichen Ursache geschuldet ist wie bei der quantalen Kurzzeitadaption, d.h. dadurch, dass die Adaption im Endeffekt vom Stimulus gesteuert wird, ergibt sich eine Verzerrung des Stimulussignals.

[20] Siehe das Insert in der linken Hälfte der obigen Abbildung: Einzelne Pulsabfolgen der beiden Einzelneuronen sind sehr variabel trotz eines homogenen Populationssignals. Da STDP auf wiederkehrende Pulsabfolgen von Neuronen zueinander anspricht, ist hier zumindest keine ‚Großsignal'-Lernrichtung vorgegeben.

[21] Für die Simulation mit konstanten, zufällig verteilten synaptischen Gewichten, korrespondierend mit der linken spektralen Darstellung, ist es natürlich irrelevant, ob die ersten 4 Sekunden weggelassen werden. Um die Vergleichbarkeit der spektralen Darstellungen in Abbildung IV.8 zu gewährleisten, wurde jedoch das Vorgehen bei der Simulation identisch gewählt (um z.B. dieselbe spektrale Auflösung zu erreichen).

### IV.1.4 Abschlussbetrachtung und Anwendungen des neuronalen Noise Shaping

Gemeinsam ist den in den vorherigen Abschnitten angeführten Verfahren, die durch Adaption den Rauschpegel senken und/oder das Signal verstärken, dass bei ihnen die neuronale Hintergrundaktivität eher als notwendiges Übel hingenommen wird, das entsprechend reduziert/verschoben werden muss. Parallel dazu existiert eine gegenteilige Sicht der Hintergrundaktivität, die diese als notwendig ansieht, um das Membranpotential eines Neurons konstant möglichst knapp unter der Feuerschwelle zu halten [Destexhe03] bzw. ein Neuron nach einem Aktionspotential und der darauf folgenden Hyperpolarisierung schnell wieder in einen ‚aktiven', feuerbereiten Zustand zu heben. Das Hintergrundrauschen kontert damit zum Einen das Refraktär-(Erholungs-)verhalten des Neurons, zum Anderen die permanente Membranentladung durch die Leckströme. Dieses Phänomen ist teilweise auch aus den Messdaten zum typischen Membranpotential von (biologischen) Kortexneuronen in Abbildung II.5 (rechts) entnehmbar. Ein Spannungspegel unterhalb des Ruhepotentials (entsprechend einer Hyperpolarisierung) wird nach einem Aktionspotential entweder gar nicht erreicht oder nur für sehr kurze Zeit gehalten, der durchschnittliche Spannungspegel entspricht eher einer leichten Depolarisierung. Der postulierte Grund für dieses rauschbasierte Verhalten ist, ein Neuron präzise auf einen eingehenden Stimulus reagieren zu lassen, da bei einer konstant knapp unter der Feuerschwelle gehaltenen Membran wenige eingehende Pulse benötigt werden, um das Neuron feuern zu lassen (=erhöhte Sensitivität). Außerdem wird durch die Verkürzung der notwendigen Integrationszeit ein schnelleres Reagieren des Neurons und damit eine rapidere Signalverarbeitung ermöglicht [Destexhe03].
Eine Auswertung des Frequenzspektrums der Gesamt-Populationspulsfolge kann auch hier als zusätzliches Analysewerkzeug neben der ISI-Auswertung im Zeitbereich dienen. Beispielsweise können Burst-Ereignisse gezielt Neuronen des Netzwerks durch die entsprechend zeitlich begrenzte erhöhte Aktivität in einen aktiven, feuerbereiten Zustand heben [Hermann79]. Als Umkehrschluss zum letzten Absatz aus Abschnitt II.2.2, kann gefolgert werden, dass eine entsprechende lokale Häufung von kurzen ISIs (bei ansonsten unveränderter ISI-Verteilung) zu einem Anstieg von niederfrequenten Anteilen im Spektrum der Pulse einer Population führt. Dies wird belegt durch die Spektrumsdarstellung in Abbildung IV.1. In Erweiterung des Spezialfalls Burst, bei dem die erwähnten charakteristischen Niederfrequenzkomponenten auftreten, kann auch ein allgemeines zeitlich gehäuftes Auftreten von kurzen ISIs als Anstieg der niederfrequenten Rauschamplitude im Frequenzspektrums abgelesen werden (Anhang A.1). Diese Häufung von kurzen ISIs ist ein Anzeichen dafür, dass ein Neuron sehr schnell nach einem Feuern wieder durch zahlreiche präsynaptische Pulse der umliegenden Population nahe an die Feuerschwelle herangeführt wird, d.h. in den erwähnten aktiven oder high-conductance Zustand überführt wird [Destexhe03].
Rauschen in Neuronenpopulationen erfährt offensichtlich eine differenzierte Steuerung, die über eine einfache Rauschreduktion in einigen Frequenzbändern hinausgeht. Dies ist in Grundzügen beispielsweise in dem komplexen Spektrumsverlauf der biologischen Messungen in Abbildung II.14 sichtbar. Eine Synthese aus den Konzepten zu Rauschunterdrückung und aktivem Rauscheinsatz könnte etwa wie folgt aussehen: Das Rauschen wird bei niedrigen Frequenzen erhöht, um eine schnelle Signalübertragung zu ermöglichen, d.h. die Übertragung hoher Frequenzen wird dadurch verbessert gemäß dem obigen ‚high-conductance'-Gedanken. Wie in Abbildung IV.6 dargelegt, ist dies einer der Effekte, den die quantale Kurzzeitadaption auf inhibitorisch gekoppelte Populationen hat. Im mittleren Bereich erfolgt eine allgemeine Reduktion, um durch ein niedriges Grundrauschen gute Übertragung zu gewährleisten. Dies geht einher mit dem Effekt, dass bei der quantalen Adaption im mittleren Frequenzbereich linear übertragen wird,

d.h. hier muss ein gutes SNR allein durch Rauschverringerung sichergestellt werden. Die Übertragung hochfrequenter Stimuli wird zusätzlich durch die Amplitudenüberhöhung der PTP sichergestellt, transiente Vorgänge/Signale scheinen damit eine Sonderstellung einzunehmen. Die Weiterleitung hochfrequenter/transienter Vorgänge kann außerdem noch durch STDP-Veränderungen der Netzwerkkopplung begünstigt werden (Abbildung IV.8). Für niedrige Stimuli-Frequenzen wird das erhöhte Rauschen durch die quantale Adaption gekontert, die für eine verstärkte postsynaptische Antwort und damit für eine Rekonstruktion des SNR sorgt.
Die in den vorhergehenden Abschnitten gewonnenen Erkenntnisse zur aktiven und passiven Informationsübertragung unter Rauscheinfluss sind einerseits eher unter neurotheoretischen Aspekten interessant. Es wurde aber auch die Ähnlichkeit mit konventionellem Noise Shaping in Bezug auf Strukturen und Verarbeitungsfunktion schon wiederholt für technische Adaptionen ins Auge gefasst, z.B. wurde von Marienborg et al. ein entsprechender ASIC entworfen [Marienborg02]. Vorteilhaft gegenüber konventionellen technischen Lösungen scheint dabei vor allem die parallele, verteilte und fehlertolerante Natur der neuronalen Rauschbeeinflussung. Ein Aufbau aus einer hohen Anzahl parallel arbeitender, stereotyper Einzelbausteine würde beispielsweise den Entwurfsaufwand gegenüber konventionellen Architekturen deutlich verringern. Ähnliche Parallelstrukturen werden in konventionellen AD-Wandlern bereits eingesetzt, wobei deren zeitversetzte Kopplung eine lineare Skalierung des SNR mit der Kanalanzahl $N$ ermöglicht, im Gegensatz zu ungekoppelten Wandlern, deren SNR nur mit $N^{1/2}$ zunimmt [Poorfard97]. Dieser lineare Zusammenhang zwischen SNR und $N$ wurde für inhibitorisch gekoppelte neuronale Netze ebenfalls bestätigt [Spiridon99], wobei deren Grundelemente und Kopplungsrealisierung wesentlich weniger komplex sind und deshalb dort ein deutlich höherer Grad an Parallelität verwendet werden könnte. In früheren Arbeiten wurde versucht, durch stärkeres Abweichen von der klassischen Synapsen/Neuronen-Struktur den Grad der Rauschverringerung zu erhöhen, etwa durch eine Weiterkopplung des Membranpotentials [Mayr05b], vergleichbar mit der Integratorweiterkopplung in konventionellen MASH-Architekturen [Norsworthy96]. Zur Verbesserung der technischen Einsetzbarkeit könnten zusätzlich einige der in den letzten Abschnitten diskutierten neuronalen Adaptionen verwendet werden. Eine Vereinfachung der PTP-Adaption wurde bereits erfolgreich in früheren Arbeiten zur Signalüberhöhung bzw. SNR-Verbesserung eingesetzt [Mayr05c], hier könnte untersucht werden, was für Auswirkungen eine stärkere Anlehnung an das biologische Vorbild hat. Die amplitudenabhängige Verringerung der Inhibition bei der quantalen Kurzzeitadaption kann ebenfalls zur Erhöhung der Signalamplitude beitragen. Zur statischen Verbesserung der Übertragungseigenschaften des Netzwerks könnte der Ansatz aus [Mayr04] zur Optimierung der Kopplungsgewichte um die zuletzt geschilderte STDP-Adaption erweitert werden. Hier wäre zu untersuchen, inwieweit die in Abbildung IV.8 beobachteten Verbesserungen statischer oder dynamischer Natur sind, d.h. ob STDP als reine Substitution der in [Mayr04] verwendeten genetischen Algorithmen dienen kann, mit einmaligem Anlernen der Gewichte, oder ob eine Adaption zur Laufzeit nötig ist. Verschiedene Ansätze aus der Literatur zur Stimulusrekonstruktion aus neuronalen Pulsfolgen [Gabbiani99, Schrauwen03] könnten verwendet werden, um entsprechende Dezimationsfilter zu bauen, die als letzte Stufe eines derartigen neuronalen AD-Wandlers benötigt werden.

## IV.2 Pulsbasiertes Local Orientation Coding

In biologischen und technischen bildverarbeitenden Systemen kommt der ersten Stufe dieser Bildverarbeitung besondere Bedeutung zu, da hier zum Einen die Wandlung der einkommenden Bildinformation in eine vom System verarbeitbare Form, d.h. Pulse oder Bitwerte, stattfindet. Zum Anderen werden die für die weitere Verarbeitung relevanten Informationen extrahiert, z.B. in der Retina die LoG-Charakteristiken. Auch in der technischen Bildverarbeitung wird in der Regel in nachfolgenden Stufen nicht die unbearbeitete Bildinformation benötigt, da die Zielsetzung eine Bildanalyse unter Anwendungsgesichtspunkten ist, nicht eine Bildwiedergabe. Für einen Bildsensor einer optischen Computermaus sind z.B. signifikante Ecken, Kanten, etc. interessant [Tabbone95], die im nächsten Bild mit entsprechender Translation wiedergefunden werden können, um die Mausbewegung zu berechnen. In dieser ersten Stufe muss also eine anwendungsorientierte Informationsverdichtung und -codierung stattfinden, um die Datenrate für höhere Verarbeitungsstufen zu reduzieren und die Daten/Bildinformation in für die weitere Verarbeitung relevanter Form bereitstellen zu können [Zitova99]. Durch diese Vorverarbeitung wird die Verarbeitungsgeschwindigkeit in nachfolgenden Stufen erhöht und ihre Komplexität reduziert [Mayr01].
Bildoperatoren, die für eine bestimmte Anwendung die o.a. Aufgaben leisten und günstig in einer Mixed-Signal VLSI-Implementierung zu realisieren sind, sind meist sehr spezifisch auf eine bestimmte Anwendung optimiert, so dass ein solcher Schaltkreis nur für diese Anwendung eingesetzt werden kann [Morie01]. Ein in Software implementierter Bildoperator kann demgegenüber für neue Anwendungen einfach neu parametrisiert oder umprogrammiert werden [Zitova99], wobei hier die Software und zugehörige Hardware mit der vollen Datenrate der vom Sensor gelieferten Bildinformation arbeiten muss, eine Entlastung dieser symbolischen Verarbeitungsstufen durch Vorselektion kann in diesem Fall nicht stattfinden. Ein Kompromiss zwischen Flexibilität und effizienter Informationsextraktion würde ähnlich der Retina (oder der unteren Stufen des V1) aufgebaut sein, d.h. ein solcher Bildoperator müsste lokale Bildzusammenhänge wie Grauwertverläufe, Kanten und Kontraste extrahieren, die signifikante Information über diese lokalen Strukturen enthalten, jedoch allgemein genug sind, um für verschiedene Anwendung einsetzbar zu sein. Nachfolgende Stufen würden dann die für die jeweilige Anwendung interessanten Strukturen aus der Ausgabe des Bildoperators selektieren und weiterverarbeiten. Ein zusätzlicher Vorteil eines solchen vorverarbeitenden Bildsensors ist der ihm zur Verfügung stehende Dynamikbereich, da er mit der vollen Bandbreite der Sensoren arbeiten kann, nicht einer begrenzten, quantisierten Variante wie ein softwarebasierter Bildoperator.

### IV.2.1 Grundvariante des Local Orientation Coding

Ein Bildoperator, der die oben genannten Anforderungen erfüllt, ist das Local Orientation Coding (LOC). Basierend auf einer Software-Version [Goerick94], wurde der Operator in [König02, Mayr06b] weiterentwickelt, um seine Realisierbarkeit als Mixed-Signal-VLSI-Schaltung zu erhöhen, u.a. durch Eliminierung globaler Abhängigkeiten und Erhöhung der Robustheit. Das Verhalten des LOC-Operators wird von folgenden Gleichungen beschrieben:

$$\varepsilon_{m,n}(i,j)=\begin{cases}k(i,j), & b(m+i,n+j)\le b(m,n)-t(m,n)\\ 0, & else\end{cases} \qquad \textbf{(IV.14)}$$

$$(i,j)\in\{(0,-1),(-1,0),(1,0),(0,1)\}\ \textit{für}\ N_4$$

$$(i,j)\in\{(-1,-1),(-1,1),(1,-1),(1,1)\cup N_4\}\ \textit{für}\ N_8$$

Ausgehend von einer 3*3 Nachbarschaft, wird der Grauwert des mittleren Pixels $b(m,n)$ mit seinen Nachbarn $b(m+i,n+j)$ verglichen und das Ergebnis in dem richtungsabhängigen Orientierungskoeffizienten $k(i,j)$ codiert, wobei $(m,n)$ die absoluten Koordinaten des Mittenpixels darstellen und $(i,j)$ die Relativkoordinaten der Nachbarschaft. Vom mittleren Pixel $b(m,n)$ wird vor dem Vergleich noch ein ortsabhängiger Schwellwert $t(m,n)$ abgezogen, um die Rauschanfälligkeit des Operators zu vermindern. Dieser berechnet sich wie folgt:

$$G(m,n)=\sum_{i,j}\left[b(m+i,n+j)*\frac{1}{Z}e^{-\frac{i^2+j^2}{2\sigma^2}}\right] \qquad \textbf{(IV.15)}$$

Eine Gaußsche Glättung $G(m,n)$ der Bildgrauwerte mit Einzugsbereich $\sigma$ wird mit der Summe der Maskenkoeffizienten $Z$ normiert:

$$Z=\sum_i\sum_j e^{-\frac{i^2+j^2}{2\sigma^2}} \qquad \textbf{(IV.16)}$$

Im Unterschied zu Gleichung (IV.14) bezeichnet (*i,j*) in beiden obigen Gleichungen den Einzugsbereich der Gaußmaske, also bei großem $\sigma$ u.U. den gesamten Bildbereich. Der Schwellwert *t*(*m,n*) wird aus der absoluten Differenz zwischen dem Grauwert des mittleren Pixels und der Gaußschen Faltung gewonnen, skaliert um einen Parameter *C*:

$$t(m,n) = C * \left| b(m,n) - G(m,n) \right| \qquad \textbf{(IV.17)}$$

Dieser Schwellwert stellt somit ein Maß dafür dar, wie stark das mittlere Pixel vom durchschnittlichen lokalen Grauwert abweicht. Durch Subtraktion von *t*(*m,n*) in Gleichung (IV.14) wird also erreicht, dass in einem Bildbereich mit geringen Grauwertvariationen auch kleine Grauwertunterschiede als signifikant betrachtet werden, während in Bereichen mit starken Unterschieden in der Helligkeit nur große Abweichungen vom Mittenpixel entsprechende Antworten *k*(*i,j*) generieren. Die Skalierung mit *C* dient dazu, die generierten LOC-Merkmale an verschiedene Anwendungen anzupassen, da damit der Charakter der extrahierten Merkmale beeinflusst werden kann. (siehe auch Abbildung IV.13)

$$b'(m,n) = \sum_{i,j} \varepsilon_{m,n}(i,j) \qquad \textbf{(IV.18)}$$

Die Orientierungskoeffizienten *k*(*i,j*) werden dann aufsummiert und geben als Merkmalszahl *b'*(*m,n*) Informationen über die Strukturen im Grauwertverlauf in der Nachbarschaft des Pixels *b*(*m,n*) wieder. Um die Eindeutigkeit der Zuordnung *b'*(*m,n*) und *k*(*i,j*) zu gewährleisten, wird eine binäre Skalierung der Antworten *k*(*i,j*) gewählt. (Abbildung IV.9)

$$\begin{pmatrix} 0 & 1 & 0 \\ 2 & R & 4 \\ 0 & 8 & 0 \end{pmatrix} \begin{pmatrix} 1 & 2 & 4 \\ 8 & R & 16 \\ 32 & 64 & 128 \end{pmatrix} \begin{pmatrix} 1 & 0 & 2 \\ 0 & R & 0 \\ 4 & 0 & 8 \end{pmatrix} \quad n \uparrow \rightarrow m$$

**Abbildung IV.9: Orientierungskoeffizienten, (v.l.n.r.) Standard N4 bzw. N8 und diagonale N4 Nachbarschaft**

Abbildung IV.10 gibt die möglichen Masken des LOC-Operators für eine 4-fach Nachbarschaft und die zugehörigen Merkmalszahlen *b'*(*m,n*) wieder.

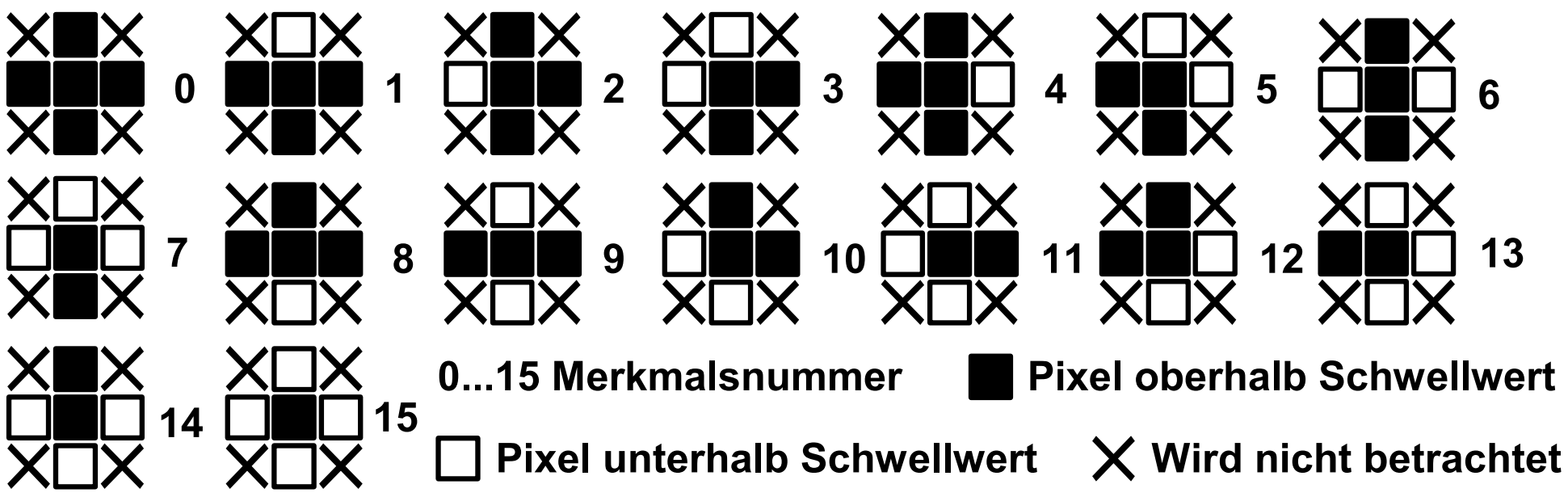

**Abbildung IV.10.: Alle möglichen Resultate für eine 4-fach Nachbarschaft und zugehörige Merkmalsnummern (entsprechend der Summe der Orientierungskoeffizienten)**

Wie in Abbildung IV.10 zu sehen, wird durch die zu den Merkmalszahlen gehörigen Masken ein weiter Bereich lokaler Grauwertstrukturen oder Texturen repräsentiert. Die Relevanz solcher lokaler Strukturen für die Analyse großflächiger Merkmale im Bild ist in Abbildung IV.11 dokumentiert.

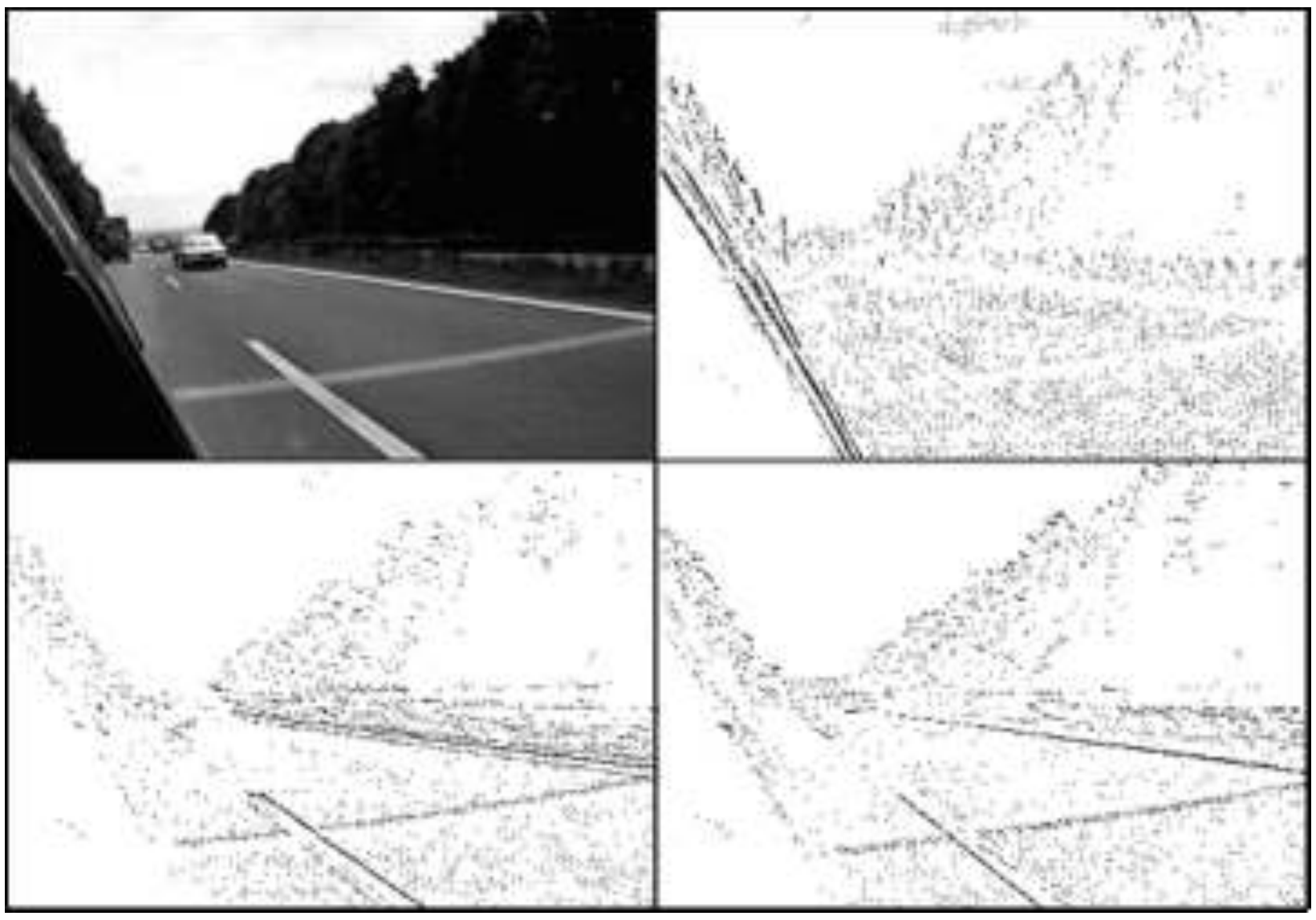

**Abbildung IV.11: Beispiele für verschiedene LOC-Resultate, Orientierungscodierung nach Abbildung IV.9. Links oben Originalbild, daneben Ergebnis für Merkmal 14, links unten Merkmal 10, rechts unten Merkmal 5**

Obwohl der LOC-Operator nur einen Zusammenhang bei extrem kleinräumigen Strukturen feststellen kann, werden dadurch auch Strukturen wie der Mittelstreifen oder Seitenstreifen der Straße zuverlässig extrahiert, da sich diese durch eine Häufung entsprechender lokaler Strukturen oder Texturen entlang des großflächigen Merkmals auszeichnen. Hierbei wird sogar die Richtung des Hell-Dunkel-Kontrastes berücksichtigt (siehe den Mittelstreifen der Straße in Abbildung IV.11, Teilbilder unten links und rechts). Bei einer Kantenextraktion ist unter anderem der Winkel dieser Kante von Interesse, der zwar aufgrund der kleinen Nachbarschaft des LOC-Operators von einer einzelnen LOC-Maske nur grob genähert werden kann. Dadurch, dass eine Kante aber von mehreren LOC-Masken (Merkmalsnummern) angezeigt wird, wie z.B. die Seitenlinie, kann zwischen den Fundamentalwinkeln der einzelnen Masken interpoliert werden. Grundsätzlich zeigt die obige Abbildung, dass vom LOC-Operator relevante Bildinformation extrahiert wird, die sich für nachfolgende, objektorientierte Bildverarbeitungsstufen eignet. Dabei wird durch diese Vorselektion gleichzeitig die Datenrate erheblich reduziert, da Bereiche mit geringem Informationsgehalt, wie z.B. der Himmel oder das Autoinnere am linken Bildrand, sehr wenige LOC-Antworten generieren.

Eine VLSI-Variante des Sensors wurde entworfen[22] [König02, Mayr05a, Mayr06b], die mit Hilfe eines Diffusionsnetzwerks eine lokale Glättung der Pixelströme durchführt und die verschiedenen Rechenoperationen wie Addition, Subtraktion, Skalierung und Absolutwert durch translineare Maschen und Stromspiegel realisiert. Der Vergleich des bearbeiteten Pixelstroms des mittleren Pixels mit den Nachbarn wird parallel ausgeführt, wodurch eine hohe Verarbeitungsgeschwindigkeit gewährleistet ist.

Der hohe Anteil an analogen Schaltungen in der o.a. Implementierung, bei dem die Digitalwandlung erst am Ende der Verarbeitung stattfindet, führt leider entweder zu einem hohen Flächenbedarf für zuverlässige Analogschaltungen, oder zu erheblichen Fehlern gerade im kleinen Strombereich, da hier z.B. für Stromspiegel große Gateflächen nötig wären. Die Beleuchtungsunabhängigkeit der LOC-Merkmale, die eigentlich durch den Vergleich, also die Auswertung relativer Grauwertinformation sichergestellt wird, wird hierdurch in Frage gestellt, da ein vergleichbares Funktionieren der Schaltungen bei kleinen und großen Photoströmen nicht gewährleistet werden kann. Großzügiger dimensionierte Analogschaltungen führen hier jedoch nicht unbedingt zum Ziel, da dann vor allem bei kleinen Photoströmen die Verarbeitungsgeschwindigkeit stark abnimmt. Zusätzlich können Analogschaltungen die Sub-Mikrometer-

[22] Der VLSI-Entwurf des Operators und die Fertigung des Schaltkreises wurden von der Deutschen Forschungsgemeinschaft im Rahmen der DFG-VIVA-Förderung (SPP 1076, Az. Ko 1255/4-1 und Ko 1255/4-2) finanziert.

Skalierung moderner CMOS-Technologien erfahrungsgemäß nicht vollständig ausnützen, so dass Portierungen zu kleineren Technologien keine nennenswerte Verkleinerung der analogen Verarbeitungsstufen mit sich bringen. Dieser Variante abträglich ist außerdem die Tatsache, dass die Portierung von Mixed-Signal-Schaltungen zu unterschiedlichen Technologien überproportional mehr Aufwand erfordert, je größer der Anteil der Analogschaltungen ist, da diese i.d.R. neu entworfen werden müssen.
In den nächsten Abschnitten soll deshalb eine alternative Implementierung des LOC entworfen werden, der in einer sehr frühen Stufe die Bildinformation als robustes, pseudo-digitales Signal vorliegen hat und dieses in einer neuro-inspirierten Arbeitsweise dazu verwendet, LOC-ähnliche Bildmerkmale zu extrahieren.

### IV.2.2 Pulsbasiertes LOC, Herleitung und Beschreibung

Biologisch inspiriert wird die pulsbasierte Variante des LOC-Operators (PLOC) unter topologischen und verarbeitungstechnischen Aspekten von den in Abschnitt II.2.5 diskutierten quasi-digitalen Pulsverrechnungen. Um mit diesen sinnvolle Bildverarbeitung ausführen zu können, wird für die Bildcodierung und Informationsextraktion eine vereinfachte Version der in Abschnitt II.2.3 eingeführten Rangordnungscodierung verwendet, angewandt auf eine linear pulsgewandelte Form des Eingangsbildes ($\lambda_{Pixel}{\sim}I_{photo}$, siehe auch Anhang C.1). Als ‚Template' dient dabei [Shamir04], d.h. es wird aus einzelnen pulsenden Pixelzellen und deren relativer Pulsstatistik zueinander eine Hypothese über den zugrunde liegenden Stimulus aufgestellt; mithin wird eine Bildanalyse auf bestimmte durch die Phasenlage codierte Merkmale durchgeführt. In der Praxis sieht dies so aus, dass im Gegensatz zum LOC-Operator kein analoger Vergleich der Pixelströme von Mittenpixel und Nachbarn stattfindet, sondern das Auftreten von mindestens einem Puls des jeweiligen Nachbarpixels während eines ISIs des Mittenpixels registriert wird und daraus am Ende von jedem ISI des Mittenpixels ähnlich wie in Abbildung IV.10 ein Maß für die lokale Struktur entsteht [Mayr05d, Mayr07d].
Wenn dieser Prozess über einen längeren Zeitraum unter dem Aspekt der Feuerraten betrachtet wird, zeigt der PLOC-Operator ein frequenzabhängiges Schwellwertverhalten, d.h. bei einer Frequenz des Nachbarpixels größer oder gleich der des Mittenpixels wird das entsprechende Bit gesetzt, in einem ähnlich einseitigen Vergleich wie beim LOC. Für eine Pulsrate des Nachbarpixels kleiner der des Mittenpixels gibt es sowohl ISIs, in denen das entsprechende Bit gesetzt wird, als auch Ausfälle, es muss also eine differenziertere Betrachtung angestellt werden.
Bei einer Periodendauer $T_2$ des (Nachbar-)Pixels 2 und unbekannter Phasenlage von Pixel 2 relativ zu (Mitten-)Pixel 1 ergibt sich für die Wahrscheinlichkeit, mit der ein Puls von Pixel 2 in einer Zeit $t$ nach dem Puls von Pixel 1 auftritt, folgende Dichtefunktion:

$$f_{T_2}(t) = \begin{cases} \dfrac{1}{T_2} & \text{für} \quad 0 \le t < T_2 \\ 0 & \text{sonst} \end{cases} \tag{IV.19}$$

Die Dichtefunktion vom Zeitpunkt Null bis zum Ende von $T_1$ ist dann in Abhängigkeit des Verhältnisses der Periodendauern/Pulsraten von Pixel 1 und 2:

$$P(0 \le t \le T_1) = \int_0^{T_1} f_{T_2}(t)dt = \begin{cases} \dfrac{T_1}{T_2} \; bzw \; \dfrac{\lambda_2}{\lambda_1} & \text{für} \quad T_2 > T_1 \\ 1 \quad \text{für} \quad T_2 \le T_1 \end{cases} \tag{IV.20}$$

Wie oft ein Feature gewählt wird, hängt damit vom Verhältnis der Pulsraten ab[23]. Die Häufigkeit der einzelnen Puls-/Bitmuster relativ zum Mittenpixel ergibt sich entsprechend über das Produkt der jeweiligen Einzelwahrscheinlichkeiten für „Puls" oder „nicht Puls" im Intervall zwischen zwei

[23] Die obige Annahme wird auch über das ‚sicher auftretende Merkmal' für $\lambda_2 \ge \lambda_1$ bestätigt

Pulsen des Mittenpixels. Für eine beispielhafte Pulsratenverteilung rund um ein Mittenpixel ist dies im Folgenden aufgeführt [Mayr07d]:

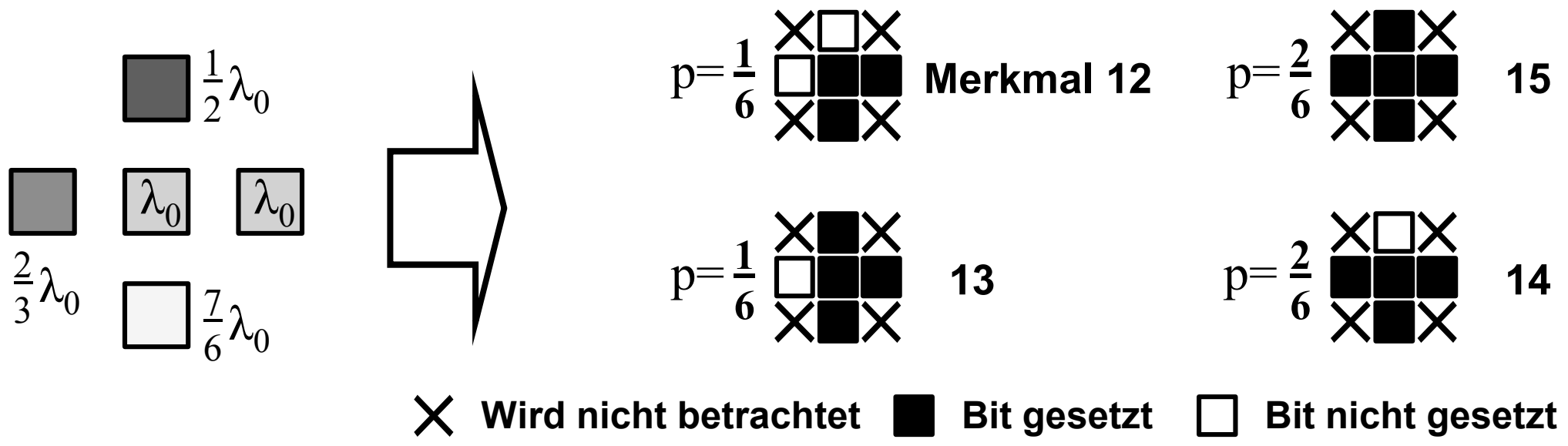

**Abbildung IV.12.: Anteil von gemeldeten PLOC-Merkmalen als Funktion der Nachbarpixelfrequenzen relativ zum Mittenpixel[24]**

Die Größe $\lambda_0$ stellt dabei eine beliebige Normierung dar, die sich durch den Quotient in Gleichung (IV.20) wieder kürzt. Anhand des Merkmals 13 soll kurz das Zustandekommen der Wahrscheinlichkeit erläutert werden: Sowohl das rechte als auch das untere Pixel haben eine Pulsrate größer oder gleich dem Mittenpixel, d.h. sie werden auf jeden Fall einen Puls für jedes ISI des Mittenpixels erzeugen. Das obere Pixel erzeugt im Mittel für jedes zweite ISI des Mittenpixels einen Puls. Dass ein Puls des rechten Pixels ausfällt, wie für Merkmal 13 benötigt, tritt für ein Drittel der ISIs des Mittenpixels auf. Wenn davon ausgegangen wird, dass die Anfangsphasenlagen der Pixel relativ zueinander unkorreliert waren, ergibt sich die Gesamthäufigkeit für Merkmal 13 aus der Überlagerung der entsprechenden Einzelpixel bzw. deren gesetzter Bits im jeweiligen ISI, und damit zu 1/3*1/2=1/6. Die Merkmalsnummern werden hier mit demselben Koeffizientenschema wie in Abbildung IV.9 (links) vergeben, gesetzte Bits/Pulse addieren den Einzelkoeffizienten zur Gesamtmerkmalsnummer des Mittenpixels. Wenn über den Beobachtungszeitraum konstante Periodendauern für die einzelnen Pixel angenommen werden, wechseln sich alle o.a. Bitmuster mit einer festen Wiederkehrdauer ab. Es kann damit auch für kleine Beobachtungszeiten angenommen werden, dass die relative Häufigkeit der einzelnen Bitmuster in guter Näherung den obigen Wahrscheinlichkeiten entspricht. Wie aus Abbildung IV.14 (r.o.) ersichtlich, führt die beschriebene Auswertung von Reihenfolgen/Phasenlagen noch nicht zu aussagekräftigen Bildmerkmalen, da durch die hohe lokale Variabilität der Grauwerte in natürlichen Bildern in gewissem Sinne ‚jedes' Merkmal an ‚jeder' Stelle mindestens einmal gemeldet wird. Deshalb wird die Analogie zu dem in [Shamir04] beschriebenen Populationscode weitergeführt, d.h. es werden zusätzlich zu der o.a. Statistik über Phasenlagen zwei weitere statistische Auswertungen entwickelt.

Die erste Auswertung ist eine Signifikanzbewertung der einzelnen aufgefundenen Merkmale. An jeder Stelle im Bild werden über einen Beobachtungszeitraum T insgesamt λ*T einzelne Merkmale gemeldet. Eine Normierung der jeweiligen Anzahl $N_k$ eines einzelnen Merkmals $k$ auf die Gesamtanzahl an Merkmalen in einem Pixel *(m,n)* trägt in Verbindung mit einem entsprechenden Signifikanz-Schwellwert $\theta_S$ dazu bei, einzelne, eher zufällige Merkmale von häufiger gemeldeten, systematischen zu unterscheiden:

$$b_k^{'}(m,n) = \begin{cases} 1 & für \quad \dfrac{N_k(m,n)}{\sum_i N_i(m,n)} \geq \theta_S \\ 0 & sonst \end{cases} \qquad damit \quad b'(m,n) = \begin{pmatrix} b_1^{'}(m,n) \\ b_2^{'}(m,n) \\ \vdots \\ b_k^{'}(m,n) \end{pmatrix} \tag{IV.21}$$

[24] Die Merkmalsnummern von LOC und PLOC sind nicht identisch, da im LOC ein Koeffizient/Bit gesetzt wird für eine Abweichung unter den Schwellwert, während bei PLOC ein Nachbarpixel im selben Takt pulsen muss, d.h. nicht zu stark abweichen darf. LOC- und PLOC-Merkmalsnummern lassen sich jedoch eineindeutig ineinander überführen.

Die Antwort dieses Vergleichs wird in einer dreidimensionalen Matrixstruktur abgelegt, d.h. es wird nicht wie bei den *b'(m,n)* aus Gleichung (IV.18) nur das dominante Merkmal an den jeweiligen Koordinaten weitergegeben. Aufgrund der wechselnden relativen Phasenlagen können beim PLOC mehrere Merkmale an denselben Bildkoordinaten über dem Schwellwert liegen, damit existiert für jede Bildkoordinate ein Antwortvektor, in dem die einzelnen Merkmale abgelegt sind; oder mit anderen Worten eine Anzahl Teilbilder, in denen jeweils die Antwort des PLOC-Operators für ein einzelnes Merkmal eingetragen ist. Damit ließen sich beispielsweise für die Grauwertverteilung aus Abbildung IV.12 bei einem Schwellwert $\theta_S$ von 0,2 die gemeldeten Merkmale auf 14 und 15 beschränken. Es wird ein ähnliches Verhalten erzeugt wie beim LOC-Operator für eine Variation des Schwellwertes *t(m,n)* über die Skalierung *C* (Gleichung (IV.17)). Die folgende Adaption des Beispiels aus Abbildung IV.12 für den LOC verdeutlicht dies:

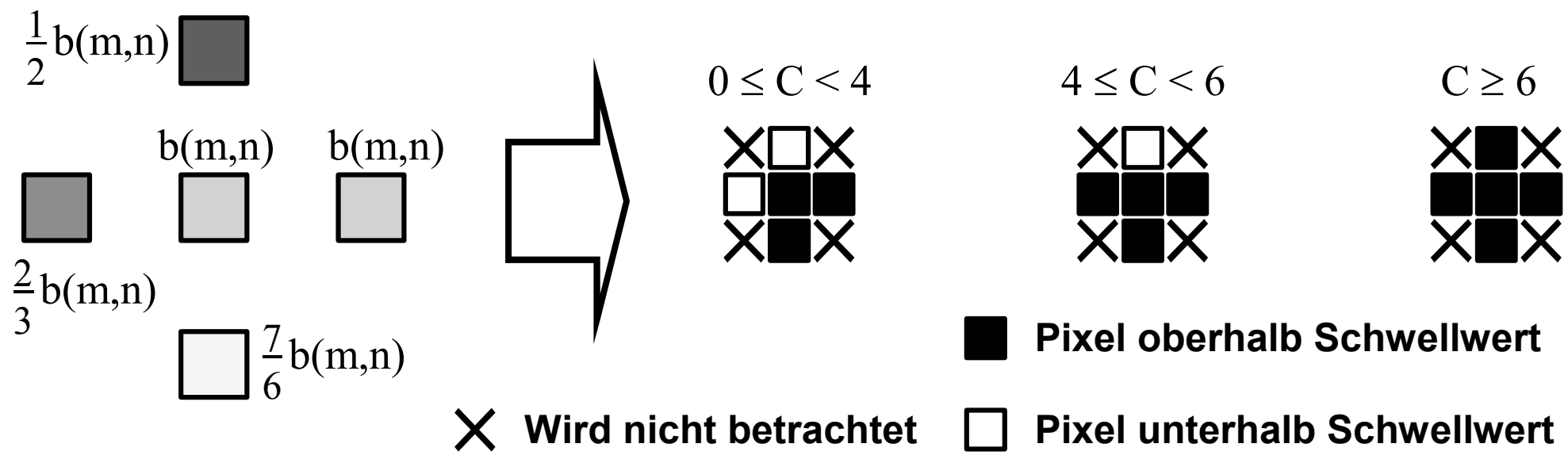

**Abbildung IV.13.: Auswirkung des Koeffizienten C im LOC-Operator**

Es wurden folgende Annahmen getroffen: Die Gaußglättung aus Gleichung (IV.15) wurde durch eine Faltung mit einer diskreten 3*3 Pixel Binomialmaske ersetzt [Jähne05 (Abschnitt 10.4)], die Pixel in den Ecken haben denselben Grauwert *b(m,n)* wie das Mittenpixel. Damit ergeben sich wie oben gezeigt die entsprechenden Werte *C*, für die das linke und das obere Pixel als zur Maske zugehörig oder disjunkt angenommen werden. Wie bereits in [Mayr06b] für die Skalierung C beim LOC Operator erwähnt, kann $\theta_S$ dazu eingesetzt werden, die vom PLOC gelieferten Merkmale an die Anwendung anzupassen, wobei hier der Einstellbereich durch die vorher erfolgte Normierung auf 0...1 festgelegt ist. Werte über 0,5 geben nur das dominanteste Merkmal an einer Stelle wieder. Die Einstellung ist ähnlich robust wie beim LOC, beispielsweise ergeben ±15% Schwankungen bei der Verteilung aus Abbildung IV.12 und einem nominalen $\theta_S$ von 0,2 keinen Unterschied in der Zusammensetzung der über Gleichung (IV.21) weitergemeldeten Merkmale. Trotz dieser Ähnlichkeit können zwischen *C* und $\theta_S$ zwei wesentliche Unterschiede ausgemacht werden:

Zum Einen findet das, was beim LOC in sequentiellen Schritten durch eine entsprechende externe Variierung von C erreicht werden kann, beim PLOC wie oben angeführt gleichzeitig statt, es wird die Menge aller lokal dominanten Merkmale bis $\theta_S$ extrahiert, nicht nur das jeweils signifikanteste. Außerdem wird beim PLOC eine Gesamtbeurteilung der lokalen Nachbarschaft durchgeführt, d.h. es wird nicht nur wie beim LOC ausgehend vom Mittenpixel analysiert, sondern durch die variierende relative Phasenlage auch die Nachbarpixel zueinander. Dies ist der Grund, warum z.B. Merkmal 13 aus Abbildung IV.12 beim PLOC für diese Grauwertverteilung auftaucht, beim LOC jedoch nicht.

Eine weitere statistische Auswertung zur Verbesserung der Qualität der PLOC-Merkmale kann unter Zuhilfenahme spatialer Abhängigkeiten erfolgen. PLOC-Merkmale, die Indikatoren für makroskopische Bildzusammenhänge sind, sollten sich durch eine gewisse lokale Häufung von Merkmalen eines Typs auszeichnen, benachbarte Pixel also zumindest teilweise korrelierte Merkmalsvektoren *b(m,n)* liefern. Es wird ein Korrelationsmaß definiert, das die Größe der Merkmals-Schnittmenge benachbarter Pixel mit der Größe der zugehörigen Vereinigungsmenge normiert und mit einem Schwellwert $\theta_{korr}$ bewertet:

$$A(m+i,n+j)=\begin{cases}1 & \text{für } \dfrac{\sum_k \left[b_k^{'}(m,n)\cap b_k^{'}(m+i,n+j)\right]}{\sum_k \left[b_k^{'}(m,n)\cup b_k^{'}(m+i,n+j)\right]}\geq \theta_{korr} \\ 0 & sonst\end{cases} \tag{IV.22}$$

Die Relativkoordinaten (i,j) permutieren dabei über die komplette 8-fach Nachbarschaft (siehe Gleichung (IV.14)). Die Menge *k* an einzelnen Merkmalen, über welche die Korrelation betrachtet wird, muss nicht zwangsläufig alle PLOC-Merkmale enthalten, sondern kann in Abhängigkeit von der Problemstellung angepasst werden. Beispielsweise würde *k* für die Auffindung der oberen Kante des Mittelstreifens in Abbildung IV.15 die zu den beiden oberen und dem linken unteren Teilbild gehörigen Merkmale enthalten. Aus den paarweisen Korrelationsentscheidungen $A(m+i,n+j)$ für das Mittenpixel und einen einzelnen Nachbarpixel $b(m+i,n+j)$ wird in einem zweiten Schritt eine Gesamtantwort $b_{korr}(m,n)$ für das jeweilige Mittenpixel berechnet:

$$b_{korr}(m,n)=\begin{cases}1 & \text{für } \sum_{i,j} A(m+i,n+j)\geq N_{korr} \\ 0 & sonst\end{cases} \tag{IV.23}$$

Die Anzahl der entsprechend Gleichung (IV.22) in der 8-fach Nachbarschaft korrelierten Pixel werden aufsummiert, und wenn diese gleich oder größer als eine Mindestanzahl korrelierter Pixel $N_{korr}$ ist, wird die Antwort der Korrelationsnachbearbeitung $b_{korr}(m,n)$ gesetzt.

### IV.2.3 Simulationsergebnisse zu Merkmalfindung und Klassifiziererеinsatz

Im folgenden werden einige Simulationsergebnisse für die diskutierten PLOC-Ausprägungen wiedergegeben und die Auswirkung der beiden o.a. Nachbearbeitungen der Merkmale diskutiert. In Abbildung IV.14 wird eine Straßenszene pulsgewandelt und der PLOC-Operator in der 32-Merkmal-Variante darauf angewandt. Alle Pixel, bei denen das Merkmal 7 mindestens einmal gezählt wurde, sind schwarz markiert. Bei einer Normierung auf die Gesamtmerkmalanzahl gemäß Gleichung (IV.21) und Verwendung eines Schwellwerts $\theta_S$ von 0,1 ergibt sich das Teilbild unten links, eine positive Antwort für $b_k'(m,n)$ ist wieder schwarz markiert:

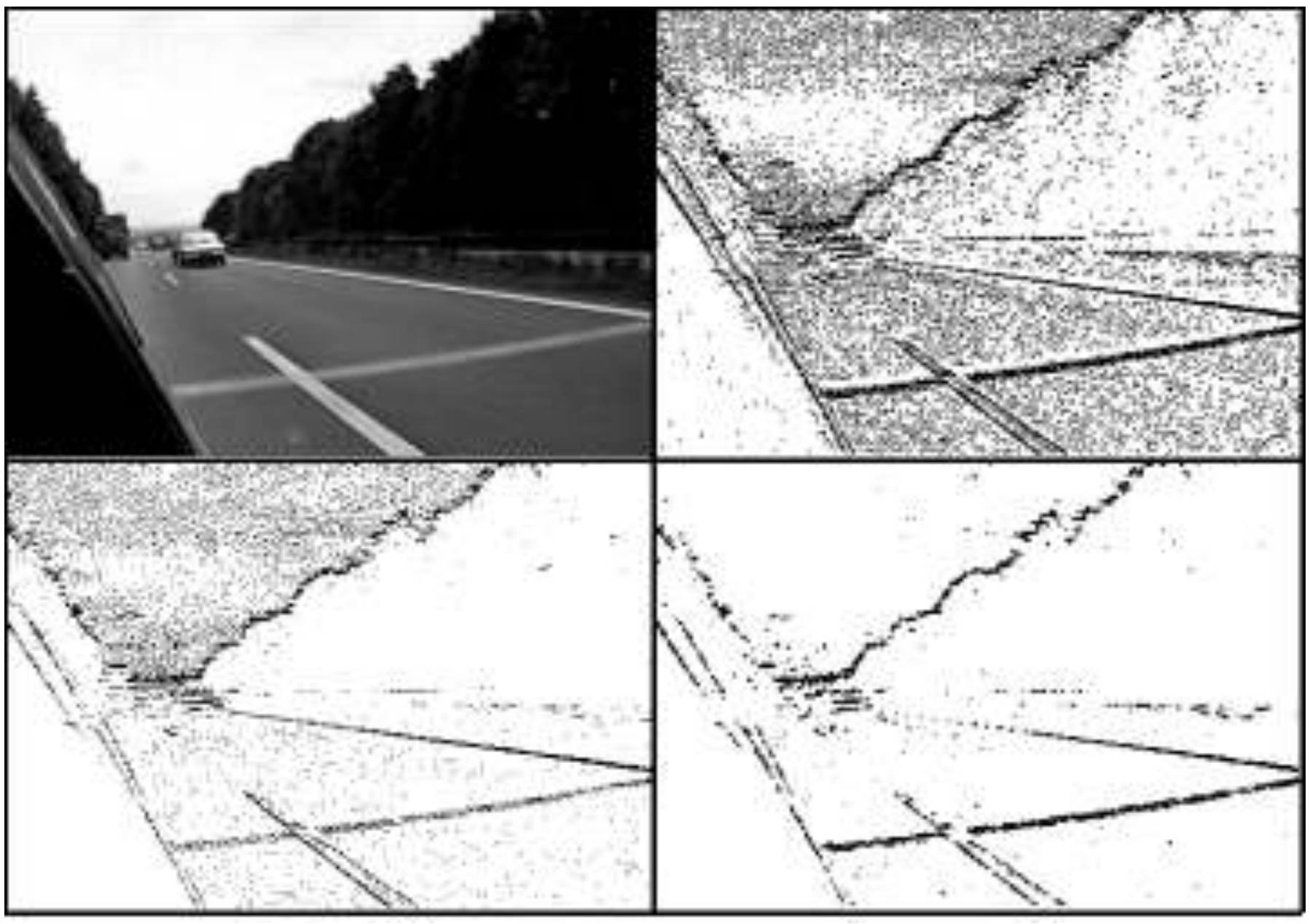

**Abbildung IV.14: Originalbild und PLOC mit verschiedenen Nachbearbeitungen [Mayr05d]**

Eine parallele Betrachtung des $\theta_S$-bewerteten PLOC und des äquivalenten Merkmals 5 im LOC Beispiel aus Abbildung IV.11 ergibt eine vergleichbare Herausarbeitung von Bildmerkmalen. Der PLOC-Operator ist etwas besser darin, Rauschen in den dunklen Teilen des Bildes zu unterdrücken, etwa in der Randbegrünung oder dem Straßenbelag. Allerdings werden für helle Bildbereiche vom PLOC wesentlich mehr fehlerhafte Merkmale als von LOC geliefert, beispielsweise im Bereich des Himmels. Eine Anwendung des Korrelationsoperators ergibt noch mal eine deutliche Verbesserung der Merkmalsqualität (Abbildung IV.14 unten rechts). Vor allem großflächige Bildstrukturen wie die Straßenmarkierung, aber auch Details wie das Auto werden wesentlich prägnanter herausgearbeitet. Interessant ist hierbei auch die Robustheit der Merkmalsextraktion gegenüber absoluten Grauwertpegeln und Kantenkontrast. Im obigen Beispiel wird der Korrelationsoperator nur auf ein einziges Merkmal angewendet, d.h. die Summe über die betrachtete Merkmalsmenge $k$ aus Gleichung (IV.22) reduziert sich auf die Überprüfung, ob dasselbe Merkmal im jeweiligen Nachbarpixel auch auftritt. Der Korrelationsschwellwert $\theta_{korr}$ ist damit unkritisch[25], $0<\theta_{korr}<1$. Der Schwellwert $N_{korr}$ für die Anzahl an Nachbarn, die dasselbe Merkmal aufweisen müssen, liegt bei fünf, d.h. das betrachtete Pixel und eine Mehrheit der zugehörigen 8fach-Nachbarschaft müssen Merkmal 7 zeigen. In Abbildung IV.15 sind Beispiele für weitere Merkmale aufgeführt, die unter Verwendung der geschilderten reduzierten Korrelationsnachbearbeitung erzeugt wurden:

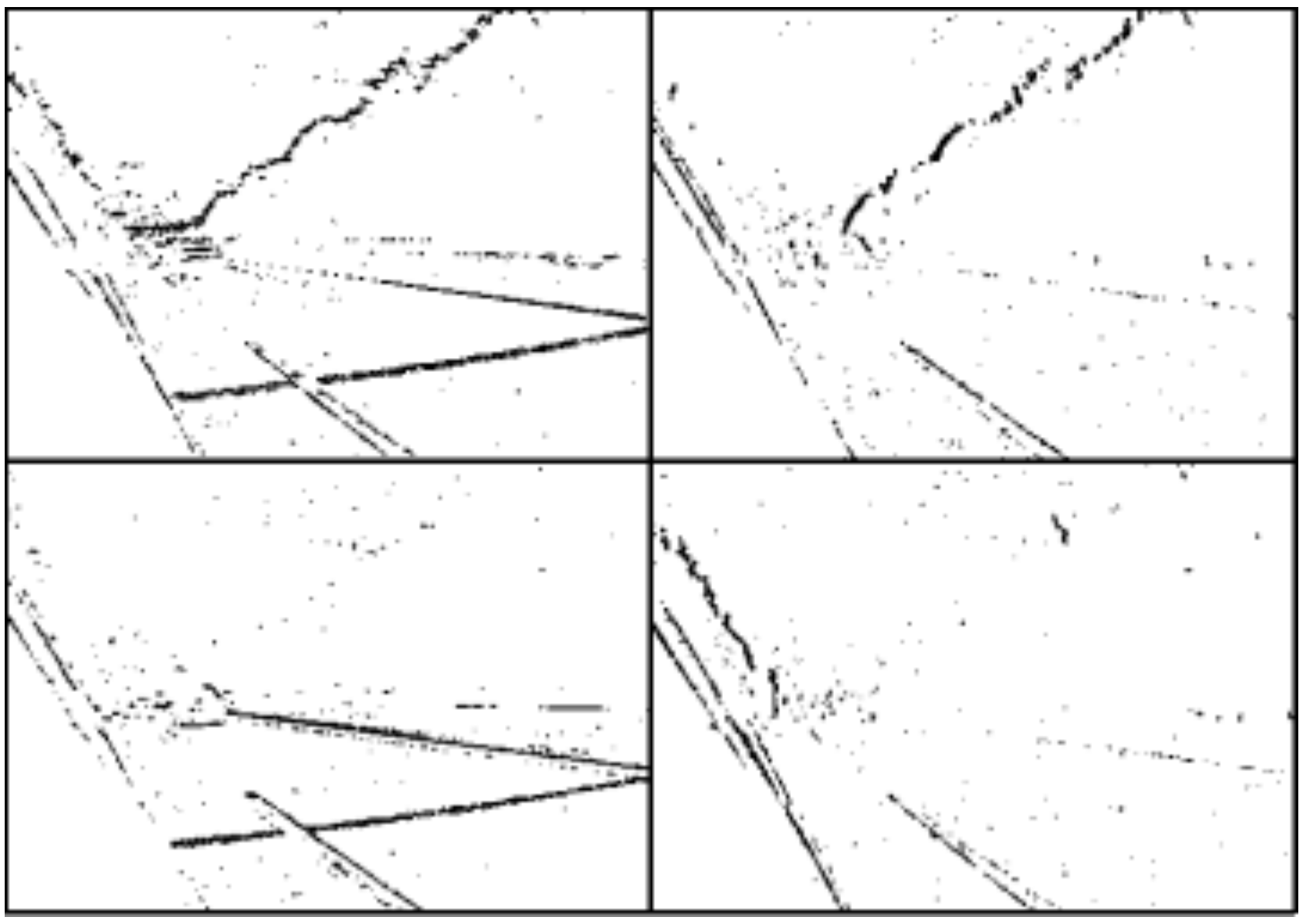

**Abbildung IV.15: verschiedene Merkmale für PLOC mit Nachbarschaftsnachbearbeitung**

Einige der makroskopischen Merkmale treten dabei für mehrere PLOC-Merkmale auf, wie z.B. die Oberkante des rechten Waldstreifens in den beiden oberen Bildern. Andere Merkmale wie der Mittelstreifen liefern leicht unterschiedliche Antworten je nach PLOC-Merkmal, es treten linke und rechte Kante des Mittelstreifens auf oder nur eine von beiden. In einer symbolischen Nachbearbeitung über mehrere Merkmalsbilder kann, wie eingangs für den LOC beschrieben, beispielsweise über Interpolation zwischen mehreren Fundamentalwinkeln der PLOC-Masken eine Feinanalyse der Kantenorientierung durchgeführt werden oder die Kontrastrichtung extrahiert werden. Wie aus Abbildung IV.15 ersichtlich, eignen sich die nachbearbeiteten PLOC-Merkmale sehr gut als Eingangsgrößen einer abstrahierteren Bildanalyse, ohne dass sie sich wie andere hardwaregebundene Verarbeitungen bereits zu sehr auf bestimmte Bildcharakteristiken einschränken. Gleichzeitig werden weite Bildbereiche ohne signifikante Bildinformationen ausgeblendet, d.h. durch diese PLOC-Vorselektion wird die rechnerische Last für nachfolgende Stufen deutlich verringert.

[25] Anhang B.1.2 enthält Beispiele, bei denen $\theta_{korr}$ für Korrelationen über größere Merkmalsmengen eine differenziertere Rolle zukommt.

LOC/PLOC-Merkmale können auch ohne symbolische (Nach-)Bearbeitung direkt für Bildanalysen verwendet werden. In [König02, Mayr06b] wird eine Klassifikationsaufgabe vorgestellt, bei der Augenpartien und zufällige Gesichtsausschnitte unterschieden werden sollen. Der PLOC-Operator wurde auf diese Testbench angewendet, wobei dieselbe Vorgehensweise verwendet wurde wie in [Mayr06b], d.h. ein RNN Klassifizierer wurde mit linken Augenpartien trainiert und mir rechts/links gespiegelten rechten Augenpartien getestet. Der Merkmalsvektor, welcher der Klassifikation zu Grunde liegt, ist ein Histogramm über die jeweilige Anzahl an Merkmalen einer Nummer in dem zu klassifizierenden Bildausschnitt. Es wurde dabei nicht die volle N8 Nachbarschaft mit ihren 256 Merkmalen verwendet, sondern zum Einen die Merkmale des N4 wie in Abbildung IV.9, zusätzlich wurden weitere 16 diagonale Merkmale eingesetzt, die aus einer 45° nach links gedrehten N4 Nachbarschaft bestehen (Abbildung IV.9 rechtes Teilbild). Auf diese 32 Merkmale wurde eine automatisierte Merkmalsselektion angewendet, die Merkmale bezüglich ihres Beitrags zur Klassentrennung auswählt. Durch die so entstandene Reduktion in der Anzahl der Merkmale tendiert der RNN-Klassifizierer weniger zum Auswendiglernen des Trainingssets und damit zur Erhöhung der Robustheit der Klassifikation. Aus demselben Grund wurde von Anfang an mit dem oben erwähnten Satz von 32 Merkmalen gearbeitet, da sowohl Merkmalselektion als auch Klassifizierer auf der N8 Nachbarschaft nicht genügend verallgemeinern. Beispiele für Augenpartien, zugehörige Histogramme und Darstellung der 2D-projizierten Trainings- und Testklassengebiete sind im Folgenden wiedergegeben:

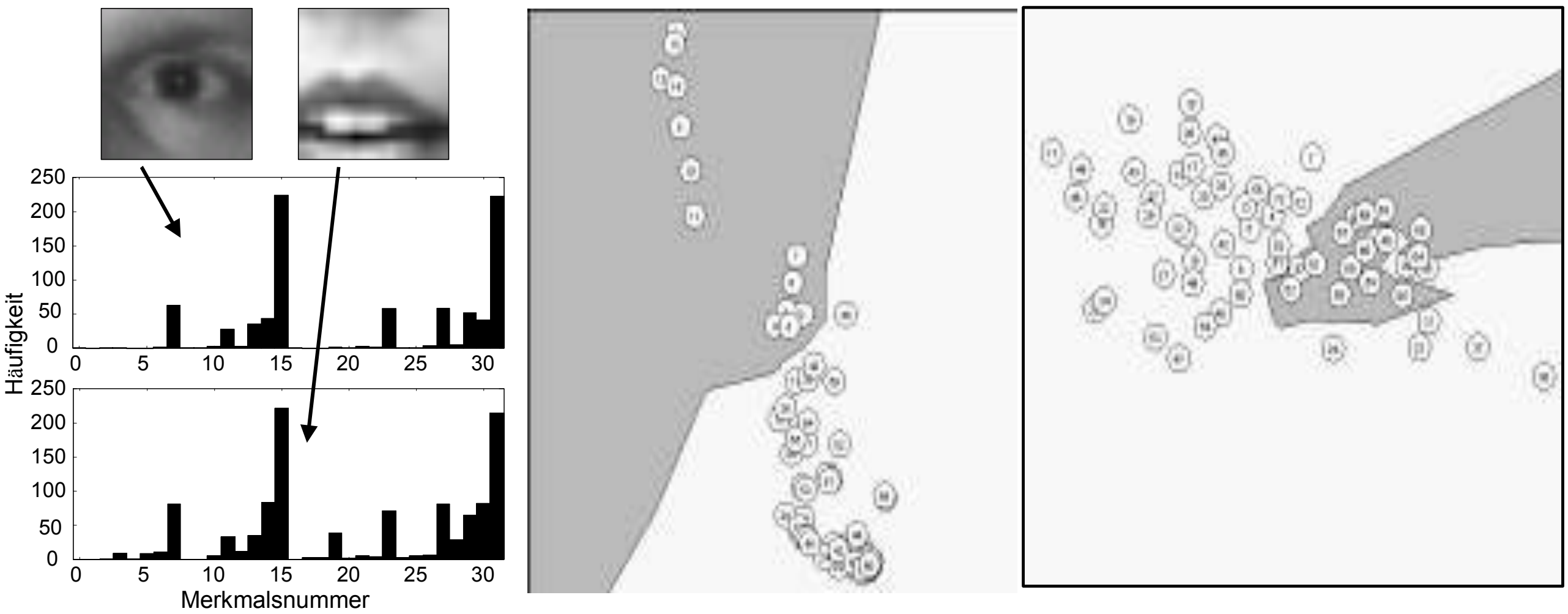

**Abbildung IV.16.: PLOC-Beispielanwendung Augenklassifikation [Mayr07d], Beispiele für Bildausschnitte und Merkmalshistogramme, sowie die Trainings- und Testklassengebiete (v.l.n.r), $\theta_S$ = 0,1, keine Korrelationsnachbearbeitung**

Als allgemeine Erkenntnis über verschiedene Permutationen der Merkmalsselektion für diese Anwendung lässt sich sagen: Es werden für die Klassifikation überwiegend Merkmale aus der originalen N4 Nachbarschaft verwendet, diagonale Merkmale scheinen für die Augenerkennung nicht so wichtig zu sein. Dies ist vermutlich der Tatsache geschuldet, dass vertikale PLOC-Merkmale sich besser eignen, um diagonale Bildstrukturen zu untersuchen, z.B. Merkmal 5 für eine diagonale Kante (beispielsweise der rechte obere Augenrand, siehe auch das linke obere Beispielbild für eine Augenpartie in Abbildung IV.16).

Diese Tests wurden im Rahmen der QuickCog Softwareumgebung durchgeführt, einer Toolbox zum Aufbau von Bildverarbeitungs- und Erkennungssystemen [König98]. Die Klassifikationsaufgabe wird vollständig gelöst, wobei der Testmerkmalsraum in der rechten Hälfte der oberen Abbildung einige Bildausschnitte aufweist, die sehr nah an der Klassentrennlinie liegen. Generell fällt jedoch auf, dass die beiden Merkmalsräume eine größere Selektivität und einfachere Klassentrennlinien aufweisen als der LOC-Operator [Mayr06b]. Vermutlich ist dies mit der multimodalen Analyse im PLOC zu erklären, die jeweils mehrere prägnante lokale Merkmale ausgibt und damit eine differenziertere Bildanalyse erlaubt. Für die Histogrammdarstellung im linken Drittel von Abbildung IV.16 wurden zwei Bildausschnitte gewählt, die im Merkmalsraum sehr dicht beieinander liegen, d.h. unter PLOC-Gesichtspunkten ähnliche Bildstrukturen enthalten. Dieses Beispiel gibt einen Fall wieder, in dem die diagonale N4-Variante wichtig für eine korrekte Klassifikation ist. Wie in den Merkmalshistogrammen zu sehen, zeigt Merkmal 19 (=Diagonalmerkmal 3) signifikante Unterschiede zwischen der Augenpartie und dem Mund, was auf die waagrechten Strukturen der Zähne zurückzuführen ist. Im Rahmen dieser Klassifikationsaufgabe können beide auseinandergehalten werden, aber wie beispielsweise im Abbildung B.1 im Anhang ersichtlich, kann die Ähnlichkeit zwischen Mund- und Augenpartien für LOC und PLOC zu Fehlklassifikationen führen. Dies scheint jedoch die einzige Schwäche des Operators in der vorliegenden Aufgabe zu sein, andere Gesichtsbereiche liegen im PLOC-Merkmalsraum deutlich weiter von den Charakteristiken einer Augenpartie entfernt und können damit robust unterschieden werden.

### IV.2.4 Technischer und neuronaler Ausblick

Eine Realisierung des PLOC-Operators ließe sich mit den pulsenden Pixelzellen (Anhang C.1) und einer kleinen Anzahl von Digitalbausteinen sehr einfach durchführen. In zweiter Stufe hinter den zur Merkmalsaquirierung geschalteten RS-Flipflops würden Zwischenspeicher die Bitmuster des vergangenen Taktes bis zum Auslesen über einen zentralen Bus gespeichert halten, wobei die Weitergabe des Bitmusters von den RS-Flipflops und das Zurücksetzen der Flipflops jeweils zum Puls des Mittelpixels erfolgt. Die folgende Abbildung verdeutlicht das Verarbeitungsschema und die benötigten Hardware-Elemente für eine Nächster-Nachbar-PLOC-Zelle, d.h. es findet eine Analyse über die horizontalen und vertikalen Nachbarn wie in Abbildung IV.12 statt:

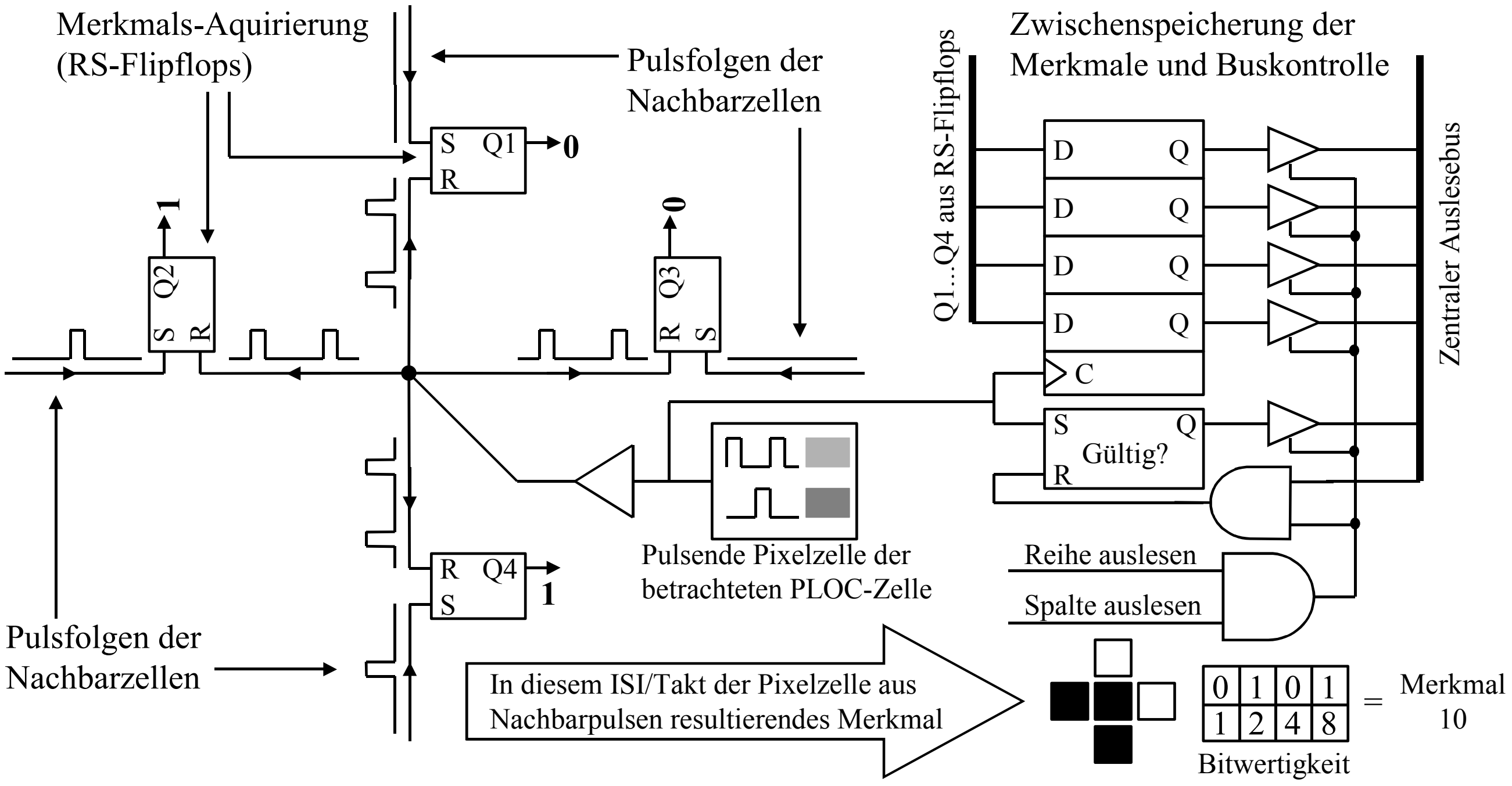

**Abbildung IV.17: Beispielimplementierung des PLOC-Operators**

Das Rücksetzen wird über einen zusätzlichen Puffer leicht verzögert ausgeführt, um eine sichere Übernahme der Merkmale in die D-Flipflops zu gewährleisten. Bustreiberbausteine und eine Zeilen- und Spaltenauswahl vervollständigen die Implementierung. Da die Merkmale asynchron mit der Pixelfrequenz erzeugt werden, bietet sich als einfachstes Ausleseschema ein kontinuierliches Scanprinzip mit der maximalen Pixelfrequenz an[26]. In diesem Fall wird zusätzlich ein Bit zur Gültigkeitsbewertung benötigt, das von der Pixelzelle im selben Zug gesetzt wird wie die Zwischenspeicher, um anzuzeigen, dass es sich beim gespeicherten Wert um ein noch nicht ausgelesenes Merkmal handelt. Während des Auslesens wird dieses Bit über die Zeilen- und Spaltenauswahl sowie eine globale Löschleitung zurückgesetzt.

[26] Im Sinne der in Abschnitt II.4 diskutierten Energieaufnahme für globale/verteilte Takte wären Alternativen zum Scanprinzip für leistungsoptimierte PLOC-Implementierungen überlegenswert.

Die Normierung der jeweiligen Anzahl einzelner Merkmale auf die Gesamtmerkmalszahl ließe sich einfach in Merkmalsakkumulatoren am Rand der Matrix durchführen, indem immer eine 2er-Potenz von Gesamtmerkmalen abgewartet wird und sich die Normierung/Division der Einzelmerkmale damit auf eine Shift-Operation reduziert. Die geshifteten Werte können dann direkt mit einem global angewendeten Schwellwert $\theta_S$ bewertet und ausgelesen werden. Weitere Verarbeitung wie beispielsweise die Korrelationsanalyse kann dann flexibel in Software implementiert werden.

Eine Flächenabschätzung der PLOC-Zelle für die komplette N8 Nachbarschaft in einer 0,13 µm CMOS Technologie lautet wie folgt: Digitale Grundbausteine 8 D-FF, 9 RS-FF, 2 AND, 9 Tristate Treiber sowie 1 Puffer, bei einem Füllfaktor für die Digitallogik von 80% ergibt dies eine Fläche von 872µm$^2$. Zusätzliche 175µm$^2$ werden für die pulsende Pixelzelle benötigt[27], damit ergibt sich eine Kantenlänge von ca. 32µm*32µm für die vollständige Zelle ohne Bus. Eine vergleichbare Implementierung des analogen LOC-Operators [Mayr06b] in einem 0,6µm Prozess resultierte in eine Kantenlänge von 83µm*80µm, mithin ein Flächenunterschied um den Faktor 6-7.

Für eine Implementierung, aber auch unter neuronalen Gesichtspunkten ist von Interesse, inwieweit Rauschen die Identifizierung von Merkmalen behindert, dadurch dass beispielsweise ein Taktjitter des Nachbarpixels trotz insgesamt stabiler Grundfrequenz sporadische Pulsverschiebungen verursacht, die Merkmalsartefakte auslösen. Eine entsprechende Abwandlung von Gleichung (IV.19) für einen zusätzlichen Jittereinfluss wird in Anhang A.3 hergeleitet. Beispielhaft soll untersucht werden, wie oft bei identischen Grundperioden des Mitten- und Nachbarschaftpixels $T_1$ und $T_2$ das sichere Ereignis eines gesetzten Bits ausfällt. Aus Gleichung (A.39), Formel (1) lässt sich die Wahrscheinlichkeit ermitteln, mit der kein Bit gesetzt wird:

$$P(t \le 0 \cup t \ge T_2) = 2\int_0^{T_1} \frac{(t+T_j)}{2T_2T_j^2}\,dt = \frac{1}{3T_2T_j^2}\left[(t+T_j)^3\right]_{-T_j}^{0} = \frac{T_j}{3T_2} \qquad \textbf{(IV.24)}$$

Da die Dichtefunktion symmetrisch ist, wird die Wahrscheinlichkeit aus dem doppelten Wert des rechtsseitigen Teilintervalls ermittelt, das außerhalb des Intervalls $T_1(=T_2)$ liegt. Eine (biologische) Feuerrate $\lambda$ von 50s$^{-1}$ bzw. ein $T_2$ von 20 ms bei einem Jitter von 1 ms [Kretzberg01] resultiert in einer Fehlerwahrscheinlichkeit von 1,7%. Die Relationen zwischen Jitter und Pulsfrequenz liegen auch für technische Implementierungen in ähnlicher Relation zueinander oder sind sogar besser (z.B. 10kHz Pixelfrequenz und 2-4µs Jitter des Integrators). Jitter-bedingte Fehler, bei denen nicht existente Merkmale extrahiert werden, können damit für technisch relevante Rauschamplitude durch ein entsprechendes $\theta_S$ beseitigt werden. Die am wahrscheinlichsten auftretenden Fehler werden jedoch Verschiebungen der relativen Häufigkeit von einzelnen (legitimen) Merkmalen zueinander sein, welche in der weiteren Auswertung aufgrund des egalisierenden Schwellwertes keine Rolle spielen.

Die für den PLOC gezeigte Tendenz, trotz einer unklaren Phasenlage zu Beginn der Beobachtungszeit vorzugsweise dominante lokale Merkmale zu liefern, kann auch von biologischer Relevanz sein:

In [VanRullen01, VanRullen05] wird ein Rank Order Coding analysiert, bei dem die ‚Time to first Spike' als relevante Größe ausgewertet wird, ohne dass hierbei der dafür benötigte Rücksetzungsmechanismus biologisch motiviert wird. Da das betrachtete System (Retina bis V1-Eingang) abgesehen von möglichen Sakkaden zwar Rückkopplung, aber kein Rücksetzen kennt, ist die Existenz eines Reset zumindest fraglich. Aus Abbildung IV.12 und Gleichung (IV.20) ist dagegen die Tendenz der für den PLOC gewählten Analyse ersichtlich, bereits über wenige Pulsperioden/ISIs ohne Rücksetzungsvorbedingung eine zuverlässige Schätzung dominanter Phasenlagen zu ermöglichen. Dendritische Verarbeitung der Retinainformation im Kortex könnte auf gleiche Weise signifikante Bildmerkmale extrahieren, ohne einen Rücksetzungsmechanismus zu benötigen. Für eine Population von Retinaneuronen, die ähnlich geartete Bildmerkmale codieren, jedoch aufgrund unterschiedlicher Anfangsbedingungen dekorreliert sind, könnte über statistische Analysen wie in Gleichung (IV.23) sogar bereits für eine einzige Pulsperiode/ISI das dominante Merkmal geschätzt werden. Biologische Mechanismen hierzu liefert beispielsweise die Merkmalsverkettung über Synchronisation im MT-Bereich des visuellen Kortex [Gerstner99].

[27] Diese Fläche wurde aus Photodiode und analogem Schaltungsteil gemäß Abbildung C.1 gebildet (10,6µm*16,5µm), der in Abbildung C.2 ersichtliche digitale Steuerungsteil wird bereits von der Digitalschaltung der PLOC-Zelle beinhaltet.

# V Fast Analog Computing with Emergent Transient States - FACETS

Das EU-Projekt FACETS hat sich zum Ziel gesetzt, durch Emulierung biologischer Verarbeitungsprinzipien in Simulation und Hardware, basierend auf einer großen Anzahl von Neuronen ($\cong 10^6$) und komplexer Vernetzung mit ca. $10^9$ Synapsen, emergente biologische Verarbeitungsmechanismen und Netzwerkverhalten zu erforschen. Zusätzlich sollen die so gewonnenen Kenntnisse und die geschaffene Hardware dazu verwendet werden, nicht-Turing-basierte Informationsverarbeitungsprinzipien zu charakterisieren und zu erproben [Meier04]. Hierbei soll insbesondere in Bezug auf verteilte und hochparallele Verarbeitung von Informationen (wie in Abschnitt II.4 dargelegt) von seriellen Turing-Prinzipien Abstand genommen werden. Das Vorbild der Modellierung bildet der V1-Bereich des visuellen Systems des Säugetiers, der in Abschnitt I.3 bereits phänomenologisch beschrieben wurde. Im Rahmen der Modellierung sollen aus aktueller neurobiologischer Forschung abgeleitete Adaptionsregeln, v.a. Spike Timing Dependent Plasticity (STDP) zum Einsatz kommen. Die Electronic Vision Group (Uni Heidelberg) und der Lehrstuhl von Prof. Schüffny kooperieren im Rahmen von FACETS beim Aufbau eines waferbasierten neuromorphen Systems mit den oben angegebenen Größenordnungen an Neuronen und Synapsen. Ähnlich wie bei dem in [Schemmel06] beschriebenen IC werden in dem Waferscale System (im Folgenden Stage 2 genannt) die neuronalen Dynamiken deutlich beschleunigt ablaufen, um (Langzeit-)Lernvorgänge beobachten zu können. Über einen Beschleunigungsfaktor von $10^4$ bei einer mittleren Neuronenpulsrate von 10Hz entsteht eine durchschnittliche postsynaptische Pulsrate von 100kHz, und daraus resultierend beträchtliche Anforderungen an die Pulskommunikation bei der projektierten Neuronen- bzw. Synapsenanzahl. Um verschiedenste Netzwerktopologien und –verhalten abbilden zu können, sollen die einzelnen Baublöcke in weiten Grenzen konfigurierbar sein.

Da FACETS zum Zeitpunkt dieser Niederschrift nur etwas mehr als die Hälfte der Projektzeit durchschritten hat, sind viele der hier geschilderten Entwurfsaspekte als Momentaufnahme zu verstehen. Es wird im Folgenden versucht, das Projekt im Zusammenhang zu schildern um einen Gesamteindruck zu liefern, wobei verschiedene, vom Autor maßgeblich mitentworfene Bereiche mehr Gewicht erhalten. Um den Anteil des Autors an den in diesem Kapitel geschilderten Arbeiten einordnen zu können, kann eine Einteilung der Urheberschaft der bisher geleisteten Arbeiten wie folgt stattfinden:

Der Systementwurf der Layer2 Pulskommunikation[28] und Studien für Layer1 wurden von H. Eisenreich, dem Autor und S. Henker ausgeführt, teilweise basierend auf der IP aus [Schemmel04]. Der Autor war dabei vor allem mit der Ausrichtung des Entwurfs an neuronalen Gesichtspunkten betraut, um die Relevanz der Hardware in Bezug auf neurobiologische Erkenntnisse von Projektpartnern sicherzustellen. Zur entsprechenden Bewertung des Entwurfs wurde von J. Partzsch und dem Autor eine Datenbank aufgebaut, in der neuronale Benchmarks von projektinternen Neurowissenschaftlern für Systemsimulationen des Hardwareentwurfs aufgearbeitet wurden [Partzsch07b]. Die Arbeiten zu Mapping und Konfigurationserzeugung erfolgten in Zusammenarbeit mit Matthias Ehrlich und Karsten Wendt [Mayr07b, Wendt07], unter Beteiligung von J. Fieres (Heidelberg). Ziel ist es dabei, die Benchmarks und spätere Forschungsanwendungen möglichst detailgetreu auf die Systemsimulation und die fertige Hardware abzubilden. Die Grundkonzepte in diesen Bereichen und die zugehörige Bewertung der Hardware unter neuronalen Gesichtspunkten während des Designprozesses wurden vom Autor entworfen [Mayr06c]. Studien zu den Baublöcken für Neuronen und Synapsen, zur Waferscale-Integration, sowie zum Gesamtsystementwurf wurden von der Electronic Vision Group um J. Schemmel (Heidelberg) ausgeführt. Am Dresdner Lehrstuhl wurde von M. Ander, H. Eisenreich, S. Scholze und G. Ellguth ein IC-Prototyp für die Pulskommunikation über lange Reichweiten entwickelt [Ehrlich07].

---

[28] Siehe Abschnitt V.3.3.

# V.1 Spike Timing Dependent Plasticity - STDP

Das bekannteste Postulat zur Funktionsweise neuronaler Lernvorgänge wurde 1949 von Hebb aufgestellt [Hebb49] (siehe auch Abschnitt II.2.7). Traditionell wurde das darin beschriebene Synapsenverhalten immer auf Pulsratenmodelle angewandt. In der neurobiologischen Forschung wurden ebenfall ratenbasierte Plastizitätsregeln gefunden, die sich biologisch motivieren lassen und verschiedene Phänomene neuronaler Adaption gut beschreiben [Bienenstock82]. Neuere Forschungen geben jedoch starke Indizien dafür, dass Adaption und Verarbeitung in neuralen Geweben auf einzelnen, reproduzierbaren Pulsmustern aufbauen [Aronov03, Gutkin03, VanRullen01, VanRullen05]. In den Jahren 1997/98 wurden von verschiedenen Forschergruppen Experimente durchgeführt, die sich mit Langzeitlernvorgängen an Synapsen in Abhängigkeit von prä- und postsynaptischen Pulsmustern befassten [Markram97, Bi98]. Eine bedeutende Erkenntnis dieser Experimente war, dass die Plastizität von Synapsen von der Differenz zwischen dem Zeitpunkt des prä- und postsynaptischen Pulses innerhalb eines Zeitfensters von ca. 0-60ms abhängig ist und damit eine Hebbsche Adaption realisiert, die auf einzelnen Pulsen aufbaut [Song01]. Die Gesamtheit dieser Erscheinungen wurde als Spike Timing Dependent Plasticity bezeichet. Da STDP im Rahmen der FACETS-Hardware die primär interessierende Adaptionsregel darstellt, wird in den nächsten Abschnitten das entsprechende Netzwerkverhalten näher diskutiert[29].

### V.1.1 Induktion aus biologischem Synapsenverhalten

Von Bi und Poo [Bi98] und Markram et al. [Markram97] wurde das Adaptionsverhalten von exzitatorischen Neuronen des Hippocampus in Abhängigkeit des Zeitintervalls $t_{post}$-$t_{prä}$ zwischen dem präsynaptischen Puls an einer Synapse und dem Puls des nachgeschalteten Neurons gemessen:

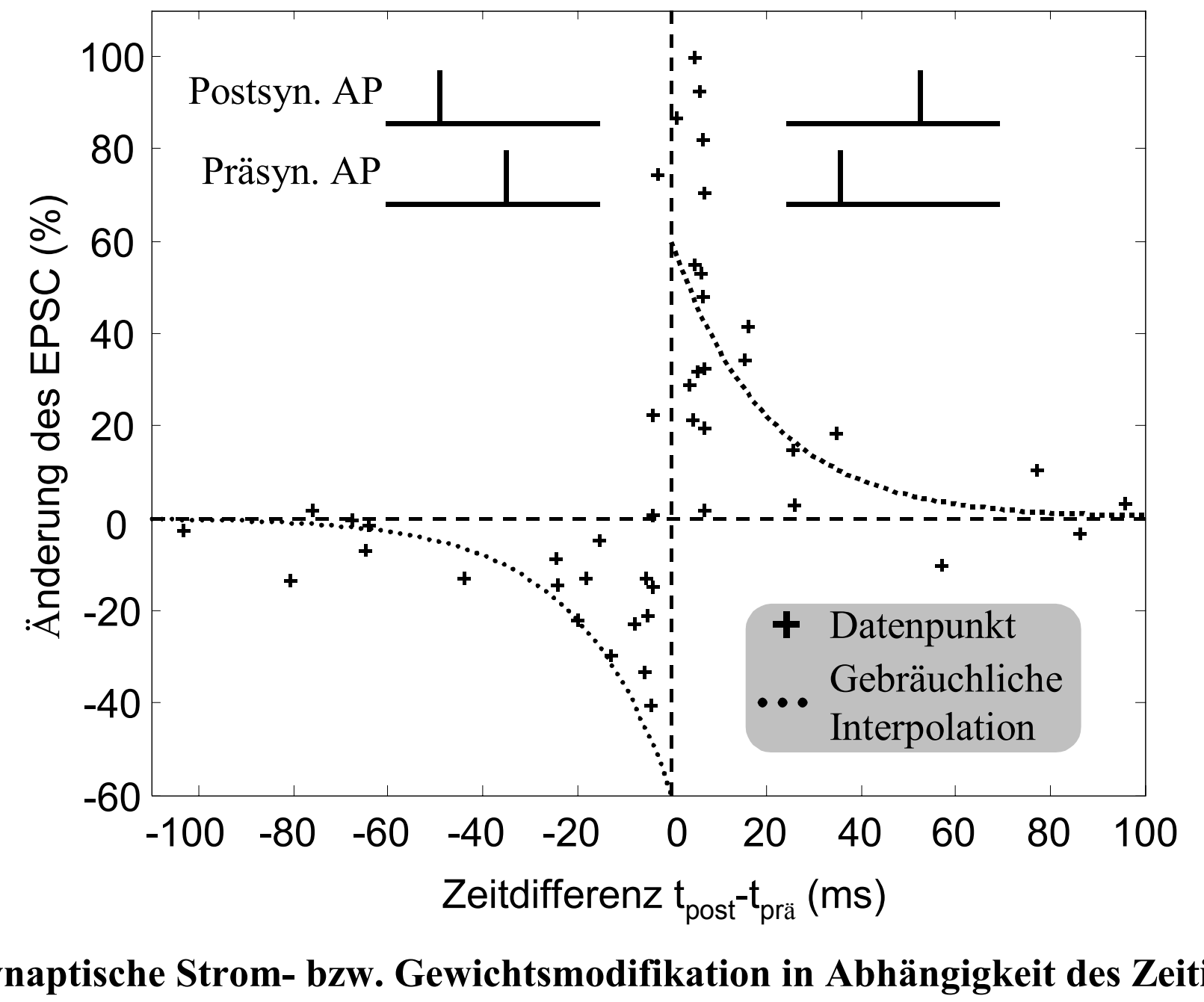

**Abbildung V.1.: Synaptische Strom- bzw. Gewichtsmodifikation in Abhängigkeit des Zeitintervalls zwischen prä- und postsynaptischem Puls [Bi98], und gängige Näherung nach [Delorme01]**

Jedes der obigen Kreuze steht für ein Experiment, bei dem eine bestimmte zeitliche Abfolge aus prä- und postsynaptischem Puls einmal pro Sekunde an einer Synapse erzeugt wurde, über einen

[29] Wie eingangs des Kapitels erwähnt, ist der Lehrstuhl im Rahmen von FACETS vor allem mit der Realisierung der Pulskommunikation und dem Systementwurf betraut, während Neuronen und Synapsen (und damit die STDP-Adaption) von Schemmel et al. entworfen werden. Da die stattfindende Adaption jedoch aufgrund ihrer pulsbasierten Natur auch für den Entwurf der Pulskommunikation und die Bewertung des Systementwurfs unter neurobiologischen Gesichtspunkten starke Relevanz hat, wird hier eine erweiterte Einführung in die Thematik gegeben.

Versuchszeitraum von 60s. Prä- und postsynaptisches Neuron wurden jeweils durch entsprechende Konstantstromeinspeisung zum Pulsen gebracht. 20 min nach dem Experiment wurde die Veränderung des postsynaptischen Stroms (EPSC), der durch einen präsynaptischen Puls ausgelöst wird, gegenüber dem Strom vor dem Experiment gemessen. Wenn diese Veränderungen über der Zeitdifferenz $t_{post}$-$t_{prä}$ aufgetragen werden, ergibt sich eine Verstärkung (Long Term Potentation, LTP), wenn der postsynaptische Puls innerhalb eines Zeitfensters von ca. 60 ms nach dem präsynaptischen Puls stattfindet. Umgekehrt verringert sich der EPSC, die sogenannte Long Term Depression (LTD), wenn der postsynaptische Puls vor dem präsynaptischen ausgelöst wird. Die rechte Hälfte des Graphen lässt sich über Hebb erklären, da der präsynaptische Puls in diesem Fall zum Auslösen des postsynaptischen Pulses beigetragen hat. Die linke Hälfte des Graphen scheint der Erhöhung der Energieeffizienz zu dienen, da ein kurz nach dem postsynaptischen Puls ausgelöster EPSC durch die Hyperpolarisierung des postsynaptischen Neurons keine Wirkung hat und damit nur unnötig die Ionenkanäle aktiviert. Zusätzlich dient die starke Unstetigkeit zwischen LTP und LTD vermutlich dazu, die zeitliche Selektivität der Neuronen zu erhöhen [Delorme01, Kepecs02]. Jedes Neuron versucht, seine Synapsen so zu trimmen, dass möglichst kurz nach einem präsynaptischen Puls der entsprechende postsynaptische ausgelöst wird, der anzeigt, dass ein bestimmtes Pulsmuster am Eingang festgestellt wurde [Song01]. Eine gängige mathematische Näherung der Messwerte aus Abbildung V.1 erfolgt mit e-Funktionen in Abhängigkeit der Zeitdifferenz $t_{post}$-$t_{prä}$ [Delorme01]:

$$dW = \begin{cases} \eta(1-W)e^{-\frac{t_{post}-t_{prä}}{\tau}} & \text{für} \quad t_{post} > t_{prä} \\ -\eta W e^{-\frac{t_{prä}-t_{post}}{\tau}} & \text{für} \quad t_{post} \le t_{prä} \end{cases} \qquad \text{(V.1)}$$

Die Veränderung des EPSC wird dabei als synaptisches Gewicht $W$ interpretiert (siehe Abschnitt II.1.2), welches sich bezogen auf das obige Versuchsprotokoll bei einem einzelnen Zusammentreffen von prä- und postsynaptischem Puls um ca. 1-2% verändert. $W$ bezeichnet dabei eine einheitenlose Größe zwischen 0 und 1 wie in Abschnitt II.1.2 eingeführt. Die Änderung des Gewichts pro Pulsereignis wird über die Lernrate $\eta$ gesteuert, ähnlich wie in Gleichung (III.1). Das Zeitfenster, in dem zusammentreffende Pulse für STDP-Plastizität berücksichtigt werden, kann durch $\tau$ gesteuert werden. Aus den Versuchsdaten leitet sich damit die übliche Parametrisierung von Gleichung (V.1) zu $\eta$=0,02 und $\tau$=20ms her, wenn die obige Gewichtsänderung zu jedem Pulspaar durchgeführt wird. Die gestrichelte STDP-Kurve in Abbildung V.1 wurde gemäß diesen Parametern erstellt, bei einem initialen Gewicht von 0,5. Aufgetragen ist die prozentuale Veränderung des Gewichts, skaliert auf 60 Wiederholungen.

Je nach Neuronenart ergeben sich für die jeweilige Form des STDP sehr unterschiedliche Verläufe:

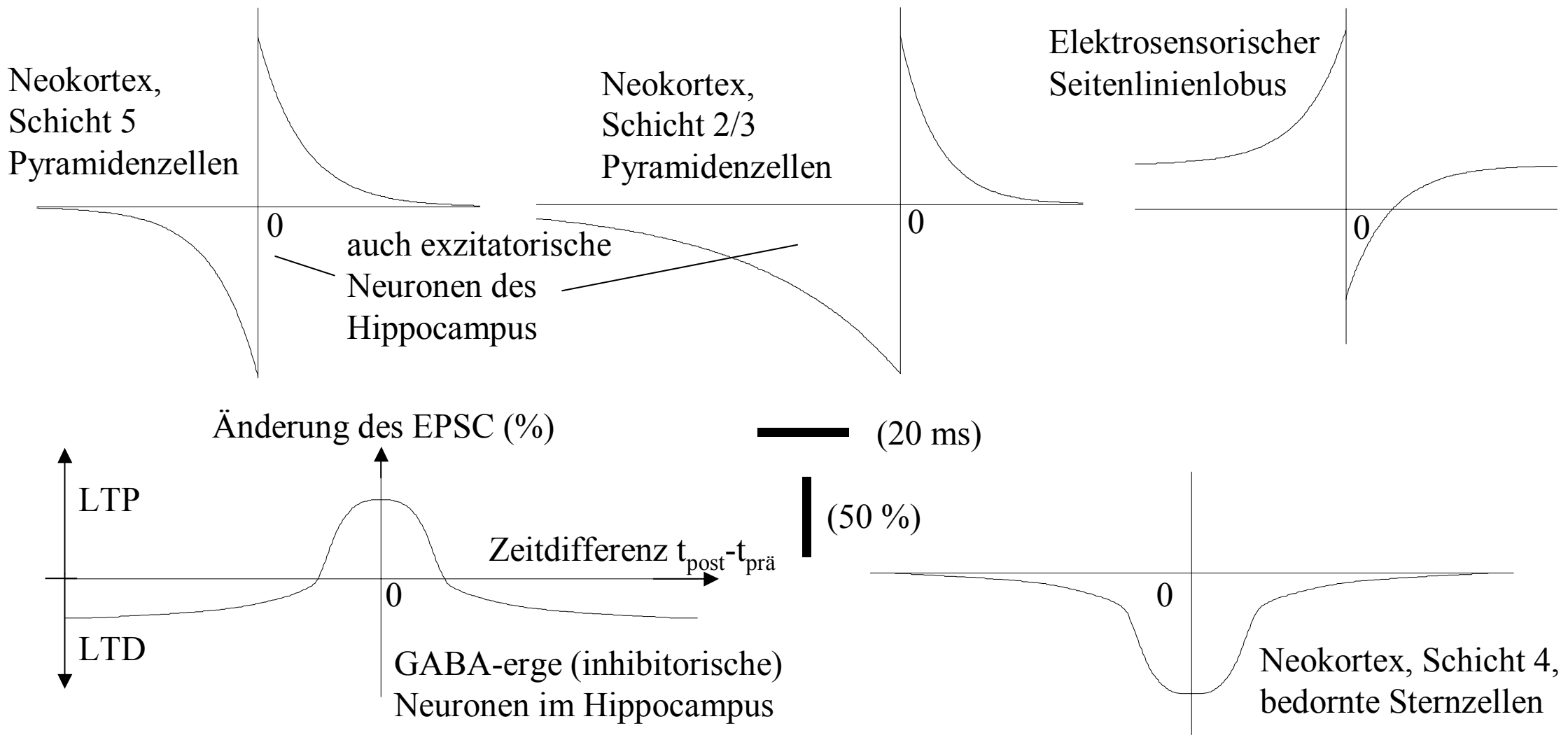

**Abbildung V.2.: STDP-Ausprägungen in unterschiedlichen Hirnarealen, nach [Abbott00, Bell97]**

Der Verlauf des STDP lässt dabei wieder Rückschlüsse auf das entsprechende Verhalten bzw. die Verarbeitungsfunktion zu [Kepecs02]. Eine größere Fläche unter der LTD-Kurve erhöht beispielsweise die Fähigkeit des Netzwerks, die mittlere Pulsfrequenz konstant zu halten [Abbott00]. Die Anti-Hebbsche Form beim elektrosensorischen Seitenlinienlobus eines elektrischen Fisches dient dazu, bereits gelerntes in der Verarbeitung auszublenden, so dass nur noch als neu erkannte Muster weitergemeldet werden [Bell97]. Die Form des STDP bei einem GABA-ergen Neuron im Hippocampus erinnert eher an klassische Hebb-Interpretationen, die von einer Verstärkung des Synapsengewichts bei korrelierter prä- und postsynaptischer Aktivität ausgehen, ohne Betrachtung der zeitlichen Abfolge. Übersichten zu verschiedenen Aspekten von STDP-Adaption finden sich in [Abbott00, Kepecs02].

### V.1.2 Auswirkungen der Lernregel

Einer der ersten Effekte, die mit auf STDP aufbauenden neuronalen Netzen nachgebildet werden konnten, ist das Entstehen der in Abschnitt I.3.3 eingeführten rezeptiven Felder. Stark vereinfachte Nachbildungen des V1 wurden dabei mit den Ausgängen einer Retinasimulation verbunden, welche mit wechselnden Naturszenen stimuliert wurde. Es entwickelten sich die bekannten Gabor-ähnlichen Filtercharakteristiken [Delorme01]. Weiterführende Tests etablierten den Zusammenhang zwischen lateraler Inhibition und Selektivität der rezeptiven Felder [Delorme03a]. In einer reinen Feedforward-Struktur bilden STDP-Synapsen zwar unweigerlich eine Winkel/Orientierungsselektivität aus, aber nur die lateralen Verbindungen sorgen für einen Wettbewerb zwischen Neuronen und damit für eine Eliminierung der Redundanz [Kepecs02, Delorme03a]. Wenn zusätzlich die aus dem V1 bekannte Rückkopplung zwischen Schichten [Binzegger04] mit in das Modell einbezogen wird, entsteht eine regelmäßige Abdeckung des Bildbereiches mit rezeptiven Feldern [Song01] ähnlich der im V1 gefundenen Kolumnen-organisation [Hubel68 (Fig. 9)]. Da im Rahmen von FACETS in erster Linie die Informations-verarbeitung im V1 detaillierter als bisher erforscht werden soll, stellen entsprechende Netzwerke sicher eine der Hauptanwendungen des späteren Stage 2 Systems dar. Um verschiedene Effekte der Hardware speziell in Bezug auf V1-Modelle charakterisieren zu können und ein Gefühl dafür zu entwickeln, was entsprechende Netzwerke beispielsweise an Konfiguration oder Puls-kommunikation für Anforderungen stellen, wurden extensive Systemsimulationen durchgeführt. Im Folgenden sollen anhand ausgewählter Simulationsergebnisse (ohne Berücksichtigung von Hardwareeffekten) einige der oben angeführten Charakteristiken von STDP in Bezug auf visuelle Informationsverarbeitung untermauert werden:

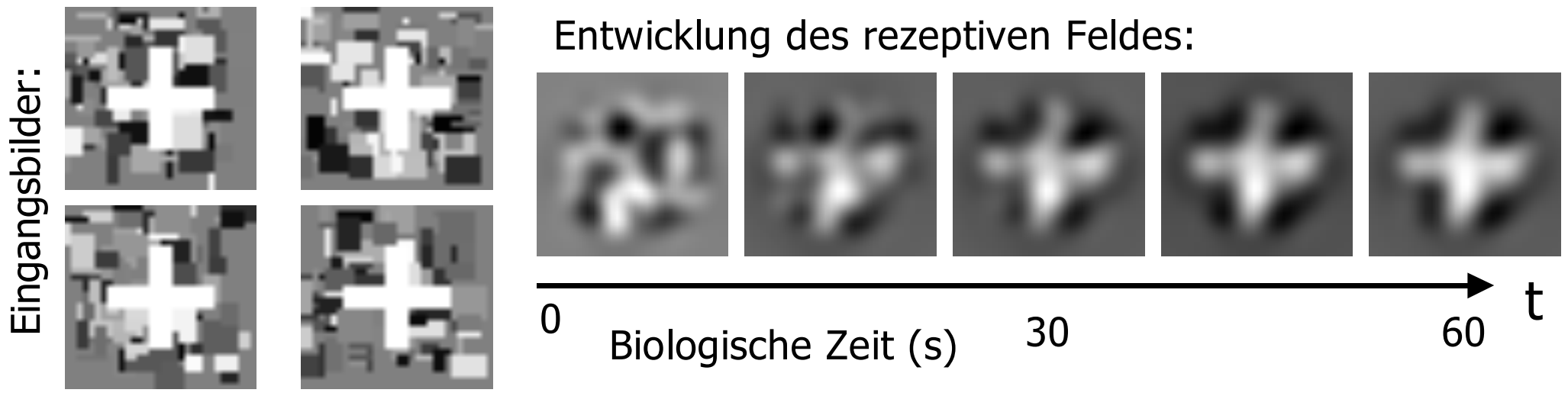

**Abbildung V.3.: Zeitliche Entwicklung des Aufbaus rezeptiver Felder**[30]

Im obigen Fall wurde ein einzelnes LIAF-Neuron[31] mit STDP adaptierenden Synapsen an die Ganglienzellen einer vereinfachten Form der in [Wohrer06] beschriebenen Retina angeschlossen. Der Retinanachbildung wurde dann im 150ms Rhythmus in zufälliger Reihenfolge mit den 4 rechts gezeigten Eingangsbildern belegt, wodurch auf den STDP-Synapsen pulscodierte DoG-

[30] Die Rekonstruktion der rezeptiven Felder erfolgte, indem die DoG-Charakteristiken des Retinamodells an ihren jeweiligen Koordinaten mit den synaptischen Gewichten multipliziert wurden und daraus die Gesamtfaltungsmaske bzw. das Empfindlichkeitsprofil des Neurons berechnet wurde.

[31] Parametern wie in Abschnitt IV.1.2 mit realistischem Schwellwert.

Interpretationen der Eingangsbilder übertragen werden [Warland97] (siehe auch Abschnitt I.3.1). In der obigen Abbildung ist deutlich die Fähigkeit von STDP zu sehen, wiederkehrende Phasenabfolgen zwischen eigenem (postsynaptischen) Puls und eingehenden (präsynaptischen) Pulsen zu lernen. Zufällige Bildanteile werden ausgeblendet, während die den regelmäßigen Phasenabfolgen zugrunde liegenden deterministischen Bildanteile in den synaptischen Gewichten gespeichert werden. In [Partzsch07a] wird gezeigt, dass ein entsprechender Aufbau von Gaborfiltern aus einer regelmäßigen Abdeckung des Originalbildes mit DoG-Masken auch für eine konventionelle Bildfilterung günstig ist. Rechenoperationen, die sonst für jede (leicht verschobene oder gedrehte) Gabormaske neu ausgeführt werden müssten, lassen sich über die DoGs effizient bündeln (siehe auch Anhang B.2.1).

Die nächste Simulation besteht aus einer Vervielfältigung des obigen Versuchs, d.h. es wurden mehrere Neuronen über STDP-Synapsen jeweils an alle Ganglienzellen der Retina gekoppelt. Die Neuronen sind untereinander mit konstanten inhibitorischen Gewichten vollvernetzt verbunden, um eine Selektion zwischen den rezeptiven Feldern zu ermöglichen [Delorme03a]. Die vier Eingangsbilder auf der rechten Seite werden im selben zeitlichen Schema wie oben der Retina zugeführt.

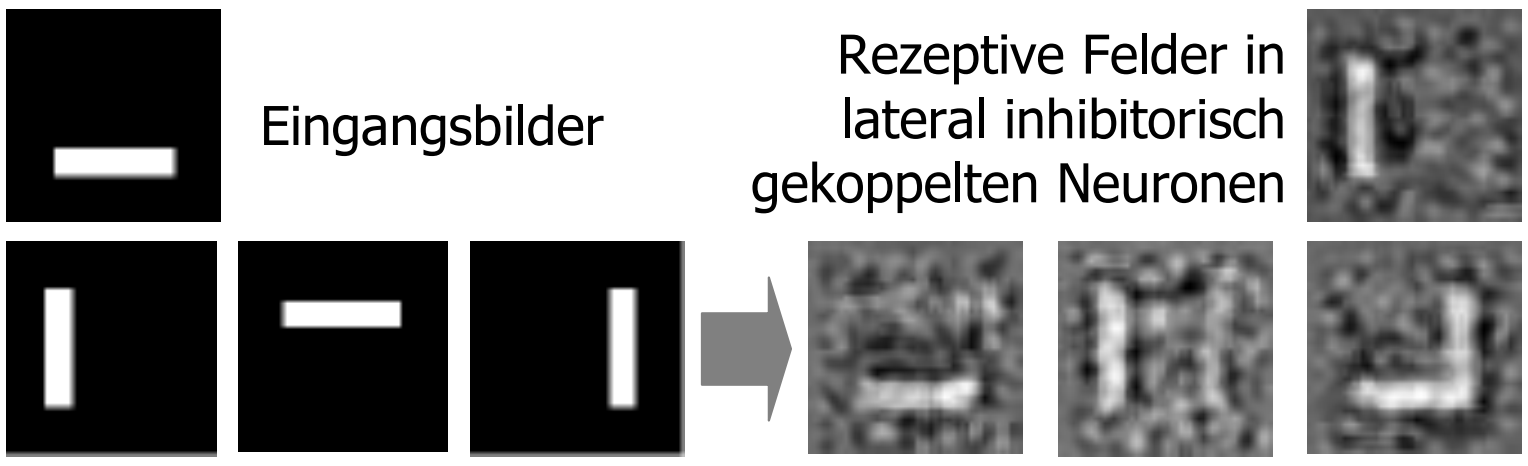

**Abbildung V.4.: Synthese von gemeinsamen Bildstrukturen und Konkurrenz zwischen Zellen**

Eine Auswahl der hierbei entstehenden rezeptiven Felder zeigt, dass sich Neuronen jeweils auf ein bestimmtes Muster konzentrieren, welches von den anderen Neuronen nicht repliziert wird. Teilweise gibt es dabei Neuronen, die sich auf ein einzelnes der Eingangsbilder spezialisieren (links unten und rechts oben), während andere Neuronen eine Synthese aus den verschiedenen Eingangsbildern erzeugen, die etwa auf den rechten und unteren Balken gleichermassen reagiert (Rezeptive Felder, rechts unten). In dem in der Mitte gezeigten rezeptiven Feld scheint dominierend der linke Balken vertreten zu sein, zusätzlich jedoch auch der obere und der rechte Balken. Entsprechende Experimente, die Lernen von einfachen Mustern über LTP und LTD in Neuronenkulturen in-vitro demonstrieren, wurden von Ruaro et al. [Ruaro05] durchgeführt. Zellen des Hippocampus (mit inhärenter Plastizität wie in Abbildung V.2 oben Mitte und links) wurden dabei unvernetzt in Petrischalen aufgebracht. Spontan entstehende Netzwerke ähnlich Abbildung II.8 wurden dann über wiederholte Stimulation mit einem Elektrodenarray auf das Erkennen von Kanten und Winkeln trainiert.

In Erweiterung der STDP-Lernregel aus Gleichung (V.1) wird in der obigen Simulation zusätzlich ein Abklingen des Gewichtes ähnlich wie in Gleichung (III.1) eingeführt. Da die in Abbildung V.4 verwendeten Eingangsmuster im Gegensatz zu natürlichen Bildern und zu den in Abbildung V.3 verwendeten einen gleichförmigen Hintergrund besitzen, meldet eine DoG-Faltung aus diesen Bildbereichen keine Antwort. Dies führt dazu, dass dort aufgrund fehlender präsynaptischer Pulse keine Adaption stattfindet, d.h. die synaptischen Gewichte ohne diesen Abklingterm in ihrer ursprünglichen Verteilung verharren würden. Das Abklingen synaptischer Gewichte um einen Faktor von ca. $\gamma=10^{-4}$ pro postsynaptischem Puls führt dazu, dass sich rezeptive Felder bei entsprechender Pulslage durch die relativ gesehen wesentlich größere Lernrate $\eta$ zwar zielgerichtet entwickeln können, jedoch periodisch aufgefrischt werden müssen, um nicht verloren zu gehen. In biologisch realistischeren Netzwerken mit spontaner präsynaptischer Aktivität (d.h. Aktivität ohne Eingangssignal am Netzwerk) kann ein ähnlicher Effekt dadurch erreicht werden, dass die Fläche unter der LTD-Kurve größer ist als unter der LTP-Kurve, dass also für unkorrelierte Aktivität Synapsen tendenziell eher abgeschwächt werden [Abbott00].

In einem mehrschichtigen Aufbau dieses Netzes lässt sich die zunehmende Komplexität der rezeptiven Felder demonstrieren. Nur die erste Schicht ist in diesem Fall mit dem Retinamodell verbunden, während nachfolgende Schichten jeweils aus allen Neuronen der darunterliegenden Schicht über STDP-Synapsen Signale erhalten und die Neuronen innerhalb jeder Schicht konstant inhibitorisch gekoppelt sind. Abbildung V.5 zeigt Beispiele für entsprechende rezeptive Felder aus einem dreischichtigen Netzaufbau[32]:

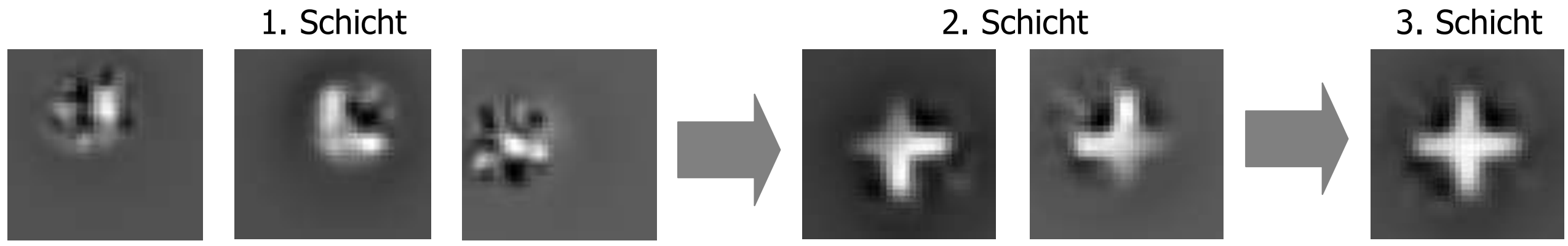

**Abbildung V.5.: Wachsende Komplexität rezeptiver Felder in mehrschichtigen STDP-Netzwerken (Eingangsbilder und –prozedere wie in Abbildung V.3)**

Weitere Eigenschaften von neuronalen Netzen, die mit STDP-behafteten Synapsen gekoppelt sind, zeigt die folgende Übersicht:

- In [Nowotny03] werden mit in-vitro Versuchen und Simulation die Synchronisationseigenschaften von STDP untersucht. Dieser Effekt kommt dadurch zustande, dass durch STDP das synaptische Gewicht dahingehend verändert wird, das postsynaptische Neuron direkt nach dem präsynaptischen feuern zu lassen.
- Die temporale Natur von STDP kann auch verwendet werden, eine präsynaptische Pulssequenz zu lernen. Mechanismen dieser Art werden für den auditiven Kortex beim Erkennen von Audiomustern postuliert [Kepecs02].
- Die oben beschriebenen rezeptiven Felder können über die Eigenschaft des Sequenzlernens auf bewegte rezeptive Felder [Kandel95] erweitert werden, d.h. ein Neuron reagiert nur auf eine Gaborcharakteristik, die mit einer bestimmten Richtung und Geschwindigkeit über das Bild bewegt wird [Senn02, Buchs02].
- STDP führt in einem zufällig verbundenen Netzwerk zur spontanen Entstehung von Neuronengruppen, in welchen die Neuronen in einer festgelegten Sequenz feuern [Izhikevich04a]. Es lassen sich dabei starke Parallelen zur Entstehung von Erinnerungsvermögen in Form von Synfire Chains entdecken [Durstewitz00].
- Eine Anwendung von STDP zur Kompensation von Streuungen in der Hardware findet sich in [Cameron05][33].
- Die Verbesserung von Signalübertragungseigenschaften durch STDP wurde in Abschnitt IV.1.3 gezeigt.
- Arbeiten zur STDP-gesteuerten Selbstkorrektur eines Netzwerks nach einer Beschädigung sind in [Hopfield04, Song01] enthalten. Durch beschädigte Synapsen wird nach dortiger Erkenntnis der Feuerzeitpunkt des postsynaptischen Neurons leicht nach hinten verschoben. STDP verstärkt dann zum Einen die bestehenden Synapsen und korrigiert damit den Feuerzeitpunkt des Neurons. Zusätzlich werden Synapsen, die ähnliche Information wie die beschädigten übertragen, durch STDP in den Vordergrund geholt und damit die Funktion des Neurons wiederhergestellt.
- Abbott et al. [Abbott00] betrachten die Eigenschaft der Selbstregulierung der Gewichtsverteilung und Populationsaktivität eines STDP-Netzwerk. Basis ist dabei die neurobiologische These, dass ‚echtes' STDP nur die Wahrscheinlichkeit des ersten Spikes verändert, welcher als Reaktion auf einen Stimulus ausgelöst wird [Legenstein05]. Damit erhält ein durch die Plastizität bevorzugter Stimulus einen entsprechend verstärkten Einfluss auf den EPSC, ohne die Gesamtnetzwerkrate signifikant zu verändern[34].

[32] In diesem Fall wurde die Rekonstruktion hierarchisch ausgeführt, d.h. die rezeptiven Felder der mit der Retina verbundenen Schicht wurden wie oben beschrieben rekonstruiert, während darüberliegende Felder iterativ aus der Multiplikation der rezeptiven Felder der vorhergehenden Schicht mit dem jeweiligen Synapsengewicht gewonnen wurden.

[33] Für eine eingehendere Diskussion der dortigen Anwendung, siehe Abschnitt V.1.4.

### V.1.3 Aktuelle STDP-Forschung

Wie in Abschnitt V.1.1 erwähnt, wurde STDP zuerst gemäß Gleichung (V.1) bzw. Abbildung V.2 eingeführt, d.h. mit einer festen Form der synaptischen Modifikation nur in Abhängigkeit des Intervalls zwischen prä- und postsynaptischem Puls. Nach der obenstehende Auflistung haben sich Ausprägungen dieser statischen Plastizitätsregel als sehr erfolgreich dabei erwiesen, verschiedenste neuronale Mechanismen simulativ nachzustellen. Jedoch wurde bereits in sehr frühen Arbeiten ersichtlich, dass eine STDP-Regel nach [Delorme01] nicht das ganze Spektrum an gemessenen Daten erklären kann. So wurde etwa in [Markram97] eine Frequenzabhängigkeit der synaptischen Plastizität festgestellt, mithin lässt sich das ‚klassische' STDP-Verhalten an den dort untersuchten Synapsen nur für ein Auftreten eines entsprechenden Pulspaares öfter als $10s^{-1}$ finden. Froemke und Dan [Froemke02] zeigen in einer Reihe von in-vitro Experimenten mit Dreifachpuls-kombinationen (post-prä-post bzw. prä-post-prä) sowie verschiedenen Vierfachkombinationen, dass die biologisch gemessene plastische EPSC-Entwicklung sich teilweise gegenteilig zur obigen STDP-Regel verhält. In derselben Arbeit wird ein neues STDP-Modell postuliert, welches hauptsächlich aus einer Überlagerung von synaptischer Kurzzeitadaption [Markram98] und der klassischen Form des STDP besteht. Dieses Modell erklärt plastische Veränderungen durch die Dreifachkombinationen deutlich besser als das bisherige, zusätzlich wurde es anhand von im V1 aufgenommenen prä- und postsynaptischen Pulsfolgen getestet, ebenfalls mit gutem Ergebnis. Die Abhandlung in [Kepecs02] zeigt weitere Belege aus neurobiologischen Messungen, für die das konventionelle STDP nicht als Erklärung ausreicht. Die Autoren gehen einen Schritt weiter als in [Froemke02], statt der Überlagerung von statischem STDP und Kurzzeitadaption wird eine Metaplastizität des STDP postuliert, d.h. die STDP-Kennkurven aus Abbildung V.2 passen sich dynamisch an verschiedene Zustandsgrößen des Neurons an. Ein entsprechendes neurobiologisches Modell wird in [Saudargiene04] vorgestellt, bei dem die Langzeitplastizität einer Synapse vor allem vom Verlauf lokaler Zustandsvariablen abhängig ist. Dazu gehören der momentane Leitwert von bestimmten Typen synaptischer Rezeptoren (siehe Abbildung II.6), sowie der Verlauf und Absolutwert des lokalen postsynaptischen Membranpotentials (das entsprechend durch den EPSC des präsynaptischen Pulses verändert wird). Saudargiene et al. untersuchen das Plastizitätsverhalten unter verschiedenen biologischen Randbedingungen und verifizieren ihr Modell mit neurobiologischen Messdaten, u.a. aus [Song01, Froemke02]. Eine gegenüber der letzten Variante etwas vereinfachte Form von Metaplastizität findet sich in [Senn02]. Dabei wird der grundsätzliche Verlauf der STDP-Kurve aus Abbildung V.1 beibehalten, jedoch in einer Abwandlung des gleitenden Schwellwertes der BCM-Plastizität [Bienenstock82] das Flächenverhältnis zwischen LTP und LTD in Abhängigkeit von der postsynaptischen Aktivität angepasst. Diese Art der Plastizität wird ebenfalls mit gutem Erfolg zur Nachsimulation der Ergebnisse aus [Markram97, Froemke02] herangezogen. Mithin gibt es noch kein anerkanntes Modell für ‚komplexes' STDP, da unterschiedliche neurobiologische Vorgänge auf Zeitskalen im 1-10ms Bereich [Saudargiene04], 10-500ms [Froemke02], als auch im Minutenbereich [Senn02] jeweils zur Erklärung der Daten aus [Froemke02, Markram97, Song01 u.a.] ausreichend erscheinen.

Ein weiteres aktuelles Forschungsgebiet ist neben der oben beschriebenen Erweiterung von STDP auf komplexere Zusammenhänge die Beeinflussung des Lernvorgangs, wie er etwa in der Natur durch Neurotransmitter wie Dopamin ausgelöst wird [Izhikevich07]. Nachdem STDP auf dem Gebiet des unüberwachten Lernens bereits große Erfolge verzeichnen konnte (siehe Abschnitt V.1.2), gibt es aus neuerer Forschung mehrere Ansätze, STDP-Netzwerke zielgerichtet auf ein

[34] Eine entsprechende Modulation der ‚initial release probability' unterstützt auch die Argumentation in [Delorme01], dass STDP sehr sensibel auf den ersten Spike reagiert und damit für schnelle Signalübertragung und –verarbeitung sorgt.

bestimmtes Verhalten hin zu steuern [Legenstein05, Izhikevich07]. Eine Übersicht hierzu sowie Vergleiche mit anderen überwachten Lernverfahren bei pulsgekoppelten neuronalen Netzen finden sich in [Kasinski06].
Legenstein et al. untersuchen in einem wegweisenden Beitrag [Legenstein05] die Konvergenz des Lernens bei einem STDP-Netzwerk im Vergleich zu dem beim Perzeptron [Zhang00] aufgestellten Theorem und stellen entsprechende Konvergenzkriterien auf. Im zweiten Teil dieser Publikation werden die Ergebnisse durch Simulationen untermauert und auf komplexere Neuronen- und Synapsentypen erweitert (LIAF-Neuronen, Synapsen weisen STDP und quantale Kurzzeitadaption auf). Das überwachte Lernen wird durch einen zusätzlichen Strompuls[35] auf das postsynaptische Neuron zum Zeitpunkt des gewünschten postsynaptischen Aktionspotentials vorgegeben:

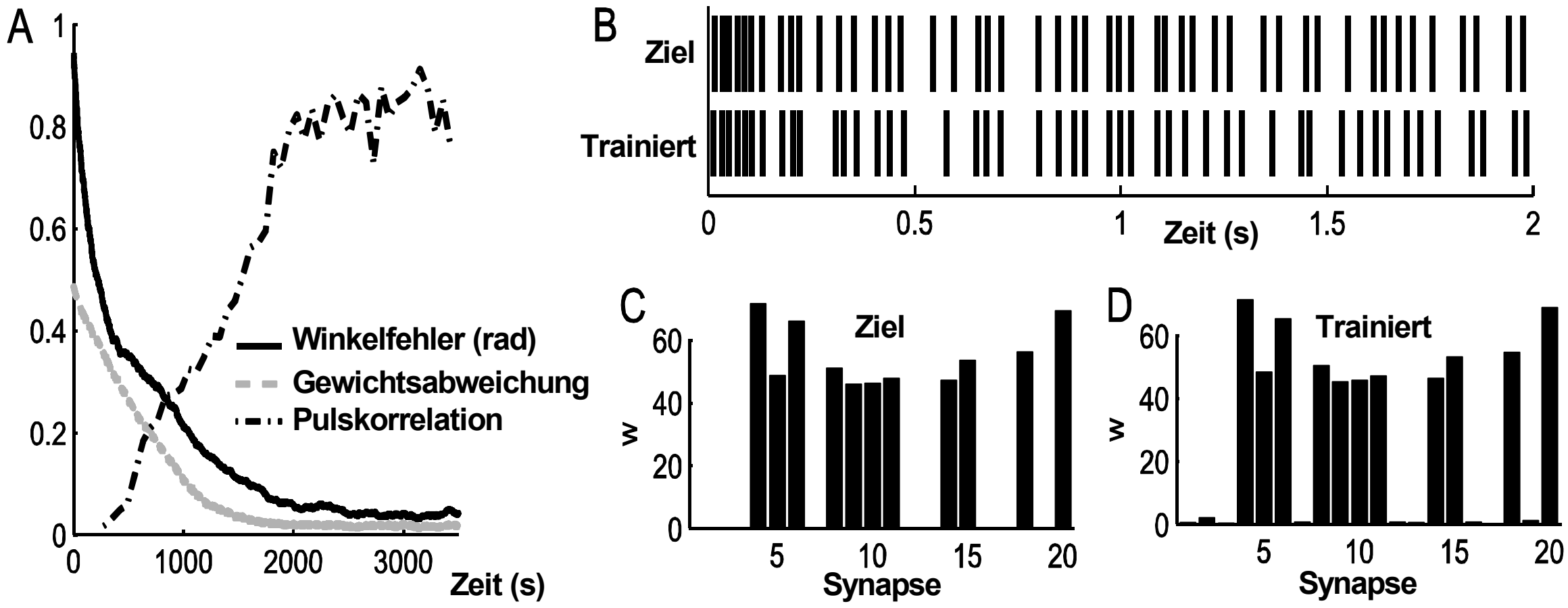

**Abbildung V.6.: Kenndaten für das gesteuerte Lernen einer Transformation zwischen prä- und postsynaptischer Pulsfolge, nach [Legenstein05]**

In dem obigen Experiment werden 100 Synapsen an einem LIAF-Neuron (10 konst. inhibitorisch, 90 konst. exzitatorisch) mit gaussverteilten Gewichten zwischen 22 und 86 mit einem zufälligen binären Vektor multipliziert, der jede Synapse entweder an- oder ausschaltet. An den Eingang der Synapsen werden 100 unkorrelierte Poisson Pulsfolgen angelegt, die über die an- und abgeschalteten Synapsen bzw. deren Gewichtswerte einen Zielvektor in Form von Ausgangspulsen des LIAF-Neurons erzeugen. Für das überwachte Lernen der Transformation zwischen prä- und postsynaptischen Pulsfolgen werden die konstanten Gewichte der exzitatorischen Synapsen mit einer STDP Adaption ausgetauscht und ihre Anfangswerte zufällig verteilt. Danach liegen wieder 3600s lang die Poisson Pulsfolgen an, während die Ziel-Pulszeiten der postsynaptischen Pulsfolge über Strompulse auf das LIAF-Neuron vorgegeben wurden. Abbildung V.6 (B) zeigt einen Vergleich zwischen trainierter und Zielpulsfolge am Ende der Trainingszeit. Die Korrelation zwischen beiden Pulsfolgen während der Trainingszeit kann dem gestrichelten schwarzen Graphen in Abbildung V.6 (A) entnommen werden. Die STDP-Adaption ‚erkennt' bei bereits relativ gut gelernten Gewichten keine deutliche Zielrichtung mehr und neigt zum Oszillieren. In Abbildung V.6 (C)&(D) sind zufällig ausgewählte korrespondierende synaptische Gewichtswerte des originalen Gewichtsvektors und des Vektors nach dem STDP Training zu sehen. Die restlichen Graphen in Abbildung V.6 (A) bewerten den trainierten Gewichtsvektor im Vergleich zum Ziel, zum Einen über den Betragsabstand zwischen beiden (graue Kurve, normiert auf den Zielvektor), sowie den Winkel zwischen beiden Vektoren (durchgezogene schwarze Kurve, in Vielfachen von $\pi$).
In [Izhikevich07] wird simulativ untersucht, wie STDP zu klassischem Konditionierungstraining beiträgt. Unüberwachtes STDP nimmt dabei an jedem Neuron während eines Versuches Adaptionen gemäß den anliegenden Pulsen vor. Dabei werden langsame synaptische Vorgänge gestartet, welche auf ein externes Belohnungssignal hin diese Gewichtsänderungen verankern.

[35] 0,2ms@1µA, sehr viel kleiner als eine Membranaufladung bis zum Schwellwert, d.h. das Neuron wird nur leicht zum Feuern hingeführt, jedoch kein Ausgangspuls durch den Strompuls erzeugt.

Mehrere Sekunden nach den entsprechenden STDP-Adaptionen während des Versuchs wird aufgrund der Belohnung Dopamin ausgeschüttet, welches alle STDP-Adaptionen verankert, die ungefähr während der Versuchszeit stattgefunden haben. Zufällige Adaptionen, die nicht mit der Erfüllung der Aufgabe während des Versuchs zu tun haben, gleichen sich bei mehreren Wiederholungen des Versuchs wieder aus, während Pulsfolgen, die mit der Erfüllung der Aufgabe korreliert sind, durch das wiederholte STDP-Lernen gelernt und danach durch das Dopamin verankert werden.

### V.1.4 STDP-Schaltkreise

Da im Rahmen von FACETS am Lehrstuhl vor allem die schaltungstechnische Realisierung entsprechender neuronaler Netzwerke mit STDP-Adaption durchgeführt werden soll, erscheint ein Überblick über bisherige Forschung auf diesem Gebiet angebracht. Trotz der relativ kurzen Zeit, in der STDP als Adaption in der wissenschaftlichen Gemeinde diskutiert wird [Kepecs02], gibt es dazu bereits einige Ansätze, die sich auf drei Forschungsgruppen verteilen, Indiveri et al. [Indiveri06, Muir05], Murray et al. [Bofill-i-Petit04, Cameron05], sowie Schemmel et al. [Schemmel04, Schemmel06]. Vereinzelt werden auch Anstrengungen auf diesem Gebiet von Forschungsgruppen unternommen, die weniger an neuronalen Gesichtspunkten interessiert sind, sich jedoch vom STDP-Einsatz die Lösung einer Problemstellung ihres spezifischen Gebietes erhoffen, etwa Koickal et al. in der Geruchsklassifikation [Koickal06].
Ein Teil der VLSI-Implementierungen von STDP-basierten neuronalen Netzen sind explorativ orientiert, d.h. zur reinen Erforschung der Netzwerkdynamiken und prinzipiellen Realisierbarkeit von verschiedenen Formen der STDP-Adaption in CMOS-Schaltungen. Murray et al. beispielsweise konstruierten einen IC mit 5 Neuronen und jeweils 6 STDP-Synapsen in einer 4-1 Feedforward-Konfiguration, welcher lernt, Korrelationen zwischen Eingangspulsfolgen zu extrahieren [Bofill-i-Petit04]. In [Indiveri06] wird ein IC vorgestellt, der 32 Neuronen mit jeweils 8 Synapsen in frei konfigurierbarer Anordnung zum allgemeinen Test von Netzwerkdynamiken enthält. Die beiden genannten Realisierungen beschäftigen sich mit relativ kleinen Netzwerken, welche in biologischer Echtzeit betrieben werden. Einen anderen Weg geht der in [Schemmel04, Schemmel06] entwickelte, auf die Neuroforschung ausgerichtete IC. Da STDP eine „langsame" Lernregel ist, bei der entsprechende Strukturveränderungen auf Zeitskalen im Stundenbereich von Interesse sind, finden dort die synaptischen Dynamiken um den Faktor $10^4$ beschleunigt statt. Dadurch kann die Experimentzeit wesentlich verkürzt bzw. mehr Experimente im selben Zeitraum durchgeführt werden. Dieser IC ist außerdem wesentlich größer als die bisher vorgestellten, mit 384 Neuronen bzw. 384*256 Synapsen. Er nähert sich somit der synaptischen Ausfächerung von Neuronen im visuellen Kortex an [Binzegger04] und ermöglicht entsprechend großflächige Experimente. Auf die in [Schemmel04, Schemmel06] vorgestellte Neuroarchitektur wird in Abschnitt V.3.1 genauer eingegangen, da diese Schaltungen den Vorläufer der neuronalen Bestandteile der FACETS-Hardware darstellen.
Neuromorphe ICs mit STDP-Adaption werden auch in konkreten Aufgabenstellungen eingesetzt. In [Cameron05] wird ein Konzept vorgestellt, mit dem matchingbasierte Zeitfehler in einem Pixelfeld zur räumlichen Tiefenmessung ausgeglichen werden sollen. Das Prinzip basiert darauf, matchingbasierte Zeitfehler in pulsenden Pixelsensoren über ihr Kopplungsgewicht zur tiefenberechnenden Stufe auszugleichen. Wenn für ein bestimmtes Eingangsmuster die entsprechende Zielabfolge von Pulsen für die Tiefenschätzung bekannt ist, kann diese über STDP auch für die tatsächliche Abfolge gelernt und damit der Fehler ausgeglichen werden. Weitere Anwendungen finden sich im Bereich der Geruchsanalyse und –klassifizierung. In [Muir05] wird der IC aus [Indiveri06] verwendet, um einen Teil des Riechkolbens (Bulbus Olfactorius) nachzubilden. Der Netzwerkaufbau wird über die Konfiguration des ICs gesteuert, während die Pulswandlung von Gerüchen sowie die Bewertung der Netzwerkaktivität softwarebasiert erfolgt[36]. Eine Weiterentwicklung dieser Anwendung zeigt [Koickal06], dort erfolgt gebündelt auf einem IC der komplette Aufbau eines Riechkolbens, angefangen bei pulswandelnden Geruchssensoren, über STDP-verarbeitende Neuroschaltungen bis hin zur olfaktorischen Klassifikation auf Basis der neuronalen Verarbeitung. Dies stellt zugleich den ersten Vorläufer dafür dar, auf STDP beruhende neuromorphe IC-Schaltungen in kommerziellen Produkten einzusetzen.

# V.2 Weitere FACETS-relevante neuronale Adaptionsregeln

## V.2.1 Kurzzeitadaptionen

Als Erweiterung der STDP-Plastizität finden im FACETS-Projekt auch diverse neuronale und synaptische Adaptionen Beachtung, die auf Zeitskalen im Millisekunden- bis Sekundenbereich agieren. Einige dieser Adaptionen und ihre Auswirkungen auf das Netzwerkverhalten wurden bereits in den Abschnitte II.1.2, IV.1.2 und IV.1.4 vorgestellt. In korrespondierender laufender Forschung [Häusler07] finden beispielsweise die quantale Kurzzeitadaption [Markram98] und Posttetanic Potentation [Koch99] Beachtung. Da in der FACETS-Hardware die Synapsen als Gewichtswerte realisiert werden, entfällt die neurobiologisch authentische Modellierung über $p$ und $q$. Wie in Abschnitt IV.1.2 argumentiert wird, kann die $p$-Adaption über PTP aber für eine Neuronenpopulation mit zumindest teilweise korrelierten Eingangssignalen auch als diskret quantisierter Gewichtswert betrachtet werden. Die Ausschüttungsmenge $q$ beeinflusst direkt den EPSC bzw. IPSC und ist folglich sofort als Gewichtswert interpretierbar.
Die obigen Effekte können als dynamisch modulierte Leitwerte mit entsprechendem Verhalten und Zeitkonstanten modelliert werden [Indiveri03]. Entsprechende gesteuerte Leitwerte sind bereits in der Vorgängerversion der Hardware Bestandteil der Synapsen [Schemmel06]. Mit ihnen wird dort eine weitere synaptische Eigenschaft emuliert, die sogenannten „High-Conductance-States“ [Destexhe03]. Dabei ändert sich das elektrische Verhalten eines Dendriten in Abhängigkeit der Aktivität an der ihm vorgeschalteten Synapse, wodurch die Dendriten aktive Signalverarbeitung an den EPSCs/IPSCs auf dem Weg zur Membran des Neurons vornehmen. Beispielsweise reagieren Neuronen in diesem Zustand schneller und sensibler auf die nahe an der Soma liegenden Synapsen [Destexhe03].
Zusätzlich zu den obigen verteilten, synapsenspezifischen Adaptionen existiert bei Neuronen auch zentral auf ihrer Membran eine Kurzzeitplastizität, die sogenannte „Spike Frequency Adaptation (SFA)“ [Partridge76]. Diese weist ein ähnlich transitives Verhalten auf wie der Teilbereich der quantalen Adaption bei hohen Frequenzen. Primär handelt es sich dabei um eine Erschöpfung der Ionenkanäle in der Membran, so dass für einen eingehenden Stimulusstrom progressiv weniger Aktionspotentiale ausgelöst werden.

## V.2.2 Bienenstock-Cooper-Munroe Adaption

Die Bienenstock-Cooper-Munroe (BCM) Regel [Bienenstock82, Bear95] wurde ursprünglich zur Modellierung verschiedener Effekte in der Entstehung rezeptiver Felder im visuellen Kortex von Katzen eingeführt. Vornehmlich sollte damit der gegenseitige Wettbewerb von Neuronen nachvollzogen werden, der entsteht, wenn Augen selektiv abgedeckt werden und damit die eingehende Information an den Neuronen starken Schwankungen unterliegt. Im Einklang mit den zur Entstehungszeit von BCM gebräuchlichen Modellen neuronaler Codes wurde die ursprüngliche Plastizitätsregel ratenbasiert eingeführt. Die Pulsraten $\lambda(t)$ repräsentieren dabei als zeitlich veränderlicher skalarer Wert das gleitende Mittel der Pulsrate in einem Zeitraum weniger

[36] Die momentane Arbeit in Bezug auf den in [Schemmel04, Schemmel06] vorgestellten IC zielt ebenfalls darauf ab, über eine entsprechende Netzwerkkonfiguration und externe Versorgung mit Stimuli diverse Beispielanwendungen auf den IC abzubilden.

Sekunden. Eingehende Pulsraten werden in diesem Modell an den Synapsen mit den synaptischen Gewichtswerten multipliziert und aufsummiert. Die Bewertung dieser Summe mittels einer sigmoiden Übertragungsfunktion gemäß Gleichung (II.18) ergibt die zugehörige postsynaptische Antwort. Die eigentliche BCM-Regel beschreibt das Verhalten der Gewichte über die Zeit in Abhängigkeit der prä- und postsynaptische Pulsrate[37]:

$$dW = \Phi\left(\lambda_{post}(t) - \theta_M\right) \cdot \lambda_{prä}(t) - \gamma \cdot W \qquad \textbf{(V.2)}$$

Eine wichtige Eigenschaft der Plastizitätsregel ist, dass sich die Richtung des Lernens (LTP/LTD) in Abhängigkeit der postsynaptischen Pulsrate $\lambda_{post}$ und eines Schwellwertes $\theta_M$ umkehrt. Die Funktion $\Phi$ stellt eine nichtlineare Bewertung dieser Differenz dar, sie muss in ihrem Wertebereich an $W$ angepasst sein. Der Schwellwert $\theta_M$ passt sich als „sliding threshold" an die mittlere postsynaptische Aktivität an und sorgt damit für eine Balance zwischen LTP und LTD bzw. für eine Regulierung der mittleren Netzwerkaktivität [Bear95, Koch99 (Abschnitt 13.5.3)]. Weitere Elemente der Lernregel sind die Abhängigkeit der Lerngeschwindigkeit von der präsynaptischen Pulsrate $\lambda_{prä}$ und ein konstantes Abklingen des Gewichtes mit der Geschwindigkeit $\gamma$ durch den zweiten Summanden.

Im Kontext von FACETS ist BCM vor allem deshalb interessant, weil es wie STDP ebenfalls Lernverhalten von Neuronen im visuellen Kortex bei der Entstehung rezeptiver Felder modelliert. Sowohl bezüglich biologischer Wirkmechanismen als auch experimenteller Nachweise ist BCM deutlich besser etabliert [Badoual06, Koch99 (Abschnitt 13.5.3)]. Aus diesem Grund gab es bereits frühzeitig Anstrengungen, STDP so zu parametrisieren, dass auch BCM-Effekte damit reproduziert werden können [Izhikevich03] und somit STDP als Obermenge von BCM darzustellen. Andere Ansätze tendierten dazu, Teile von BCM wie etwa den „sliding threshold" zu übernehmen, um durch die entstehende Metaplastizität Effekte erklären zu können, die durch konventionelles STDP nicht abgedeckt werden [Senn02] (siehe auch Abschnitt V.1.3). Im Gegensatz zu den etablierten Wirkmechanismen von BCM [Bear95] ist STDP in seiner konventionellen Variante wie in Gleichung (V.1) auch deshalb problematisch, da es immer eine Zeitmessung zwischen prä- und postsynaptischem Aktionspotential benötigt, welche hypothetisch über vom Soma entlang des Dendriten rücklaufende Aktionspotentiale vollzogen wird. Dieser Mechanismus wird jedoch von manchen Forschern in Frage gestellt [Saudargiene04]. Mithin existiert noch kein Gesamtmodell, das sowohl BCM-Effekte als auch STDP-Effekte geschlossen erklären kann und biologisch sinnvoll motiviert ist [Badoual06]. Es wurden, wie oben angeführt, verschiedene Versuche unternommen STDP im Hinblick auf BCM-Effekte abzuändern, jedoch gibt es sehr wenige Arbeiten, die den anderen Weg gehen, d.h. die BCM modifizieren und mit STDP-Protokollen testen [Badoual06]. Dies scheint vor allem darauf rückführbar zu sein, dass einzelpulsbasierte Protokolle wie bei STDP leicht zu Raten erweitert werden können [Izhikevich03]. Demgegenüber ist es nicht offensichtlich, wie eine ratenbezogene BCM-Formulierung aus Gleichung (V.2) für einen Test mit einzelnen Paaren aus prä- und postsynaptischem Puls angepasst werden kann.

Im Folgenden soll eine solche Anpassung getestet werden, welche ihre Inspiration aus Forschungsarbeiten bezieht, die versuchen, STDP direkt vom lokalen Membranpotential abhängig zu machen [Saudargiene04]. Das lokale Membranpotential kann in einer BCM-Formel als Ersatz für die mittlere postsynaptische Aktivität verwendet werden, wodurch eine ratenbezogene Zustandsvariable eliminiert wird. Die entstehende BCM-Adaption kann ähnlich wie in Gleichung (III.1) bzw. [Schreiter04] formuliert werden ($U_{Mem}$ entspricht dem dortigem Akkuzustand $a_k$), mit gegenüber [Schreiter04] gemäß der Langzeitplastizität angepassten Parametern:

$$dW = \eta * W * \left(U_{Mem} - \theta_M\right) * \chi\left(r_{prä}(t)\right) \qquad \textbf{(V.3)}$$

[37] Spätere Modifikationen von BCM haben teilweise versucht, neuere Erkenntnisse zur Codierung neuronaler Information zu integrieren, etwa indem die Pulsraten $\lambda(t)$ als instantane Raten interpretiert werden, d.h. als tiefpassgefilterte Puls<u>folgen</u> [Koch99 (Abschnitt 13.5.3)].

In der obigen Formel wurde in Abwandlung der klassischen BCM-Regel die Multiplikation mit der präsynaptischen Aktivität $\lambda_{prä}$ aus Gleichung (V.2) durch eine Aktivierung $\chi(r_{prä}(t))$ der Lernregel zu jedem Puls der präsynaptischen Pulsfolge $r_{prä}(t)$ ersetzt, ähnlich wie in Gleichung (III.1). Damit wird auch die zweite ratenbasierte Zustandsvariable aus der originalen BCM-Formulierung ersetzt. Diese ereignisbasierte Anwendung der Plastizitätsregel wird im Folgenden wie bei konventionellem STDP aus Gleichung (V.1) implizit vorausgesetzt, d.h. die Notation „$\cdot\chi(r_{prä}(t))$" wird weggelassen. Der Abklingterm wird ebenfalls vernachlässigt. Des Weiteren wird die BCM-Regel für diese Analysen als statisch angesehen, d.h. es existiert keine zeitliche Anpassung des Schwellwertes[38].

Der Schwellwert der BCM-Regel wird gleich $\theta_M$=-65mV gewählt. Dies kann zum Einen aus einer konsequenten Anwendung der Prinzipien in [Bienenstock82] begründet werden. Dort wird $\theta_M$ als mittlere postsynaptische Aktivität über einen längeren Zeitraum angesehen. Wenn $U_{Mem}$ als Zustandsvariable der postsynaptischen Aktivität interpretiert wird, ergibt sich $\theta_M$ als mittlere Membranspannung bzw. als Ruhepotential (für den zugehörigen Zahlenwert siehe Abbildung V.7 C, mittleres Membranpotential am linken bzw. rechten Rand). Zum Anderen kann die Wahl von $\theta_M$ auch neurobiologisch motiviert werden. Der BCM-Schwellwert wird vermutlich durch die Aktivität spannungsgesteuerte $Ca^{2+}$-Kanäle bereitgestellt [Bear95], diese haben bei ca. –65mV einen „Pol" [Koch99 (Fig. 9.3)]. Existierende Calcium-Aktivität wird durch Spannungen unterhalb dieser Schwellspannung bis ca. –90mV linear gedämpft, bei inaktiven Ca2+-Kanälen werden diese oberhalb –65mV bis ca. –40mV linear aktiviert. Dieser lineare Zusammenhang zwischen Calcium-Aktivität und Membranspannung wird in Gleichung (V.3) auch widergespiegelt durch die Wahl einer linearen Funktion für die Wertung $\Phi$ des Einflusses von $U_{Mem}$ auf die synaptische Modifikation. Diese Gleichsetzung von $Ca^{2+}$-Fluss und synaptischer Plastizität ist dadurch gerechtfertigt, dass Calcium-Aktivität einen direkten Einfluss auf die Plastizität ausübt[39] [Koch99 (Abschnitt 13.3), Kandel95 (Kapitel 15)].

Um die modifizierte BCM-Regel in einer STDP-Notation aufschreiben zu können, muss die Plastizitätsänderung in Abhängigkeit des Zeitabstands zum postsynaptischen Puls bekannt sein, statt als Funktion der Membranspannung. Mithin muss ein Zusammenhang zwischen $U_{Mem}$ und der Zeitdauer jeweils zu dem davor und danach liegenden AP hergestellt werden. Dazu werden die gesamten Messdaten, aus denen beispielsweise die Pulsfolge aus Abbildung II.5 entnommen wurde, in diskreten Intervallen abgetastet und als Membranspannung relativ zu den Zeiten der Aktionspotentiale notiert (jeweils nur vor- bzw. rückbezüglich zum naheliegendsten AP):

[38] Die im Weiteren zur Analyse verwendeten biologischen Messdaten aus [Piwkowska07] beziehen sich auf Intervalle im 10s-100s Bereich, während der „sliding threshold" $\theta_M$ Anpassungszeiten von mehreren Tagen hat [Bear95], so dass er in guter Näherung relativ zu den Messdaten als konstant angenommen werden kann.

[39] Bei STDP werden auch Anstrengungen unternommen, dieses direkt auf Calcium basierend aufzubauen [Badoual06].

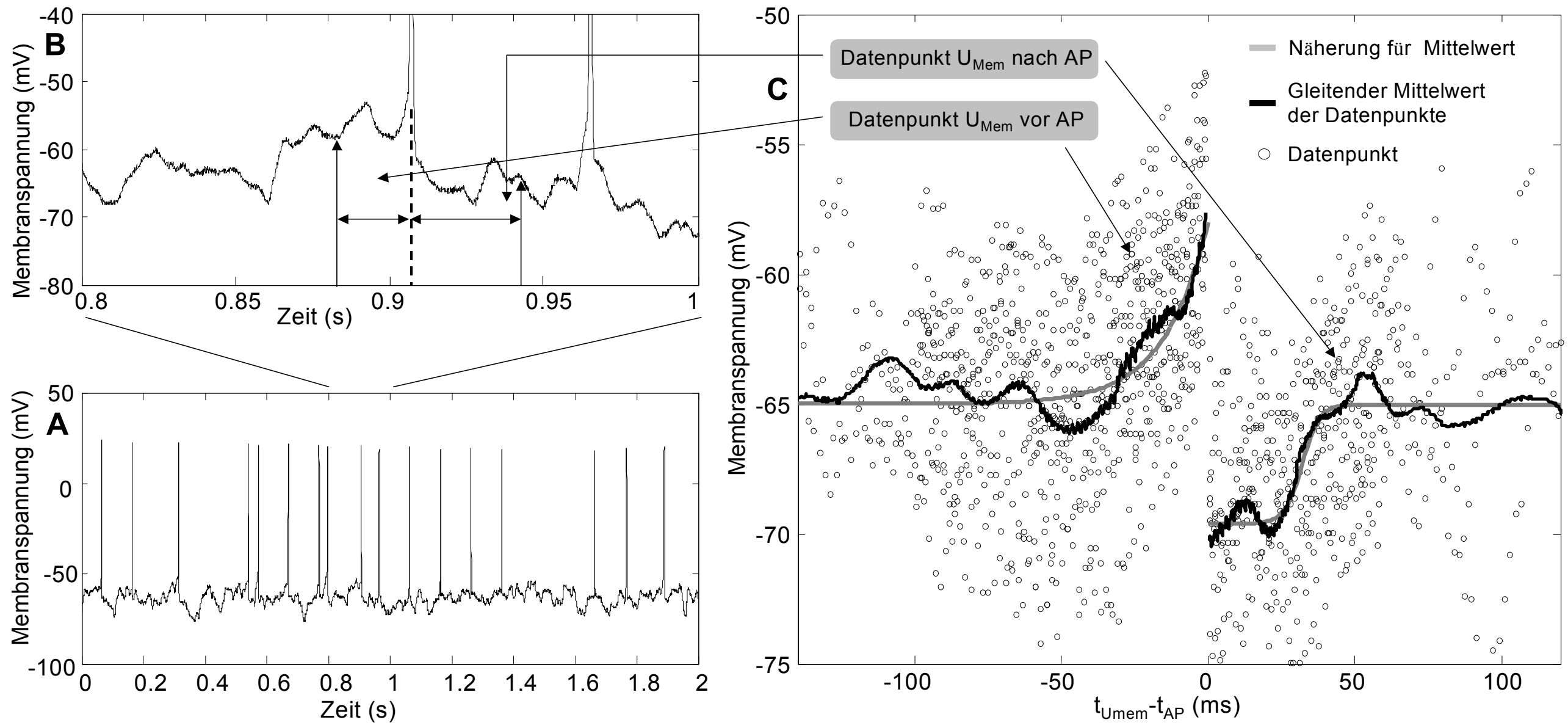

**Abbildung V.7.: (A) Beispiel für die zur Modellierung verwendeten Messungen der Membranspannung, (B) Ausschnitt der Messung mit eingezeichneten Datenpunkten und (C) Membranspannung als Funktion der Zeit vor/nach einem AP, Datenpunkte, gleitender Mittelwert und Näherung**

In weiter zeitlicher Entfernung von Aktionspotentialen nimmt $U_{Mem}$ im Mittel einen Wert von ca. –65mV an (schwarze Kurve in Abbildung V.7 (C)). Auf ein Aktionspotential zu, d.h. $-50ms < t_{Umem}-t_{AP} < 0$, nimmt die Membranspannung exponentiell zu, bis zu einem Wert, bei dem die Dynamiken der Ionenkanäle zum Tragen kommen und ein AP entsteht (siehe Abschnitt II.1.1). Nach einem Aktionspotential lässt sich ein Verharren bei einer mittleren Spannung von ca. –70mV für eine Zeitdauer von 25ms beobachten, bedingt durch das Refraktärverhalten, mit anschließender Rückkehr zum Ruhepotential. Die Refraktärzeit der vorliegenden kortikalen Regular Spiking Neuronen aus [Piwkowska07] scheint (unter den dort verwendeten Rahmenbedingungen) mit ca. 50ms laut Abbildung II.5 (links) sehr viel länger zu sein als die ca. 3-5ms von Neuronen des Hippocampus [Kandel95 (Abb. 14.11)]. In Gestalt der grauen Kurven in Abbildung V.7 (C) wurden empirisch Funktionen erstellt, die den mittleren Verlauf des Membranpotentials vor und nach einem AP annähern. Für den Verlauf von $U_{Mem}$ auf ein Aktionspotential zu ergibt sich:

$$U_{Mem} = 7mV * e^{\frac{\Delta t}{15ms}} - 65mV \qquad \textbf{(V.4)}$$

Eine ähnliche Näherung kann für die Zeit nach einem AP durchgeführt werden, wobei das Refraktärverhalten von $U_{Mem}$ durch eine Tangens Hyperbolicus Funktion berücksichtigt wird:

$$U_{Mem} = 2{,}3mV * \tanh\left(\frac{\Delta t - 32ms}{6ms}\right) - 67{,}3mV \qquad \textbf{(V.5)}$$

Der in Abbildung V.7 (C) zu Tage tretende Zusammenhang zwischen Membranspannung und $\Delta t$ kann nun zur Transformation der BCM-Plastizität in eine STDP-Notation verwendet werden. Dabei sei nochmals erwähnt, dass die obigen Überlegungen nur der Herstellung einer Korrespondenz zwischen STDP und BCM dienen, jedoch der Wirkmechanismus von BCM gemäß Gleichung (V.3) wie angegeben spannungs- und nicht zeitpunktsabhängig postuliert wird. Für die folgende Plastizitätsberechnung wird hypothetisch angenommen, dass zu einem bestimmten Zeitpunkt relativ zum postsynaptischen Aktionspotential ein präsynaptischer Puls eintrifft. Um diesen Zeitpunkt werden dann die naheliegendsten 30 Datenpunkte aus Abbildung V.7 (C) entnommen und anhand dieser Membranpotentiale nach Gleichung (V.3) jeweils die einzelne prozentuale synaptische Gewichtsmodifikation berechnet. Alle 30 Modifikationen werden multiplikativ verknüpft und ergeben die gesamte EPSC-/Gewichtsänderung. Die bei diesem Prozess entstehende, gegenüber Abbildung V.7 (C) um den Faktor 30 reduzierte Zahl an Datenpunkten ist in Abbildung V.8 (A) wiedergegeben:

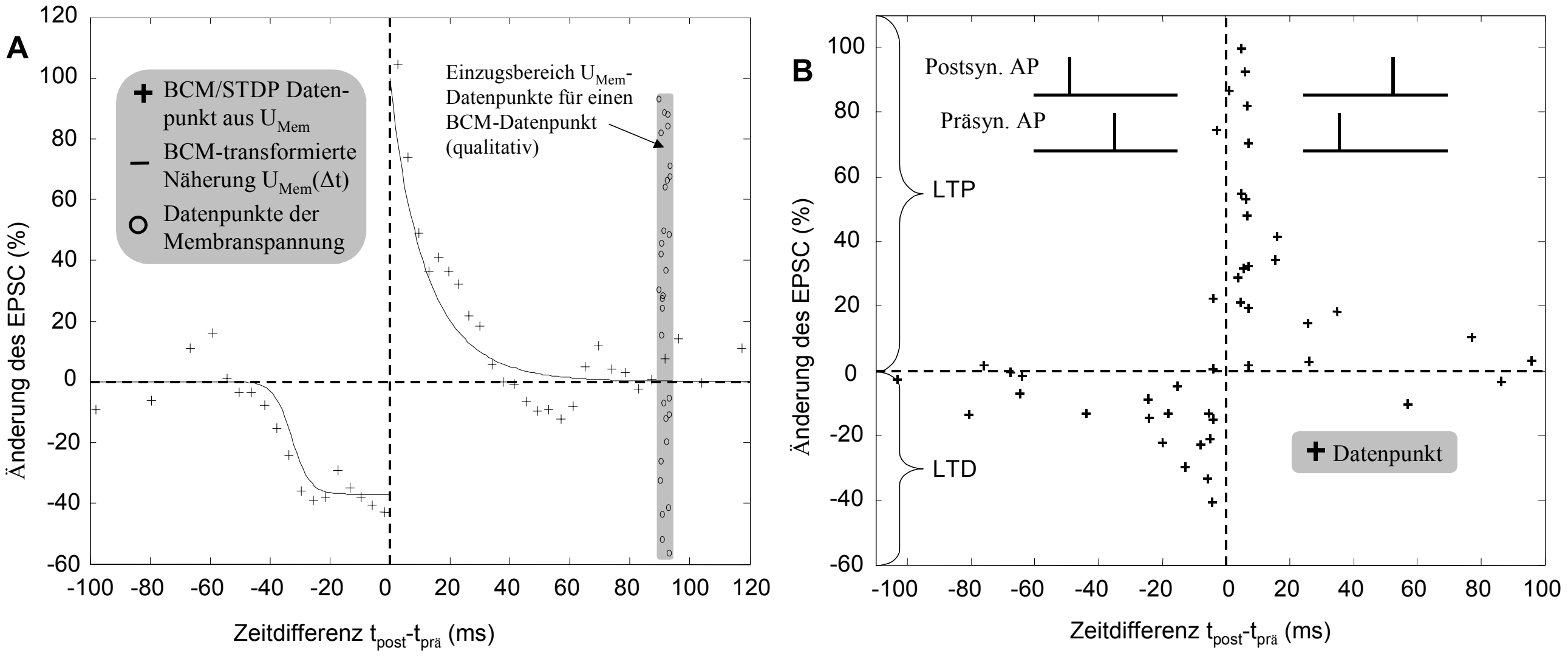

**Abbildung V.8.: (A) BCM-Plastizität nach Gleichung (V.3), angewendet auf die Datenpunkte und Näherungskurven aus Abbildung V.7, in STDP-Notation überführt; (B) zum Vergleich konventionelle STDP-Daten aus Abbildung V.1 bzw. [Bi98]**

Das beschriebene Protokoll wurde gewählt, um möglichst genau die Versuchsbedingungen in Bi et al. [Bi98] zu replizieren, bei denen die EPSC-Veränderung nach 60 Pulspaaren ausgewertet wurde. Da der Einzugsbereich für 60 Datenpunkte aus Abbildung V.7 bereits eine Bandbreite von ca. ±6ms ausmachen würde und damit evtl. die zeitliche Präzision der Transformation beeinträchtigt, wurden die berücksichtigten Datenpunkte um den Faktor 2 reduziert. Damit ist sichergestellt, dass für ein gewähltes $\Delta t=t_{post}-t_{prä}$ nur zeitlich eng benachbarte Membranspannungen berücksichtigt werden. Die Normierungskonstante bzw. Lernrate η wurde empirisch zu $^1/_{(300mV)}$ gewählt, um aus den 30 Datenpunkten eine mit den 60 Pulspaar-Wiederholungen in Abbildung V.8 (B) vergleichbare Größenordnungen an EPSC-Änderung zu erhalten.
Sowohl bei LTP als auch bei LTD zeigen die Datenpunkte eine sehr gute Übereinstimmung mit den klassischen STDP-Experimenten. Auf der LTD-Seite von Abbildung V.8 (A) kann ein leichter Einfluss der Refraktärzeit ausgemacht werden, jedoch liegen alle Datenpunkte des STDP aus [Bi98] in Abbildung V.8 (B) auch ohne Berücksichtigung von unterschiedlichen Refraktärzeiten innerhalb der statistischen Bandbreite der oben angewendeten BCM-Plastizitätsregel. Die in der rechten Hälfte von Abbildung V.7 sichtbare Streuung des Membranpotentials ließe beispielsweise auch bei kleinen $\Delta t$ eine Verteilung an Membranpotentialen zu, bei denen beinahe keine synaptische Gewichtsänderung auftritt, vergleichbar mit einigen der Datenpunkte nahe des Ursprungs in Abbildung V.8 (B). Ein weiterer Beleg für ein qualitatives STDP-Verhalten der modifizierten BCM-Regel kann über die Näherungen für den Membranspannungsverlauf aus den Gleichungen (V.4) und (V.5) erfolgen. Wenn der Zusammenhang zwischen $U_{Mem}$ und Zeit vor einem Aktionspotential aus Gleichung (V.4) über Gleichung (V.3) in eine synaptische Gewichtsänderung überführt wird, ergibt sich:

$$dW = \eta * W * e^{-\frac{\Delta t}{15ms}} \qquad \textbf{(V.6)}$$

Dieser LTP-Anteil der Plastizität ist bis auf eine leicht unterschiedliche Zeitkonstante identisch mit Gleichung (V.1). In der jetzigen spannungsbasierten Formel ist allerdings noch keine Gewichtsbegrenzung nach oben eingebaut, das Gewicht kann jedoch durch die Multiplikation mit W zumindest nicht negativ werden. Für eine geschlossene BCM-Formel, die nicht wie Gleichung (V.1) separat für LTP und LTD notiert wird, müsste eine Gewichtssättigung nach beiden Seiten beispielsweise über eine tanh-Funktion eingeführt werden. Ein „Erschöpfungszustand“ in der synaptischen Ausschüttung, d.h. ein maximales synaptisches Gewicht, ist sicher biologisch sinnvoll, kann an dieser Stelle jedoch noch nicht ausreichend in einem BCM-Kontext begründet werden und wird deshalb vernachlässigt. Eine Transformation wie oben kann auch für $U_{Mem}$ nach einem Aktionspotential aus Gleichung (V.5) durchgeführt werden, wodurch sich die LTD-Seite der Adaption ergibt:

$$dW = 2{,}3mV * \eta * \left[ \tanh\left( \frac{-\Delta t - 32ms}{6ms} \right) - 1 \right] \qquad \textbf{(V.7)}$$

Diese weist ebenfalls große Ähnlichkeit mit der graphischen Darstellung der LTD-Hälfte von Gleichung (V.1) auf, jedoch wird im Gegensatz zu der dortigen Darstellung durch die explizit berücksichtigte Refraktärzeit ein Sättigungsplateau erreicht. Es existiert mithin für kurz nach einem postsynaptischen Puls eintreffende Aktionspotentiale keine exponentielle Gewichts-veränderung mehr. Bei einem Blick auf die Originaldaten aus [Bi98] in Abbildung V.8 (B) ist zumindest fraglich, ob der zweifellos für die LTP-Hälfte vorhandene exponentielle Zusammenhang von Zeitdifferenz und Gewichtsveränderung einfach für die LTD-Seite übernommen werden kann, d.h. ob Gleichung (V.7) oder die untere Zeile von Gleichung (V.1) die bessere Näherung für eine $\Delta t$-bezogene Plastizität darstellt. Graphisch dargestellt sind Gleichung (V.6) und (V.7) in Form der durchgezogenen Kurven in Abbildung V.8 (A), wobei wie bei der oben beschriebenen Transformation der Datenpunkte die Kurven entsprechend für 30 Wiederholungen gewichtet wurden.

Es wurde in den o.a. Passagen gezeigt, dass die $U_{Mem}$-bezogene BCM-Plastizität aus Gleichung (V.3) STDP-Effekte nachvollziehen kann, d.h. für ein Protokoll aus einzelnen Aktionspotentialen sinnvolle Ergebnisse liefert. Jedoch müssen natürlich auch mit der modifizierten BCM-Formel die allgemein bei Langzeitplastizität experimentell nachgewiesenen pulsratenabhängigen Effekte [Bear95, Markram97, Senn02] reproduziert werden. Insbesondere sollte ein auf die postsynaptische Rate bezogener Schwellwert vorhanden sein, der LTD bei geringen Pulsraten von LTP bei hohen Pulsraten trennt [Koch99 (Fig. 13.7), Bear95]. Ein möglicher Mechanismus für diese frequenzabhängig unterschiedliche Plastizität wird bei einer membranbasierten Formel durch die Leckströme bereitgestellt. Bei Verwendung eines IAF-Modells ohne Leckstrom für das Neuron ist das mittlere Membranpotential unabhängig von der Rate, nur der zeitliche Anstieg ändert sich. Wenn ein Leckstrom eingeführt wird, tendiert das Membranpotential vor allem bei niedriger Aktivität dazu, überproportional viel Zeit im unteren Spannungsbereich zu verbringen [Koch99 (Abschnitt 14.3)] (siehe auch Abbildung II.12 und zugehörige Textpassagen), wodurch LTD dort dominiert. Weitere Forschung ist nötig um festzustellen, ob sich aus dem Spannungsschwellwert in Gleichung (V.3) über die realistischeren Leckströme eines HH-Modells ein Ratenschwellwert gemäß Gleichung (V.2) ableiten lässt. Außerdem müssen in Erweiterung des oben untersuchten pulspaarbasierten STDP weitere biologische Experimente, wie etwa die Triple-Puls-Protokolle [Froemke02, Senn02] mit der modifizierten BCM-Regel nachgestellt werden, um ihre allgemeine Aussagekraft zu testen.

Im Hinblick auf FACETS ist eine Einarbeitung in BCM zum Einen dahingehend relevant, dass evtl. auch Aufgaben aus diesem Bereich auf die Hardware zukommen. Zumindest die Parameter-umschreibung von BCM nach STDP laut [Izhikevich03] wird in die Benchmark-Algorithmen integriert sein. Zusätzlich soll in der Software ein Rahmen dafür geschaffen werden, das generelle Versuchsprozedere von BCM-Simulationen zu integrieren, etwa indem neben puls- auch ratenbasierte Stimuli und Auswertungen realisiert werden.

Ein weitergehender Vorschlag, welcher auf der in diesem Abschnitt durchgeführten Analyse fußt, ist, möglicherweise statt STDP für die FACETS Hardware die hier hergeleitete modifizierte BCM Regel zu verwenden. In Form der Plastizitätsregel aus Gleichung (III.1) existieren bereits VLSI-Schaltungen [Schreiter04], die etwas einfacher aufgebaut sind als vergleichbare STDP-Schaltungen [Bofill-i-Petit04, Indiveri06]. Insbesondere können zwei Zustandsvariablen bzw. deren Kapazitäten eingespart werden, die im bisherigen STDP-Modell zur Speicherung von jeweils $t_{post}$-$t_{prä}$ bzw. $t_{prä}$-$t_{post}$ verwendet werden (siehe auch Abschnitt V.3.1). Die einzige Zustandsvariable, die BCM in der vorliegenden Realisierung benötigt, ist das Membranpotential, welches ohnehin als Teil des Neuronenmodells vorliegt. Dieses kann in effizienten analogen Rechenschaltungen dann mit einem Schwellwert belegt und mit $\Phi$ bewertet werden, um Lernvorgänge nach der BCM-Regel auszuführen und, wie oben gezeigt, auch STDP-Lernen zu reproduzieren.

# V.3 Systemaufbau

Um die zu Anfang genannten Größenordnungen an Neuronen und Synapsen physisch zu realisieren und die zugehörigen hohen Kommunikationsbandbreiten bei beschleunigtem Netzwerkbetrieb bereitzustellen, wurde bereits in der Anfangsphase des Projektes ein Aufbau als Waferscale-System ins Auge gefasst [Meier04]. Hierbei werden die einzelnen Dies eines Wafers nicht voneinander getrennt und separat gebondet, sondern direkt auf dem Wafer über zusätzliche Metalllagen nachträglich miteinander verbunden. Im Vergleich zu neuronaler VLSI, die aus einzelnen ICs über ereignisbasierte Datenpakete (vgl. Abschnitt III.3.2) miteinander gekoppelt werden [Lin06], lassen sich über das Postprocessing wesentlich höhere Leitungsdichten und kürzere Verbindungslängen erreichen, was die Datenrate und damit die mögliche Netzwerkpulsrate deutlich erhöht. Abbildung V.9 zeigt den prinzipiellen Aufbau des Waferscale-Systems (Namenskonvention: Stage 2 Hardware):

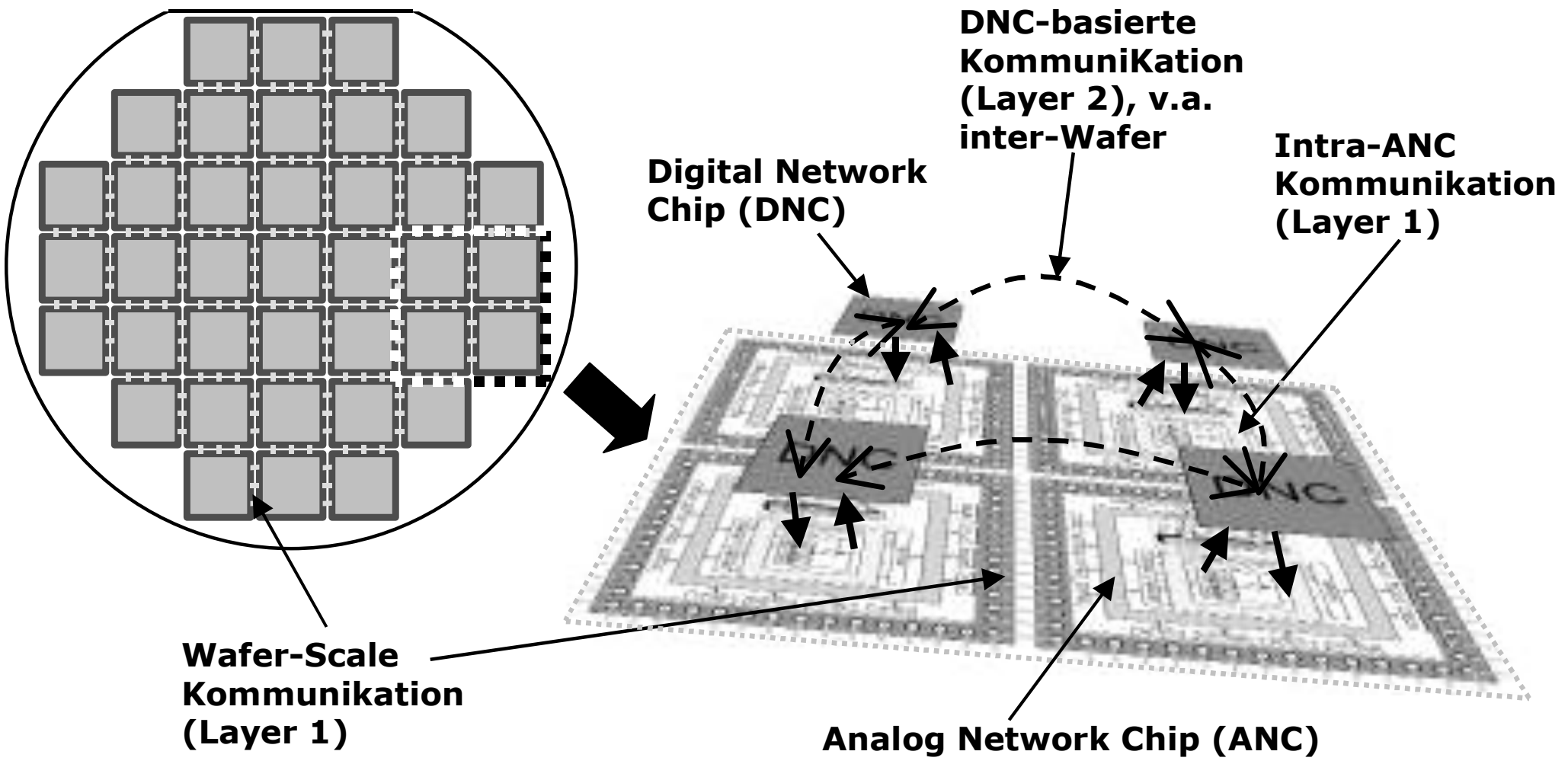

**Abbildung V.9: Prinzipübersicht des Systemaufbaus der FACETS Stage 2 Hardware**

Auf dem Wafer sind dabei die sogenannten Analog Network Chips (ANC) angeordnet, welche Arrays aus Neuronen und Synapsen enthalten, sowie Schnittstellen zu den benachbarten ANCs und zur externen Kommunikation. Die Kommunikation innerhalb eines ANCs sowie zwischen ANCs auf dem selben Wafer erfolgt über statisch geschaltete Busse mittels eines ‚Layer1' genannten asynchronen Adressprotokolls. Wenn dessen Ressourcen aufgebraucht sind oder Pulse zu anderen Wafern gesendet werden sollen, werden sie nach Adresse und Zeitpunkt codiert und zu den außerhalb des Wafers liegenden Digital Network Chips (DNC) übertragen. Der DNC nimmt diese Pulspakete entgegen, erweitert sie um eine Zieladresse entsprechend der zu realisierenden Netztopologie und versendet sie. Am Ziel werden die Pakete dekodiert, in Pulse zurückgewandelt und in ihre Zielsynapse (an einem der Neuronen des Ziel-ANC) eingespeist. Diese paketbasierte Kommunikation wird als ‚Layer2' bezeichnet.
Um einen möglichst fehlerfreien Entwurf des Stage 2 Systems sicherzustellen, werden im Vorfeld weite Teile bereits als einzelne Prototypen gefertigt, so etwa ein von A. Srowig (Heidelberg) entworfenes Verfahren zur analogen Parameterspeicherung für die Neuronen und Synapsen über Floating Gates [Ehrlich07]. Tests der elektrischen Qualität des Postprocessing werden über Wafer mit reinen Metallstrukturen in verschiedenen Layoutvarianten durchgeführt. Verkleinerte Versionen des DNC und ANC werden in Multi-Project-Wafer(MPW)-Runs hergestellt und getestet.

Hierbei müssen Skalierungseffekte beachtet werden, um eine problemlose Migration zur vergrößerten Variante im Stage 2 System sicherzustellen.
Es sei an dieser Stelle nochmals darauf verwiesen, dass der im Folgenden geschilderte Entwurfsstand nur eine Momentaufnahme mit Stand ca. Juni 2007 darstellt und damit alle Übersichten und zugehörigen Parameter temporären Charakter haben.

### V.3.1 High Input Count Analog Neural Network - HICANN

Der HICANN stellt den ersten, verkleinerten Prototyp eines ANC dar, entworfen von J. Schemmel (Heidelberg) [Schemmel07]. Wie oben erwähnt, enthält er primär den neuromorphen Teil des Stage 2 Systems, d.h. die Neuronen, Synapsen und die zugehörige Plastizitätssteuerung. Zusätzliche Komponenten sind die Layer1-Verdrahtung mit ihren Konfigurationsspeichern, sowie Interfaceschaltungen zur externen Kommunikation (Layer2, Schreiben der Parameter, etc.):

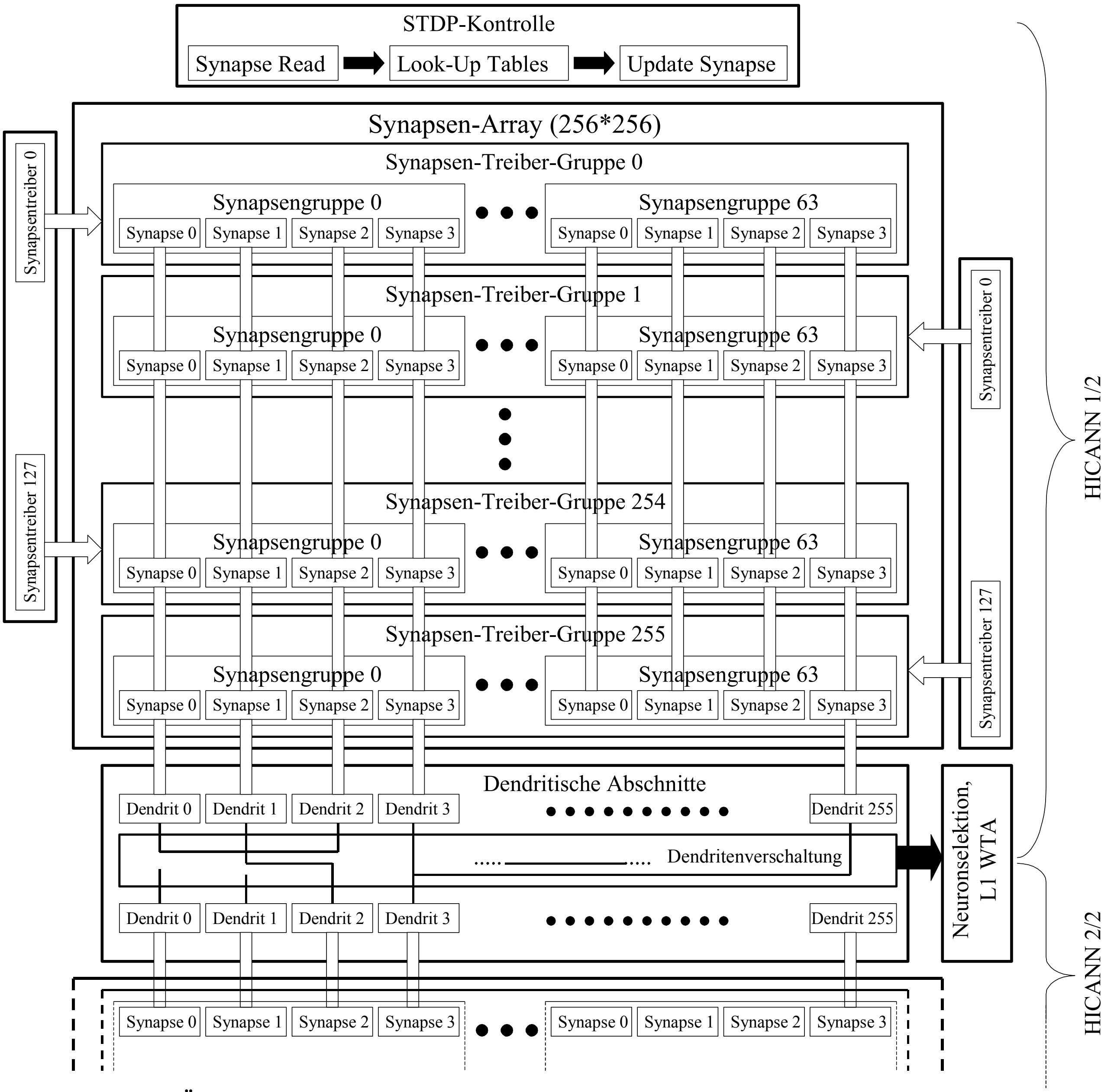

**Abbildung V.10.: Übersicht der Baublöcke des HICANN, Neuronen, Synapsen und Schnittstellen [Schemmel07]**

Jeder der Synapsentreiber an beiden Rändern entnimmt aus dem vorbeilaufenden Layer1-Bus eine Layer1 Sammelleitung mit 64 möglichen Pulsadressen bzw. –quellen. Da über Layer1 digitale Signale übertragen werden, müssen die einzelnen präsynaptischen Pulse in den Synapsentreibern wieder in analoge Spannungsverläufe umgewandelt werden [Schemmel04]. Je nach konfigurierter Zielsynapse werden dann die Pulse an eine der Synapsengruppen der zum Synapsentreiber gehörigen Synapsentreibergruppe weitergegeben. Da die Leitungsanzahl über der Synapsenmatrix limitiert ist, erhalten immer 4 Synapsen denselben präsynaptischen Puls, resultierend in 64 Synapsengruppen bzw. 256 Synapsen pro Zeile bzw. pro Synapsentreiber. Dies stellt nur eine kleine Einschränkung dar, da die Synapsen einer Synapsengruppe an verschiedene Neuronen geschaltet werden können. Von den Synapsentreibern werden jeweils 128 Layer1-Sammelleitungen des links und rechts der Matrix verlaufenden Layer1-Busses abgegriffen. Die Synapsen einer Spalte können damit konfigurierbar Pulsereignisse aus 256 verschiedenen Layer1-Sammelleitungen abgreifen bzw. aus 256*64 individuellen Pulsquellen. Die 256 Synapsen einer Spalte sind an einen gemeinsamen sogenannten dendritischen Abschnitt angeschlossen, welcher ein eigenständiges Neuron mit Leitwerten und Pulsgenerator darstellt. Das den dendritischen Abschnitten zugrunde liegende Neuronenmodell wird zur Zeit ähnlich entworfen wie in [Schemmel04], d.h. ein HH-Modell mit reduzierter Anzahl Zustandsvariablen gemäß [Destexhe97]. Andere mögliche Neuronenmodelle, vor allem im Hinblick auf möglichst gute Wiedergabe der Verhältnisse im V1, finden sich in [Izhikevich04b].

Die dendritischen Abschnitte werden im untersten Baublock konfigurierbar verbunden, um entsprechende Zusammenschaltungen an Synapsen zu ermöglichen. Als Extremfälle können damit aus einem HICANN (beide Hälften) 512 Neuronen mit jeweils 256 eingehenden Synapsen oder 8 Neuronen mit jeweils 16384 Synapsen konfiguriert werden. Diese Bandbreite an synaptischer Ausfächerung (mit über die dendritische Verschaltung realisierbaren Zwischenlösungen) deckt die im V1 Bereich gefundenen neurobiologischen Daten ab [Binzegger04]. Es ist geplant, zwei Möglichkeiten für diese Verschaltung zu realisieren: Punktneuronen können gebildet werden, in dem alle Pulsgeneratoren bis auf einen deaktiviert werden, so dass von den restlichen dendritischen Abschnitten nur die Kapazität und die aktiven und passiven Leitwerte zum Verhalten beitragen. Alternativ kann, wie im biologischen Vorbild, jeder dendritische Abschnitt bei Erreichen seines Schwellwertes einen Puls auslösen, der sich dann zu den weiteren dendritischen Abschnitten gemäß der Verschaltung fortpflanzt. In der untersten Ebene wird dann über Multiplexer einer der dendritischen Abschnitte als ‚Soma' ausgewählt. Dessen Puls wird zum Einen an die Synapsen zurückgesendet und trägt dort zur Messung des Prä-Post/Post-Prä-Intervalls der STDP Adaption bei. Zum Anderen wird der Puls mittels WTA zur Kommunikation über L1 vorbereitet und/oder dessen Zeitpunkt digitalisiert zum Versenden über L2 bzw. zur externen Analyse. Der HICANN ist symmetrisch ausgeführt, d.h. die eben beschriebene Synapsenmatrix und ihre Beschaltung ist unterhalb der dendritischen Abschnitte nochmals angeordnet (In Abbildung V.10 angedeutet durch HICANN1/2 bzw. HICANN 2/2).

STDP wird in den Synapsen über zwei durch Kondensatoren realisierte Zustandsvariablen durchgeführt, die jeweils entweder durch prä- oder durch postsynaptische Pulse geladen werden und diese Ladung über einen konstanten Leitwert wieder verlieren. Zum Zeitpunkt des jeweiligen komplementären Pulses werden die Kondensatoren ausgelesen und das in der Zustandsvariable repräsentierte Zeitintervall als Index in einen programmierbaren Look-Up-Table (LUT) verwendet, welcher die zugehörige Gewichtsänderung enthält. Das neue Gewicht wird dann wieder in einen digitalen Gewichtsspeicher in der Synapse abgelegt. Dies ermöglicht die Implementierung einer festen, programmierbaren STDP-Kurve [Schemmel04, Schemmel06] mit Charakteristiken aus Abbildung V.2, d.h. stetige und unstetige Plastizität, beliebige Wechsel von LTD nach LTP und umgekehrt, symmetrische und antisymmetrische STDP-Kurven. Da jeweils die 1024 Synapsen der zu einer Synapsengruppe gehörigen Spalten einen eigenen LUT besitzt, ist regional unterschiedliches STDP [Kepecs02] machbar. Über entsprechende Parametrisierung der LUTs kann außerdem ein BCM-ähnliches Verhalten erzeugt werden [Izhikevich03].

Durch die Konfigurierbarkeit der LUTs ist eine langsame Metaplastizität umsetzbar, die sich an extern messbaren Daten orientiert (d.h. vor allem prä- und postsynaptischen Pulsen)[40]. Somit kann beispielsweise die sliding-threshold-STDP-Variante aus [Senn02] realisiert werden. Plastizität, die sich in Abhängigkeit von lokalen Zustandsvariablen ändert [Saudargiene04] kann mit dem im Moment projektierten System nicht implementiert werden. In der aktuellen Entwurfsphase des HICANN wird zusätzlich das Einbeziehen diverser Arten von Kurzzeitplastizität (siehe Abschnitt V.2.1) diskutiert. Je nach dem hierbei erreichten endgültigen Stand ist evtl. das erweiterte STDP-Überlagerungsmodell aus [Froemke02] realisierbar (siehe Abschnitt V.1.3).
Von großem Interesse ist bei den diskutierten Kurzzeitadaptionen besonders die quantale Adaption [Markram98] da diese Bestandteil einer der innerhalb FACETS verwendeten Benchmarks ist [Häusler07] (siehe auch Abbildung V.14). Präsynaptische Adaptionen dieser Art ließen sich in den Synapsentreibern realisieren, bei denen je eingehender Pulsquelle eine analoge Zustandsvariable mit entsprechender Zeitkonstante einen Mittelwert der präsynaptischen Aktivität bildet. Bei der Rekonstruktion des analogen Pulses wird dann die Amplitude als eine Funktion dieser Aktivität moduliert. Postsynaptische Adaptionen wie etwa SFA [Partridge76] können wie die bereits implementierten „High-Conductance-States“ [Schemmel06] über Leitwerte an der Membran realisiert werden, die wie oben durch mittelwertbildende Zustandsvariablen von der postsynaptischen Pulsrate gesteuert werden. Eine derartige Hardware-Modellierung von SFA als gesteuerter Leitwert direkt an der Membran ist neurobiologisch realistisch [Partridge76] und technisch leicht durchzuführen [Indiveri03]. Brette und Gerstner beschreiben in [Brette05] ein „adaptive exponential IAF“-Neuronenmodell, das mit wenigen konfigurierbaren Parametern eine große Bandbreite der oben angeführten Einzelaspekte des Neuronenverhaltens nachbilden kann. Dieses Modell stellt die Basis des jetzigen Neuronenentwurfs im HICANN dar [Schemmel07]. Überwachtes STDP wie in Abschnitt V.1.3 vorgestellt, kann in der Form aus [Legenstein05] ebenfalls eingesetzt werden. Da Pulse zu beliebigen Zeitpunkten als externer Stimulus über Layer1 an Synapsen angelegt werden können, ist die Steuerung durch zusätzliche Stromimpulse zu gewünschten Feuerzeitpunkten gemäß [Legenstein05] leicht realisierbar. Die Steuerungssynapse würde bei deaktivierter Adaption mit festem Gewicht betrieben werden, wodurch sich der dadurch ausgelöste Beitrag zur Membranaufladung sehr genau quantisieren lässt.
Der endgültige ANC wird aus 8 HICANNs in einer 2x4 Anordnung bestehen. Diese Aufteilung ermöglicht, einzelne Prototypen innerhalb eines kostengünstigen MPW-Runs zu erstellen und gleichzeitig eine Validierung des gesamten späteren ANC zu erreichen. Der modulare Aufbau aus gleichartigen HICANNs beinhaltet aber auch, dass das Layer1 Routing zwischen den HICANNs auf einem ANC identisch mit dem über IC-Grenzen hinweg ist, d.h. das Layout der Busse zwischen HICANNs auf demselben Die kopiert das spätere Postprocessing (siehe Abbildung V.12 und Abbildung V.13). Die projektierten Abmessungen eines HICANN sind 5mm*10mm, oder 20mm*20mm für den endgültigen ANC.

### V.3.2 Pulskommunikation intra-Wafer: Layer 1

Wie erwähnt, sollen die sogenannten Layer1 Verbindungen für den Großteil der Pulskommunikation innerhalb eines Wafers aufkommen und damit auch maßgeblich am Aufbau der Netztopologien beteiligt sein. Dadurch, dass in der HICANN-Architektur die Synapsen in einem Baublock mit den Neuronen untergebracht sind, besteht für die zugehörigen ‚Dendriten' nur die Konfigurationsmöglichkeit über die Verschaltung dendritischer Abschnitte. Alle anderen, komplexeren oder weitreichenderen Verbindungen müssen über die ‚Axone' des Layer1 erledigt werden, d.h. den geschalteten Verbindungen zwischen Neuronen eines HICANNs und Synapsen eines anderen HICANNs. Der Aufbau des Layer1 ist zeitkontinuierlich und adressbasiert. Pulse aus 64 Quellen (Neuronen) werden von einer Winner-Take-All (WTA) Schaltung und einem nachgeschalteten Prioritätsencoder in Pulsadressen codiert, die asynchron sofort nach ihrer Codierung auf der zugehörigen Layer1 Sammelleitung übertragen werden. Kollidierende Pulse werden vom WTA nicht verworfen, sondern zeitverzögert weitergegeben.

[40] Zustandsvariablen wie das Membranpotential sind zwar zu Testzwecken auch extern messbar, jedoch nur für einzelne, ausgewählte dendritische Abschnitte. Die Übertragung in einem für externe Adaption nötigen Rahmen ist durch die dort im Vergleich zur Pulsdigitalisierung höhere Auflösung (zeitlich und Amplitude) technisch unmöglich.

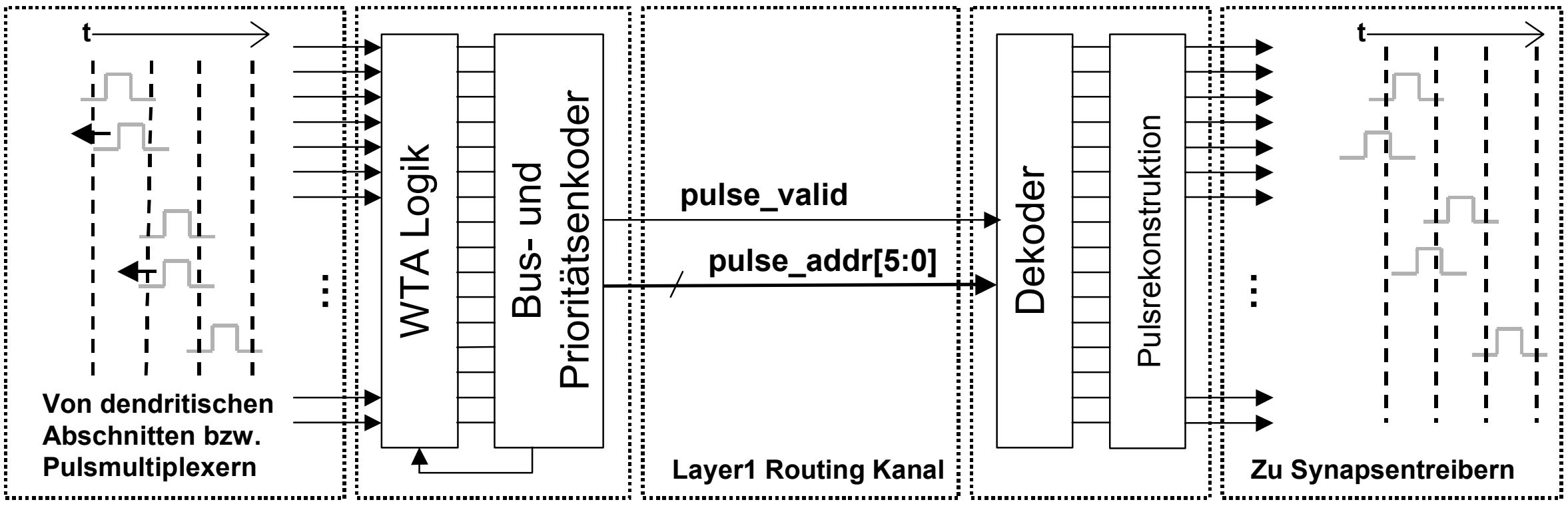

**Abbildung V.11.: Prinzipschaltbild von WTA und Decoder an der Schnittstelle zwischen Dendriten und L1**

Die ursprüngliche Planung sah dabei für Layer1 eine Parallelübertragung dieser Adressen auf 7 Leitungen vor, 6 Adressleitungen und ein Validsignal. Die Layer1-Busse hätten demnach aus N(Anzahl der Busse)*7 Leitungen bestanden. Gegenwärtig zeichnet sich hier ein Wechsel des Designs auf asynchrone serielle Übertragung ab, mit einer Signalleitung und schirmenden Masseleitungen. Unabhängig von der Realisierungsvariante signalisiert eine anliegende Adresse über die Bits auf der Leitung gleichzeitig sowohl die Existenz eines Pulses als auch das zugehörige Quellneuron. Über Busse dieser Layer1 Sammelleitungen und konfigurierbare Verschaltung gelangt die Adresse zum Ziel-HICANN, wo sie an den Synapsentreibern wieder in analoge Pulse auf 64 Leitungen zurückgewandelt wird. Einen Überblick der Busstruktur des L1 gibt die folgende Abbildung:

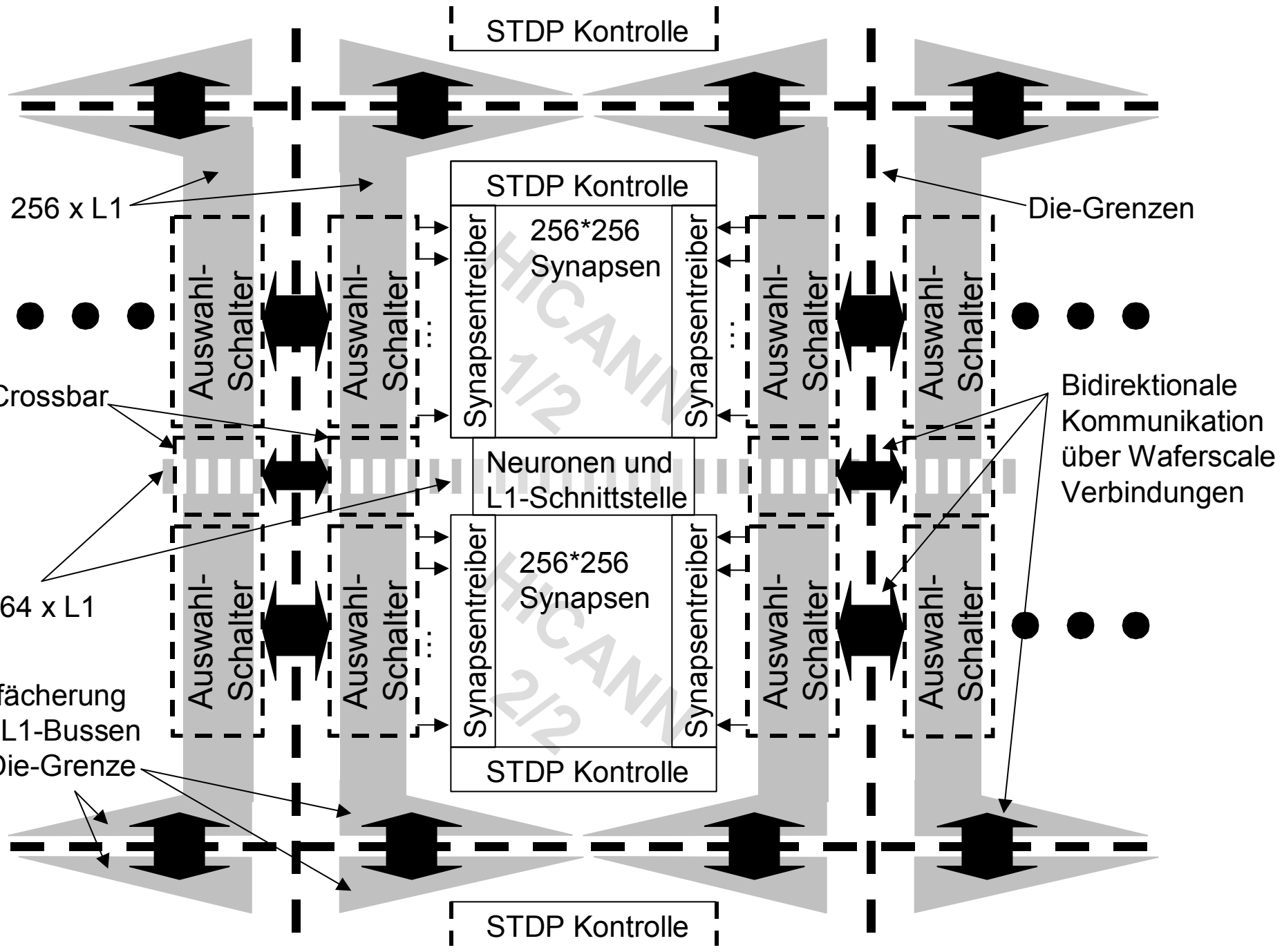

**Abbildung V.12.: Busstruktur des Layer 1, mit Crossbars, Auswahlschaltern an der HICANN-Matrix und Waferscale Verbindungen [Schemmel07, Ehrlich07]**

Die Strichlinie gibt die Grenzen zwischen HICANNs bzw. die Die-Grenze zwischen ANCs an. Entlang des linken und rechten Randes des HICANN führt vertikal jeweils ein Bündel von 256 Layer1 Sammelleitungen. Parallel zu den beiden Hälften des HICANN befinden sich

Auswahlschalter, die aus den 256 vorbeiführenden jeweils eine Layer1 Sammelleitung pro Synapsentreiber auswählen. Die Matrix der Auswahlschalter ist dabei dünn besetzt, d.h. nicht jeder Synapsentreiber hat Zugriff auf alle 256 Leitungen. Die genaue Besetzung der Matrix ist eines der Optimierungsprobleme des Mapping in Abschnitt V.5, da deren Aufbau maßgeblich die Art der abbildbaren Netztopologien beeinflusst. Über Waferscale-Verbindungen haben die Synapsentreiber neben dem Zugriff auf den eigenen 256er Layer1 Bus auch die Möglichkeit, entsprechende Sammelleitungen des Layer1 Bus auf dem Nachbar-Die auszuwählen. Die zwischen den beiden HICANN-Hälften liegenden dendritischen Abschnitte werden wie bei Abbildung V.10 beschrieben als Pulsquellen ausgewählt und dann über die Codierung aus Abbildung V.11 in ein Layer1 Signal umgewandelt. Je nach der aus den dendritischen Abschnitten realisierten Anzahl an Neuronen (8 bis 512) können damit bis zu 8 Layer1 Sammelleitungen belegt werden. Über den dendritischen Abschnitten geht ein 64fach Layer1 Bus entlang, der diese Pulsquellen sammelt und zusätzlich horizontale Verdrahtungsressourcen bereitstellt. An den Kreuzungsstellen mit den vertikalen Layer1 Bussen befinden sich Crossbars zur konfigurierbaren Verschaltung der Busse[41]. An den oberen und unteren HICANN-Grenzen werden die vertikalen Busse aufgefächert, um einen Waferscale-konformen Leitungsabstand zu erreichen. An den HICANN-Grenzen können die horizontalen und vertikalen Busse unterbrochen werden, so dass die Busse auch abschnittsweise verwendbar sind. Die einzelnen Abschnitte der L1-Verbindungen sind selektiv in beiden Richtungen verwendbar, d.h. Quellen- und Senkenende sind nicht fest vorgeschrieben. Mit einer anfänglichen Konfiguration werden sie jeweils auf eine Richtung festgelegt.
Der HICANN-Prototyp wird bereits diese Layer1-Busstruktur aufweisen, erweitert um Multiplexer an den Die-Grenzen, mit denen einzelne Layer1 Sammelleitungen aus den Bussen herausgegriffen werden können. Dies ist notwendig, da für den Prototyp nur serienmäßige gebondete Pads mit einem Mindestabstand zueinander von ca. 60µm bereitstehen. Im Vergleich dazu erlauben die späteren Waferscale-Verbindungen Leitungen im Abstand von ca. 4-8µm [Ehrlich07].

### V.3.3 Langreichweitenkommunikation: Layer 2

Für den Aufbau eines Stage 2 Systems aus mehreren Wafern reicht die Layer1 Kommunikation nicht aus. Zusätzlich muß die Möglichkeit bestehen, zur Verhaltensanalyse in hoher Bandbreite Pulsereignisse außerhalb des Wafers auslesen zu können, sowie Eingangssignale für das Netz bereitstellen zu können. Außerdem müssen Möglichkeiten bestehen, mit denen die Konfigurationen der verschiedenen Crossbars, Auswahlschalter, Buszuweisungen sowie die Parameterspeicher der Synapsen und Neuronen beschrieben werden können. Die Gesamtheit dieser Kommunikationsressourcen wird als Layer2 bezeichnet. Das Rückgrat dieser Kommunikation bildet eine serielle synchrone digitale Datenübertragung über Low Voltage Differential Signalling (LVDS) Leitungen mit mehreren Gbit/s pro Leitung [Scholze07]. Ein Teil der entsprechenden Schnittstelle ist auf den HICANNs integriert, wobei Layer2 hier direkt mit Layer1 interagiert, d.h. Ereignisse auf dem vertikalen Layer1 Bus (siehe Abbildung V.12) werden mit einer Zeitmarke und weiterer Identifikation versehen und als digitales Datenpaket versendet. Eintreffende Ereignisse werden analog wieder in das Layer1-Protokoll zurückverwandelt und dann über Layer1 zu ihrem Zielort geroutet. Außerhalb des Wafers werden die LVDS-Leitungen mit den DNCs verbunden (siehe Abbildung V.9), welche die weitere Versendung übernehmen:

[41] Vereinzelt werden in der Literatur bereits ähnliche Konzepte verfolgt, neuronale Untereinheiten auf neuromorphen VLSI-ICs durch verteilte Crossbars zu einer Gesamtfunktionalität zu verschalten, etwa in [Eickhoff06]

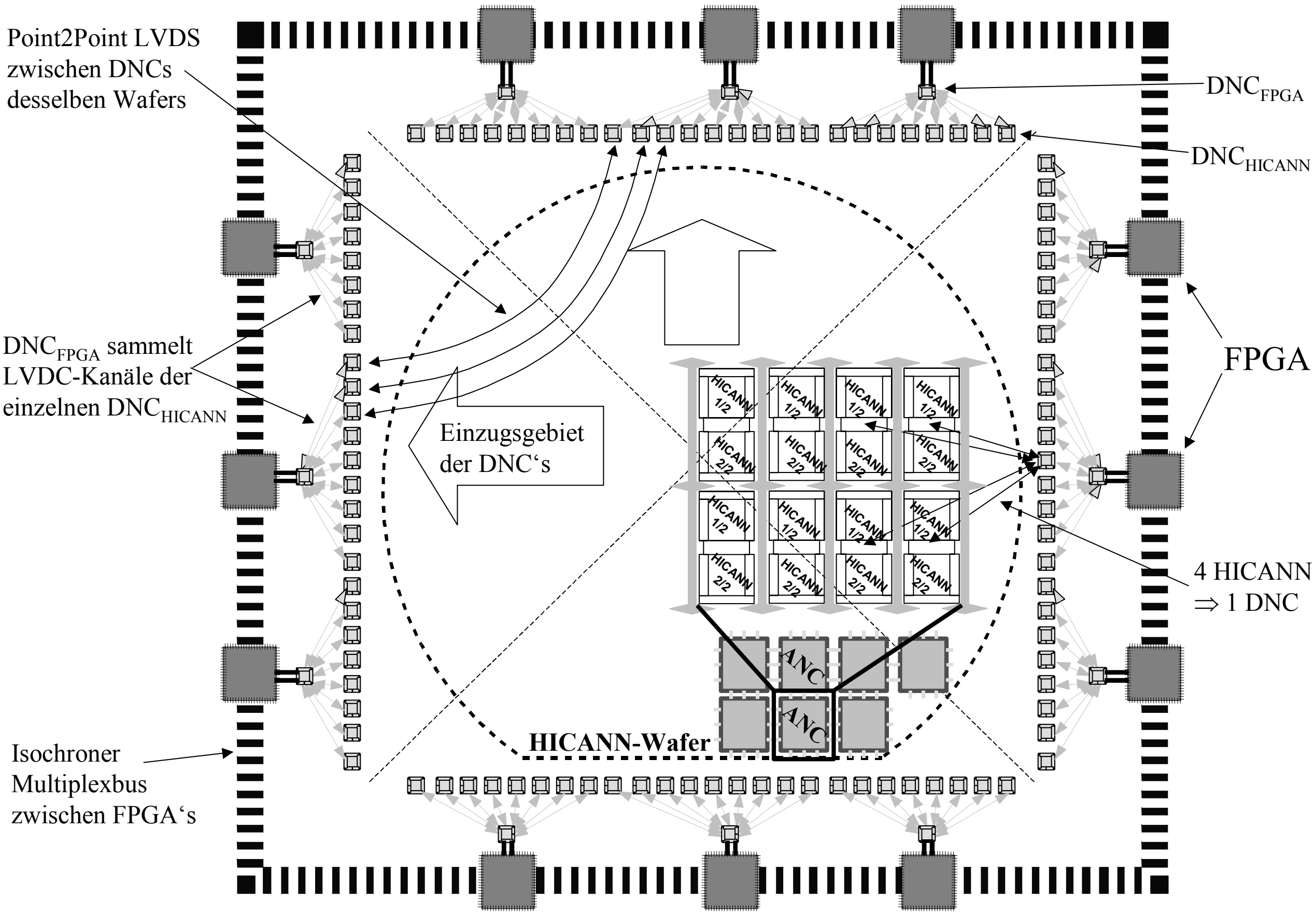

**Abbildung V.13.: Struktur der Layer2-Pulskommunikation [Scholze07]**

Die DNCs beinhalten LUTs, in denen Ziele der gerouteten Pulse abgelegt sind, und diverse PLLs und Zustandsautomaten zur Herstellung der elektrischen und logischen Kommunikation [Ehrlich07]. Identische DNCs werden an zwei verschiedenen Stellen in der Kommunikationshierarchie eingesetzt, gekennzeichet durch $DNC_{HICANN}$ und $DNC_{FPGA}$. Die $DNC_{HICANN}$ sind jeweils mit 4 HICANNs verbunden und stellen damit das Gegenstück zu den LVDS Transmittern auf den HICANNs her. Eingehende Pulse können entweder direkt wieder zu einem der anderen 3 HICANNs zurückgegeben werden (Routing in der Nachbarschaft). Alternativ dazu existieren Point2Point Direktverbindungen zu anderen DNCs desselben Wafers, mit denen mit entfernten Bereichen des Wafers kommuniziert werden kann, die wegen aufgebrauchter Layer1 Ressourcen nicht direkt erreicht werden können. Für das Auslesen von Pulsen und/oder eine Kommunikation mit anderen Wafern wird das Pulspaket an die $DNC_{FPGA}$ weitergegeben. Diese fassen die Kommunikationskanäle mehrerer $DNC_{HICANN}$ zusammen und stellen eine Schnittstelle zu kommerziellen FPGAs her. In den FPGAs erfolgt ein Wechsel des Busprotokolls, statt des Versandes einzelner Datenpakete über feste Leitungen wie bei den DNCs sind alle FPGAs über einen Bus miteinander verbunden, der im Zeitmultiplex betrieben wird, orientiert an dem in [Fieres04] verwendeten Aufbau. Die FPGAs stellen Standardschnittstellen wie Firewire bereit, über die dann PC-basiert das Netzwerk konfiguriert, gesteuert und ausgelesen wird.

# V.4 Benchmarks für die Systemsimulation

Da die eben beschriebene Stage 2 Hardware in FACETS als Forschungswerkzeug zur Unterstützung der simulationsbasierten neuronalen Verhaltensmodellierung verwendet werden soll, muss bereits in der Entwurfsphase sichergestellt werden, dass eine möglichst große Kongruenz zwischen Hardware und darauf abzubildenden Netzwerken besteht. Modelle verschiedener Forschungsgruppen innerhalb FACETS wurden deshalb als ‚Benchmarks' aufbereitet, die möglichst repräsentativ für den späteren Einsatz der Stage 2 Hardware sein sollen [Partzsch07b].

Die Benchmarks werden in der Entwurfsphase als Konfiguration für Systemsimulationen der Stage 2 Hardware verwendet und dienen damit zur Identifizierung von z.B. Engpässen in der Kommunikation oder neurospezifischen Designdefiziten[42]. Jede Benchmark besteht idealerweise aus der Netztopologie, Parametrisierung der zugehörigen Neuronen, Synapsen und axonalen Verbindungen (Plastizität, axonale Verzögerungen, etc), sowie aus den Stimuli, mit denen das Netzwerk versorgt wird und einer Methode zur Bewertung der Netzwerkausgabe. Dabei ist im Gegensatz zu den originalen simulativen Experimenten weniger das tatsächliche Ergebnis des Experiments wichtig, sondern die Korrelation zwischen diesem Ergebnis und der Ausgabe der Stage 2 Hardware[43], um die Qualität der Abbildung der Benchmark auf die Hardware beurteilen zu können. Es muss mithin nicht nur eine qualitative, sondern auch eine quantitative Bewertung des Netzwerkausgangs/verhaltens möglich sein. Beim Entwurf der Stage 2 Hardware wird hier immer ein Kompromiss nötig sein, da sich die Netztopologien und Parametrisierungen teilweise stark zwischen den verschiedenen Benchmarks unterscheiden. Das Waferscale System soll außerdem möglichst auch für zukünftige Modelle nutzbar sein und kann deshalb nicht so rigide festgelegt werden wie etwa eine auf wenige Zielanwendungen optimierte Architektur [Bofill-i-Petit04, Schreiter04]. Aus diesem Grund wird die Stimuluserzeugung auch außerhalb der Hardware durchgeführt und als fertig generiertes Pulsmuster ähnlich wie in [Muir05] über die vorhandenen Pulskommunikationsmöglichkeiten eingespeist, statt z.B. in Form von Pixelsensoren direkt integriert zu sein [Morie01] (siehe auch Kapitel III).

### V.4.1 Typischer Input/Output von Verarbeitungsaufgaben

Verschiedene Möglichkeiten, ein neuronales Netz mit Stimuli zu versorgen, wurden bereits in den letzten Kapiteln vorgestellt, etwa über Poisson-generierte Pulsfolgen (Abschnitt II.2.1). Für die dort untersuchten Ratencodes wurde dabei die Rate $\lambda$ als im Beobachtungszeitraum konstant angenommen. Eine Erweiterung stellen modulierte Poisson-Generatoren dar, d.h. eine Pulsfolge wird gemäß Gleichung (II.16), erzeugt, aber $\lambda=\lambda(t)$ gewählt. Für eine sinusförmige Modulation erzeugt dies Pulsfolgen ähnlich Abbildung II.15. In [Vogels05] werden rechteck- sinus- und rampenmodulierte Poisson-Pulsfolgen verwendet, um Signalübertragung durch ein mehrschichtiges Netzwerk zu testen und veränderlich Pulsraten für eine XOR Ratenverarbeitung bereitzustellen. Eine ähnliche Möglichkeit, die jedoch je nach Neuronenmodell etwas deterministischere Pulsfolgen erzeugt, ist, den Stimulus wie in Abschnitten IV.1 als Strom auf ein (simuliertes) Netz zu geben, dessen Ausgangspulse dann als Eingang für die Stage 2 Hardware verwendet werden. Als weitere neurobiologische Verfeinerung eines derart vorgeschalteten Netzes kann ein komplettes Simulationsmodell etwa einer Retina verwendet werden [Wohrer06], das seinerseits mit dem Stimulus (z.B. Bildfolge) konfrontiert wird, und diesen über Modelle der retinalen Zellen DoG-filtert und pulscodiert (Abschnitt I.3.1, Verwendung z.B. als Eingangssignal für die STDP-Nachbildung rezeptiver Felder in Abschnitt V.1.2). Eine sehr detaillierte Stimulusgenerierung könnte es sogar notwendig machen, von einer Vorberechnung abzusehen und den Stimulus zur Laufzeit zu generieren, um Rückkopplungen des Netzwerks auf das Stimulusmodell zu berücksichtigen, etwa bei einer Einbeziehung der V1-Kontrolle des LGN [Freeman02] (siehe auch Abbildung I.5). Es existieren daneben noch einige nicht neurobiologisch verankerte Ansätze, neuronale Netze für Verarbeitungsaufgaben mit Pulsstimuli zu versorgen, etwa eine SNR-optimierte direkte Codierung eines Stimulus in eine Pulsfolge [Schrauwen03] für Liquid-Computing Anwendungen. Manchmal enthalten die Stimuli solcher informationstheoretischer

[42] Fehler im Plastizitätsverhalten durch zu starke Quantisierung der Gewichte, fehlende Parametrisierungs-möglichkeiten, etc.

[43] Mit Stage 2 Hardware ist in diesem Kontext der jetzige Entwurfsstand bzw. ein Simulationsmodell gemeint, wobei die für Benchmarks und Konfiguration entwickelten Softwareroutinen hardwarenahe Ausgabeformate besitzen und später zur Ansteuerung des realen Waferscale Systems verwendet werden sollen. Die Namenskonvention, eine Hardware-Simulation (!) als ‚Hardware' zu bezeichnen, wird eingeführt als Unterscheidung zu den Neurosimulationen, aus denen die Benchmarks gewonnen werden. Diese werden auf speziellen Softwarewerkzeugen ausgeführt, die zur Emulierung neurobiologischen Verhaltens geschrieben wurden.

Ansätze keine Information per se, sondern werden nur als Template etwa zur Klassifizierungsleistung eines Netzwerks verwendet [Häusler07], oder sie definieren eine synaptisch zu erlernende Transformation zwischen eingehender und resultierender Pulsfolge (Abbildung V.6).
Die Stage 2 Hardware kennt im Betrieb deutlich andere Randbedingungen als die Simulation, wie etwa diskrete Verzögerungszeiten, quantisierte Zustandsvariablen oder abweichende synaptische Plastizitätsparameter, da diese von mehreren Synapsen in den Synapsentreibern zusammengefasst werden (siehe Abbildung V.10). Das Verhalten einer Benchmark in der Hardware wird demnach in der Regel sowohl quantitativ als auch qualitativ von einer Neurosimulation abweichen. Dies passiert beispielsweise wenn axonale Verzögerungszeiten leicht (quantitativ!) abweichen, und sich bei einer STDP-Adaption um den Feuerzeitpunkt des postsynaptischen Neurons herum daraus LTP statt LTD und damit eine qualitative Veränderung ergibt. Inwieweit dies Auswirkung auf die durch neurobiologische Simulationen etablierte Verarbeitungsfunktion hat, muss anhand des Netzwerkverhaltens überprüft werden. Die einfachste Bewertung stellt dabei für ratenbasierte Verarbeitungsfunktionen eine Zählung der Pulse eines festgelegten Ausgangsneurons in einem Beobachtungsintervall dar, ähnlich wie in Abschnitt II.2.1 (Liegt die korrekte Anzahl Pulse im Intervall?, mit Konfidenzgrenzen). Die Erweiterung dieses Verfahrens auf ein Populationssignal mehrerer Neuronen wird in Abschnitt II.2.3 (erster Absatz) eingeführt. Mit derartigen Verfahren kann z.B. die XOR-Verarbeitung von Eingangspulsströmen wie in [Vogels05] überprüft werden. Ein adaptives Filter wie bei Gabbiani und Metzner [Gabbiani99] erlaubt die Rekonstruktion eines in einer Pulsfolge codierten Stimulus. Eine Stimulustransformation bzw. Übertragung durch das Netzwerk kann damit quantitativ analysiert werden.
Wenn als Bewertungsmaßstab die resultierende Pulsfolge aus der Simulation vorgegeben wird [Häusler07], kann in Anlehnung an [Legenstein05, Mayr05c (Fig. 5)], über beide Pulsfolgen für jeden Puls ein gaussförmiges Profil eingesetzt werden:

$$r_{Ziel}(t)=\sum_i \delta(t-t_i) \quad , \quad r_{\sigma,Ziel}(t)=\frac{1}{\sqrt{2\pi}\sigma}\sum_i e^{\frac{(t-t_i)}{2\sigma^2}} \qquad \textbf{(V.8)}$$

Ausgangspunkt ist eine Definition der Hardware- und Zielpulsfolge ähnlich wie in Gleichung (A.1), wobei in Folge jeder Diracimpuls durch eine Gaußglocke mit Standardabweichung σ an den Pulszeitpunkten $t_i$ ersetzt wird[44]. Ein Integral über das Produkt beider Pulsfolgen in einem Beobachtungszeitraum, i.d.R. die Länge der vorgegebenen Pulsfolgen, ergibt ein ‚weiches' Maß für die Korrelation zwischen beiden Pulsfolgen:

$$C_{HW,Ziel}=\int_{T_0}^{T_1}\sum_i e^{\frac{(t-t_i)}{2\sigma^2}}\sum_i e^{\frac{(t-t_j)}{2\sigma^2}}\,dt \qquad \textbf{(V.9)}$$

Im Integral bezeichnet $t_i$ die Pulszeitpunkte der Zielpulsfolge und $t_j$ die Zeitpunkte der bei der Stage 2 Simulation entstehenden Pulsfolge. Durch Normierung des Korrelation zwischen Hardwarepulsfolge und Vorgabe $C_{HW,Ziel}$ auf die Autokorrelation der Zielpulsfolge entsteht eine Bewertungsmöglichkeit für die Ähnlichkeit beider Pulsfolgen $\overline{C}_{HW,Ziel}= C_{HW,Ziel}/C_{Ziel,Ziel}$. Etwaiger zeitlicher Versatz der einzelnen Pulse wird dabei je nach Wahl des Standardabweichung des Gaußprofils unterschiedlich stark bestraft, komplett fehlende Pulse verringern die Korrelation ungefähr um den Prozentsatz des einzelnen Pulses relativ zur Gesamtanzahl der Pulse. Es ist in diesem Zusammenhang sinnvoll, aus identischen (simulativen) Experimenten mehrmals die Zielpulsfolge zu erzeugen und diese über die obige Korrelation zu bewerten, um mit Hinblick auf die Stage 2 Implementierung desselben Experiments den Ergebnisspielraum abschätzen zu können.
Da die Kommunikationsressourcen des Stage 2 Systems vor allem auf die Weitergabe von Pulsen optimiert sind, stellt eine pulsbasierte Methode die beste Wahl zur Analyse des Netzwerkverhaltens dar. Für die Rekonstruktion rezeptiver Felder bei V1-Nachbildungen können dabei die bewährten

[44] Für den Grenzwert $\sigma \rightarrow 0$ sind beide Definitionen gleichwertig. Die ‚Aufweichung' mit realistischen σ erzeugt ein differenzierteres Korrelationsmaß.

Methoden aus der Biologie Verwendung finden [Hubel68, Jones87a]. Diese sind jedoch relativ zeitintensiv, es werden viele Ausgangspulse benötigt, außerdem sind die mathematischen Methoden nur zur Bewertung statischer RFs geeignet[45]. Für ein Netzwerk mit plastischer Adaption ist jedoch primär die Entwicklung der Gewichte über den Simulationszeitraum interessant. In Verallgemeinerung der obigen Methoden zur Rekonstruktion rezeptiver Felder wurden deshalb verschiedene Möglichkeiten getestet, aus der Puls-I/O-Relation eines Neurons und seiner Synapsen die Gewichtsverteilung der Synapsen zu rekonstruieren [Partzsch07b]. Somit können allgemeine plastische Vorgänge im (mit Stage 2 emulierten) neuronalen Gewebe untersucht werden, ohne Bezug zur speziellen Verarbeitungsfunktion.

### V.4.2 Netztopologien

Die Kommunikationsressourcen der Stage 2 Hardware können maßgeblich darüber definiert werden, welche Arten von Netztopologien mit ihnen umsetzbar sind. Hier gibt es eine große Spannweite von mehr oder weniger biologisch orientierten Strukturen, stochastischen und deterministischen Netzwerkelementen, etc. Im Folgenden sollen in einer kurzen Übersicht Beispiele für die verschiedenen Kategorien gegeben werden. Die erste Variante ist aus [Häusler07] entnommen, sie besteht aus einer Sammlung einzelner Neuronenpopulationen, die untereinander mit bestimmten Wahrscheinlichkeiten Synapsen ausbilden. Das zugehörige Modell in der linken Hälfte von Abbildung V.14 versucht, grundlegende Charakteristiken des visuellen Kortex nachzustellen, so etwa die relativen Populationsgrößen zueinander, den Anteil an exzitatorischen bzw. inhibitorischen Neuronen und die Verbindungswahrscheinlichkeiten [Häusler07]. Eine Orientierung erfolgt dabei an Messdaten von Thomson und Kollegen [Thomson03]. Die Netztopologie konzentriert sich auf die Schichten 2-5 des visuellen Kortex, wobei Schichten 2 und 3 wegen ihrer starken Vernetztheit bzw. undeutlicher Grenzen [Binzegger04] als eine Neuronenpopulation angesehen werden:

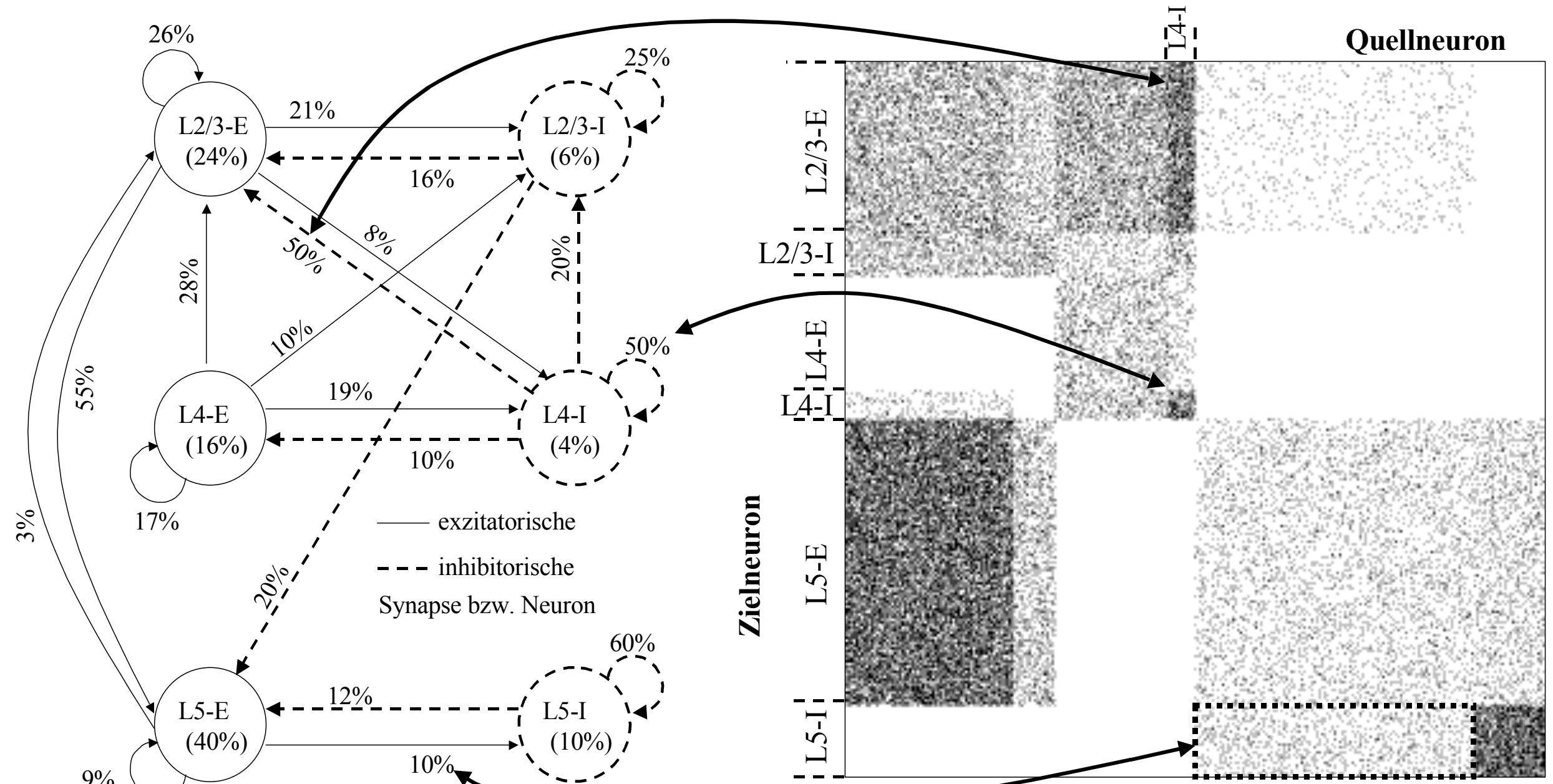

**Abbildung V.14.: Netztopologie aus [Häusler07], schichtenbasiertes Modell mit aus der Biologie entlehnten Verbindungswahrscheinlichkeiten, Aufbau (links, ohne I/O-Kanäle, nur interne Verbindungen) und resultierende Verbindungsmatrix[46] für 550 Neuronen Gesamtpopulation**

[45] Alternativ bietet sich (zumindest für die Systemsimulation der Hardware) eine Rekonstruktion der rezeptiven Felder aus den synaptischen Gewichten an, wie bei Abbildung V.3 und V.5 beschrieben.

[46] Wie am Anfang von Abschnitt V.3.2 ausgeführt, bestehen die zwischen den Neuronen in der Stage 2 Hardware gebildeten Verbindungen unter neurobiologischer Sicht aus Axonen, während Dendriten und Synapsen relativ starr miteinander gekoppelt sind. Da jedoch jede dieser Verbindungen aus der Abfolge ‚Axon-Synapse-Dendrit' besteht und

Die Pfeile im Modell geben jeweils die Richtung der Synapse (prä auf post) und die Wahrscheinlichkeit für jedes individuelle Neuron einer Subpopulation an, eine entsprechende Verbindung auszubilden. In Schicht 2/3 liegen dabei 30% der Neuronen, in Schicht 4 20% und in Schicht 5 50%, wobei die Aufteilung in exzitatorische und inhibitorische Population in jeder Schicht im Verhältnis 4:1 stattfindet.
Wenn die Neuronen der Gesamtpopulation von Schicht 2/3 bis 5 durchnummeriert werden, jeweils die exzitatorische Unterpopulation zuerst, und die synaptischen Verbindungen als Matrixeinträge verwendet werden, mit jeweils dem Quellneuron als Spalte und Zielneuron als Zeile, erhält man eine zu der Netztopologie gehörige zweidimensionale Verbindungsmatrix. Wie die Pfeile in Abbildung V.14 anzeigen, lassen sich dabei die verschiedenen synaptischen Verbindungswahrscheinlichkeiten als unterschiedlich dicht besetzte Rechtecke in der Matrix wieder finden, mit Kantenlängen jeweils entsprechend der Größe der Quell- bzw. Zielneuronenpopulation. In Abschnitt V.5 wird die weitere Bedeutung von Verbindungsmatrizen im FACETS-Kontext ausgeführt.
Wie oben erwähnt, liegt Modellen des visuelle Kortex meist dessen physische, dreidimensionale Gestalt zugrunde, wobei räumlich begrenzte biologische Messdaten (meist aus der Verbindungsstruktur mehrerer 10 bis 100 Neuronen) hochgerechnet werden, um Simulationen mit größeren Anzahlen an Neuronen und Synapsen durchzuführen. Im vorhergehenden Modell wurde außer den auf diesen Messdaten beruhenden Verbindungswahrscheinlichkeiten noch die (makroskopisch erkennbare) vertikale Schichtstruktur des Kortex berücksichtigt. In einer weiteren Verfeinerung derartiger Modelle wird die horizontale Einteilung in voneinander abgesetzte Kolumnen berücksichtigt. Ein Beispiel für ein solches Netz ist in [Djurfeldt05] enthalten. Es existieren drei Typen von Neuronen: Pyramidenzellen (py), normal-pulsende Zellen (engl. regular spiking cell, RS) und Korbzellen (engl. basket cell, ba). Das Netz ist vor allem horizontal stark strukturiert, mit einer hierarchischen Einteilung in sogenannte Mini- und Makrokolumnen [Shepherd04]. Jede Minikolumne besteht aus 30 stark miteinander vernetzten Pyramidenzellen und 2 RS-Zellen, die Schichteinteilung innerhalb der Minikolumne [Binzegger04, Kandel95] wird bei diesen Verbindungen vernachlässigt. Die RS-Zellen erhalten gleichverteilte Verbindungen mit einer Wahrscheinlichkeit von wenigen Prozent von allen im Netz vorhandenen Pyramidenzellen. Jeweils 100 Minikolumnen und 100 Korbzellen bilden eine Makrokolumne. Die folgende Abbildung illustriert die Struktur des Netzes:

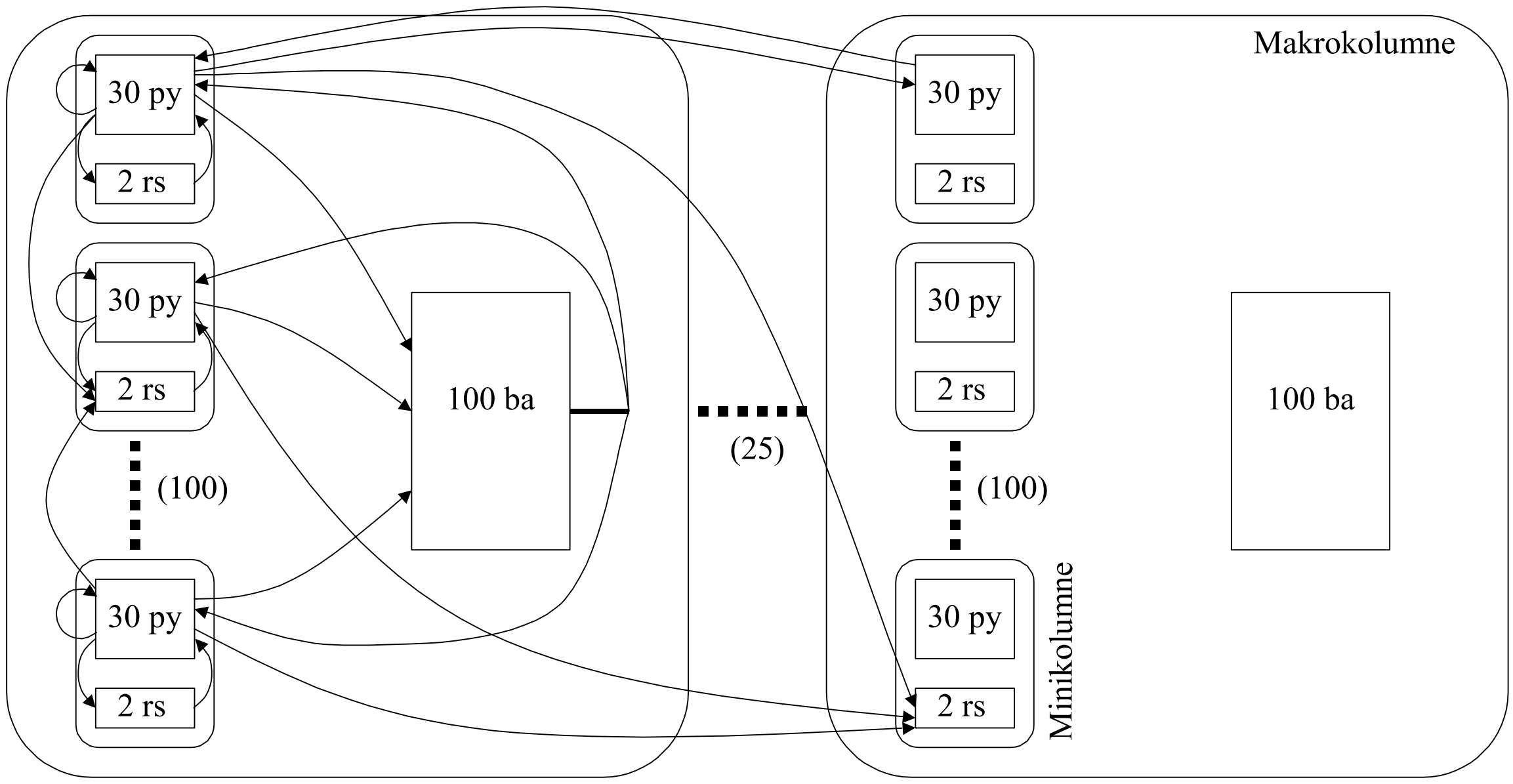

**Abbildung V.15.: Kolumnenorientierte Netzstruktur des V1-Modells der KTH Stockholm [Djurfeldt05]**

es gebräuchlich ist, die Synapse als eigentliche Verbindung zwischen Neuronen zu sehen, werden die Einträge in der Verbindungsmatrix als ‚Synapsen' bezeichnet.

Jede Korbzelle erhält eingangsseitig Verbindungen von zufälligen Pyramidenzellen derselben Makrokolumne. Ihr Ausgang ist in hoher Dichte mit den Pyramidenzellen der Makrokolumne vernetzt. Die 25 Makrokolumnen dieses V1-Modells sind, wie oben erwähnt, über die kolumnenübergreifenden Verbindungen von Pyramidenzellen zu RS-Zellen miteinander verbunden. Zusätzlich besteht, wie oben in Abbildung V.15 angedeutet, eine dichte reziproke Vernetzung zwischen jeweils den Pyramidenzellen von korrespondierenden Minikolumnen. Das Modell hat in der als Benchmark vorliegenden Dimensionierung 75000 Pyramidenzellen, 5000 RS-Zellen und 2500 Korbzellen und damit insgesamt 82500 Neuronen mit $27{,}5*10^6$ Synapsen.
Da Formen von STDP an vielen weiteren Stellen des Gehirns gefunden wurden (siehe Abbildung V.2), erschöpft sich natürlich die Einsatzfähigkeit der Stage 2 Hardware nicht in der Nachbildung des visuellen Kortex. Eine mögliche Anwendung/Forschungsgebiet, bei dem STDP bereits erfolgreich eingesetzt wird [Koickal06, Muir05], ist die Modellierung des olfaktorischen Kortex bzw. Riechkolbens. Aufgrund der Freiheitsgrade in der Rekonfigurierbarkeit der Stage 2 Hardware können generell viele der aktuellen Forschungen bzgl. sensorischer Subsysteme des Kortex unterstützt werden. Vorraussetzung ist nur, dass dort als Modell für die Langzeitplastizität Varianten pulsbasierter Plastizität zum Einsatz kommen [Abbott00, Bell97, Muir05]. Die Modellierung innerhalb Stage 2 wird auch insoweit vereinfacht, als diese Subsysteme insgesamt ähnliche Strukturen aufweisen [Shepherd04].
In [Izhikevich04a] wird eine weitere möglicherweise Stage 2-relevante Architektur vorgestellt, ein generisches Topologiemodell für mikroskopische Kortexausschnitte, mit dem allgemein die Entstehung von Gedächtnis in kortikalem Gewebe nachbildet werden kann. Es wird topologisch vor allem die räumliche Struktur von Axonen und Dendriten sehr genau modelliert, mit resultierender small-world Verbindungscharakteristik [Blinder05]. Makroskopische Strukturen wurden größtenteils vernachlässigt, was durch das kleine simulierte Kortexvolumen begründet wird. Axonale und dendritische Verzögerungszeiten bei myelinisierten und unmyelinisierten Verbindungen werden hingegen sehr genau modelliert. Die Autoren versuchen, über diese Verzögerungen in Verbindung mit STDP-Adaption den selbstorganisierenden Aufbau von Synfire-Chains nachzuvollziehen. Entsprechende Kopplungsstrukturen ergeben sich dabei ähnlich zufällig wie in [Vogels05], d.h. durch die größtenteils zufällige Verdrahtung entstehen zwangsläufig Rückkopplungsstrukturen, diese werden dann durch STDP verstärkt. Die Anregung des Netzwerks erfolgt trivialerweise über einen rauschenden Membranstrom, der für pulsende Hintergrundaktivität sorgt, welche wiederum das spontane Entstehen von Synfire-Chains begünstigt.

# V.5 Mapping und Konfigurationserzeugung

Durch das ‚Mapping' (Zuordnung) soll ein vorgegebenes Simulationsmodell oder ein aus neuronalem Gewebe rekonstruiertes Netz auf die Stage 2 Hardware abgebildet werden. Die in Abbildung V.14 eingeführte Verbindungsmatrix einer Topologie kann als zweidimensionale Repräsentation einer eigentlich dreidimensionalen biologischen Struktur dienen. Dies ist von Vorteil bei der Abbildung der Netztopologie auf die (inhärent) zweidimensionale Stage 2 Hardware [Mayr07b]:

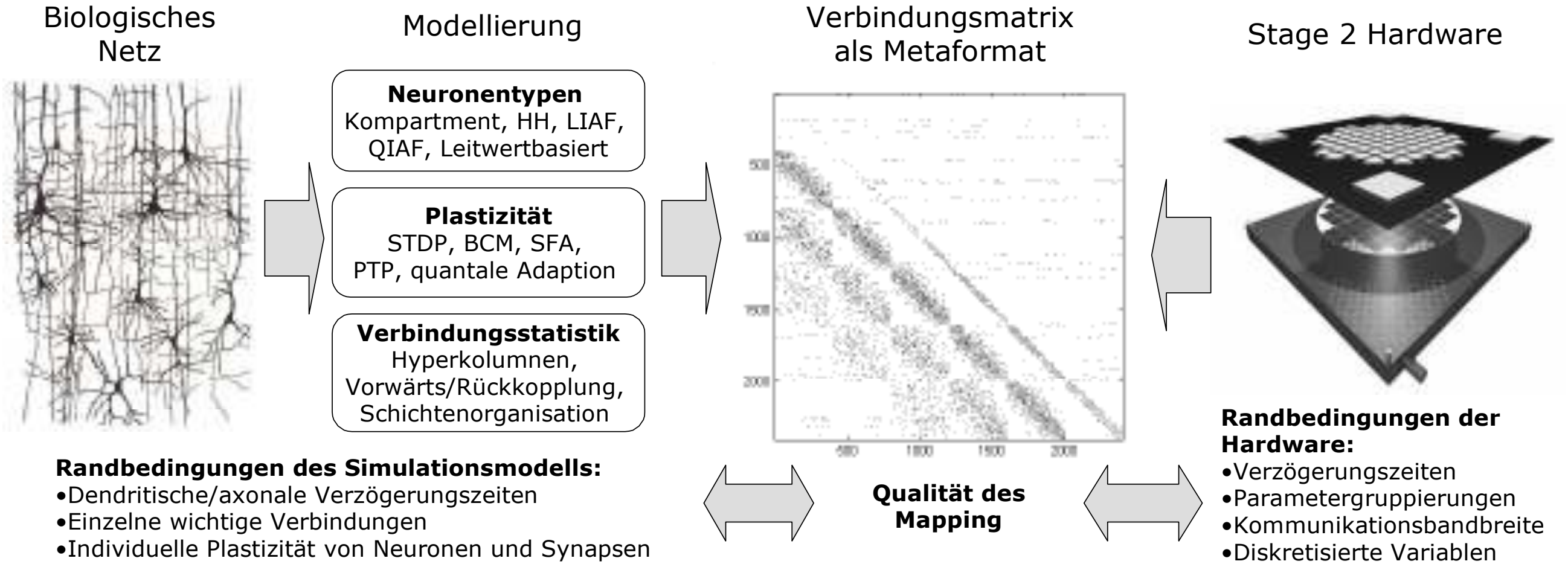

**Abbildung V.16: Topologie-Mapping-Fluss von Biologie über Simulation zur Hardware [Mayr06c] (Biologisches Netz aus [Cajal09], Stage 2 Darstellung aus [Meier04], die Verbindungsmatrix stellt eine Synthese der V1-Modelle aus [Djurfeldt05, Häusler07] dar)**

Wie in den letzten Abschnitten ausgeführt, werden für die auf Stage 2 abzubildende Benchmark die biologischen bzw. simulativen Grundlagen zusammengefasst und per Software aufbereitet. Die Neuronenmodelle der Simulation werden in Parametersätze umgeschrieben, welche für die dendritischen Abschnitte des HICANN möglichst ähnliches Verhalten erzeugen. Gleiches erfolgt mit den verschiedenen Formen der synaptischen Plastizität, die auf ihre Repräsentation durch entsprechende Parameter bzw. Konfigurationen der Hardware übertragen werden. Die Umschreibung der Parametersätze von Vorgabe in Hardwaremodell ergibt sich dabei aus Simulationen der einzelnen Baugruppen, bei denen korrespondierende Parametersätze bzw. Umschreibungsvorschriften erarbeitet werden. Für Umschreibungen wird der jeweilige Parametersatz des Hardwaremodells ermittelt, bei dem sich für einen Vektor an Eingangspulsfolgen der geringste Fehler im Vergleich der Ausgangspulsfolgen von Vorgabe und Hardwarerealisierung ergibt (Gleichung (V.9)). Der Parametersatz der Vorgabe kann dann mit der so ermittelten Korrespondenz auf den Parametersatz der Hardware umgeschrieben werden. Datenbanken dieser Parametersatz-Korrespondenzen bilden die Grundlage dafür, in Systemsimulationen die verschiedenen Neuronen- und Synapsenmodelle der simulativen Vorgaben auf das Hardwaremodell umzuschreiben. Teilweise lassen sich korrespondierende Einzelparameter auch direkt ineinander umschreiben. Ein Beispiel hierfür wäre die Umschreibung eines Membranableitwiderstands, der zum Einen der erhöhten Geschwindigkeit des Hardwaremodells angepasst werden muss und außerdem im Hardwaremodell nicht direkt, sondern etwa über eine Gatespannung hergestellt wird, mit entsprechend nichtlinearem Zusammenhang zwischen Stellgröße und gewünschtem Leitwert [Partzsch07b].

Verschiedene zusätzliche Randbedingungen werden ebenfalls in die abzubildenden Datensätze integriert, etwa Vorgaben für axonale Verzögerungszeiten oder unterschiedliche Prioritäten für synaptische Verbindungen. Zuletzt wird das Topologiemodell für das Netz der Benchmark in eine zweidimensionale Verbindungsmatrix umgeschrieben, wobei an jeden Eintrag der Matrix die zugehörigen Parameter oder Randbedingungen angefügt sind. Aus der gegenteiligen Richtung erfolgt die Aufarbeitung der Stage 2 Hardware, ebenfalls mit ihren Randbedingungen und ihrer (konfigurierbaren) Netzwerktopologie. Wie gut diese dann im Mappingprozess zum Überlappen gebracht werden können, bestimmt die Qualität des Mappings und damit in der späteren simulationsunterstützenden Anwendung der Stage 2 Hardware auch den Nutzen der damit durchgeführten Forschung.

Zur Beurteilung des Mapping und der neuromorphen Eigenschaften des aktuellen Hardwareentwurfs wird in einer Systemsimulation für das Gesamtsystem eine Verhaltensbeurteilung durchgeführt. Dabei wird ‚Ähnliches Verhalten' abhängig von der jeweiligen Benchmark in Zusammenarbeit mit der Forschungsgruppe definiert, welche die Benchmark zur Verfügung gestellt hat. Es kommen die in Abschnitt V.4.1 diskutierten Verfahren zur Bewertung der Ausgabe der Systemsimulation zum Einsatz. Diese bestehen entweder aus den ursprünglich für die Benchmark definierten Auswertungen oder aus einer Ähnlichkeitsbeurteilung von Pulsausgaben gemäß Gleichung (V.9) zwischen der Originalsimulation und der Hardwareimplementierung. Quantitative Werte für eine Verhaltensähnlichkeit werden gemäß den angeführten Beurteilungsverfahren momentan mit den relevanten Forschungsgruppen vereinbart. Aus der detaillierten Analyse derartiger Systemsimulationen ergeben sich wieder Forderungen an Mappingtool und/oder Hardwareentwurf. Wenn etwa während des Mappings Verbindungen weggelassen wurden, die sich in der Systemsimulation als wichtig für das Netzwerkverhalten erweisen, müssen beispielsweise die Crossbars geändert werden, um diese Verbindungen zuzulassen.

### V.5.1 Topologieprojektion

Das Mapping ist in mehrere Hierarchiestufen unterteilt [Mayr07b], beginnend mit einem Stage 2 System aus mehreren Wafern (inter-Wafer), der Zuordnung auf einem Wafer (intra-Wafer) und als letzter Stufe die Konfiguration der einzelnen ANCs und DNCs (Die-Level). Stellvertretend für die beiden obersten Mappingstufen wird der Prozess in Abbildung V.17 im Detail für die intra-Wafer Ebene veranschaulicht:

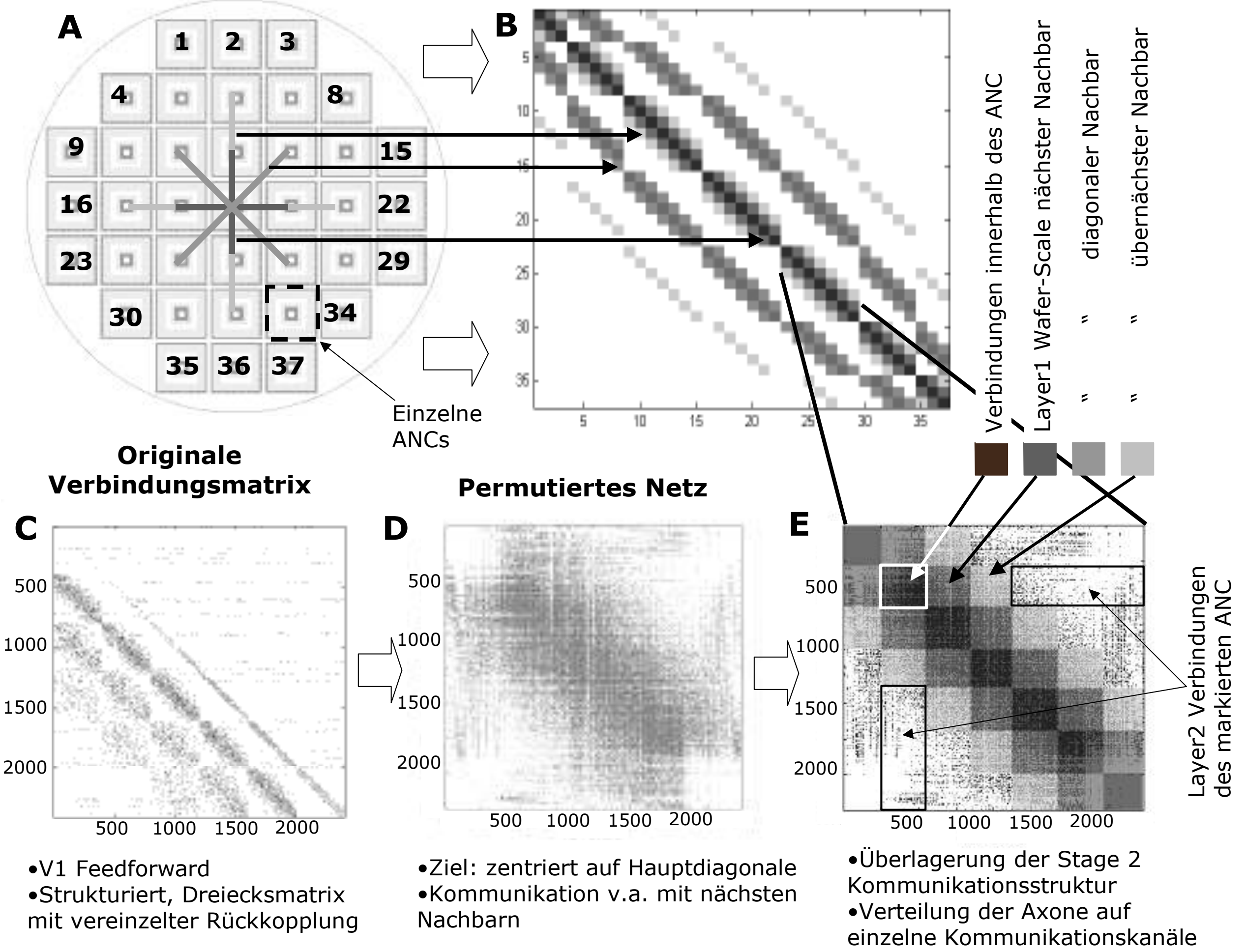

**Abbildung V.17: Praktisches Beispiel für die Herangehensweise eines (intra-Wafer) Mapping-Algorithmus [Mayr06c], Abbildung der Waferstruktur (obere Zeile, A&B), optimierende Permutation des Benchmark-Netzes und Abbildung auf einen Ausschnitt des Wafers (untere Zeile, C-E)**

In der oberen Abbildungszeile ist zu sehen, wie sich eine spezifische Die-Verteilung auf dem Wafer und eine festgelegte Reichweite der Layer1 Verbindungen in der Verbindungsmatrix widerspiegelt [Mayr06c]. Im vorliegenden Fall wird von einer Reichweite für Layer1 von 2 ANC-Grenzen bzw. 3 ANCs ausgegangen, d.h. es wird angenommen, dass jede weiterreichende Verbindung über Layer2 realisiert werden muss. Resultierende Verbindungen sind demnach nächster Nachbar, diagonaler Nachbar, und verlängerter nächster Nachbar. Jedes der Quadrate in der Verbindungsmatrix (Abbildung V.17 B) repräsentiert ähnlich wie in Abbildung V.14 die Verbindungen einer Untergruppe von Neuronen zu einer anderen Untergruppe. Die schwarzen Quadrate auf der Hauptdiagonalen entsprechen dabei den Verbindungen/Synapsen, die Neuronen eines ANC innerhalb dieses ANC ausbilden. In der unteren Zeile von Abbildung V.17 wird mit der Verbindungsmatrix eines generischen V1-Modells begonnen. Basierend auf einer oder mehrerer Zielfunktionen werden die Zuordnungsvektoren (Spalte und Zeile) zwischen Neuronen des Simulationsmodells und dem Neuronenindex der Verbindungsmatrix durchpermutiert. Ziel ist dabei, die zu den Neuronen gehörigen Synapseneinträge in der Matrix möglichst vorteilhaft auf die in Abbildung V.17 B abgeleitete Verbindungsstruktur der ANCs abzubilden. Für ein Netz mittlerer Größe mit 2000 Neuronen, welches auf einen Teil des Wafers abgebildet wird, liefert als grundlegendster Ansatz eine Zentrierung auf die Hauptdiagonale bereits gute Ergebnisse (Abbildung V.17 D). In der vorliegenden Anwendung können beispielsweise zusätzlich zur Zentrierung noch (Neben-)Zielfunktionen einbezogen werden, etwa eine akkurate Abbildung der Neuronen- und Synapsenparametersätze. Die Korrespondenzen für eine gute Übereinstimmung des Hardwareverhaltens und der simulativen Vorgabe ergeben sich zwar aus den vorherigen Abschnitt geschilderten Einzelsimulationen, jedoch können von dieser idealen Umschreibung während des Mappingprozesses Abweichung auftreten. Dies ergibt sich aus der Tatsache, dass im Hardwareentwurf nicht für alle Neuronen und Synapsen eigene Konfigurationsspeicher existieren, so dass Neuronen und Synapsen mit ähnlichen Parametersätzen zusammengefasst werden. Es werden somit Konfigurationsspeicher gemeinsam benutzt, wodurch sich zwar eine kompakte Hardwareabbildung ergibt, jedoch Abweichungen im Verhalten unausweichlich sind. Mittels einer Systemsimulation kann dann beurteilt werden, ob die Abweichungen im Rahmen der spezifischen Benchmark zulässig sind oder mehr Priorität auf die Einhaltung der Parameterwerte gelegt wird, was durch eine geringere Auslastung der Hardware erkauft wird.

Algorithmen zum Permutieren von Zuordnungsvektoren unter mehreren Zielfunktionen können teilweise aus der Literatur adaptiert werden, sie finden dort Anwendung bei Ablauf/Terminoptimierungen als Multiobjective Combinatorial Optimization (MOCO) [Jaszkiewicz02, Wendt07].

Um die für diese Netztopologie nötige synaptische Ausfächerung von 2000 Synapsen pro Neuron zu erreichen, werden die dendritischen Abschnitte der HICANNs eines ANC so miteinander verschalten, dass insgesamt 360 Neuronen pro ANC entstehen (Abbildung V.10). Ein Teil der ANC-Verbindungsmatrix aus Abbildung V.17 B wird dann auf die zentrierte Neuronen-Verbindungsmatrix überlagert. Wie eben erwähnt, ändert sich dabei der Index von der ANC-Nummerierung in Abbildung V.17 B auf eine Neuronennummerierung mit 360 Neuronen pro ANC in Abbildung V.17 E (entspricht der Seitenlänge der einzelnen Quadrate). Welche Verbindungen über welchen Kanal realisiert werden, ergibt sich durch den entsprechenden Teil der ANC-Verbindungsmatrix, in den eine synaptische Verbindung fällt. Die Zentrierung der synaptischen Verbindungen auf die Hauptdiagonale hatte damit anschaulich den Effekt, zum Einen möglichst viele Verbindungen bzw. deren Pulskommunikation jeweils innerhalb eines ANC zu konzentrieren. Zum Anderen kann ein Großteil der übrigen Kommunikation über den Nachbarradius der Waferscale Layer1-Busse realisiert werden. Synaptische Verbindungen, die nicht in eines der markierten Quadrate fallen, müssen über die DNC-basierte Layer2-Kommunikation realisiert werden.

Das oben illustrierte intra-Wafer Mapping wurde als Beispiel für die beiden obersten Mapping-Stufen gewählt, da das inter-Wafer Mapping sehr ähnlich aufgebaut ist. Die Zuordnung der Neuronen auf einzelne Wafer folgt dabei auch dem Ziel, die Kommunikation zwischen den Wafern so weit wie möglich zu reduzieren. Die inter-Wafer Verbindungsmatrix ist einfacher aufgebaut als jene des intra-Wafer Mapping in Abbildung V.17 B, es wird nur auf die Hauptdiagonale optimiert.

Die Kommunikation zwischen den Wafern hat keine vergleichbare unterschiedliche Wertigkeit wie die Kommunikation zwischen den ANCs im obigen Beispiel.

### V.5.2 Mapping und Konfiguration

Nachdem die Zuordnung der Neuronen zu einzelnen ANCs bzw. deren HICANNs erfolgt ist, muss als letzte (und wichtigste) Stufe des Mapping eine Abbildung der Einzelelemente der simulativen Vorgabe auf die Hardware erfolgen, d.h. der Neuronen, Synapsen und ihrer zugehörigen Parametersätze. Diese Abbildung kann allgemein wie folgt formuliert werden[47] [Wendt07]:

$$m: H_p \rightarrow B_p \quad , \quad H_p \subseteq H \quad B_p \subseteq B \qquad \textbf{(V.10)}$$

*H* repräsentiert die Menge aller Hardwareelemente und *B* die Menge aller biologischen Elemente. $H_p$ ist die Menge aller Hardwareelemente aus dem gesamten Stage 2 System, welche für das Mapping der simulativen Vorgabe verwendet werden. $B_p$ ist die Untermenge von Elementen aus *B*, die erfolgreich in die Hardware abgebildet werden kann. Dies ergibt zugleich das wichtigste Optimierungskriterium für das Mapping:

$$l_B : B_p \rightarrow R, \quad l_B(B \setminus B_p) \Rightarrow 0 \qquad \textbf{(V.11)}$$

Dabei wird versucht, den Verlust $l_B$ an nicht abbildbaren biologischen Komponenten minimal zu halten[48]. Da eine pareto-optimale Front [Jaszkiewicz02] aller Lösungen des Mappingproblems in diesem Kontext zu wenig Aussagen über die Qualität des Mapping zulassen würde, werden die einzelnen Verlustelemente in $l_B$ additiv zu einem skalaren Maß zusammengefasst [Wendt07, Mayr05c]. Diese Zusammenfassung erfolgt gewichtet, mit empirisch ermittelten Faktoren je nach Relevanz der einzelnen Elemente für das korrekte Funktionieren der Benchmark. In der Reihenfolge absteigender Wichtigkeit der Beurteilungskriterien des Mapping folgt als nächstes die Übereinstimmung $c_B$ der Parameter und anderer Nebenbedingungen zwischen Vorgabe und auf Stage 2 abgebildeter Benchmark:

$$c_B : (H_p \rightarrow B_p) \rightarrow R, \quad c_B(m) \Rightarrow \max. \qquad \textbf{(V.12)}$$

Nebenbedingungen könnten z.B. vorgegebene Verzögerungen von axonalen Verbindungen sein. Die in $c_B$ enthaltenen Einzelkorrelationen zwischen Parametersätzen werden ähnlich wie oben additiv zusammengefasst, mit dem Optimierungsziel einer Maximierung der Gesamtkorrelation. Als letztes Optimierungskriterium wird eine möglichst effiziente Ausnutzung der Hardware unter vorgegebener Abbildung *m* und Gesamthardware *H* berücksichtigt:

$$c_H : (H_p \rightarrow B_p, H) \rightarrow R, \quad c_H(m, H) \Rightarrow \max. \qquad \textbf{(V.13)}$$

Hier bestehen deutliche Freiheitsgrade, da aufgrund der Konfigurierbarkeit von Stage 2 die Abbildung der Vorgabe auf die Hardware nicht eindeutig ist, d.h. sich bei gleich bleibender Güte für die Kriterien aus den Gleichungen (V.11) und (V.12) unterschiedliche Realisierungen derselben Vorgabe finden lassen. Da Stage 2 einen Kompromiss aus verschiedenen Benchmarkvorgaben und Randbedingungen der Hardware darstellt, wird keine der Benchmarks (und auch keine der Vorgaben aus zukünftigen Einsätzen in der Forschung) sich optimal auf die Hardware abbilden lassen, gewisse Teile der Stage 2 Ressourcen werden ungenutzt bleiben. Ein einfaches Beispiel wäre hier eine Gruppe von zehn Synapsen, welche einen bestimmten Parametersatz unterschiedlich vom Rest der Synapsen verwirklichen müssen. Dafür würden im HICANN ein Synapsentreiber mit diesen Parametern initialisiert, jedoch nur zehn der 256 so konfigurierten Synapsen verwendet. Die

[47] Die Abbildung ist in der Richtung von Hardware auf Biologie notiert, da davon ausgegangen wird, dass ein biologisches Element immer einem oder mehreren Hardwareelementen entspricht, und damit mathematisch gesehen nur in dieser Richtung von einer Abbildung gesprochen werden kann. Der Begriff der Abbildung wird im Folgenden dennoch in etwas erweiterter Bedeutung auch für die entgegengesetzte Richtung verwendet.

[48] Das Symbol ‚⇒' bedeutet in diesem Fall die Optimierungsrichtung der Bewertungsfunktion, d.h. welchen Wert $l_B$ in (V.11) für ein optimales Mapping annehmen muss.

Optimierung aus Gleichung (V.13) ist dahingehend interessant, eine synaptische und neuronale Auslastung im Bereich 50-80% zu erreichen, welche nötig ist, um auf technisch machbaren Stage 2 Systemgrößen von maximal mehreren zehn Wafern wirklich Netze in der am Anfang des Kapitels erwähnten Größenordnung realisieren zu können.

Zur Systemmodellierung wird eine Graphendarstellung der simulativen Vorgabe und der Stage 2 Hardware verwendet. Die Knoten $V_G$ repräsentieren die Elemente und Parameter der Modelle, die gerichteten Kanten $E_G$ die topologischen Eigenschaften sowie die semantische Zuordnung der Knoten zueinander. Abbildung V.18 zeigt beispielhaft die Modellierung eines neuronalen Netzwerks aus Neuronen (N) und Synapsen (S) als Graph mit Erhaltung der Topologie und Einführung spezieller Parameterknoten für die Parameter (P). Links daneben wird auf gleiche Weise die HICANN-Struktur auf einen Graphen projiziert, ebenfalls mit Neuronen (N), verbunden mit einem synaptischen Feld (S), dem Layer1 Bussystem (L1), bestehend aus Kodierern (WTA), Auswahlschaltern (AS) und Crossbars (CB):

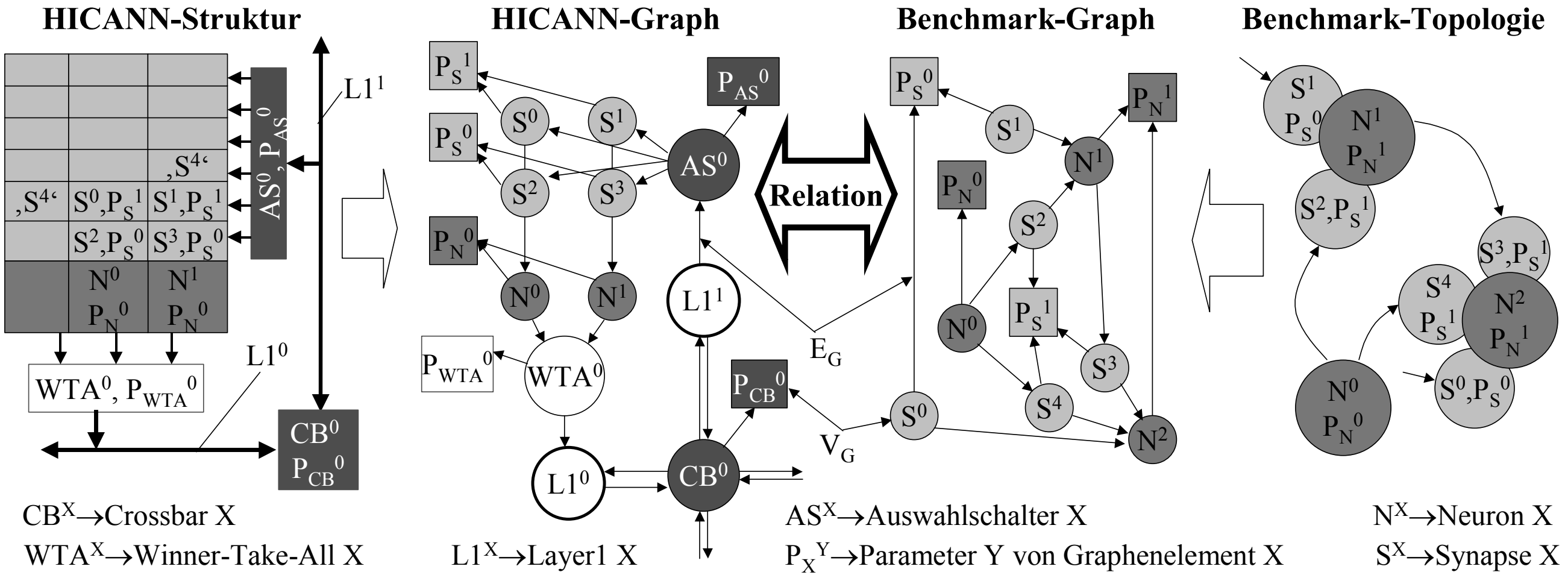

**Abbildung V.18.: Graphenbasiertes Matching als letzter Schritt des Mappings von neuronaler Vorgabe auf die Stage 2 Architektur, Abbildung von Einzelelementen [Wendt07]**

Resultierend daraus verallgemeinert sich die Abbildung $m$ zu einer Graphenrelation $r_m$ aus einer Menge von Tupeln:

$$r_m = \begin{cases} (b_1, h_{11}, \cdots, h_{1m}) \\ \cdots \\ (b_n, h_{n1}, \cdots, h_{nm}) \end{cases} \quad mit \quad \begin{matrix} b_1, \ldots b_n \in V_B \\ h_{i1}, \ldots h_{im} \in V_H \end{matrix} \qquad \textbf{(V.14)}$$

$V_B$ repräsentiert jeweils die Menge aller Knoten der biologischen und $V_H$ die der Hardware-Graphenmodelle. Die Auflistungen ($h_{i1},...,h_{im}$), ... beziehen sich jeweils auf die Menge aller Elemente des HICANN-Graphen und stellen eine Auswahl dar, die nötig ist, um ein spezifisches Element $b_i$ des Benchmark-Graphen zu realisieren. Die Relation $r_m$ ist nicht disjunkt, d.h. biologische Elemente können mehreren Hardwarekomponenten zugewiesen sein und umgekehrt. Ein Beispiel für eine nicht eindeutige Graphenrelation wäre die dendritische/axonale Verzögerung, die im Modell meistens auf Grundlage dreidimensionaler Netzstrukturen als einzelner Zahlenwert vorgegeben wird [Koch99 (Abschnitt 6.5.1)]. Jedoch stellt dies in der Hardware die Summe der Verzögerungen aus allen Kommunikationselementen von Layer1 und Layer2 dar [Mayr07b], über die diese Verbindung geschaltet ist, etwa für Layer1 WTA-Codierung, Busübertragung und –auffächerung am Ziel, für Layer2 die (paketbasierte) Übertragung und evtl. Umcodierung oder Routing-FIFOs (siehe [Scholze07] sowie Abschnitte V.3.2 und V.3.3 bzw. Abbildung III.13). Einige Beispiel-Tupel der Relation (V.14) zu den Graphen aus Abbildung V.18 lauten wie folgt:

$$r_m = \begin{cases} \left(N^1, N^0\right) \\ \left(N^2, N^1\right) \\ \left(P_N^1, P_N^0\right) \\ \left(P_N^1, P_N^0\right) \\ \vdots \end{cases} \quad \cdots \quad r_m = \begin{cases} \vdots \\ \left(P_S^0, P_S^0\right) \\ \left(P_S^1, P_S^1\right) \\ \left(S^3, WTA^0, P_{WTA}^0, L1^0, CB^0, P_{CB}^0, L1^1, AS^0, P_{AS}^0, P_S^1, S^1\right) \\ \left(S^1, \cdots, L1^1, AS^0, P_{AS}^0, P_S^0, S^2\right) \end{cases} \qquad \textbf{(V.15)}$$

Die folgende Abbildung verdeutlicht die Zuordnung der Parametersätze, Synapsen und Neuronen in der obigen Relation. An der linken Seite befindet sich der Benchmark-Graph aus Abbildung V.18, daneben ist derselbe Graph mit der Indexzuordnung dargestellt, die sich beim Mapping des Graphen auf die Hardware ergeben, auf der rechten Seite dann das zugehörige Abbild der Hardware.

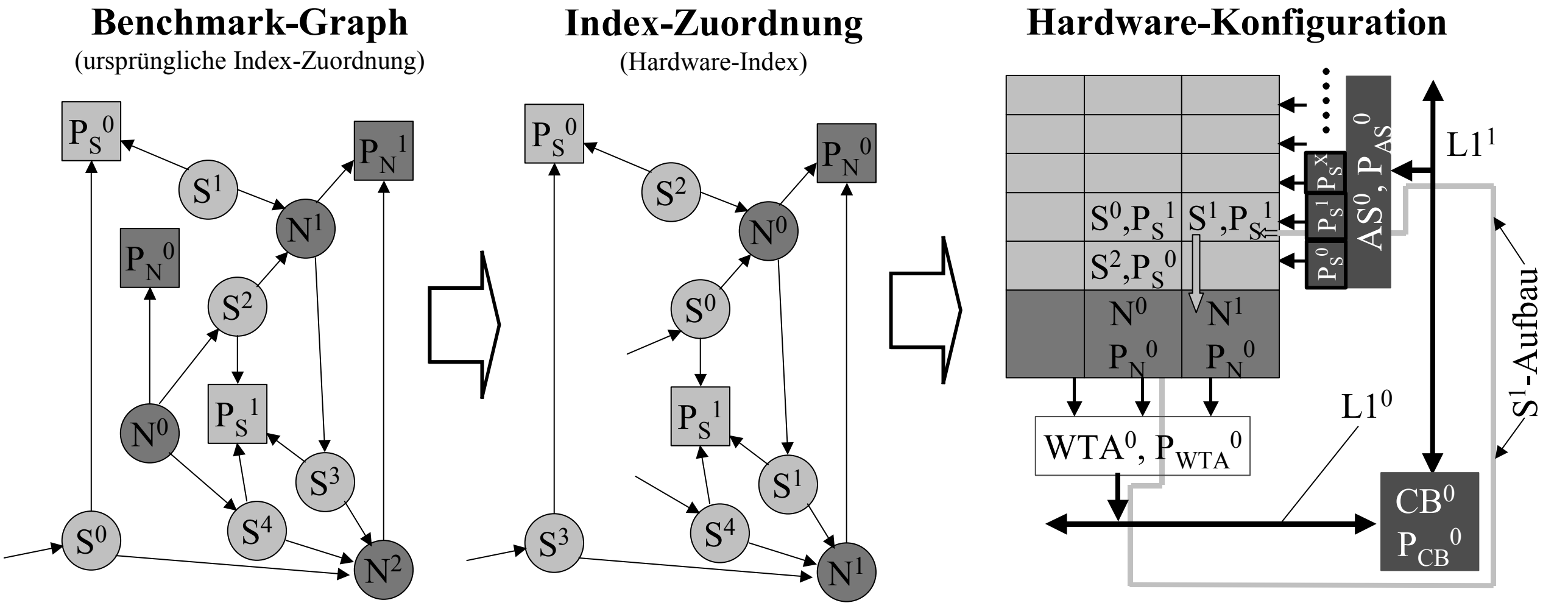

**Abbildung V.19.:Zuordnung der Indexe des Benchmarkgraphen zu den Hardwareelementen und Routing-basierte Realisierung einer Beispielsynapse**

Grau unterlegt ist in der Hardwaredarstellung das Zustandekommen der Synapse $S^1$ wiedergegeben. Ein Ausgangspuls von Neuron $N^0$ wird über $WTA^0$, den Layer1-Bus $L1^0$, die Crossbar $C^0$, Layer1-Bus $L1^1$ und den Auswahlschalter $AS^0$ auf den Eingang der Synapse $S^1$ gelegt (Gleichung (V.15)). Da alle Synapsen einer Spalte an dem unter der Spalte liegenden dendritischen Abschnitt bzw. Neuron anliegen (siehe Abbildung V.10), liegt der Ausgang von Synapse $S^1$ wie von dem mittleren Graphen in der obigen Darstellung vorgegeben an Neuron $N^1$. Für die Synapse $S^2$ würde ein externes Pulssignal eines Nachbar-HICANN über weitere L1 Ressourcen herangeführt werden. Wie aus Abbildung V.19 ersichtlich, wird der Parametersatz bzw. die Konfiguration von Synapsen für eine komplette Synapsenzeile von den am Rand gelegenen Synapsentreibern festgelegt. Unter Einbeziehung der bis jetzt erfolgten Konfiguration kann Synapse $S^4$ aus dem Benchmark-Graphen auf zwei verschiedene Arten realisiert werden:

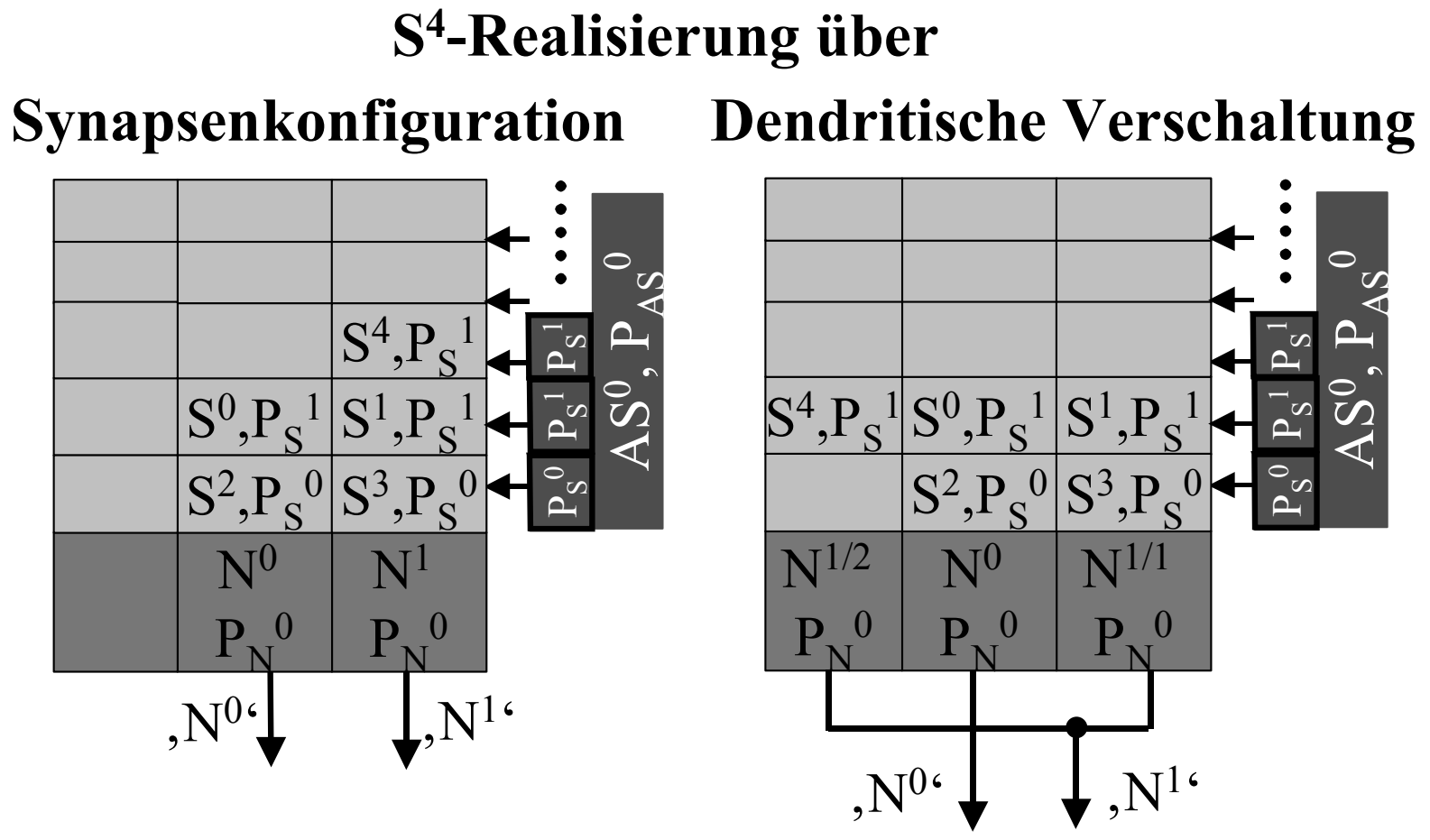

**Abbildung V.20.: Realisierung der Synapse $S^4$ mit Parametersatz $P_S{}^1$ über die Nutzung einer nicht verwendeten Synapse (links) oder über dendritische Verschaltung (rechts)**

Zum Einen kann die Synapsenzeile über den bereits verwendeten Zeilen auch mit $P_S{}^1$ konfiguriert werden, so dass $S^4$ über $S^1$ zu liegen kommt. Dies setzt voraus, dass die über $N^1$ liegende Synapsenspalte noch nicht durch die restlichen an $N^1$ anliegenden Synapsen ausgelastet ist. Vorteil hierbei wäre die Einsparung von Neuronen/dendritischen Abschnitten in dem betrachteten HICANN links von $N^0$. Alternativ kann der bereits konfigurierte Parametersatz $P_S{}^1$ in der zweituntersten Zeile der Synapsenmatrix auch für $S^4$ verwendet werden, indem die links von $S^0$ liegende Synapse als $S^4$ verwendet wird. Der unterhalb von $S^4$ liegende dendritische Abschnitt muss dann mit Hilfe der dendritischen Verschaltung (siehe Abbildung V.10) mit dem eigentlichen Neuron $N^1$ verschalten werden. Die Auslastung der Synapsen wäre durch die über $N^{1/2}$ ungenutzt liegenden restlichen Synapsen nicht optimal. Jedoch stellt dies eine Möglichkeit dar, $S^4$ korrekt zu realisieren, wenn über $N^{1/1}$ bereits alle Synapsen verwendet werden, jedoch in diesem HICANN noch ungenutzte dendritische Abschnitte vorhanden sind. Beide Varianten führen zur selben neuronalen Funktionalität, so dass die Wahl der Realisierung mittels Gleichung (V.13) hinsichtlich der besten Auslastung der Hardware erfolgt (unter Einbeziehung der sich aus der restlichen Topologie ergebenden Randbedingungen).
Praktisch wird das Graphenmatching über ein softwarebasiertes Stage 2-Äquivalent durchgeführt, welches Variablen für die einzelnen Graphenelemente beinhaltet. Dieses Modell ist nicht als Simulation ausführbar, jedoch auch nicht komplett statisch, da Wechselbeziehungen der Graphenelemente untereinander mit einbezogen sind. Beispiele hierfür wären etwa Layer1 Busabschnitte, die in einem HICANN nicht mehr zur Verfügung stehen, da sie bereits zur Kommunikation der angrenzenden HICANNs untereinander eingesetzt werden. Die Variablen dieses Modells werden in einem inkrementellen Graphenmatching mit Elementen des Topologie-Graphen aus Abbildung V.18 belegt. Falls nicht die gesamte Menge an Elementen geschrieben werden kann, welche in den höheren Mappingstufen diesem HICANN zugewiesen wurden, erfolgt eine Rückverweisung auf die zugehörige höhere Stufe des Algorithmus. In [Wendt07] findet sich ein Überblick über verschiedene Algorithmen, mit denen die in diesem Abschnitt abstrakt formulierten Mappingziele praktisch erreicht werden können. Wenn auf dieser untersten Stufe des Mapping für alle Teilgraphen der Benchmark Abbildungen auf die HICANN-Graphen gefunden wurden, ist das Mapping abgeschlossen. Die Inhalte der oben angesprochenen Variablen des Stage 2-Äquivalent können direkt oder mit geringfügiger Konvertierung als Speicherinhalte der Systemsimulation oder der späteren (realen) Hardware verwendet werden.

## V.6 Aktueller Stand und weiterer Entwurf

Der momentane Entwurfsstand der verschiedenen Hardware- und Softwarekomponenten, die das Stage 2 System ausmachen, stellt sich wie folgt dar: Auf der Hardwareseite wurde ein DNC

Prototyp[49] entworfen, der zur Zeit vermessen wird. Des Weiteren existieren Testschaltungen zu Floating Gate Parameterspeichern[50] und zu den Waferscale Verbindungen [Ehrlich07]. Ein HICANN Prototyp befindet sich momentan im der ersten Entwurfsphase. Wie in den Abschnitte V.4 und V.5 beschrieben, tragen neben der Hardware auch umfangreiche Softwaremodule zur Funktion von Stage 2 bei. Es existieren parametrisierbare Simulationen, mit denen Einflüsse von Gewichtsquantisierung, Pulsverzögerungen, Parameterabweichung auf das neuronale Verhalten untersucht werden können (siehe Abschnitt V.1.2). Zusätzlich wurde ein Modell für das gesamte Stage 2 System in Hardwarebeschreibungssprachen erstellt, welches vor allem zur Unterstützung des Entwurf der Pulskommunikation eingesetzt wird [Scholze07], indem Kombinationen aus Layer1 und 2 mit Populationen von Poissonpulsquellen getestet werden. In nächster Zeit sollen diese beiden Ansätze zu einem hardwarenahen Systemmodell konvergieren, mit dem dann groß angelegte Benchmark-Simulationen durchgeführt werden. Eine umfangreiche Benchmark-Datenbank wurde aufgebaut. Diese ist permanenten Anpassungen an aktuelle Forschungsergebnisse der Projektpartner unterworfen [Partzsch07b]. Für einen späteren Einsatz von Stage 2 in der neuronalen Forschung geben damit die Projektpartner über entsprechende Benchmarks explizit oder implizit ihre Anforderungen an Stage 2 bekannt.

Zu Projektanfang wurde angedacht, eine integrierte Entwurfsumgebung für Stage 2 zu erstellen, welche quantitative (z.B. Anzahl der dendritischen Abschnitte) und qualitative Änderungen (z.B. Lage und Aufbau der Layer1-Crossbars) parametrisierbar in den Entwurf einfließen lassen würde [Ehrlich07]. Simulationsmodell und Hardware wären dann skriptbasiert jeweils nach dieser Parametrisierung erstellt worden. Dies hat sich als zu ambitioniert herausgestellt, da die Leistungsdaten der Stage 2 Hardware hinsichtlich Neuronen- und Synapsendichte zu hoch sind, als dass sie mit einem Mixed-Signal Place-and-Route-Werkzeug erreichbar wären. Im Handentwurf sind für Neuronen und Synapsen, aber auch für die Crossbars mit ihrem verteilten Konfigurationsspeicher größere Packungsdichten erreichbar. Zusätzlich liegt die Syntheselaufzeit für ein derart großes, inhomogenes System aus verschiedensten Grundbausteinen mit sehr engen Randbedingungen (Platz, Verbindungslängen, etc.) im Tage- bis Wochenbereich und damit zu hoch für eine sinnvolle Designexploration. Der Entwurf der Stage 2 Hardware erfolgt mithin von Hand (mit Unterstützung durch konventionelle Synthesewerkzeuge).

Im Modell für die Systemsimulation wird ein gemischter Ansatz verfolgt, so können kleinere Designänderungen, etwa die Verbindungsbesetzung einer Crossbar, über Konfigurationsdateien angepasst werden. Größere Änderungen, wie etwa die Lage einer Crossbar im Gesamtdesign müssen direkt im Quelltext der ausführbaren Systembeschreibung von Hand angepasst werden. Da die neuronale Bewertung des derart modifizierten Entwurfs einen wesentlichen Zeitfaktor im Designprozess darstellt, sollte zumindest das Abbilden von verschiedenen Benchmarks in schneller Folge möglich sein. Zu diesem Zweck kann das Mappingtool leicht an veränderte Hardware angepasst werden [Wendt07, Mayr07b], so dass zumindest das Pro und Kontra eines neuen Entwurfs unter neuronalen Gesichtspunkten rasch ermittelt werden kann. Zu dieser Bewertung zählt beispielsweise, ob die vorgegebene Netzwerktopologie mit den vorhandenen Layer1 und Layer2 Kommunikationsressourcen überhaupt realisierbar ist. Falls dieses reine Topologiemapping erfolgreich (oder mit vernachlässigbaren Verlusten) durchgeführt werden konnte, wird eine Systemsimulation durchgeführt, deren Ergebnis mit den in Abschnitt V.4.1 beschriebenen Methoden beurteilt werden kann. Vernachlässigbare Verluste beurteilen sich Benchmark-spezifisch mittels Systemsimulationen, wie unterhalb Abbildung V.16 angesprochen. Als realen Probelauf des Mappings wird dieses im Moment dahingehend modifiziert, Benchmarks auch auf das in [Schemmel06] beschriebene System abzubilden.

[49] Zur Erzeugung der CLK-Signale für die verschiedenen Taktdomänen des sogenannten Minilink-IC wurde für Takte kleiner 600MHz auf einen digitalen IP-Core zurückgegriffen [Eisenreich07], für GHz-Takte wurden PLLs analog handentworfen.

[50] Wichtig war hierbei die Integration von analogen Floating-Gate Speichern in die konventionelle 180nm CMOS-Technologie, die zum Entwurf des Gesamtsystems verwendet wird. Die Werteinschreibung erfolgt über Gate-Tunnelströme [Ehrlich07].

Zusätzlich befinden sich mehrere Softwarewerkzeuge im Aufbau, die verschiedene Aspekte des Mapping oder der Systemsimulation optisch aufarbeiten und dreidimensional darstellen (siehe Anhang B.2.2). Aspekte des Entwurfsprozesses können damit visuell beurteilt werden. Diese topologische Bewertung ermöglicht insbesondere gegenüber einer reinen Analyse der Netzwerkausgabe die räumliche Lokalisierung von Fehlern/Abweichungen. Wenn diese Analyse auf die biologische Topologie projiziert wird, können beispielsweise nicht darstellbare lokale Plastizitätsvariationen festgestellt werden. Bei einer Projektion der Fehler auf die Stage 2 Topologie können z.B. zu klein dimensionierte Parameterspeicher lokalisiert werden. Der Entwurfsfluss stellt sich damit wie folgt dar:

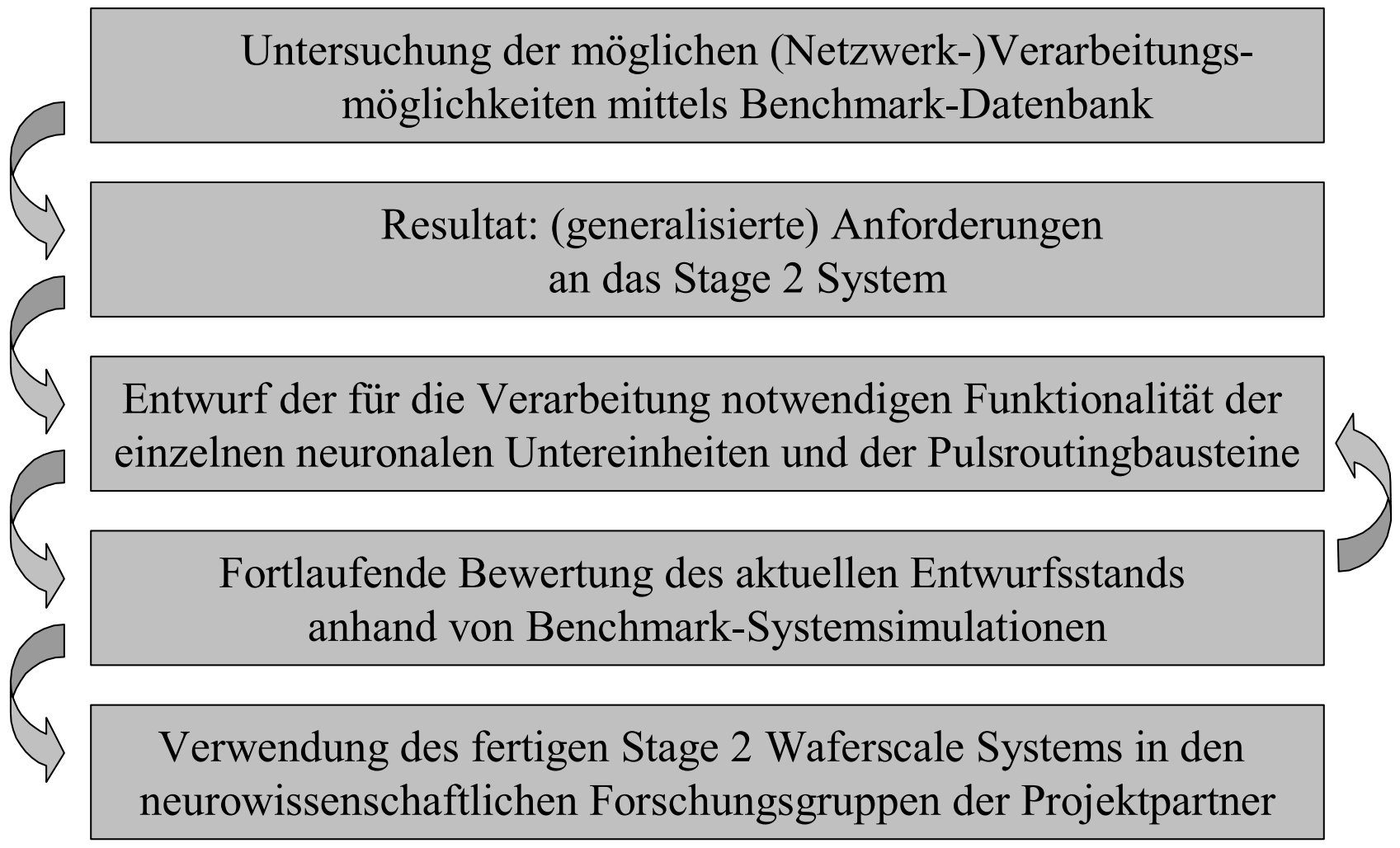

**Abbildung V.21.: Ablauf und Ziele des Stage 2 System Entwurfsflusses**

Das FACETS-Projekt hat ungefähr die Hälfte seiner Laufzeit erreicht und befindet sich derzeit im 3. und 4. Unterpunkt des obigen Entwurfsflusses. Eine abschließende Bewertung ist deshalb nicht möglich, jedoch zeigt die obige Zusammenfassung den bis jetzt erreichten Entwurfsstand auf und gibt etwa mittels der in Entwicklung befindlichen Entwurfswerkzeuge und der bis jetzt erfolgten Erprobung von Teilkonzepten in Mini-ASICs eine realistische Gangrichtung vor, mit der das Gesamtprojekt zu einem erfolgreichen Abschluss gebracht werden kann.

# VI Zusammenfassung der Arbeit und Perspektiven

Nosce te ipsum oder im griechischen Original „γνῶθι σεαυτόν“ (Gnôthi seautón) bezeichnet ein angeborenes Bedürfnis des Menschen, sich selbst zu erforschen und zu verstehen. Die Kenntnis des eigenen Gehirns und der darin stattfindenden Vorgänge ist seit Jahrtausenden als wichtiger Teilbereich fest in diesem Bedürfnis verankert [Finger01]. Diese philosophische Motivation für die Erforschung neurobiologischer Vorgänge wurde in den letzten Jahrzehnten um eine praktische Komponente erweitert, mithin das Verständnis der entsprechenden Mechanismen zur Verwendung innerhalb technischer Aufgabenstellungen.

In diesem Zusammenhang beschäftigt sich die vorliegende Arbeit mit der Realisierung von „Bildverarbeitungsalgorithmen mittels pulsgekoppelter Netze“. Die dabei anstehende Adaption von Prinzipien der Informationsverarbeitung des biologischen Vorbilds beinhaltet zum Einen die Übernahme von Einzelaspekte (evtl. mit konventionellen Prinzipien kombiniert), um technische Aufgabenstellungen in neuartiger, verbesserter Weise zu lösen. Einzelaspekte in diesem Sinne sind z.B. Verarbeitungsfunktionen von Untereinheiten, etwa die verschiedenartigen Plastizitätsvorgänge bei Neuronen und Synapsen, oder spezielle Topologien der Vernetzung. Zum Anderen wird auch versucht, komplette algorithmische Strukturen zu analysieren und zu übernehmen, um Aufgabenstellungen zu lösen, für die noch keine umfassende konventionelle Lösung existiert. In der Bildverarbeitung wäre dies beispielsweise eine rotations- skalierungs- und translationsinvariante Objekterkennung, wie sie vom visuellen Kortex in Säugetieren leicht vollzogen wird. Zur Nachbildung dieser Algorithmen bzw. Einzelaspekte müssen weiterhin Schaltungen und Systeme entworfen werden, welche sie technisch zum Einsatz bringen. Abweichungen vom biologischen Vorbild sind in diesem Kontext unausweichlich, weshalb nach der technischen Adaption gezielt die Invarianz der gewünschten Verarbeitungsfunktion gegenüber den Modifikationen verifiziert werden muss. Im Rahmen dieser Gesamtaufgabenstellung können zwei Wissenschaftsgebiete ausgemacht werden, für die wesentliche Beiträge geleistet wurden:

---

Die Verwendung von neuromorphen Strukturen und Prinzipien im anwendungsorientierten Einsatz stellt das erste dieser Themengebiet dar, beispielsweise in speziellen Bildsensoren. Dazu wurden in dieser Arbeit zwei neue Konzepte entworfen, zum Einen die pulsbasierte Bildfaltung mittels Mikroschaltungen und variablen Topologien in Kapitel III [Mayr06d, Mayr07c]. Der Aufbau von komplexen Bildfilterfunktionen aus generischen neuronalen Mikroschaltungen einfacher Funktionalität wurde gezeigt, sowie Algorithmen zur Generierung der Netztopologie für beliebige Faltungsmasken erstellt. Die Mikroschaltung und weitere Funktionalität zum Aufbau dieser Faltungsmasken wurde im Sinne einer neuartigen ‚neuronalen FPGA' realisiert. Globales Pulsrouting ermöglicht den Aufbau beliebiger Netztopologien auf dem IC. Dies steht im Kontrast zu üblichen neuromorphen ICs, welche entweder mit festen Topologien aufgebaut sind oder die Vernetzung extern implementieren. Neu entwickelte Softwarewerkzeuge erstellen die Konfiguration der neuronale FPGA, d.h. es werden Netztopologien und die den Mikroschaltungen zugehörigen Parameter auf die FPGA abgebildet und damit die Verarbeitungs-/Filterfunktion realisiert.

Des Weiteren wurde in Form von pulsbasierter Texturanalyse ein neuer neuromorpher Bildoperator entwickelt (Abschnitt IV.2 [Mayr05d, Mayr07d]), das sogenannte Pulsed Local Orientation Coding. Dieses belegt in Simulationen seine Überlegenheit gegenüber ähnlichen konventionellen Operatoren. Extrapolationen für eine technische Implementierung des Operators zeigen, dass er hinsichtlich Fläche und Technologieportierbarkeit deutliche Vorteile gegenüber einer Mixed-Signal Implementierung von vergleichbaren Operatoren hat. Eine mathematische Fehleranalyse wurde durchgeführt, welche die Robustheit des Operators gegenüber Pulsjitter zeigt[51].

[51] Eine reduzierte Form der Texturanalyse, die optische Computermaus, stellt gleichzeitig das bekannteste Beispiel für einen kommerziellen Einsatz von neuromorphen Prinzipien in Bildsensoren dar [Giles01].

Andere derzeitige Beispiele auf diesem Gebiet benutzen AER-Codes, um Bildinformationen selektiv, d.h. bandbreiten- und energiesparend darzustellen [Posch07]. Weitere Vorteile von AER-Bildsensoren sind erweiterter Dynamikbereich und vereinfachte, variable AD-Wandlung durch die Pulsrepräsentation des analogen Pixelstroms. Die Fortführung dieses Gedankens hin zu allgemeiner Signalverarbeitung führt zum Entwurf pulsbasierter digitaler und gemischt analog/digitaler Signalverarbeitung, mit der Zielsetzung paralleler, fehlertoleranter und leistungsoptimierter Systeme [Giles01]. Motiviert wird dies durch die gleichlaufende Vorgabe bei biologischen Systemen, d.h. die Realisierung von redundanz- und energieoptimierter, fehlertoleranter Signalübertragung [Laughlin03]. Einen wichtigen Baustein solcher Systeme stellt die Art der pulsbasierten Informationscodierung dar. Pulsbasierte Codes werden in der Literatur nur vereinzelt und meist biologiebezogen miteinander verglichen. In Verallgemeinerung dieser Ansätze wurde in Abschnitt II.2 ein signaltheoretischer Vergleich aller gebräuchlichen Codes hinsichtlich Fehlertoleranz und der jeweiligen Informationsdichte hergeleitet. Damit wird Hardwareentwicklern ein Werkzeug gegeben, entsprechende Architekturentscheidungen treffen zu können.
Aus der Biologie entlehnte Soft-Fail Fähigkeiten werden in zunehmendem Maße auch für die konventionelle Halbleiterfertigung relevant als einzige Methode, um neuartige, inhärent fehlerbehaftete Hardware hoher Packungsdichte sinnvoll einzusetzen [Eickhoff06, Türel05]. Ein Aufbau von höheren Verarbeitungsstufen aus einfachen, identischen Grundelementen in konfigurierbarer Topologie stellt eine dieser Möglichkeiten dar, neuronale Prinzipien auf fehlertolerante Architekturen anzuwenden. Neuartige Entwürfe und Realisierungen von derartigen Systemkonzepten wurden in den Kapiteln III und V erstellt.

---

Ein weiteres Gebiet der neuronalen Forschung, auf dem zur Zeit viele Hardwareressourcen zum Einsatz gebracht werden, sind biologienahe Nachbildungen neuronaler Strukturen. Sie werden primär als Werkzeuge für die Grundlagenforschung eingesetzt. Entsprechende softwarebasierte Ansätze auf Hochleistungsrechnern sind z.B. Blue Brain [Markram06] und die an der KTH Stockholm unternommene V1-Modellierung [Djurfeldt05].
Arbeiten des Autors auf diesem Gebiet beschäftigten sich mit der Analyse biologischer neuronaler Verarbeitung im Frequenzbereich. Zur Untersuchung von neuronalen Verarbeitungsfunktionen finden konventionell verschiedenste statistische Ansätze Einsatz. Mittels verschiedener Simulationen, mathematischer Analysen und der Untersuchung biologischer Messdaten wird jedoch belegt, dass eine Untersuchung im Frequenzbereich wichtige Aspekte neuronaler Verarbeitungsfunktionen aufzeigen kann (Abschnitt IV.1.1). Als Analyseinstrument wurde in diesem Zusammenhang die (nach bestem Wissen des Autors) erste geschlossene Lösung für die Transformation von Poisson-Pulsfolgen in den Frequenzbereich hergeleitet (Abschnitte II.2.2 und A.1). Bisherige Analysen im Frequenzbereich wurden nur für statische neuronale Netze durchgeführt, ohne Berücksichtigung von Plastizität [Spiridon99, Mayr05c]. In Kontrast dazu wurde in Abschnitt IV.1.2 eine simulative und mathematische Untersuchung der quantalen Kurzzeitadaption im Frequenzbereich durchgeführt. Dabei wurde insbesondere für die Übertragung modulierter Pulsfolgen eine Erhöhung des Signal-Rausch-Abstandes (SNR) gegenüber äquivalenten konstanten Pulsfolgen festgestellt. In einer Verallgemeinerung dieses Ansatzes für über längeren Zeitraum wirkende Lernvorgänge wurden die Auswirkungen von STDP im Frequenzbereich simulativ untersucht (Abschnitt IV.1.3). Dabei wurde festgestellt, dass Langzeitadaptionen wie STDP dazu beitragen, gezielt den Hintergrundrauschpegel zu vermindern und damit auch das SNR vor allem bei hohen Signalfrequenzen zu erhöhen. Dies stellt einen Vorgang ähnlich der Optimierung der Koeffizienten in der Modulatorschleife beim konventionellen Delta-Sigma-Modulator dar.
In einer weiteren Untersuchung wurde eine alternative Implementierung der STDP-Plastizitätsregel entwickelt. Insbesondere wurde gezeigt, dass eine von der Membranspannung abhängige modifizierte BCM-Regel maßgebliche Teile der biologischen Messdaten nachbilden kann, auf deren Grundlage ursprünglich STDP postuliert wurde (Abschnitt V.2.2). Die neu geschaffene Variante der BCM-Regel hat dabei den Vorteil, nur auf lokalen Zustandsvariablen aufzubauen, im

Gegensatz zu den Zeitmessungen zwischen entfernten Pulsereignissen bei STDP, wodurch sich ihre Hardwarerealisierung vereinfachen würde.
In vielen Forschungsgruppen wird derzeit auch an Hardwareimplementierungen von forschungsorientierten neuronalen Nachbildungen gearbeitet, die versuchen, schneller, größer und/oder kostengünstiger zu sein als Software-Simulatoren [Eickhoff06]. Bekannte Personen/Projekte auf diesem Gebiet sind beispielsweise v. d. Malsburg, Indiveri, Douglas et. al. mit DAISY [Indiveri06], Boahen et. al. [Lin06], sowie Meier, Gerstner, Maass, et. al. im Rahmen des FACETS-Projektes [Meier04]. In dieser Arbeit wurden für FACETS Konzepte erstellt zur Realisierung der Routing-Hardware, der Implementierung biologisch realistischer Simulationen auf dem endgültigen Wafer-Scale System, und zur entsprechenden biologieorientierten Verifikation der Gesamt-Hardware, siehe Kapitel V und [Ehrlich07, Mayr06c, Mayr07b, Wendt07]. Die Bewertung des Hardwareentwurfs unter dem Gesichtspunkte der späteren neuronalen Simulationsaufgaben wurde in bisherigen Ansätzen nicht in dieser Weise verfolgt. Für einen sinnvollen Einsatz derartiger Hardwaresimulatoren in neurobiologischer Forschung scheint jedoch ein entsprechendes biologieorientiertes Entwurfs- und Validierungskonzept unerlässlich[52]. Im Rahmen der Validierung gewonnene Erkenntnisse wurden verschiedentlich wieder im Entwurf der FACETS Hardware berücksichtigt, etwa beim Aufbau des pulsbasierten Kommunikationsnetzes.

---

Die vorliegende Arbeit nähert sich ihrem Thema somit auf mehreren Ebenen, von der Entwicklung direkt einsetzbarer Bildoperatoren, über eine Taxonomie neuronaler Verarbeitung als Anhaltspunkte für künftige Hardwareentwürfe bis hin zu VLSI-basierter neuronaler Grundlagenforschung, mit der die Basis für die nächste Generation an neuromorphen Schaltungen gelegt wird. Unmittelbares Ziel ist hier der Einsatz neuronale Verarbeitungsprinzipien im technischen Alltag.
Demgegenüber stellt die so genannte „Wetware“, also eine Verbindung von biologischen Neuronen und technischen Komponenten, ein weiteres interessantes, jedoch nicht so offensichtliches Anwendungsgebiet von neuronalen Schaltungen und Nachbildungen dar. Diese werden einerseits wie die o.a. Simulationen für Analysezwecke eingesetzt, beispielsweise als Zellkulturen in Petrischalen, mit denen über entsprechende Eingangssignale, chemische und elektrische Steuerung und Messungen eine Verarbeitungseinheit aufgebaut wird. Mit dieser kann etwa Bildverarbeitung [Ruaro05] oder eine Robotersteuerung [Potter03] realisiert werden.
Im medizinischen Bereich wird „Wetware“ bereits experimentell für Prothesen angewandt, um beschädigte Teile des Nervensystems durch technische Emulierungen zu ersetzen. Wie in verschiedenen Kapiteln dieser Arbeit angesprochen, liegt das meiste gesicherte Wissen über neuronale Verarbeitungsprozesse für Hirnareale vor, für die sich eine messbare I/O-Relation definieren lässt. Deshalb gibt es die am weitesten fortgeschrittenen „Neuro“-Prothesen auf den Gebieten der Sensorik, v.a. des visuellen Pfades [Weiland06], und der Aktorik [Santhanam06].
Forschungsvorhaben wie FACETS dienen in diesem Kontext auch dazu, über biologische Messungen und Modellierung bessere Kenntnisse über höhere Verarbeitungsstufen zu erlangen und damit ultimativ den Bau entsprechender Prothesen, etwa bei einer V1- oder Sehnervschädigung, zu ermöglichen [Meier04].

[52] Teilweise wurde die o.a. Zielsetzung auch bereits mit dem Router-Schaltkreis aus Kapitel III verfolgt [Mayr06a]. Auch hier sollte die forschungsorientierte Untersuchung von Verarbeitungsmodalitäten in Mikroschaltungen durch den ASIC-Einsatz gegenüber den Softwaresimulationen beschleunigt werden. Simulative Modelle sowohl des Einzelelement- als auch des Netzwerkverhaltens wurden mit möglichst hoher Übereinstimmung auf eine konfigurierbare Hardwareplattform abgebildet.